%% file: thesis.tex
\documentclass[a4paper,12pt,numbered,print,index]{Classes/PhDThesisPSnPDF}

\usepackage{amsthm,integrable,stmaryrd,xparse} 
\usepackage{twistor}

\usepackage{filecontents}
\usepackage{natbib}
\usepackage{bibentry}
\nobibliography*

\input{Preamble/preamble}

\input{thesis-info}

\ifdefineAbstract
\fi

\ifdefineChapter
\fi

\counterwithout{footnote}{chapter}

\usepackage[utf8]{inputenc}
\usepackage{babel}
\usepackage[T1]{fontenc}
\usepackage[autostyle]{csquotes}
   \MakeAutoQuote{‘}{’}

\begin{document}

\frontmatter

\maketitle

\include{Dedication/dedication}
\include{Declaration/declaration}
\include{Acknowledgement/acknowledgement}
\include{Abstract/abstract}


\tableofcontents

\listoffigures


\nomenclature[z]{AdS}{Anti-de Sitter space}
\nomenclature[z]{CCA}{Celestial chiral algebra}
\nomenclature[z]{CFT}{Conformal field theory}
\nomenclature[z]{dS}{de Sitter space}
\nomenclature[z]{IR}{Infrared}
\nomenclature[z]{KK}{Kaluza-Klein}
\nomenclature[z]{OPE}{Operator product expansion}
\nomenclature[z]{QFT}{Quantum field theory}
\nomenclature[z]{QG}{Quantum gravity}
\nomenclature[z]{RR}{Ramond-Ramond}
\nomenclature[z]{SDYM}{Self-dual Yang-Mills theory}
\nomenclature[z]{SDGR}{Self-dual gravity}
\nomenclature[z]{UV}{Ultraviolet}
\nomenclature[z]{VEV}{Vacuum expectation value}
\nomenclature[z]{zrm}{Zero-rest-mass}

\printnomenclature


\mainmatter

\include{Chapter1/chapter1}
\include{Chapter3/chapter3}

\include{Chapter2/chapter2}

\include{Chapter4/chapter4}
\include{Chapter5/chapter5}

\include{Chapter6/chapter6}

\include{Chapter7/chapter7}


\begin{spacing}{1.10}


\bibliographystyle{JHEP}
\cleardoublepage
\bibliography{thesis} 



\end{spacing}


\begin{appendices} 

\include{Appendix1/appendix1}

\include{Appendix2/appendix2}

\end{appendices}

\printthesisindex 

\end{document}

%% file: thesis-info.tex
\title{Celestial Chiral Algebras
and Self-Dual Gravity}

\subtitle{}

\author{Simon Heuveline}

\dept{Department of Applied Mathematics and Theoretical Physics}

\university{University of Cambridge}
\crest{\includegraphics[width=0.33\textwidth]{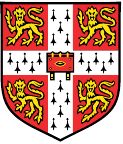}}

\degreetitle{Doctor of Philosophy}

\college{St. John's College}

\subject{LaTeX} \keywords{{LaTeX} {PhD Thesis} {Mathematics} {University of
Cambridge}}

%% file: Dedication/dedication.tex

\begin{dedication} 

I would like to dedicate this thesis to my mother.

\end{dedication}

%% file: Declaration/declaration.tex

\begin{declaration}

This thesis is partly based on the papers \cite{Bu:2022iak, Bittleston:2023bzp, Bittleston:2024rqe, Bogna:2024gnt}. The original research for this thesis was conducted
by the author as a postgraduate student at the Department of Applied Mathematics and
Theoretical Physics, University of Cambridge
under the supervision of David Skinner. The material in Chapter \ref{chapter2} is based on work done
in collaboration with Wei Bu and David Skinner and insights from Roland Bittleston \cite{Bu:2022iak}. Chapter \ref{chapter4} is based on research done in collaboration with Roland Bittleston and David Skinner \cite{Bittleston:2023bzp}. Chapter  \ref{chapter5} is based on research done in collaboration with Roland Bittleston, Giuseppe Bogna, Adam Kmec, Lionel Mason, and David Skinner \cite{Bittleston:2023bzp}. Chapter \ref{chapter6} is partly based on research done in collaboration with Giuseppe Bogna \cite{Bogna:2024gnt}.

\medskip

This thesis is the result of my own work and includes nothing which is the outcome of work done in collaboration except as declared above and specified in the text. It is not substantially the same as any work that has already been submitted, or, is being concurrently submitted, for any degree, diploma or other qualification at the University of Cambridge or any other University or similar institution except as declared in the preface and specified in the text. It does not exceed the prescribed word limit for the relevant Degree Committee.

\end{declaration}

%% file: Acknowledgement/acknowledgement.tex

\begin{acknowledgements}      

Firstly, I would like to thank my supervisor David Skinner for his continued guidance, support and patience throughout my PhD. His insights and feedback have been critical in shaping the direction of this work. I am very grateful for all the hours he spent with me in front of a blackboard going through calculations in detail and all the inspiring bigger-picture physics chats we had on walks from DAMPT to the city centre.

Secondly, I would like to thank Roland Bittleston, who has been much more than just a collaborator over the years. 
His dedication and patience in working with me and explaining things to me have been a constant source of knowledge and inspiration, without which this work would not have been possible. 

I am also incredibly
grateful for my other collaborators Giuseppe Bogna, Wei Bu, Kasia Budzik, Adam Kmec,  Lionel Mason, Chris Pope, Surya Raghavendran, and Atul Sharma, for their patience and hard work, and for sharing so much of their wisdom with me. 

I had the privilege to be able to engage in mathematics and physics discussions with many further people whose insights contributed to this thesis. I would like to thank every single one of them!


I would like to thank St. John's College, Cambridge,  the STFC HEP Theory Consolidated grant ST/T000694/1, and the  Simons Collaboration on Celestial Holography for providing me with financial support with minimal
administrative overhead.

Moreover, I would like to thank Tim Adamo, Sean Hartnoll, and David Skinner for providing me with helpful comments on the draft of this thesis and encouraging me to upload this thesis on arXiv.

Finally, I would like to thank my family,  my brother, my father, my mother, and Yuki, as well as all my wonderful friends for their constant love and support that kept me going throughout the years of my PhD. Particularly, I would like to thank Natalie Parham for an exceptional amount of such love and support. Every single person that has been nice to me over the years contributed to this work and I am incredibly grateful that there are way too many such people to start naming here.

\end{acknowledgements}

%% file: Abstract/abstract.tex
\begin{abstract}


Celestial holography suggests, among other things, that collinear singularities of graviton scattering amplitudes are described by the OPEs of some putative dual CFT. One of the great successes has been the insight that this duality is true at tree-level which led to the discovery of new infinite dimensional symmetry algebras of tree-level amplitudes in flat space closely related to $w_{1+\infty}$. This thesis studies these celestial chiral algebras in the light of twistor theory and derives tree-level deformations thereof induced by non-trivial background geometries that solve some form of the self-dual Einstein equations.

After an elaborate introduction, we begin by reviewing how holomorphic collinear singularities of gravity and gauge theory amplitudes in a certain basis are reminiscent of OPEs in a $2$-dimensional CFT. Then, we discuss how a non-commutative $\mathbb{R}^4$-background deforms these celestial OPEs in an interesting way.
 The following chapter reviews some basic twistor theory and various actions on twistor space and spacetime that describe self-dual gravity and self-dual Yang-Mills theory at the classical level.

In the following chapter, we give a detailed analysis of celestial symmetries in an asymptotically (locally) Euclidean space, Eguchi-Hanson space, that solves the equations of self-dual Einstein gravity. This deformation arises naturally from a backreaction on twistor space analogous to parts of Burns holography, the top-down construction of Costello, Paquette and Sharma and we will highlight similarities and differences.  We explain how the deformed celestial OPEs are closely related to certain chiral algebras, from now on referred to as celestial chiral algebras, supported on twistor lines.

In the final two chapters, we discuss the presence of a non-zero cosmological constant, which has many subtleties. Twistor theory allows us to also include a cosmological constant in the self-dual Einstein equations, and after reviewing the relevant background material, we discuss how the cosmological constant deforms the gravitational celestial chiral algebra. This gives an independent derivation of a deformed algebra previously found by Taylor and Zhu. Repeating the twistorial backrection in the presence of a cosmological constant leads us to self-dual limits of  Pleba\'{n}ski-Demia\'{n}ski black hole metrics. From their twistor perspective we derive a two-parameter deformation which generalises both the Eguchi-Hanson and cosmological constant deformations we previously discussed and in a sense interpolates between them.

\end{abstract}

%% file: Chapter1/chapter1.tex

\chapter{Introduction}  
\label{chapter1}

In this chapter, we will set the scene for the rest of this thesis by motivating the study of \emph{self-dual gravity} and \emph{celestial chiral algebras} from various different angles. For the purpose of motivation we will not be mathematically precise in this section and aim for physical motivation. Some but not all\footnote{After all, mathematically defining a generic quantum field theory is a famously hard task.} of the concepts, particularly those in section \ref{section1.4} and section   \ref{newsection1.4}, could be stated in a mathematically precise language. 

\section{Quantum gravity and black holes} 
\label{section1.1}

Quantum physics and Einstein's theory of relativity are the pillars of our understanding of the universe. Their respective mathematical frameworks, Quantum Field Theory
(QFT) and General Relativity (GR), have provided us with various predictions over the years that have been experimentally verified to stunning accuracies. 
The Standard Model of particle physics, which is based on quantum field theory, describes the electromagnetic and nuclear forces at subatomic distances. General Relativity on the other hand accurately predicts the orbits of celestial bodies, the expansion of the universe, and the formation of black holes from the collapse of massive stars. 

Although the two theories are individually believed to be true within their range of validity (very small and very large distance scales, respectively) it is not known how to combine the two into a unified framework of \emph{quantum gravity}. If the scattering of two charged particles is considered in the Standard Model, their interaction can be computed to arbitrarily high energy and precision, assuming that one has enough computing power at hand to solve complicated yet finite integrals. However, when computing the same interaction between two particles under gravity, then the resulting integrals diverge eventually when considering increasingly high energies, no matter how we try to regulate them at low energies. Theories with this property are called \emph{non-renormalizable} and gravity is perhaps the most important example of a non-renormalizable quantum field theory.


Simply ignoring these problems with quantizing gravity and treating the two theories independently is not quite possible on a fundamental level for a variety of reasons. First of all, classical gravity is described by a field, the metric tensor, that is \emph{universal} which means that every field in the Standard Model has to couple to the gravitational field. There is no such thing as a particle that is uncharged under gravity. As we consider energies beyond the \emph{Planck scale}
\be
M_P=\sqrt{\frac{\hbar c}{G}}\approx 1.22\times 10^{29} \frac{\text{GeV}}{c^2}\,,
\ee
the strength of gravity is expected to become comparable with the other forces. Hence, without a renormalizable description of quantum gravity our understanding of physics breaks down beyond the energy scale set by $M_P$. Equivalently, our understanding of physics breaks down at length scales smaller than the \emph{Planck length}
\be
\ell_P=\sqrt{\frac{\hbar G}{c^3}}\approx 1.62 \times 10^{-35} \text{m}\,.
\ee

A simple thought experiment also signifies that it is necessary to quantize the gravitational field. Let us consider a non-relativistic particle of mass $m>0$ with a wave function that is sharply peaked at two points $x_1,x_2$, and approximately modelled by
\be
|\psi\rangle\approx \tfrac{1}{\sqrt{2}}\big(|x_1\rangle+|x_2\rangle\big)\,.
\ee
An experiment to determine the position of the particle will result, with equal probability, either it being found at position $x_1$ or $x_2$. Its mass $m>0$ means that depending on the outcome of the measurement, it will source a gravitational field $\vec{g}_1$ or $\vec{g}_2$ centred at $x_1$ or $x_2$ respectively. So, performing some gravitational experiments allows one to deduce whether the gravitational field is given by $\vec{g}_1$ or $\vec{g}_2$ and hence the location of the particle. 
What was the gravitational field before such a measurement? 
Since different positions lead to different gravitational fields, the gravitational field must be approximately entangled with the position of
the particle. 
Hence, the wavefunction of the total system must be approximately of the form
\be
|\Psi\rangle\approx \tfrac{1}{\sqrt{2}}\big(|x_1\rangle\otimes|\vec{g}_1\rangle+|x_2\rangle\otimes|\vec{g}_2\rangle\big)\,.
\ee
It follows that the gravitational field should be of quantum mechanical nature itself.

There are many simplified toy models of quantum gravity which can be quantised in a controlled way. Studying such toy models in various dimensions has been one of the most successful ways to gain new insights about quantum gravity such as its holographic nature. In section \ref{section1.3} we will discuss that \emph{self-dual
gravity}, which is the topic of this thesis, is one of the very few such toy models in $4$ spacetime dimensions.

\subsection*{Black Holes}

\begin{figure}[t!]
	\begin{center}
 	\includegraphics[scale=0.25]{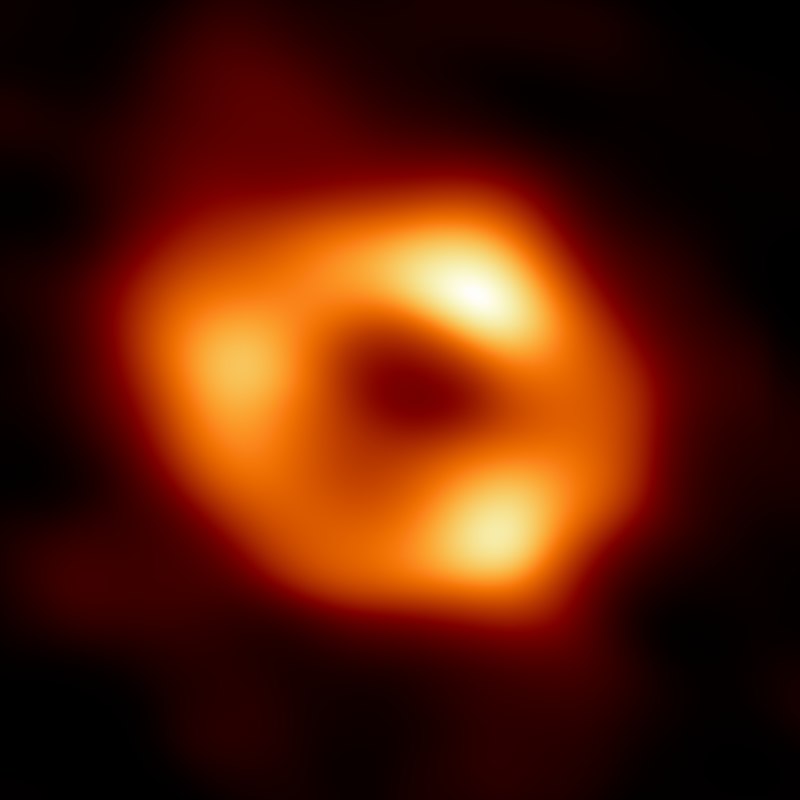}
	\caption{\emph{The image released by the Event Horizon Telescope Collaboration in 2022 \cite{EventHorizonTelescope:2022wkp} leaves no doubt that there exists a black hole in the centre of the Milky way. 
    When the first black hole solution \eqref{eq:Schwarzschild} was published by Schwarzschild in 1916 \cite{Schwarzschild:1916uq}, black holes were generally not considered to be objects of astrophysical relevance.}} \label{fig:Saggitarius}
	\end{center}
\end{figure}

A further manifestation of the relevance of quantum gravity in our universe is played by the existence of black holes such as Sagittarius A* in the centre of our galaxy (see figure \ref{fig:Saggitarius}).  
Black holes are spacetimes, which solve the Einstein field equations with the property that there exists a region of spacetime from which 'nothing can escape', not even light \cite{Hawking:1973uf}.

The singularity which can be found at the origin $r=0$ of black hole metrics such as the Schwarzschild metric
\be
\label{eq:Schwarzschild}
\dif s^2 = -\Big(1-\frac{2GM}{r}\Big)\dif t^2+ \Big(1-\frac{2GM}{r}\Big)^{-1}\dif r^2 +r^2(\dif \theta^2 +\sin^2(\theta)\dif \phi^2)
\ee
has the feature that the curvature becomes arbitrarily large. However, the classical description of gravity is only valid in locations of our universe in which the curvature is much smaller than the Planck scale $M_P$ so that the presence of black holes in our universe implies the presence of regions in our universe in which the effects of quantum gravity certainly become important. 

Even more peculiarly, in 1976 Hawking argued that the existence of black holes leads to the breakdown of predictability in general relativity, the black hole information paradox \cite{Hawking:1976ra, Ruffini:1971bza}. Although it is still unresolved, a lot of important progress has been made recently and many important lessons about quantum gravity more generally have been learned from it \cite{Raju:2020smc, Almheiri:2020cfm, Witten:2024upt}. We deeply hope that insights from the content of this thesis, particularly the study of \emph{self-dual black holes} and methods from \emph{twistor theory} and \emph{twisted holography}, will eventually contribute to some of the various ideas that circle the black hole information paradox, although we are not at this point yet.

Roughly speaking, the paradox is implied by a combination of the no-hair theorem, which states that classically stationary black holes in $4$ dimensions are uniquely determined by their mass, angular momentum and charge \cite{Israel:1967wq, Israel:1967za} and the existence of \emph{Hawking radiation} \cite{Hawking:1975vcx} which we will briefly discuss in the following.  While classically, black holes are not able to emit any radiation by their very definition, on the quantum level they emit radiation as blackbodies with
temperature
\be
\label{eq:HawkingT}
T=\frac{\hbar c^3}{8\pi G M k_B}\,,
\ee
where $M$ is their mass. This radiation is referred to as Hawking radiation. This means that 'black holes ain't so black' \cite{Hawking:1988qt} and that they slowly radiate all of their mass away until they fully disappear after a long yet finite time\footnote{Unless it is stabilized by charge or a
steady influx of energy.}. The aforementioned black-hole information paradox arises because after having fully radiated away, only featureless radiation in a mixed state remains. The black hole seems to have lost all of the information about the star that formed it, which can be well described by a pure state \cite{Hawking:1976ra, Raju:2020smc}. 

Beyond posing paradoxes, this Hawking temperature
plays the key role in \emph{black hole thermodynamics} \cite{Witten:2024upt}. There have been previous analogies that were drawn between black holes and thermodynamics \cite{Bardeen:1973gs}. In particular, the identification of the horizon area as being some form of entropy  was proposed by Bekenstein \cite{Bekenstein:1973ur}. 
Indeed, the \emph{Bekenstein-Hawking entropy}
\be
S_{\text{BH}}=
\frac{k_Bc^3 A}{4\hbar G}\,,
\ee
where $A$ is the area of the black hole's event horizon, obeys the expected \emph{first law of black hole mechanics} \cite{Bardeen:1973gs} 
\be
\dif M =T \dif S_{\text{BH}}
\ee
with $T$ being the Hawking temperature \eqref{eq:HawkingT}. All of this means that there is a consistent notion of entropy assigned to a black hole which scales like the size of its area rather than being \emph{extensive}, i.e. scaling like the volume of the black hole interior. 
This scaling property of $S_{\text{BH}}$ hints at the idea that all the information content of the black hole might somehow be fully described by degrees of freedom on the event horizon, i.e. the boundary of the black hole interior \cite{Susskind:1994vu, tHooft:1993dmi, Bousso:2002hon}. Let us elaborate.


\section{The holographic principle} 
\label{section1.2}

\begin{figure}[t!]
	\begin{center}
\begin{tikzpicture}[scale=0.5]
\draw [very thick] (0,0) circle (5);
\draw [very thick] (0,0) circle (5.1);
\draw [very thick] (0,0) circle (5.2);
\draw [very thick] (0,0) circle (5.3);
\filldraw[color=red!60, fill=red!5] (0,0) circle (2.5);
\node at (0,0){$\red{R}$};
\node at (2.3,2.25){\red{$\partial R$}};
\node at (5,-5){$\nwarrow$};
\node at (-5,-5){$\nearrow$};
\node at (-5,5){$\searrow$};
\node at (5,5){$\swarrow$};
\draw [->] (7.5,0)  -- (9.5,0);
\filldraw[color=black, fill=black] (16,0) circle (2.5);
\node at (18.3,2.25){$\partial R$};
\end{tikzpicture}
\caption{\emph{This image displays the Susskind process. The initial configuration (left) consists of some matter with mass $E$ in a spherically symmetric region $R$ surrounded by a very heavy shell of mass $M-E$, where $M$ is the mass of a black hole with horizon $\partial R$. The shell gets pushed inwards until a black hole with horizon $\partial R$ forms, which is the final state (right).}} \label{fig:Susskind}
\end{center}
\end{figure}
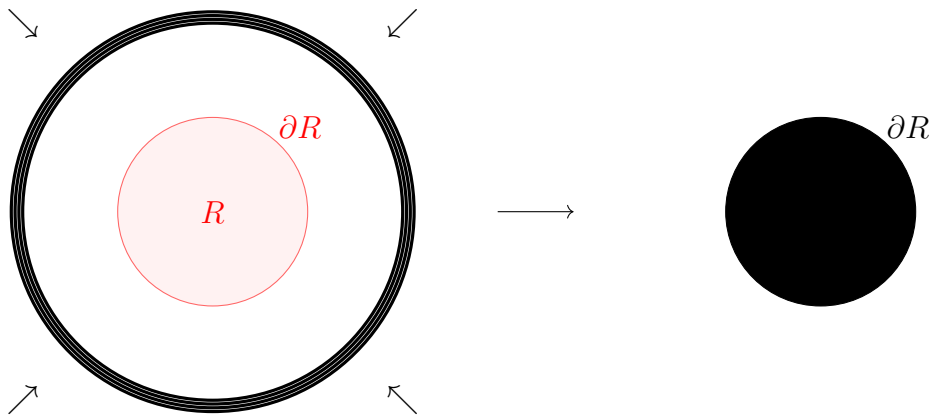

We consider an isolated matter system of mass $E$ and entropy $S_{\text{matter}}$ residing in a spherical region $R$ of a Cauchy slice of some $4$-dimensional spacetime\footnote{For details on the restrictions on the spacetime and the matter content, we refer to \cite{Bousso:2002hon} and references therein.}. We define $A=\text{vol}(\partial R)$ to be the area of the $2$-sphere that bounds $R$. For the system to be gravitationally stable, we need $E<M$, where $M$ is the mass of a black hole with event horizon $\partial R$ and surface area $A$. If we introduce a very heavy shell with mass $M-E$ around the region $R$ as displayed in figure \ref{fig:Susskind} and collapse it inwards, a black hole of mass $M$ forms.

If the massive shell is initially very far away from the region $R$, then the total initial entropy is given by
\be
S^{\text{initial}}_{\text{total}}=S_{\text{matter}}+S_{\text{shell}}\,.
\ee
On the other hand, the final state only contains a black hole with surface area $A$ and hence thermodynamic entropy\footnote{From now on, we will work in natural units.}
\be
S^{\text{final}}_{\text{total}}=S_{\text{BH}}=\frac{A}{4}\,.
\ee
By the \emph{generalised second law of thermodynamics} \cite{Bekenstein:1973ur}, 
\be
\dif S_{\text{total}}\geq 0\,,
\ee
and hence
\be
S_{\text{matter}}\leq S^{\text{initial}}_{\text{total}}\leq S^{\text{final}}_{\text{total}}=\frac{A}{4}\,.
\ee
We used that $S_{\text{shell}}\geq 0$. This gives us the \emph{spherical entropy bound}
\be
S_{\text{matter}}\leq\frac{A}{4}\,.
\ee
This bound can be saturated by putting a black hole in $R$. It is a very surprising result which implies that even if there is not a black hole in $R$, the number of degrees of freedom needed to describe a theory in $R$ scales like its surface and not its volume as would be naively expected \cite{Bousso:2002hon, Susskind:1994vu}. This led 't Hooft to the idea of the holographic principle: 'Given any
closed surface, we can represent all that happens inside it by degrees of freedom on this
surface itself' \cite{tHooft:1993dmi}. For a more precise version of this statement see \cite{Bousso:2002hon} and references therein.

\subsection*{The AdS/CFT correspondence}

As we discussed in the previous section, there is a general hope that gravitational theories on some manifold $M$, the \emph{bulk}, with boundary $\partial M$ may be described purely by degrees of freedom on $\partial M$. The first concrete realization of this holographic principle is the duality between  IIB string theory on AdS$_5$ $\times S^5$ and $\mathcal{N}=4$ super Yang-Mills
theory on $\partial_\infty$AdS$_5$ with gauge group U$(N)$ (see figure \ref{fig:AdS/CFT})  \cite{Maldacena:1997re, Witten:1998qj, DHoker:1999kzh}. 
\begin{figure}[t!]
\begin{center}
\begin{tikzpicture}
\node[draw] at (-4,-1) {IIB strings on  AdS$_5 \times S^5$};
\node[draw] at (4.5,-1) {$\mathcal{N} = 4$ SYM on $\partial_\infty$AdS$_5$};
\draw [<->] (-1.3,-1) -- (2,-1) node[midway,above]{AdS/CFT};
\draw [<->] (-1.3,-1) -- (2,-1) node[midway,below]{};
\end{tikzpicture}
\end{center}
\caption{\emph{The most prominent example of holography is the AdS$_5$/CFT$_4$ correspondence which provides a dictionary that leads to an equivalence between a gravitational bulk theory (left) and a boundary CFT (right).}}\label{fig:AdS/CFT}
\end{figure}
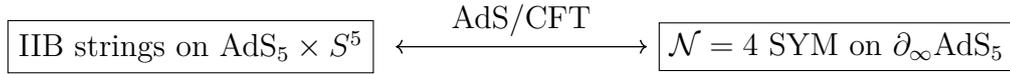$\partial_\infty$AdS$_5$ hereby denotes the \emph{boundary at infinity} of AdS$_5$ which is the boundary of its conformal compactification. This most famous example of the holographic principle is rooted in string theory. We will now discuss that the presence of an AdS$_5$-factor in the bulk geometry is not just a coincidence but it is dictated by the string theory through a \emph{backreaction} of D-branes which will play an important role in this thesis.

The AdS/CFT correspondence of figure \ref{fig:AdS/CFT} is derived by considering a stack of $N$ coincident D$3$-branes in $\mathbb{R}^{9,1}$. Fluctuations of these branes are described at low energies by
the dynamics of open strings which end on them. This dynamics in turn is given by a four-dimensional gauge
theory, specifically $\mathcal{N} = 4$ super Yang-Mills theory with U$(N)$ gauge group, on the worldvolume of the D$3$-branes. 

\begin{figure}[t!]
 \begin{center}
 \begin{tikzpicture}[scale=1]
\draw (0,0)  -- (1,2) -- (5,2) -- (4,0)--(0,0)  ;
\draw[red] (2.5,1) ellipse (1 and 0.4);
\draw[red] (2.5,1) ellipse (0.5 and 0.2);
\draw[red] (2.5,1) ellipse (0.25 and 0.1);
\node at (2.5,1){\red{$\bullet$}};

\node at (9.8,-2){$\downarrow$ \red{D-branes}};

\draw (6,0)  -- (7,2) -- (11,2) -- (10,0)--(6,0)  ;
\draw
(8.5,1) ellipse (1.25 and 0.5);
 \draw (7.75,0) .. controls (8.25,-1) and (8.25,-1.5) .. (8.25,-2);
 \draw (9.15,0) .. controls (8.65,-1) and (8.65,-1.5) .. (8.65,-2);
\draw[red] (8.23,-1.5) .. controls (8.33,-1.65) and (8.57,-1.65) .. (8.67,-1.5);
\draw[red] (8.23,-1.85) .. controls (8.33,-2) and (8.57,-2) .. (8.67,-1.85);
\draw[red] (8.07,-0.75) .. controls (8.2,-0.95) and (8.7,-0.95) .. (8.83,-0.75);
\draw[red, dashed] (8.23,-1.5) .. controls (8.33,-1.35) and (8.57,-1.35) .. (8.67,-1.5);
\draw[red, dashed] (8.23,-1.85) .. controls (8.33,-1.7) and (8.57,-1.7) .. (8.67,-1.85);
\draw[red, dashed] (8.07,-0.75) .. controls (8.2,-0.55) and (8.7,-0.55) .. (8.83,-0.75);

\end{tikzpicture}
\end{center}
\caption{\emph{This image displays backreaction in the AdS/CFT correspondence. On the left, we can see flat space before the backreaction and the location of the D-branes. On the right, we can see the backreacted geometry which is asymptotically flat but develops a throat region near the D-branes. We are of course cheating with dimensions and the AdS$_1\times S^1$ throat region we see should to be thought of as AdS$_5 \times S^5$ region where the red circles are the $S^5$-factor.}}\label{fig:Throatregion}
\end{figure}
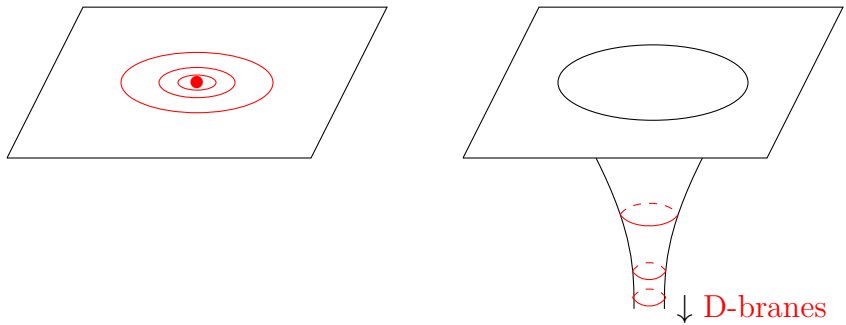

The low-energy description of closed strings is given by IIB supergravity and we can consider the same brane setup from this point of view. The $N$ D$3$-branes will backreact on the geometry (see figure \ref{fig:Throatregion}) meaning that coupling of the brane to the field content includes $N$ times a $\delta$-function source-term supported on the locus of the branes in the gravitational equations. This deformed equation has a fundamental solution which is a background metric
\bea
\label{eq:backreactionD3}
\dif s^2 =H(r)^{-1/2}\eta_{\mu \nu} \dif x^\mu \dif x^\nu +H(r)^{1/2}(\dif r^2+r^2\dif\Omega_{S^5})\,,
\eea
where
\be
H(r)=1+\frac{L^4}{r^2}
\ee
 and the coordinates are split into four Minkowski-like coordinates $x^\mu\in \mathbb{R}^{3,1}$ and hyperspherical-like coordinates $r,\Omega_{S^5}$. This metric is asymptotically flat since the backreaction from the brane on the flat $\mathbb{R}^{9,1}$ gets arbitrarily weak as we move far away from the brane. Close to the location of the branes, the backreaction is very strong and a throat region forms as depicted in figure \ref{fig:Throatregion}.
 
The AdS/CFT correspondence is concerned with closed strings in the vicinity of the brane so only the throat region of figure \ref{fig:Throatregion} will be of interest. In this \emph{near horizon limit} $r\ll L$, the metric \eqref{eq:backreactionD3} becomes
\be
\label{eq:backreactionD3nearhor}
\frac{r^2}{L^2}\eta_{\mu \nu} \dif x^\mu \dif x^\nu + \frac{L^2}{r^2}\dif r^2+l^2\dif \Omega_{S^5}\,,
\ee
which is the metric of AdS$_5 \times S^5$. 
Note, that the geometry \eqref{eq:backreactionD3} and its near horizon limit \eqref{eq:backreactionD3nearhor} are high dimensional analogues of the extremal Reissner-Nordström black holes in $4$ dimensions. It reads
\bea
\label{eq:extremalRN}
\dif s^2&= -H(r)^{-2}dt^2+H(r)^2(\dif r^2+r^2\dif \Omega_{S^2})\,,\\
H(r)&=1+\frac{GM}{r}\,,
\eea
and limits to AdS$_2\times S^2$ for $r\ll L$. 

The backreaction \eqref{eq:backreactionD3} and the insight that the low-energy dynamics of  the open string should match the low-energy dynamics of the closed string 
led Maldacena \cite{Maldacena:1997re} to formulate the duality of figure \eqref{fig:AdS/CFT}. There is much more evidence for this duality than we can possibly mention here, but let us briefly discuss how the most important quantities on both sides of the duality are related to each other via a \emph{holographic dictionary} \cite{Witten:1998qj, DHoker:2002nbb}.

This dictionary relates the different parameters of the theories on both sides to each other. Each theory depends on $2$ parameters. The bulk string theory depends on the string coupling $g_s$ and  the string length measured in units of the AdS radius $\ell_s/L$. The boundary gauge theory depends on the Yang-Mills coupling $g_{\text{YM}}$ and the rank of the gauge group $N$. These are related to each other by the dictionary
\bea
\label{eq:dictionaryParameters}
&\frac{1}{N}=\frac{g_s}{4\pi^2} \big(\frac{l_s}{L}\big)^2\,,\\
&g_{\text{YM}}^2N=4\pi^2 \big(\frac{L}{l_s}\big)^2\,,
\eea
where the latter quantity is often referred to as the \emph{'t Hooft coupling} $\lambda=g_{\text{YM}}^2N$. In the large $N$ limit, in which we keep the 't Hooft coupling fixed, we find a weakly coupled string theory $g_s\rightarrow 0$. Higher genus corrections in string perturbation theory correspond to finite $N$ corrections of the CFT. The value of the 't Hooft coupling is then controlled by the string length. Large 't Hooft coupling $\lambda\gg 1$, meaning a strongly coupled boundary theory,  corresponds to short strings $\frac{\ell_s}{L}\ll1$ which are approximated by the well-behaved supergravity limit of the theory. On the other hand, a weakly coupled boundary gauge theory with $\lambda\ll 1$ corresponds to long strings $\ell_s/L\gg1$ that can probe the entire AdS spacetime and generally don't have any well-behaved mathematical description\footnote{With the exception of some beautiful recent progress that considers the \emph{tensionless string} in AdS$_3$ \cite{Eberhardt:2018ouy, Eberhardt:2019ywk} and AdS$_5$ \cite{Gaberdiel:2021jrv, Gaberdiel:2021qbb}.}. This \emph{strong/weak} behaviour is a fundamental feature of the AdS/CFT correspondence and the fact that either side is always strongly coupled makes it very hard to get a mathematical handle on both sides at the same time. This will be different in \emph{twisted holography}.


\section{Twisted holography}
\label{section1.4}

Proving the AdS/CFT correspondence in its strongest form, i.e. beyond special limits of the parameters such as the tensionless limit ($N\gg 1$, $\lambda\ll1$) \cite{Eberhardt:2018ouy,Eberhardt:2019ywk}, is not a well-defined task because of the strong/weak nature of the duality we discussed previously. 

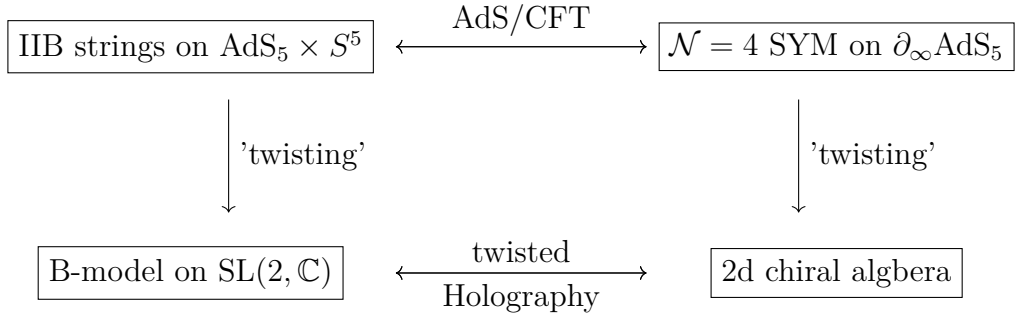
\begin{figure}[t!]
	\begin{center}
\begin{tikzpicture}
\node[draw] at (-4,-1) {IIB strings on  AdS$_5 \times S^5$};
\node[draw] at (4.5,-1) {$\mathcal{N} = 4$ SYM on $\partial_\infty$AdS$_5$};
\draw [<->] (-1.3,-1) -- (2,-1) node[midway,above]{AdS/CFT};
\draw [<->] (-1.3,-1) -- (2,-1) node[midway,below]{};

\node[draw] at (-4,-4) {B-model on SL$(2,\mathbb{C})$};
\node[draw] at (4.5,-4) {$2$d chiral algbera};
\draw [<->] (-1.3,-4) -- (2,-4) node[midway,above]{twisted};
\draw [<->] (-1.3,-4) -- (2,-4) node[midway,below]{Holography};

\draw [<-] (-3.5,-3.2) -- (-3.5,-1.7) node[midway,right]{'twisting'};
\draw [<-] (4,-3.2) -- (4,-1.7) node[midway,right]{'twisting'};

\end{tikzpicture}
\end{center}
	\caption{\emph{Twisted holography consists of a twisted subsector of the original AdS/CFT correspondence. Both sides of the duality are mathematically tractable after applying a certain twist \cite{Costello:2016mgj, Costello:2018zrm}. This is merely a schematic depiction of twisted holography and hides many important subtleties.}} \label{fig:twistedHolo}
\end{figure}

Twisted holography\footnote{\emph{Twisting} here refers to the procedure that was first introduced by Witten \cite{Witten:1988ze} and had since then had a crucial impact on the relation between quantum field theory and mathematics.} in the sense of \cite{Costello:2018zrm} is a holographic duality consisting of a simplified subector of the conventional AdS$_5$/CFT$_4$ duality (see figure \ref{fig:twistedHolo})\footnote{A twisted form of the AdS/CFT correspondence was formulated previously in \cite{Costello:2016mgj} and in \cite{Costello:2017fbo}.  There, the mathematical notion of \emph{Koszul duality} was used to define a holographic dictionary as an isomorphism between an algebras of large $N$ boundary operators and the \emph{Koszul dual} of an algebra of bulk operators \cite{Paquette:2021cij}.
}. The bulk- as well as the boundary-dynamics in twisted holography, are described by mathematically tractable theories, namely the B-model topological string theory on SL$(2,\mathbb{C})$ and a certain chiral algebra subsector of the boundary $\mathcal{N}=4$ supersymmetric Yang-Mills theory\footnote{For an actual definition of the respective theories, we refer to \cite{Costello:2018zrm}.} \cite{Lemos:2020pqv, Beem:2013sza}. 
Vaguely speaking, the topological/holomorphic nature of the B-model means that it is not possible to define the 'string length in AdS units'.

Up to certain subtleties, both sides of the duality arise as twists (in the presence of an $\Omega$-background) of the respective bulk and boundary theories of AdS$_5$/CFT$_4$, type IIB supergravity and the $\mathcal{N}=4$ supersymmetric Yang-Mills theory. $4$-dimensional $\mathcal{N}=4$ supersymmetric Yang-Mills theory has been known for a long time to admit a $2$-dimensional protected chiral algebra subsector \cite{Beem:2013sza, Lemos:2020pqv} arising from a certain twist of the $4$d theory \cite{Oh:2019bgz}. Twisting supersymmetric gauge theories more generally is a very well-studied and mathematically well-understood task \cite{Costello:2014gal}. On the other hand, topologically twisting type IIB supergravity and any supergravity theory more generally has only more recently been achieved by the work of Costello and collaborators \cite{Costello:2016mgj, Costello:2020jbh}\footnote{See \cite{Jeon:2018kec, deWit:2018dix} for another approach defining localization in supergravity.}. In supergravity theories, the supersymmetry is gauged which means that it is not possible to simply 'add' a supercharge to the BRST charge as is done in the standard twisting procedure. The resolution to this issue is that \emph{'Twisted supergravity is nothing but ordinary supergravity in an unusual background in which a bosonic ghost acquires a nontrivial vacuum
expectation value'} \cite{Costello:2020jbh}.
This has led to the conjectural identification of the left 'twisting' arrow in figure \ref{fig:twistedHolo}.

\subsection*{Backreaction in twisted holography}

Backreaction plays an important role in the AdS/CFT-correspondence as we saw in section \ref{section1.2} and it also plays an important role in twisted holography. Let us briefly discuss the simplest backreaction in the original context of twisted holography \cite{Costello:2018zrm}. In the context of standard AdS/CFT, the metric is a closed string state in the bulk and we saw the backreaction of a stack of D$3$-branes deform the metric to that of AdS$_5 \times S^5$ in the near-horizon limit \ref{fig:Throatregion}. However, the topological B-model only knows about the complex structure, instead of the metric, on the target space, which is some Calabi-Yau threefold $X$. For the remainder of this subsection, we will only consider $X=\mathbb{C}^3$. The corresponding closed string state in the B-model is given by $\beta \in \Omega^{0,1}(X, T^{1,0}X)$. This \emph{Beltrami differential} encodes deformations of the complex structure. It explicitly deforms the complex structure through
\be
\bar{\partial}\mapsto \bar{\partial}+\beta=\dif \bar{z}^i\big(\frac{\partial}{\partial \bar{z}^i}+\beta_i^j\frac{\partial}{\partial z^j}\big)\,.
\ee
For $ \bar{\partial}+\beta$ to define an integrable complex structure, it needs to obey
\be
\label{eq:BeltramiIntegrability}
0=(\bar{\partial}+\beta)^2= \bar \partial \beta + \frac{1}{2}\{\beta, \beta\} \,.
\ee
Here, $\{\ ,\ \}$ denotes the so-called \emph{Schouten bracket}\footnote{The Schouten bracket is not to be confused with the Poisson-bracket used below.} which is defined for instance in \cite{Costello:2018zrm} and won't be of further relevance to us.
Equation \eqref{eq:BeltramiIntegrability} is imposed by the \emph{BCOV-action} \cite{Bershadsky:1993cx, Costello:2015xsa} 
\be
\label{eq:BCOVaction}
S_{\text{BCOV}}[\beta]= \frac{1}{2}\int_X (\partial^{-1} \beta)\bar{\partial}\beta + \frac{1}{6}\int_X \beta^3\,.
\ee
Equation \eqref{eq:BCOVaction} is only a schematic form of the action. Contracting all the indices such that the integrand is an element of $\Omega^{3,3}(X)$ makes use of the holomorphic volume form $\Omega\in \Omega^{3,0}(X)$ and we refer to \cite{Costello:2018zrm} for the details. Note, that this means that in order to write down the action \eqref{eq:BCOVaction} it is crucial for $X$ to be Calabi-Yau which will be important below.  

Next to the integrablity condition \eqref{eq:BeltramiIntegrability}, the further condition $\partial \beta=0$ is needed for $\beta$ to be a deformation of Calabi-Yau manifolds rather than just complex manifolds and to define $\partial^{-1}\beta$. This condition $\partial\beta=0$ is imposed by hand  in analogy to the self-duality condition of the Ramond-Ramond (RR) $5$-form field strength in IIB string theory \cite{DHoker:2002nbb}. In fact, it is much more than just an analogy since $\beta$ corresponds to a certain component of the RR $5$-form field strength and $\partial^{-1}\beta$ corresponds to a certain component of the RR $4$-form itself. Even though it arises from some component of the $5$-form rather than the metric, $\beta$ is to be viewed as a gravitational field as we will motivate further below. 

To derive a holographic duality, we need to introduce D-branes. Consider a stack of $N$ coincident D$1$-branes wrapping $\mathbb{C}\subset \mathbb{C}^3$. We introduce coordinates $(z,w_1,w_2)\in \mathbb{C}^3$ such that
\be
\mathbb{C}=\{(z,w_1,w_2)\in\mathbb{C}^3\,|\,  w_1=w_2=0\}\subset\mathbb{C}^3\,.
\ee
Their presence leads to a new term in the action that couples the branes to $\beta$. The resulting action reads\footnote{In the untwisted theory, the coupling is given by the RR $4$-form being integrated over the D3-branes. Since $\partial^{-1}\beta$ rather than $\beta$ corresponds to a component of this RR $4$-form, we have to couple $\partial^{-1}\beta$ to the D$1$-brane in the twisted theory. Making use of $\Omega\in\Omega^{3,0}(X)$, $\partial^{-1}\beta$ can be viewed as a $(1,1)$ form \cite{Costello:2018zrm}.}
\be
S_{\text{BCOV}}[\beta]+N\int_\mathbb{C} \partial^{-1}\beta\,,
\ee
which leads to a new equation of motion with a source term
\be
\label{eq:sourcedIntegrability}
\bar \partial \beta + \frac{1}{2}\{\beta, \beta\} = N \delta_\mathbb{C}\,.
\ee
 In \cite{Costello:2018zrm}, it is shown that equation \eqref{eq:sourcedIntegrability} is solved by the fundamental solution
\be
\label{eq:betasource}
\beta_{\text{source}} = \frac{2N}{8 \pi^2}\frac{\bar w_1 d \bar w_2 - \bar w_2 d \bar w_1 }{||w||^4}\frac{\partial}{\partial z}\,.
\ee

What is the backreacted target space $(\mathbb{C}^3\setminus\mathbb{C},\bar{\partial}+ \beta_{\text{source}})$ resulting from $\beta_{\text{source}}$ in equation \eqref{eq:betasource}? To answer this, let us find deformed holomorphic coordinates, i.e. solutions $f(z,\bar{z},w_i,\bar{w}_i)$ to the equation
\be
\label{eq:betasourceholo}
0=(\bar{\partial}+\beta_{\text{source}})f(z,\bar{z},w_i,\bar{w}_i)\,.
\ee
Since $\beta_{\text{source}}$ does not involve any terms of the form $\frac{\partial}{\partial w_i}$,  we see immediately that $w_i$ are two solutions \eqref{eq:betasourceholo}. While $z$ can not be deformed to a holomorphic coordinate, the two combinations
\bea
u_1&=w_1z-N\frac{\bar{w}_2}{(|w_1|^2+|w_2|^2)^2} \\\
u_2&=w_2z+N\frac{\bar{w}_1}{(|w_1|^2+|w_2|^2)^2} \,,\,,
\eea
can be easily seen to be holomorphic. The four solutions $(w_1,w_2,u_1,u_2)$ are not independent. They obey the relation
\be
\label{eq:DefiningSL2}
w_1u_2-w_2u_1=N\,,
\ee
which is the defining relation of SL$(2,\mathbb{C})$. We conclude that a stack of D$1$-branes in the topological B-model wrapping $\mathbb{C}\subset \mathbb{C}^3$ backreacts the geometry from $\mathbb{C}^3\setminus\mathbb{C}$ to SL$(2,\mathbb{C})$. Since SL$(2,\mathbb{C})\cong \mathbb{H}^3\times S^3$ this space has a Euclidean AdS$_3$-factor and twisted holography can be viewed as an incarnation of AdS$_3$/CFT$_2$\footnote{We note however that SL$(2,\mathbb{C})\cong \mathbb{H}^3\times S^3$ is not an isomorphism of complex manifolds and SL$(2,\mathbb{C})$ as a complex manifold is the bulk spacetime we obtained from backreaction.} \cite{Costello:2020jbh}. This backraction is analogous to  D$3$-branes wrapping $\mathbb{R}^{3,1}\subset \mathbb{R}^{9,1}$ leading to AdS$_5 \times S^5$ in the near-horizon limit that we discussed in section \ref{section1.2}. Except the notion of a near horizon limit requires a metric which is not present in the given case so that in twisted holography the entire space consists of the throat region in figure \ref{fig:Throatregion}. 


In analogy to conventional AdS$_5$/CFT$_4$, there are further objects that can be included in the twisted holography dictionary such as giant gravitons and determinant operators (and perhaps even heavier operators of order $N^2$) \cite{Budzik:2021fyh, Budzik:2022hcd, Budzik:2023nnx}. 
Also, twisted holography has many more incarnations such as twists of AdS$_4$/CFT$_3$ or AdS$_7$/CFT$_6$ dualities which arise from twisted M-theory in the presence of M$2$-branes or  M$5$-branes \cite{Costello:2016nkh, Costello:2017fbo, Raghavendran:2021qbh}. Most importantly, for the context of this thesis, it is also possible to consider twisted holography on \emph{twistor space}\footnote{Note that it is just a coincidence that the word 'twist' appears in twisted holography and twistor theory.} $\mathbb{PT}$ rather than $\mathbb{C}^3$. We will refer to this incarnation of twisted holography as \emph{Burns holography} \cite{Costello:2022jpg, Costello:2023hmi} and discuss it after briefly mentioning some twistor theory.

\section{Twistor space, self-dual gravity and Burns holography}
\label{newsection1.4}

We will give a more detailed review of the necessary background material on twistor theory in chapter \ref{chapter3}. However, let us briefly mention some basic ideas here as well. 

The twistor space of (the complexification of) Minkowski space $\mathbb{R}^{3,1}$ is the complex $3$-dimensional manifold 
\be
\mathbb{PT}=\mathcal{O}(1)\oplus \mathcal{O}(1)\rightarrow \mathbb{CP}^1\,.
\ee
A sensible twistor space can be defined for a much more general class of $4$-manifolds equipped with a conformal class of metrics with self-dual Weyl-tensor\footnote{For a definition of $\mathcal{W}^\pm$ we refer to \cite{Mason:1991rf}.} \cite{Mason:1991rf, Atiyah:1978wi} but for the content of this section we will not attempt to do so.

Twistor theory is a longstanding program that relates holomorphic data on twistor space to conformal data on spacetime and has many incarnations for theories of scalars, gauge fields, higher spin fields and gravity in various dimensions\footnote{However, in this thesis we will only consider spacetimes of dimension $4$.}. In this thesis, we will be mostly interested in the gravitational case which was originally formulated as the content of the \emph{non-linear graviton construction}. It states that there is a one-to-one correspondence between
\begin{enumerate}
    \item Four-dimensional manifolds $M$ together with a conformal class $[g]$ of Riemannian metrics with self-dual Weyl curvature 
    $\mathcal{W}^-=0$,
   and
    \item Complex $3$-manifolds $\mathcal{PT}$ that possess at least one rational curve $\mathcal{L}_x\cong \mathbb{CP}^1$ with normal bundle $N=\mathcal{O}(1)\oplus \mathcal{O}(1)$, together with an antiholomorphic involution $\sigma:\mathcal{PT}\rightarrow \mathcal{PT}$ that acts as the antipodal map on $\mathcal{L}_x$.
\end{enumerate}
A more detailed discussion will be given in chapter \ref{chapter3} and a proof can be found in \cite{Penrose:1976jq, Penrose:1976js, Atiyah:1978wi}. More generally, such a transformation that relates some complex data on twistor space to some conformal data on spacetime is referred to as \emph{Penrose transform}.

The non-linear graviton construction essentially implies that the deformations of complex structures, which are dynamically described by the closed string sector of the B-model on twistor space are equivalent to deformations of conformal classes of self-dual metrics on spacetime. The insight that the B-model of section \ref{section1.4} can be considered on twistor space, originally with a focus on the open string sector, goes back a long time to Witten's seminal paper \cite{Witten:2003nn}. Twistor space is not Calabi-Yau and this was circumvented by supersymmetrising twistor space in the Berkovits-Witten twistor string \cite{Witten:2003nn, Berkovits:2004hg}. However, it can also be circumvented in other ways without supersymmetry as will be discussed below \cite{Costello:2021bah, Bittleston:2024efo}.

Already in this original Berkovits-Witten twistor string, an important feature of the closed-string B-model on twistor space was understood: Instead of describing any form of \emph{self-dual Einstein gravity} on spacetime, it describes some form of \emph{self-dual conformal gravity} \cite{Berkovits:2004jj}. This is reflected by the fact that the non-linear graviton construction as stated above only provides us with a conformal class $[g]$ of metrics on spacetime. There are many known problems with conformal gravity such as an unphysical fourth-order kinetic term and it is a natural question if the string theory can somehow be varied to give self-dual Einstein gravity on spacetime. The Ooguri-Vafa $\mathcal{N}=2$ string \cite{Ooguri:1990ww, Ooguri:1991fp} is believed to describe self-dual Einstein gravity on the target space but it comes with many additional complications that we will not discuss here. It is currently not known whether it admits an uplift to a string theory on twistor space. In the presence of $\mathcal{N}=8$ supersymmetry a worldsheet model that describes $\mathcal{N}=8$ supergravity and correctly computes its tree-level amplitudes has been found in \cite{Skinner:2013xp}. The data that was used in \cite{Skinner:2013xp} to break the conformal invariance on twistor space  is given by the so-called \emph{infinity twistor}. The infinity twistor allows one to pick out an explicit self-dual Einstein metric $g\in [g]$ in its conformal class and we will describe this in section \ref{chapter3} in more detail. 
Any twistor string that describes self-dual Einstein gravity has to know about this infinity twistor or break conformal invariance in some other way. Finding a twistor string theory that describes self-dual Einstein gravity on spacetime is an open problem at the time of the publication of this thesis.

Although engineering self-dual Einstein gravity from a string theory remains subtle and it is also not a theory that directly describes our universe\footnote{In fact, it is not even a unitary theory.}, it is still a well-behaved and simultaneously incredibly rich toy model. Let us briefly give an incomplete list of why we believe this to be the case:
\begin{itemize}
    \item Most importantly, self-dual conformal gravity as well as self-dual Einstein gravity are both believed to be UV finite $4$-dimensional theories of quantum gravity due to their string theory descriptions \cite{Ooguri:1991fp, Ooguri:1990ww, Costello:2021bah, Costello:2023hmi}. There are  only very few such examples\footnote{There are $\gg 10^{500}$ more such examples arising from compactifications of string theory on $\mathbb{R}^{3,1}\times \text{CY}^3$ but these are actually all $10$-dimensional theories.}.

    \item The famous 1-loop all $+$ graviton amplitude \cite{Bern:1998sv, Bern:1998xc} is present in self-dual Einstein gravity.  Both, the tree-level $-++$ and this 1-loop $+\cdots+$ amplitudes can be computed from the \emph{Chalmers-Siegel action} of self-dual gravity~\eqref{Chalmers-Siegel} and agree, at the same order of perturbation theory, with the corresponding amplitudes computed from the full Einstein-Hilbert action. However, in full gravity, these amplitudes receive further loop corrections whereas self-dual gravity is one-loop exact. Similar results exist in gauge theory and more generally, a lot about amplitudes in arbitrary helicity configurations has been learned from twistor space methods \cite{Cachazo:2012kg, Skinner:2013xp}. The rough slogan is that in perturbation theory the full physical theory is not 'too far away' from the self-dual theory through the existence of a so-called \emph{MHV-expansion}\footnote{GR does not quite admit an MHV vertex expansion in the same way as Yang-Mills \cite{Bjerrum-Bohr:2005xoa}.}. 

 \item Self-dual gravity is a classically integrable field theory which has an infinite dimensional algebra of hidden-symmetries \cite{Mason:1991rf, Dunajski:2000iq, Dunajski:2003gp, Penrose:1976jq}. These symmetries become manifest on twistor space and they are closely related to the \emph{celestial chiral algebras} that led to new symmetries of tree-level graviton amplitudes even beyond the self-dual sector \cite{Strominger:2021lvk, Guevara:2021abz, Costello:2022wso}. This twistor perspective on celestial chiral algebras has recently led to the computation of new two-loop amplitudes in non-supersymmetric QCD-like gauge theories \cite{Costello:2023vyy, Dixon:2024mzh, Dixon:2024tsb}. Similar gauge theory amplitudes were also computed on non-trivial backgrounds \cite{Bittleston:2024efo}. Although the gravitational case is more subtle, it is expected that loop-level graviton amplitudes and more general  higher loop gauge theory amplitudes can be computed from similar bootstrap methods \cite{Fernandez:2024qnu, Bittleston:2022jeq, Bittleston:2022nfr}. 
    
    \item Even though real solutions to the classical self-dual field equations only exist in Riemannian signature and $(2,2)$-signature, also referred to as \emph{Kleinian signature}, there are many classical solutions to the self-dual gravity equations that can be constructed from twistor space. These include gravitational instantons \cite{Hitchin:1979rts} such as Eguchi-Hanson space, more general ALE spaces, and the \emph{self-dual black holes} \cite{Crawley:2021auj, Crawley:2023brz, Bogna:2024gnt} that have been recently related to physical black holes \cite{Guevara:2023wlr}. The list could be extended by various other (pseudo-)hyperkähler manifolds and quaternionic Kähler manifolds, such as the \emph{Pedersen-metric}.  These metrics will play an important role in the present thesis and will be discussed in detail below.     
\end{itemize}

\subsection*{Burns holography}

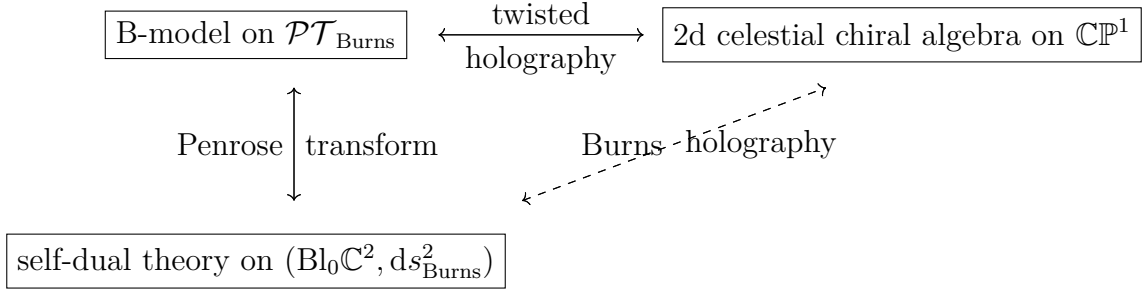
\begin{figure}[t!]
\begin{center}
\begin{tikzpicture}
\node[draw] at (-4,-1) {B-model on $\mathcal{PT}_{\text{Burns}}$};
\node[draw] at (4.5,-1) {$2$d celestial chiral algebra on $\mathbb{CP}^1$};
\draw [<->] (-1.6,-1) -- (1.1,-1) node[midway,above]{twisted};
\draw [<->] (-1.6,-1) -- (1.1,-1) node[midway,below]{holography};

\draw [<->] (-3.5,-3.2) -- (-3.5,-1.7) node[midway,left]{Penrose};
\draw [<->] (-3.5,-3.2) -- (-3.5,-1.7) node[midway,right]{transform};
\node[draw] at (-4,-4) {self-dual theory on $(\text{Bl}_0{\mathbb{C}^2}, \dif s^2_{\text{Burns}})$};
\draw [<->, dashed] (-0.5,-3.2) -- (3.5,-1.7) node[midway,left]{Burns};
\draw [<->, dashed] (-0.5,-3.2) -- (3.5,-1.7) node[midway,right]{holography};
\end{tikzpicture}
\end{center}
	\caption{\emph{Burns holography is a $2$d - $4$d duality that involves a combination of twisted holography ($2$d - $6$d) and the Penrose transform ($6$d - $4$d) which relates holomorphic data on twistor space to conformal data on spacetime. It is the first known concrete example of holography involving an asymptotically flat $4$-dimensional bulk spacetime.}} \label{fig:Burnsholo}
\end{figure}

Let us briefly review some aspects of \emph{Burns holography}, which is the name we will use to refer to the content of \cite{Costello:2022jpg, Costello:2023hmi}. It is a holographic duality that combines twisted holography with twistor theory and is summarized in figure \ref{fig:Burnsholo}. 

The B-model on twistor space $\mathbb{PT}$ of flat $\mathbb{R}^4$ is considered and although twistor space is not Calabi-Yau, with a choice of divisor it can be considered to be \emph{log Calabi-Yau}. Concretely, reference spinors $\alpha^\alpha, \beta^\alpha$ are chosen and a weightless meromorphic volume form
\be
\Omega=\frac{\epsilon_{abcd}Z^a\dif Z^b\wedge\dif Z^b\wedge \dif Z^c\wedge \dif Z^d}{4! \langle\alpha \lambda\rangle^2 \langle \lambda\beta\rangle^2}
\ee
is considered where $Z^a \in\mathbb{PT}\subset \mathbb{CP}^3$ are homogenous coordinates and $\lambda_\alpha\in\mathbb{CP}^1\subset \mathbb{PT}$ with $\lambda_\alpha\neq \alpha_\alpha$ and $\lambda_\alpha\neq \beta_\alpha$ \cite{Costello:2023hmi}. Our notation for twistors will be discussed in chapter \ref{chapter3}. A result by Pontecorvo \cite{pontecorvo1992twistor} shows that twistor spaces with such meromorphic volume forms correspond to scalar-flat Kähler geometries\footnote{Moreover, this requires $\beta=\hat{\alpha}$.} like \emph{Burns space}. While the \emph{infinity twistor} to be discussed in chapter \ref{chapter4} breaks conformal invariance in a way that singles out a desired representative $g\in [g]$ that is self-dual Einstein, the choice of reference spinors $\alpha^\alpha, \beta^\alpha$ to define a divisor breaks conformal invariance in a way that singles out a representative $g\in [g]$ that is scalar flat Kähler.

A stack of $N$ D$1$-branes wrapping the twistor line over the origin, $\mathbb{CP}^1_0$, is included.  $\mathbb{CP}^1_0$ is most simply thought of as the zero-section of  
\be
\mathbb{PT}=\mathcal{O}(1)\oplus \mathcal{O}(1)\rightarrow\mathbb{CP}^1\,,
\ee
and it can be identified with the celestial sphere of spacetime \cite{Costello:2022jpg}. Similarly to the simpler case of a $\mathbb{C}^3$ target space, the presence of D$1$-branes will result in a backreaction of the complex structure on $\mathbb{PT}\setminus\mathbb{CP}^1_0$. This backreaction deforms $\mathbb{PT}\setminus\mathbb{CP}^1_0$ to $\mathcal{PT}_{\text{Burns}}$, the twistor space of \emph{Burns space} \cite{Burns:1986, Costello:2023hmi}. Burns space is given by a Kähler-metric $\dif s^2_{\text{Burns}}$ on the blowup of $\mathbb{C}^2$ at the origin, $\text{Bl}_0{\mathbb{C}^2}$. If we placed the D$1$-branes on some other twistor line $\mathbb{CP}^1_x$ corresponding to a point $x\in\mathbb{C}^2$, then the backreacted spacetime would be $\text{Bl}_x{\mathbb{C}^2}$ and it is expected that wrapping multiple stacks of D$1$-branes along different  $\mathbb{CP}^1_{x_i}$ leads to multiple blowups on spacetime\footnote{On $\mathbb{C}^3$ the multi-centred case is well-understood \cite{Budzik:2022hcd}, however in twistor space it is more subtle.}. So Burns holography is a starting point for a concrete realization of the expectation that a certain \emph{spacetime foam} is described  by a \emph{gas of D$1$-branes} in the
topological B model on twistor space \cite{Hartnoll:2004rv}.

The Kähler scalar for the Burns metric takes the simple form
\be
K(u)=|u|^2+N\log \big(|u|^2\big)\,,
\ee
where $u^\da\in \mathbb{C}^2\setminus 0$ and $N$ is the number of $D$1-branes that controls the strength of the backreaction. This Kähler potential solves a certain fourth order PDE, the equation of motion of \emph{Mabuchi gravity} \cite{Costello:2023hmi}, everywhere away from $u^\da=0$. Solutions to this equation give scalar flat Kähler manifolds as anticipated from Pontecorvo's theorem \cite{Pontecorvo:1992twi}. Even though Burns space is scalar flat, its Ricci curvature can be explicitly seen to be non-vanishing. In fact, the metric is not even Einstein.

As we will discuss in chapter \ref{chapter5}, Burns space contains the self-dual Einstein manifold $(\mathbb{CP}^2\setminus\{\text{point}\}, \dif s^2_{\text{Fubini-Study}})$ in its conformal class but there is no mechanism to choose the corresponding conformal factor in the present context. The conformal class that gets picked from the data of the divisor will always lead to a scalar-flat Kähler manifold, i.e. a solution to the equations of Mabuchi-gravity. Beyond being stuck with Mabuchi-gravity, there is a stringent anomaly cancellation which leads to the choice $G=\text{SO}(8)$ for the gauge group of the open string sector\footnote{We should note that there are also other ways of cancelling the chiral anomalies on twistor space \cite{Costello:2022jpg, Costello:2023vyy} as will be discussed below.}. 

All of these somewhat unphysical features of the self-dual bulk theory aside, we should reiterate that Burns holography is an incredibly beautiful story which makes up the first known concrete example of a holographic duality that involves an asymptotically flat space in the bulk. The bulk theory has nice properties such as quantum integrability and it is expected to be UV finite by its very construction. Burns holography has been checked to a very impressive accuracy and the dictionary includes the matching of bulk scattering amplitudes with boundary correlation functions.  The possibility of holographic dualities in asymptotically flat spaces has been related to the infrared structures of gravity and gauge theory in more general physical theories \cite{Strominger:2017zoo}. Its study has recently been known as \emph{celestial holography} \cite{Raclariu:2021zjz, Pasterski:2021rjz, McLoughlin:2022ljp}.
Burns holography has many features which are expected from the celestial holography program as will be discussed below. In particular, the holomorphic collinear behaviour of scattering amplitudes on Burns space has been matched with corresponding OPEs of boundary correlation functions as envisioned previously in celestial holography through the figure \ref{fig:celestial} \cite{Strominger:2017zoo}.


\section{Holography in asymptotically flat spaces}
\label{section1.3}

Already a long time before the specific example of Burns holography was identified, it was a natural question whether the AdS/CFT correspondence of section \ref{section1.2} is valid beyond bulk spacetimes that include an AdS factor. After all, we do not seem to live in a universe with a negative cosmological constant. Instead, we observe a very small positive cosmological constant. Moreover, the intuition we discussed in section \ref{section1.1} seems to suggest that the holographic principle should be very generally tied to the existence of black holes in gravitational theories which does not require a negative cosmological constant to be in place. 
There has been a vast amount of work on holographic dualities in the presence of a positive cosmological constant \cite{Strominger:2001pn}, but we will not attempt to discuss any of this. Instead, we will approximate the small positive cosmological constant in our universe to be $0$ and try to see if holography can say something about gravitational theories in asymptotically flat $4$-dimensional spacetimes. This question was already studied a long time ago \cite{Susskind:1998vk, Polchinski:1999ry, Giddings:1999jq} and has recently regained a lot of attention as the centre of the \emph{celestial holography program} \cite{Pasterski:2021raf, Pasterski:2021rjz, Raclariu:2021zjz, Strominger:2017zoo, McLoughlin:2022ljp}.

\begin{figure}[t!]
\begin{center}
\begin{tikzpicture}[scale=1]
       
      \draw[color=red, thick] (-6.5,0)--(-5,1.5);
      \draw[color=red, thick] (-3.5,0)--(-5,1.5);
      \draw[color=red, thick] (-6.5,0)--(-5,-1.5);
      \draw[color=red, thick] (-3.5,0)--(-5,-1.5);
\filldraw[color=black, fill=black!15, very thick](-5,0) circle (0.5);
\draw[blue, thick] (-5.75,-0.75)--(-5.35,-0.35);
\draw[blue, thick] (-4.25,-0.75)--(-4.65,-0.35);
\draw[green, thick] (-5.75,0.75)--(-5.35,0.35);
\draw[green, thick] (-4.4,0.9)--(-4.75,0.45);
\draw[green, thick] (-4.1,0.6)--(-4.55,0.25);
      \filldraw[color=red!60, fill=red!5, very thick](0,0) circle (1.5);
      \node at (-0.8,0.8){{\color{green}$\bullet$}};
      \node at (1.2,0.3){{\color{green}$\bullet$}};
      \node at (-1,-0.4){{\color{blue}$\bullet$}};
      \node at (1,0.6){{\color{green}$\bullet$}};
     \node at (0.2,-0.7){{\color{blue}$\bullet$}};
      \node at (1.9,-1.7){$\mathbb{CP}^1_{\text{celestial}}$};
       \node at (-5,-0){S};
       \node at (-6.2,-1.5){$\mathbb{R}^{3,1}$};
       \node at (-3.7,1){\red{$\scri^+$}};
       \node at (-3.7,-0.8){\red{$\scri^-$}};
    \end{tikzpicture}
    \caption{\emph{The Penrose diagram of the conformal compactification of Minkowski space $\mathbb{R}^{3,1}$ is depicted on the left. Its boundary contains $\scri^{\pm}=\mathbb{CP}^1_{\text{celestial}}\times \mathbb{R}$, where the $\mathbb{R}$ direction is null. The S-matrix of a $4$-dimensional QFT in Minkowski space is depicted in the Penrose diagram. An S-matrix element can be rewritten to take the apparent form of a  correlator in some two-dimensional CFT on \emph{the celestial sphere} $\mathbb{CP}^1_{\text{celestial}}$ \cite{Strominger:2017zoo} as depicted on the right. Two external momenta becoming collinear corresponds to the collision of the two respective insertion points on  $\mathbb{CP}^1_{\text{celestial}}$. There are many caveats to this picture.}}\label{fig:celestial}
      \end{center}
    \end{figure}
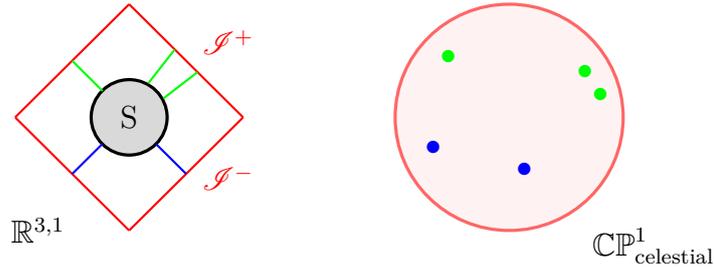

The first challenging feature is that the conformal compactification of Minkowski space has a boundary with a null direction  as displayed on the left of figure \ref{fig:celestial}. It is a topic of current debate 'where the holographically dual theory should live' 
and there are two main proposals. First of all the holographically dual theory could be a $3$-dimensional \emph{Carrollian CFT} living on 
\be
\scri=\mathbb{CP}^1_{\text{celestial}}\times \mathbb{R}\,.
\ee
Or the duality could be involve a $2$-dimensional CFT on the celestial sphere \cite{deBoer:2003vf} as displayed in figure \ref{fig:celestial}. We will refer to the latter as \emph{celestial holography} and it requires decomposing fields into modes along the null direction of $\scri$. Commonly, these are so-called \emph{conformally soft modes} and they will be discussed below.
The $3$-dimensional Carrollian approach is better suited to describe dynamics on $\scri$. For instance, consider sequential bursts of gravitational radiation. It will intersect $\scri$ at different points in the $\mathbb{R}$-direction and it is not quite clear how a theory on the celestial sphere would describe such a situation. While these two approaches look very different a priori, they have been recently suggested to be equivalent \cite{Mason:2023mti, Donnay:2022aba, Donnay:2022wvx, Bagchi:2022emh}. We will not discuss the Carrollian perspective any further.

 Naively taking the flat space limit of AdS/CFT $N\rightarrow 0$ means that there are no branes left on which a dual theory could live.  Anyway, a lot of work in the literature  has successfully implemented, in different ways, the limit
\be
\Lambda=\frac{-3}{L^2}\rightarrow 0\,,
\ee
in AdS/CFT to end up with Minkowski space in the bulk. While this led to many insights on how flat space physics arises from AdS \cite{Penedones:2010ue, Fitzpatrick:2011hu, Fitzpatrick:2011ia, Gadde:2022ghy, Marotta:2024sce, Alday:2017vkk, Hijano:2019qmi}, the limit is much more subtle on the boundary \cite{Alday:2024yyj, Lipstein:2025jfj} and it has not been possible to find a fully fledged holographic duality from such a limit.

Since flat $\mathbb{R}^n$ has vanishing curvature at every point, it is not possible to define an analogue of the string length in AdS units. Such a dimensionless ratio allowed us to consider weakly coupled regimes in the first place through \eqref{eq:dictionaryParameters}. The absence of such a scale suggests that a holographically dual theory to strings in Minkowski space is intrinsically strongly coupled. This argument can be circumvented however if an asymptotically flat spacetime with some curvature in the bulk is considered. This is an important contrast to conventional AdS/CFT with $\Lambda<0$. There, it would be very unnatural to attempt holographically understanding some complicated asymptotically AdS space (say, AdS-Schwarzschild) before considering AdS itself. For $\Lambda=0$ however, asymptotically flat spaces which have some curvature scale in the bulk might be easier to understand than Minkowski space itself.
Examples of this include Burns space and the gravitational instantons such as Eguchi-Hanson space which will be considered in the bulk of this thesis.

\subsection*{Celestial chiral algebras from splitting functions}

The most important observable of asymptotically flat theories is the S-matrix depicted on the left of figure \ref{fig:celestial}. The general difficulties with holography in asymptotically flat space  mean that a lot of literature on celestial holography has focussed on rewriting the S-matrix in a so-called \emph{conformal primary basis} and see what we can learn about a putative holographically dual $2$-dimensional \emph{celestial CFT} on general grounds. 
A massless $n$-particle amplitude depends on the external massless momenta
\be
p_i^{\alpha \da}=\lambda_i^\alpha \tilde{\lambda}_i^\da\,,\quad i=1,\dots, n
\ee
where little group scaling can be used to have $\lambda_i^\alpha\in \mathbb{CP}^1$ only defined up to scaling. This $\mathbb{CP}^1$ can be viewed as the celestial sphere. Let us use the corresponding coordinates
\be
\lambda_i^\alpha=\binom{1}{z_i}\,,\quad \tilde{\lambda}_i^\da= \omega_i \binom{1}{\tilde{z}_i}\,,
\ee
where $\omega$ is the total energy of the momentum of $p_i$ and $\omega\rightarrow0$ is called the (energetically) \emph{soft limit.}
An amplitude 
\be
\mathcal{A}(\omega_i,z_i,\tilde{z}_i)
\ee
can be transformed to \emph{celestial amplitude} by a simple Mellin-transform
\be
\label{eq:Mellin}
\mathcal{A}(\Delta_i,z_i,\tilde{z}_i)=\Bigg(\prod_{i=1}^n\int_0^\infty \dif \omega_i\, \omega_i^{\Delta_i-1}\Bigg)
\mathcal{A}(\omega_i,z_i,\tilde{z}_i)\,.
\ee
Up to important subtleties\footnote{such as the presence of a momentum conserving $\delta$-function in the amplitude \cite{Fan:2022vbz}.}, the left-hand side of equation \eqref{eq:Mellin} resembles a correlation function of a $2$-dimensional celestial CFT \cite{Pasterski:2016qvg, Pasterski:2017kqt}. The most important analogy is played by the relation between OPEs of a putative $2$-dimensional CFT and \emph{holomorphic collinear singularities} of the $4$-dimensional S-matrix. The latter arise from the singularities an amplitude develops as the momenta of two external massless states become (holomorphically) collinear, i.e. 
\be
z_{ij}=z_i-z_j\rightarrow 0\,, \quad \text{ with $\tilde{z}_i, \tilde{z}_j$ fixed}\,.
\ee
This is depicted in figure \ref{fig:celestial}. 

\begin{figure}[t!]
	\begin{center}
 	\includegraphics[scale=0.4]{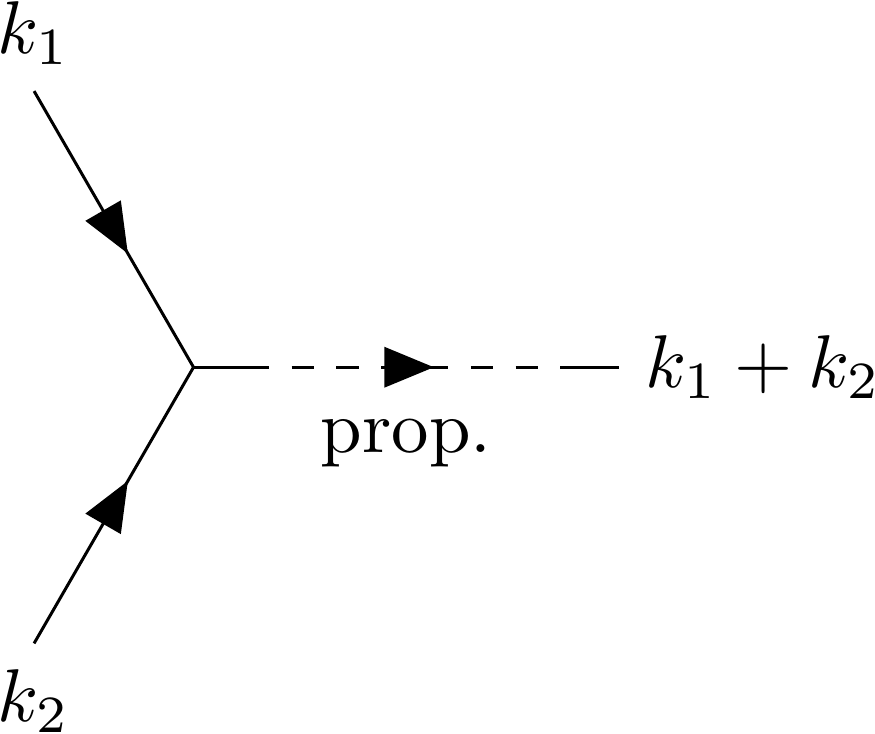}
	\caption{\emph{Tree diagram responsible for the singularity in a graviton amplitude as the momenta $k_1,k_2$ of two positive helicity external states become (holomorphically) collinear. A simple pole is generated when the propagator goes on-shell.}} \label{fig:splitting}
	\end{center}
\end{figure}

As shown long ago \cite{Weinberg:1965nx, Kulish:1970ut,Altarelli:1977zs,Berends:1987me,Mangano:1990by}, collinear singularities are governed by \emph{splitting functions} which in perturbation theory arise from the diagram in figure~\ref{fig:splitting}. 
For instance, consider a tree-level $n$-point graviton amplitude\footnote{Analogous results exist for gluon amplitudes and have similar implications. However, in the given section we will solely discuss the gravitational case.} $\mathcal{A}^n_{s_1,\dots ,s_n}(p_1,\dots,p_n)$ in the holomorphic collinear limit. 
$s_i\in\{\pm 2\}$ hereby denotes the helicity of the $i$-th external particle. $\mathcal{A}^n$ has a universal piece which factorizes as
\be
\label{eq:GravitonSplitting}
\mathcal{A}^n_{s_1,\dots ,s_n}(p_1,\dots,p_n)\xrightarrow[]{z_{ij}\rightarrow0} \sum_{s\in \{\pm 2\}} \text{Split}^s_{s_i,s_j}(p_i,p_j) \mathcal{A}^{n-1}_{s_1,\dots,s,\dots,s_n}(p_1,\dots,P , \dots,p_n)\,,
\ee
where we defined
\be
P=p_i+p_j\,,\quad \omega_P=\omega_i+\omega_j\,.
\ee

This collinear splitting displayed in figure \ref{fig:splitting} takes the form \cite{Bern:1998sv}
\be
\text{Split}^{+2}_{2,2}=\frac{-\kappa \omega_P^2}{2\omega_i\omega_j} \frac{\tilde{z}_{ij}}{z_{ij}}\,.
\ee
Carefully performing the Mellin transform \eqref{eq:Mellin} on both sides of equation \eqref{eq:GravitonSplitting}, in the case of particles $i$ and $j$ being of positive helicity, gives
\be
\label{eq:conformalprimarygravitons}
G^+_{\Delta_1}(z_1,\tilde{z}_1)G^+_{\Delta_2}(z_2,\tilde{z}_2) \sim \frac{-\kappa}{2} \frac{\tilde{z}_{ij}}{z_{ij}} B(\Delta_1-1,\Delta_2-1) \,G^+_{\Delta_1+\Delta_2}(z_2,\tilde{z}_2)
\ee
 at leading order, where $B$ denotes the Euler beta function. Equation \eqref{eq:conformalprimarygravitons} can be interpreted as a \emph{celestial OPE} between two \emph{conformal primary gravitons}\footnote{Rather than computing the S-matrix in the conventional basis of momentum eigenstates, the Mellin transform \eqref{eq:Mellin} leads to boost eigenstates \cite{Raclariu:2021zjz, Pasterski:2016qvg, Pasterski:2017kqt}. The corresponding basis is commonly referred to as conformal primary basis.} of arbitrary weight.

Up to subtleties \cite{Mitra:2024ugt}, 
the soft limit in the conformal primary basis can be implemented by the so-called \emph{conformally soft limit} \cite{Guevara:2021abz, Pate:2019lpp}
\be
\Delta\rightarrow 2,1,0,-1,\dots \,.
\ee
This limit gives rise to the conformally soft gravitons of positive helicity
\be
H^k(z, \tilde{z})=\lim_{\epsilon\rightarrow0} 
\epsilon \, G^+_{k+\epsilon}(z, \tilde{z})\,,
\ee
where  $k\in \{2,1,0,-1,\dots\}$. Expanding further in $\tilde{z}$ gives
\be
H^k(z,\tilde{z})= \sum_{m=(k-2)/2}^{(2-k)/2} \frac{H^k_m(z)}{\tilde{z}^{m+(k-2)/2}}\,,
\ee
where we took the conformal weights
\be
(h,\bar{h})=\bigg(\frac{k+2}{2},\frac{k-2}{2} \bigg)
\ee
of $H^k$ into account. After a further redefinition of the modes, a so-called light transform \cite{Strominger:2021lvk}
\be
w^p_m(z)=\frac{1}{\kappa} (p-m-1)!(p+m-1)!H^{-2p+4}_m(z)\,,
\ee
the OPEs of equation \eqref{eq:conformalprimarygravitons} take the very simple form \cite{Strominger:2021lvk, Guevara:2021abz}
\be
\label{eq:wOPE}
w^p_m(z_1) w^q_n(z_2)\sim \frac{1}{z_{12}}(m(q-1)-n(p-1))w_{m+n}^{p+q-2}(z_2)\,.
\ee
Considering individual $z$-modes 
\be
w^p_{m}(z)=\sum_{a=0}^\infty w^p_{m,a} \,z^{-a+p-3}
\ee
of the OPE \eqref{eq:wOPE} we find the Lie-algebra 
\be
\label{eq:wLieAlgebra}
[ w^p_{m,a}, w^q_{n,b}]=(m(q-1)-n(p-1))w_{m+n,a+b}^{p+q-2}\,,
\ee
which has played an important role in the celestial holography program. The structure constants of equation \eqref{eq:wLieAlgebra} are the same as the structure constants of the famous Lie-algebra $w_{1+\infty}$ \cite{Pope:1991ig, Pope:1989sr}. However, since the labels of the generator $w_{m,a}^p$ are restricted to
\be
\label{eq:indexrange}
p\in \{1,\tfrac{3}{2},2,\tfrac{5}{2},\dots\}\,, \quad m\in\{1-p,2-p,\dots,p-2, p-1\}\,, \quad a\in \mathbb{Z}+p
\ee
this is not quite $w_{1+\infty}$. In fact, it is the loop algebra of $\mathfrak{ham}(\mathbb{C}^2)$, denoted by $\mathcal{L}\mathfrak{ham}(\mathbb{C}^2)$, for reasons we will explain below\footnote{It is also often referred to as the loop algebra of the wedge subalgebra of $w_{1+\infty}$, $\mathcal{L}w_\wedge$, but this is also not quite right because importantly, the label $p$ in equation \eqref{eq:indexrange} is a half-integer. The half-integer modes can be eliminated by a $\mathbb{Z}_2$ quotient in which case we indeed have $\mathcal{L}w_\wedge\cong \mathcal{L}\mathfrak{ham}(\mathbb{C}^2/\mathbb{Z}_2)\cong \mathcal{L}\mathfrak{ham}(\mathbb{C}^2)^{\mathbb{Z}_2}$. All of these subtle differences will be very relevant in this thesis.}.

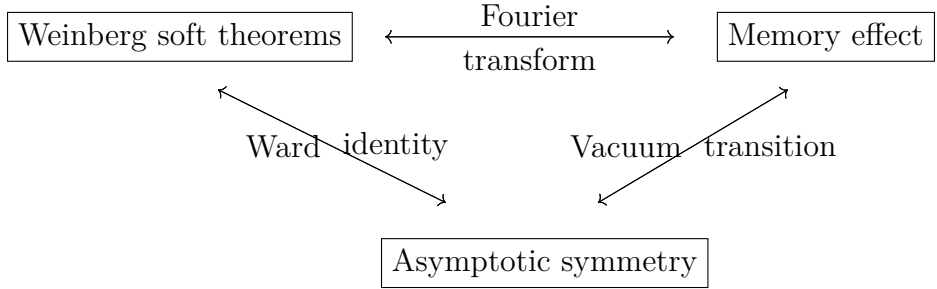
\begin{figure}[t!]
	\begin{center}
\begin{tikzpicture}
\node[draw] at (-4,-1) {Weinberg soft theorems};
\node[draw] at (4.5,-1) {Memory effect};
\draw [<->] (-1.3,-1) -- (2.5,-1) node[midway,above]{Fourier};
\draw [<->] (-1.3,-1) -- (2.5,-1) node[midway,below]{transform};

\node[draw] at (0.8,-4) {Asymptotic symmetry};

\draw [<->] (-0.5,-3.2) -- (-3.5,-1.7) node[midway,left]{Ward};
\draw [<->] (-0.5,-3.2) -- (-3.5,-1.7) node[midway,right]{identity};
\draw [<->] (1.5,-3.2) -- (4,-1.7) node[midway,left]{Vacuum};
\draw [<->] (1.5,-3.2) -- (4,-1.7) node[midway,right]{transition};

\end{tikzpicture}
\end{center}
	\caption{\emph{This infrared triangle relates three central, yet seemingly different subjects of physics. We will not discuss the individual corners and their relation in detail and refer to \cite{Strominger:2017zoo} and references therein for a detailed discussion.}} \label{fig:Triangle}
\end{figure}

The Lie-algebra \eqref{eq:wLieAlgebra} has important physical implications via so-called \emph{infrared triangles} \cite{Strominger:2017zoo}, which are striking triangular equivalence relations governing the infrared  dynamics of essentially all QFTs involving massless fields\footnote{The role of massive particles in this context is discussed in the literature \cite{Campiglia:2015kxa, Campiglia:2015qka, Himwich:2023njb} but less well established.}. It relates the three corners of soft theorems, asymptotic symmetries, and  the memory effect in a simple picture \ref{fig:Triangle} which we will not discuss in detail here. We refer to the review \cite{Strominger:2017zoo} and references therein for a detailed discussion. Next to their physical relevance, these algebras naturally arise from the twisted holography perspective and hence link to the more mathematical literature \cite{Costello:2022wso, Zeng:2023qqp}.

Roughly, the hope is that the generators $w^p_{m,a}$ lead to an infinite tower of such infrared triangles, one for each $p\in \{1,\tfrac{3}{2},2,\tfrac{5}{2},\dots\}$. Let us discuss this case by case in some more detail:
\begin{itemize}
    \item $p=1$. The elements $w^1_{0,a}$ for $a\in \mathbb{Z}$ are all central and could be consistently removed from the Lie-algebra if we wanted to. We believe that these central extensions do not play a physical role. 
    \item $p=3/2$. When interpreted as asymptotic symmetry generators, the $w^{3/2}_{m,a}$ generate certain \emph{BMS supertranslations} \cite{Strominger:2021lvk} with the four global translations given by $w^{3/2}_{\pm 1/2,\,\pm 1/2}$. Supertranslations are part of an infrared triangle \cite{Kervyn:2023adk, Strominger:2017zoo} in which the other two corners are given by the displacement memory effect \cite{Zeldovich:1974gvh} and the leading soft graviton theorem \cite{Weinberg:1965nx}. 
    
    \item $p=2$. The $w^{2}_{m,a}$ generate certain \emph{BMS superrotations} \cite{Strominger:2021lvk} when interpreted as asymptotic symmetry generators. In this case, the infrared triangle perspective was very fruitful and has led to a previously undiscovered gravitational memory effect, the \emph{spin memory effect}. There even is a proposal for measuring this spin memory effect using the Einstein telescope \cite{Nichols:2017rqr}. The triangle is completed by a subleading soft graviton theorem  in the remaining corner. This new soft theorem was shown to imply superrotation symmetry  \cite{Campiglia:2014yka, Kapec:2014opa, Kapec:2016jld}. 
    
    \item $p= 5/2$. Beyond the subleading soft graviton theorem, a further sub-subleading soft graviton theorem has been proven \cite{Cachazo:2014fwa} and there are discussions of the corresponding spacetime symmetries \cite{Campiglia:2016efb, Campiglia:2016jdj, Freidel:2021dfs}. These spacetime symmetries have been discussed to be non-local \cite{Freidel:2021dfs} which means it is certainly not as natural as the previous cases.
    
    \item $p\geq 3$. In fact,  a treatment similar to $p=5/2$ has been generalized to give rise to an infinite tower of charges \cite{Freidel:2021ytz}\footnote{These charges can also be derived from twistor space \cite{Kmec:2024nmu, Kmec:2025ftx, Seet:2024vmh}.}. However, the precise relation to the Lie-algebra \eqref{eq:wLieAlgebra} is not quite clear and again feels much less natural than the cases $p=3/2$ and $p=2$. Also, an infinite tower of soft graviton theorems is known to exist \cite{Hamada:2018vrw} and it was speculated to be related to these higher generators \cite{Guevara:2021abz}.
\end{itemize}

Although the role of the $w^p_{m,a}$ generators with $p\geq 5/2$ is not quite obvious in theories of physical interest, these generators have a very natural interpretation as symmetries of self-dual gravity through the non-linear graviton construction of section \ref{section1.3}. In fact, on twistor space, the algebra \eqref{eq:wLieAlgebra} can be obtained simply by plugging two generators
\be \label{eq:TwistorWbasis}
    w^{p}_{m,a} = \frac{(\mu^{\dot 0})^{p+m-1}(\mu^{\dot 1})^{p-m-1}}{2\lambda_0^{p-a-2}\lambda_1^{p+a-2}}\,,
\ee
 which are polynomial in $\mu^\da$ and have arbitrary integer powers in $\lambda_\alpha$, into the degenerate Poisson-bracket corresponding to the flat space infinity twistor
\be
\label{eq:Infinity}
 \{f,g\}=\epsilon^{\dot\alpha\dot\beta}\frac{\partial f}{\partial\mu^{\dot\alpha}}\frac{\partial g}{\partial\mu^{\dot\beta}}\,.
\ee
A simple calculation shows that indeed
\be
\label{eq:PoissonW}
\{w^p_{m,a},w^{q}_{n,b}\}=(m(q-1)-n(p-1))w^{p+q-2}_{m+n,a+b}\,,
\ee
which matches \eqref{eq:wLieAlgebra} and also justifies the name $\mathcal{L}\mathfrak{ham}(\mathbb{C}^2)$ for the Lie algebra \eqref{eq:PoissonW}. The reason it is natural to consider the generators \eqref{eq:TwistorWbasis} and the Poisson-bracket \eqref{eq:Infinity} is rooted in the non-linear graviton construction and was pointed out in \cite{Adamo:2021lrv, Costello:2022wso}. We will review the arguments in more detail in chapter \ref{chapter3}.

\section{Deformations of celestial chiral algebras}
\label{section1.5}

We saw that celestial holography, among other things, centers on the hope that collinear singularities of graviton scattering amplitudes are described by the OPE of a putative dual $2$-dimensional CFT \cite{Strominger:2017zoo, Raclariu:2021zjz, Pasterski:2021raf}. The fact that $\mathcal{L}\mathfrak{ham}(\mathbb{C}^2)$ arises from a tree-level graviton splitting function is one of the great successes of celestial holography and means that this duality is true at tree-level. This led to the discovery of new infinite dimensional symmetry algebras of tree-level amplitudes in flat space \cite{Strominger:2021mtt, Guevara:2021abz, McLoughlin:2022ljp}. We briefly discussed their relation to twistor space \cite{Adamo:2021lrv, Costello:2022wso} which leads to the central slogan of this thesis:

\begin{center}
\emph{The presence of the Lie-algebra $\mathcal{L}\mathfrak{ham}(\mathbb{C}^2)$, closely related to $w_{1+\infty}$, in tree-level graviton scattering on $\mathbb{R}^4$ is implied by the classical integrability of self-dual gravity.}
\end{center}

It is a natural question, whether this slogan can be generalized beyond flat $\mathbb{R}^4$ and beyond tree-level\footnote{A further natural generalization is to $\mathbb{R}^{4n}$ and other high-dimensional hyperkähler manifolds \cite{Adamo:2025mqp, Pano:2023slc}.}.

\subsection*{Quantum Corrections}

Almost the entirety of this thesis will be about tree-level results, however, we find it important to at least briefly mention some of the exciting developments around quantum corrections to celestial chiral algebras \cite{Costello:2022upu, Bittleston:2022jeq, Fernandez:2024qnu, Sahoo:2018lxl, Choi:2024ygx}. 

The Lie-algebra $w_{1+\infty}$ admits a deformation to the famous Lie-algebra $W_{1+\infty}$ \cite{Pope:1991ig}.  It was initially speculated that this $W_{1+\infty}$ might perhaps be related to quantum corrections in the bulk\footnote{This was a justified expectation since $W_{1+\infty}$ arises as a quantization of $w_{1+\infty}$ from the perspective of 2d CFTs with higher spin symmetry~\cite{Fateev:1987zh,FATEEV1987644,Zamolodchikov:1985wn}
} \cite{Strominger:2021lvk}. However, it turns out that such a deformation instead arises from a non-commutative deformation of classical self-dual gravity theory as first pointed out by the author and collaborators \cite{Bu:2022iak}. We will discuss this further in chapter \ref{chapter2}.

\begin{figure}[!t]
	\centering
 	\includegraphics[scale=0.33]{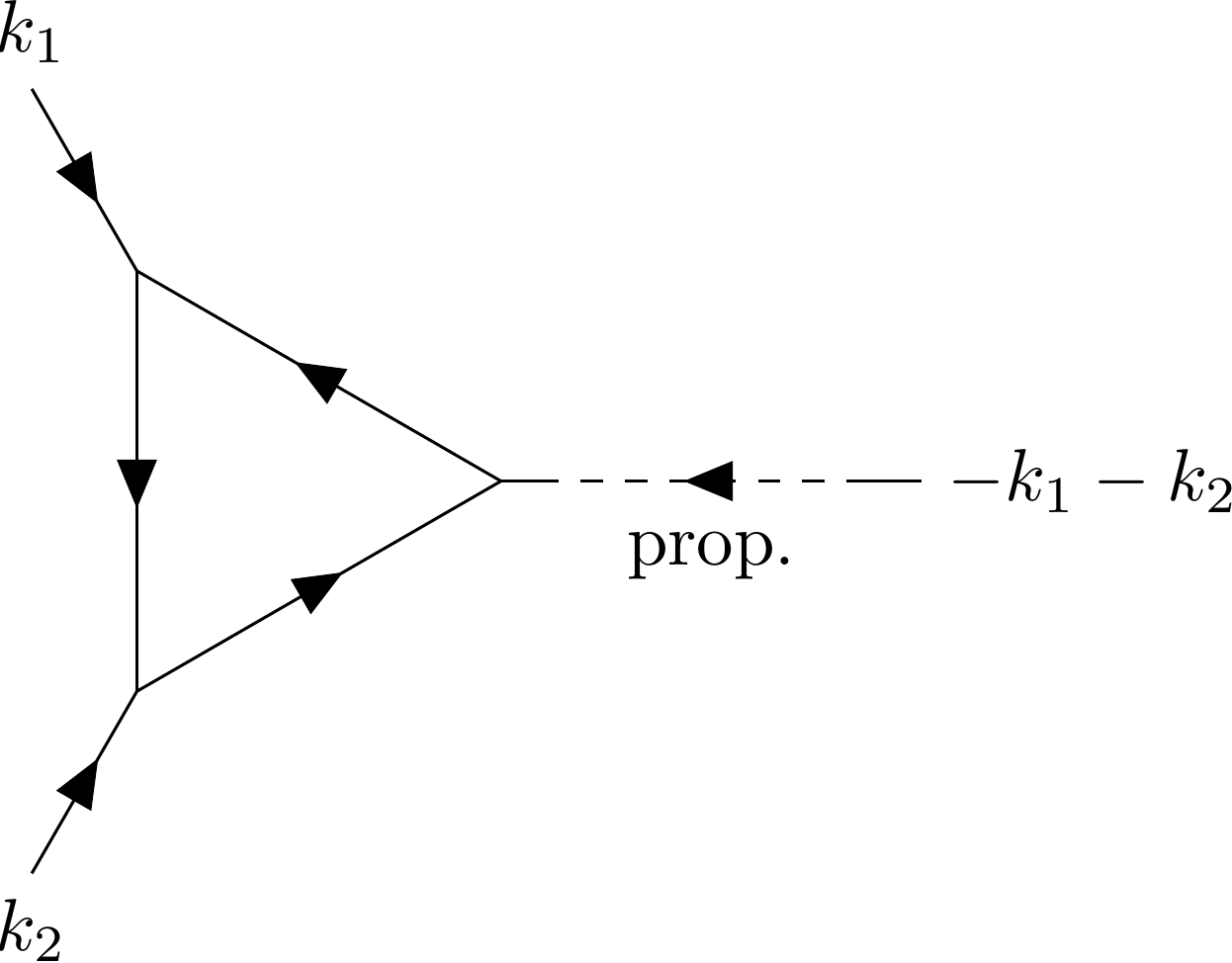}
	\caption{\emph{1-loop diagram leading to double poles in a graviton amplitude as the momenta $k_1,k_2$ of two positive helicity external states become holomorphically collinear.}} \label{fig:loopsplitting}
\end{figure}

Instead of obtaining a well-behaved deformed vertex algebra from quantum corrections, something else happens. When loop corrections are taken into account the splitting function at $1$-loop gets deformed by diagrams of the form depicted in figure \ref{fig:loopsplitting}. Through these, the graviton splitting function gets corrected by a new term which has a double pole $z_{ij}^{-2}$ \cite{Bittleston:2022jeq}\footnote{The analogous result in the Yang-Mills case was first discussed \cite{Costello:2022upu}.}.  The corresponding correction of the structure constants in \eqref{eq:wOPE} leads to a \emph{non-associative} OPE, both in the gravitational case \cite{Bittleston:2022jeq} as well as in the Yang-Mills case \cite{Costello:2022upu}. Within self-dual gravity on an undeformed $\mathbb{R}^4$ background, the diagram of figure \ref{fig:loopsplitting} will not contribute since there is no non-vanishing diagram it could feed into. This led to a result on 'perturbatively exact $w_{1+\infty}$ asymptotic symmetry of quantum self-dual gravity' \cite{Ball:2021tmb}, which is still consistent with the results on a deformed loop-level celestial chiral algebra \cite{Bittleston:2022jeq}. 

Interestingly, associativity can be restored if further fields are carefully chosen to cancel a certain anomaly on twistor space \cite{Costello:2021bah, Bittleston:2022nfr}. The twistorial descriptions of both self-dual Yang-Mills and self-dual gravity are both $6$-dimensional chiral gauge theories and suffer from a chiral anomaly \cite{Costello:2021bah, Bittleston:2022nfr}. This does not a priori mean that self-dual Yang-Mills and self-dual gravity are themselves not well-defined quantum theories on spacetime, but it means that their integrability, which is made manifest from a twistor description, gets broken at the quantum level. The associativity anomaly in the celestial chiral algebra is a manifestation of this phenomenon.

It is possible to cancel the chiral anomaly on twistor space by a topological string version of the Green-Schwarz mechanism \cite{Costello:2015xsa, Costello:2019jsy}. This requires carefully choosing a specific gauge group such as\footnote{Note the analogy to the famous $G=\text{SO}(32)$ in the original Green-Schwarz mechanism \cite{Green:1984sg}.} 
\be
G=\text{SO}(8)\,,
\ee
which was also considered in the previously discussed Burns holography setup \cite{Costello:2021bah, Costello:2023hmi}. An analogous cancellation is possible in the case of the chiral anomaly in self-dual gravity \cite{Bittleston:2022nfr}. Further ways to cancel these anomalies include supersymmetry \cite{Witten:2003nn}, and carefully chosen fermionic matter \cite{Costello:2023vyy, Bittleston:2024efo}. 

If the field content is carefully chosen to cancel the chiral anomaly, then the associativity anomaly of the corresponding celestial chiral algebra also vanishes \cite{Costello:2022upu, Bittleston:2022jeq}. This has led to consistent celestial chiral algebras at arbitrary loop orders \cite{Fernandez:2024qnu}. 
Once again, this very much suggests, that these infinite-dimensional algebras are tied to integrability. Only after we carefully choose the field content in order to obtain quantum integrable theories\footnote{Generically, self-dual theories have vanishing tree-level amplitudes but a non-vanishing $1$-loop all + amplitude \cite{Bern:1998xc, Bern:1998sv}. For carefully chosen quantum integrable theories, this $1$-loop amplitude gets cancelled and vanishes.} the celestial chiral algebras remain associative beyond the classical level. In this case, the symmetries have potential applications to the computation of loop-level amplitudes via the bootstrap method. This \emph{celestial chiral algebra bootstrap} has recently been exemplified by the computation of certain two-loop amplitudes in a QCD-like gauge theory \cite{Costello:2023vyy, Dixon:2024mzh, Dixon:2024tsb}.

There is an independent line of work on quantum corrections to soft theorems \cite{Sahoo:2018lxl,Sahoo:2020ryf, Saha:2019tub} and their symmetry interpretation via an infrared triangle of the form \ref{fig:Triangle} \cite{Choi:2024ygx, Choi:2024ajz}. Although the logarithmic divergences in these \emph{logarithmic soft theorems} seem unnatural from the chiral algebra perspective, it would be interesting to see if there is a relation between them and the consistent quantum deformations in quantum integrable theories arising from twistor space.

\subsection*{Non-trivial backgrounds}
\begin{figure}[!t]
    \centering
    \includegraphics[width=1\linewidth]{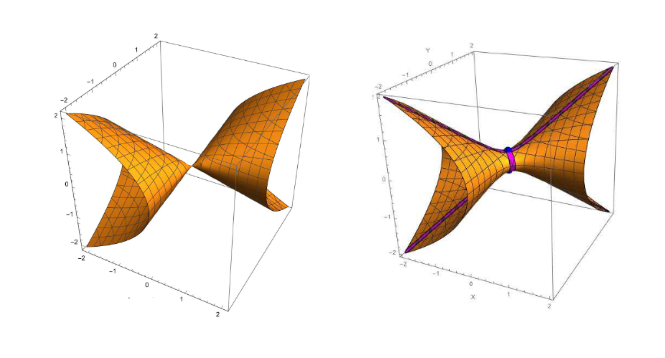}
    \caption{\emph{This figure displays the backreaction of \cite{Bittleston:2023bzp}. On the left, we can see a $2$-dimensional real slice of the flat orbifold $\mathbb{C}^2/\mathbb{Z}_2$. Inserting certain defects in analogy to D$1$-branes in Burns holography backreacts its twistor space to $\mathcal{PT}_{\text{EH}}$, the twistor space of Eguchi-Hanson space. A  $2$-dimensional real slice of Eguchi-Hanson space is depicted on the right. Notice the analogy of this figure to figure \ref{fig:Throatregion}}}
    \label{fig:EHspacePic}
\end{figure}

Non-trivial backgrounds have been studied in the celestial context for various reasons \cite{Bittleston:2023bzp, Bittleston:2024rqe, Taylor:2023ajd, Bu:2022iak, Melton:2022fsf, Costello:2022wso, Mago:2021wje,  CarrilloGonzalez:2024sto, McLoughlin:2024ldp, Adamo:2023zeh, Adamo:2024mqn, Costello:2022jpg, Costello:2023hmi, Garner:2023izn, Garner:2024tis, Gonzo:2022tjm, Melton:2023bjw, Melton:2023dee, Casali:2022fro, Stieberger:2022zyk, Fan:2022vbz, Adamo:2023fbj, Crawley:2023brz, Klisch:2025mxn, Klisch:2025pbu, Tropper:2024evi}. 
Eguchi-Hanson space \cite{Eguchi:1978xp} might not be the most natural background to consider from a physical perspective, however from a twistor perspective it is a very natural candidate. It is the simplest self-dual gravitational instanton in the class of \emph{asymptotically locally Euclidean (ALE) spaces}. All of these hyperkähler ALE spaces are known to have descriptions in terms of a twistor space \cite{Hitchin:1979rts}. The twistor space of Eguchi-Hanson
\be
\label{eq:EHTwistorspace}
\mathcal{PT}_{\text{EH}}=\{XY-Z^2=c^2(\lambda)\}\subset\text{Tot}\big(\mathcal{O}(2)\oplus \mathcal{O}(2)\oplus \mathcal{O}(2)\rightarrow \mathbb{CP}^1\big)\,,
\ee
which will be described in detail in chapter \ref{chapter4}, turns out to arise from a backreaction in \emph{holomorphic Poisson BF theory} \cite{Mason:2007ct, Bittleston:2022nfr}. Holomorphic Poisson BF theory is a twistor description of self-dual Einstein gravity with $\Lambda=0$. It is somewhat analogous to BCOV theory on twistor space describing Mabuchi gravity on spacetime \cite{Costello:2021bah}. Topologically, Eguchi-Hanson space is given by blowing up the singular point of the orbifold $\mathbb{C}^2/\mathbb{Z}_2$ and a $2$-dimensional real slice of it is depicted in figure \ref{fig:EHspacePic}.

Up to this $\mathbb{Z}_2$-quotient at infinity, it is asymptotically Euclidean and after slightly deforming the scattering states it makes sense to define scattering amplitudes on Eguchi-Hanson space. It hence also makes sense\footnote{after complexifying the external momenta} to consider the holomorphic collinear limit of such amplitudes in which again, diagrams of the form \ref{fig:splitting} dominate. In chapter \ref{chapter4}, we will explicitly compute the resulting deformed splitting function and celestial chiral algebra from a spacetime perspective. This is very much analogous to the original derivation of Strominger and collaborators \cite{Guevara:2021abz, Strominger:2021lvk}.

Then, we will also compute the deformation of the celestial chiral algebra induced by the deformed twistor space \eqref{eq:EHTwistorspace}. Although the algebra in the twistor basis looks a priori very different to the algebra in the scattering basis, the two algebras turn out to be isomorphic and an explicit isomorphism will be constructed, both in the gravitational case as well as the Yang-Mills case. A further non-commutative deformation of this Eguchi-Hanson background will be discussed.

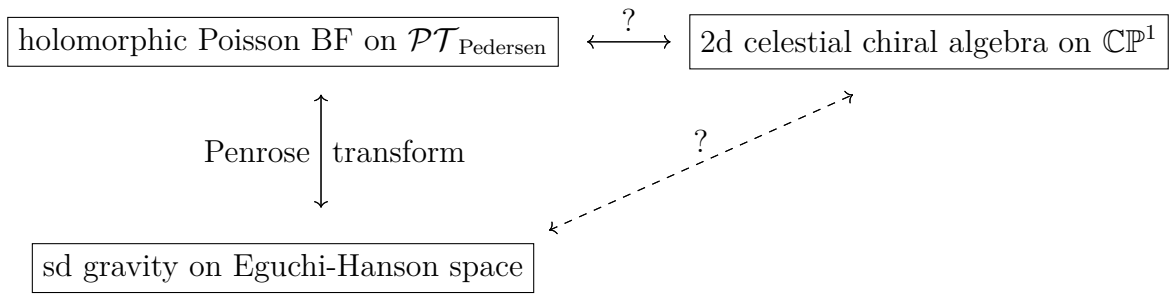
\begin{figure}[!t]
\begin{center}
\begin{tikzpicture}
\node[draw] at (-4,-1) {holomorphic Poisson BF on $\mathcal{PT}_{\text{Pedersen}}$};
\node[draw] at (4.5,-1) {$2$d celestial chiral algebra on $\mathbb{CP}^1$};
\draw [<->] (0,-1) -- (1.1,-1) node[midway,above]{?};
\draw [<->] (0,-1) -- (1.1,-1) node[midway,below]{};

\draw [<->] (-3.5,-3.2) -- (-3.5,-1.7) node[midway,left]{Penrose};
\draw [<->] (-3.5,-3.2) -- (-3.5,-1.7) node[midway,right]{transform};

\node[draw] at (-4,-4) {sd gravity on Eguchi-Hanson space};

\draw [<->, dashed] (-0.5,-3.5) -- (3.5,-1.7) node[midway,above]{?};
\draw [<->, dashed] (-0.5,-3.5) -- (3.5,-1.7) node[midway,right]{};

\end{tikzpicture}
\end{center}
	\caption{\emph{We hope that our work \cite{Bittleston:2023bzp} can be extended to a fully fledged holographic duality analogous to Burns holography. The bulk involves self-dual Einstein gravity with $\Lambda=0$ as opposed to self-dual conformal gravity in the Burns holography case.}} \label{fig:EHholography}
\end{figure}

Although this algebra arises merely from a tree-level calculation and there is a non-vanishing chiral anomaly on twistor space, our work provides a first step towards a version of Burns holography involving self-dual Einstein gravity in the bulk. Note, in particular, that Eguchi-Hanson space is Einstein, while Burns space is not. We envision that it might be possible to embed our tree-level dictionary \cite{Bittleston:2023bzp} into a fully fledged holographic duality along the lines of figure \ref{fig:EHholography}\footnote{Note the existence of a recent top-down duality involving self-dual QCD, among other backgrounds, on Eguchi-Hanson space in the bulk \cite{Costellotalk}.}.

\medskip

Trusting our relation between the celestial chiral algebras from tree splitting and the celestial chiral algebra on twistor space, we will derive further, more complicated algebras from a twistor perspective in chapter \ref{chapter5}. The degenerate Poisson bracket \eqref{eq:Infinity}, arising from the flat-space infinity has a generalization to $\Lambda\neq0$ which immediately leads to a $\Lambda$-deformed celestial chiral algebra \cite{Bittleston:2024rqe} that was previously derived without twistor methods in \cite{Taylor:2023ajd}. 

There also is a natural $\Lambda\neq 0$  generalization of the Eguchi-Hanson space backreaction. 
The backreaction on twistor space is computed using the Mason-Wolf action with $\Lambda\neq0$\footnote{For $\Lambda=0$ the Mason-Wolf action reduces to the aforementioned holomorphic Poisson BF theory.} and we find a backreacted complex manifold $\mathcal{PT}_{\text{Pedersen}}$ that has already been studied a long time ago by Pedersen \cite{pedersen1985geometry, Pedersen:1986vup}. The spacetime metric corresponding to the curved twistor space  $\mathcal{PT}_{\text{Pedersen}}$ is given by a self-dual Taub-NUT-AdS$_4$ metric that arises as a certain self-dual limit of Pleba\'{n}ski-Demia\'{n}ski black hole metrics \cite{Rodriguez:2021hks, Plebanski:1976gy}.

In Kleinian signature, this metric can be interpreted  as a so-called \emph{self-dual black hole}. In the limit $\Lambda=0$, it reduces to the well-known self-dual Taub-NUT metric. Although this self-dual Taub-NUT metric does not have a horizon in Euclidean signature, it does have a genuine horizon in Kleinian signature and it is possible to continue the metric past this horizon where the maximal continuation encounters a curvature singularity \cite{Crawley:2021auj}. This justifies the name \emph{self-dual black hole} \cite{Crawley:2023brz} for such self-dual Taub-NUT geometries. 

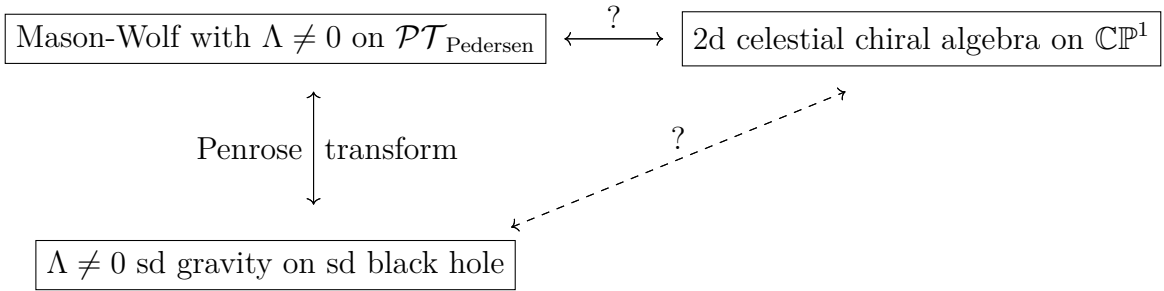
\begin{figure}[!t]
\begin{center}
\begin{tikzpicture}
\node[draw] at (-4,-1) {Mason-Wolf with $\Lambda\neq 0$ on $\mathcal{PT}_{\text{Pedersen}}$};
\node[draw] at (4.5,-1) {$2$d celestial chiral algebra on $\mathbb{CP}^1$};
\draw [<->] (-0.2,-1) -- (1.1,-1) node[midway,above]{?};
\draw [<->] (-0.2,-1) -- (1.1,-1) node[midway,below]{};

\draw [<->] (-3.5,-3.2) -- (-3.5,-1.7) node[midway,left]{Penrose};
\draw [<->] (-3.5,-3.2) -- (-3.5,-1.7) node[midway,right]{transform};

\node[draw] at (-4,-4) {$\Lambda\neq 0$  sd gravity on sd black hole};

\draw [<->, dashed] (-0.9,-3.5) -- (3.5,-1.7) node[midway,above]{?};
\draw [<->, dashed] (-0.9,-3.5) -- (3.5,-1.7) node[midway,right]{};

\end{tikzpicture}
\end{center}
	\caption{\emph{Our work \cite{Bogna:2024gnt} generalizes the previous work on Eguchi-Hanson space by considering self-dual Einstein gravity with a non-vanishing cosmological constant $\Lambda\neq 0$} in the bulk. Beyond leading to a new $2$-parameter deformation of $\mathcal{L}w_\wedge$, it also reveals a connection between Eguchi-Hanson space, Burns space and a class of \emph{self-dual black holes} \cite{Crawley:2023brz}.} \label{fig:QTNholography}
\end{figure}

Similarly to the Eguchi-Hanson case, we identify a deformed celestial chiral algebra which now depends on $2$-parameters and interpolates between the previously identified algebras. Again, the question arises, whether some picture of the form \ref{fig:QTNholography} can perhaps be engineered in string theory. Having multiple parameters is an extension of previous results that might give insights into new features of flat space holography. Very ambitiously, such new features could include a non-trivial thermodynamic phase structure related to the phase transitions previously encountered for the same self-dual Taub-NUT geometries \cite{Chamblin:1998pz, Hawking:1998ct, Martelli:2012sz}.

\section{Outline of this thesis}
\label{section1.6}
The first part of this thesis spanning chapters \ref{chapter2} and \ref{chapter3} mostly serves as a summary of known results.  Chapter \ref{chapter2} will review the original derivation of celestial chiral algebras from collinear singularities of graviton and gluon scattering amplitudes. Afterwards, a similar derivation of a deformed celestial chiral algebra is discussed. It is derived from a non-commutative deformation of self-dual gravity purely from a spacetime perspective, which is progress made by the author and his collaborators. 
Chapter \ref{chapter3} will then review basic twistor theory and how celestial chiral algebras arise from twistor space. In particular, twistor actions for self-dual gravity and self-dual Yang-Mills theory will be discussed in detail.

The second part of this thesis spanning chapter \ref{chapter4} and \ref{chapter5}, reports the main progress made by the author and his collaborators in attempting to identify holographic dualities analogous to Burns holography \cite{Costello:2022jpg, Costello:2023hmi} involving self-dual Einstein gravity in the bulk. This led to the identification of various backreacted geometries in self-dual Einstein gravity and the identifications of deformations of celestial chiral algebras from certain backreactions, non-commutative backgrounds and including a cosmological constant.

Chapter \ref{chapter4} discusses how an Eguchi-Hanson background is obtained when the twistor description of self-dual gravity is coupled to a $2$-dimensional defect wrapping a certain $\mathbb{CP}^1$, a \emph{twistor line}. This leads to a deformed twistor space from which a deformed celestial chiral algebra can be identified as the Poisson-ring of holomorphic functions. We prove explicitly that this deformed chiral algebra is isomorphic to a chiral algebra obtained from scattering amplitudes and their holomorphic collinear singularities on the curved background. This isomorphism is explicitly provided both in the case of graviton as well as gluon amplitudes. A non-commutative background can be included in order to obtain a more general $2$-parameter family of known celestial chiral algebras in the non-commutative theory.

Chapter \ref{chapter5} discusses how a non-vanishing cosmological constant can be included on twistor space and how from this perspective a deformed celestial chiral algebra arises. Then, another backreaction is performed in the presence of a cosmological constant and the deformed background is identified to be a self-dual limit of certain Pleba\'{n}ski-Demia\'{n}ski black holes. This $2$-parameter family of metrics interpolate between (a singular double cover of) Eguchi-Hanson space and AdS$_4$ and we derive a corresponding $2$-parameter deformation of $Lw_\wedge$ which interpolates between previously discussed deformed celestial chiral algebras.


\section{List of publications}
\label{section1.7}

The present thesis consists of an exposition based on established literature,
as well as new results coming from original published work by the author and his collaborators. The
original results are based on the following publications and preprints
\begin{itemize}
         \item \cite{Bu:2022iak} \bibentry{Bu:2022iak}
    \item \cite{Bittleston:2023bzp} \bibentry{Bittleston:2023bzp} 
    \item \cite{Bittleston:2024rqe} \bibentry{Bittleston:2024rqe}
    \item \cite{Bogna:2024gnt}  \bibentry{Bogna:2024gnt} 
\end{itemize}

%% file: Chapter3/chapter3.tex
\chapter{Non-commutative Deformations}
\label{chapter2}

 This chapter is loosely based on our work \cite{Bu:2022iak} and is organized as follows: At the beginning of section~\ref{sec:algebra} we review $w_{1+\infty}$ and related Lie-algebras in detail.  Our presentation partly follows the excellent review of~\cite{Pope:1991ig}, to which the reader is referred for a more comprehensive treatment.  In section \ref{section1.5} we discussed the importance of $\ham(\mathbb{C}^2)$ in celestial holography and we will see that there is a unique Lie-algebra deformation of $\mathfrak{ham}(\mathbb{C}^2)$, the Weyl algebra $\difff$.  In section~\ref{Action_section}, after briefly reviewing the Chalmers-Siegel form~\cite{Chalmers:1996rq} of the action for self-dual gravity on $\mathbb{R}^4$, we present its \emph{Moyal deformation}, which arises from switching on a non-commutative background on $\mathbb{R}^4$, and evaluate its $\overline{\rm MHV}$ 3-pt tree amplitude.
 In section~\ref{sec:OPEs}, we study the collinear splitting function in the Moyal deformed theory and obtain it in a closed form. We will then mirror the original spacetime derivation of the undeformed gravitational celestial chiral algebra \cite{Guevara:2021abz} using this deformed splitting function and obtain the Weyl-algebra. This gives a tree-level bulk interpretation of the unique Lie-algebra deformation of $\mathfrak{ham}(\mathbb{C}^2)$. 
 Within this chapter, the treatment of self-dual gravity will be elementary and no twistor theory is used. \\

\emph{Note added:} While \cite{Bu:2022iak} was being prepared, \cite{Monteiro:2022lwm} appeared on the arXiv. \cite{Monteiro:2022lwm} has some overlap with \cite{Bu:2022iak} and the present chapter.

\section{A unique deformation of $\mathfrak{ham}(\mathbb{C}^2)$}
\label{sec:UniqueDeformation}

In this section we will briefly review $w_\wedge,  w_{1+\infty}, \mathfrak{ham}(\mathbb{C}^2)$ and their relation to Poisson diffeomorphisms of the plane. We then review the deformed Lie-algebra $W_{1+\infty}$ and the more general $W(\mu)$-algebras which can be viewed as a family of deformations of $w_{\wedge}$. Since there are many different infinite-dimensional algebras at play in the celestial holography context and their nomenclature has been somewhat inconsistent through the literature, we make an effort to distinguish these algebras and point out their subtle differences\footnote{Unfortunately, this leads to the somewhat tedious name $\mathcal{L}\mathfrak{ham}(\mathbb{C}^2)$ for the gravitational algebra that was derived by Strominger \cite{Strominger:2021lvk}. In the celestial holography literature, it is often referred to as $w_{1+\infty}$ or some qualifiers thereof.}. 
We then discuss that $\mathfrak{ham}(\mathbb{C}^2)$ has a unique Lie-algebra deformation related to non-commutativity.
We refer the reader to ~\cite{Pope:1989sr,Pope:1991ig} for a comprehensive review of these topics and their relations to higher spin symmetries of 2d CFTs.

\subsection*{$ w_{1+\infty}, w_\wedge,$ and $\mathfrak{ham}(\mathbb{C}^2)$ }
\label{sec:algebra}


We briefly saw in section \ref{section1.5} how the Lie-algebra $\mathcal{L}\mathfrak{ham}(\mathbb{C}^2)$
\be
[w^p_{m,a}, w^q_{n,b}]=2(m(q-1)-n(p-1))w_{m+n,a+b}^{p+q-2}\,
\ee
arises from holomorphic collinear singularities of tree-level graviton amplitudes\footnote{The conventional factor of $2$ in the structure constants can be absorbed into the generators.}. Let us, for the sake of this chapter, forget the loop-algebra labels $a,b$ to simplify the discussion\footnote{Considering the full Lie-algebra $\mathcal{L}\mathfrak{ham}(\mathbb{C}^2)$ rather than $\mathfrak{ham}(\mathbb{C}^2)$ will be of relevance in chapter \ref{chapter5}}.  The resulting Lie-algebra $\mathfrak{ham}(\mathbb{C}^2)$ with the same structure constants
\be
\label{eq:wstructureconstants}
[ w^p_{m}, w^q_{n}]=2(m(q-1)-n(p-1))w_{m+n}^{p+q-2}\,,
\ee
is much better understood than $\mathcal{L}\mathfrak{ham}(\mathbb{C}^2)$ and in particular it is known to admit a unique Lie-algebra deformation \cite{Etingof:2023new}.
Recall that in the context of celestial holography, after disregarding the index $a$, generators $w^{p}_m$ exist for \cite{Strominger:2021lvk}
\be
 p\in\{1,\tfrac{3}{2}, 2, \tfrac{5}{2}, \dots \}, \quad m\in \{1-p,2-p,\dots, p-2,p-1\}\,,
\ee
which define $\mathfrak{ham}(\mathbb{C}^2)$.
If we restrict ourselves to indices
\be
 p\in\{1, 2, 3, \dots \}, \quad m\in \{1-p,2-p,\dots, p-2,p-1\}\,,
\ee
then the algebra is known as $w_\wedge$, the wedge subalgebra of $w_{1+\infty}$. $w_{1+\infty}$ itself is obtained from disgarding the \emph{wedge condition} $|m|\leq 1-p$ and including further modes in $m$
\be
 p\in\{1, 2, 3, \dots \}, \quad m\in \mathbb{Z}\,.
\ee
It is possible to obtain a further extension from also including negative spins 
\be
 p\in\mathbb{Z}, \quad m\in \mathbb{Z}\,,
\ee
which we will give the somewhat silly name $w_{1+\infty}^{- \text{spins}}$, since it won't be particularly important in the given context.

\begin{figure}[t!]
	\begin{center}
\begin{tikzpicture}[scale=1.85]

\filldraw[draw=red, fill=red!20] (0,1) -- (2.7,3.7) -- (-2.7,3.7)  -- cycle;
\draw[->] (-3.5,0) -- (3.5,0);
\node at (3.5,-0.2){$m$};
\draw[->] (0,-1.5) -- (0,3.7);
\draw[dashed] (-3.5,0.85) -- (3.5,0.85);

\node at (-0.2,3.85){$p$};
\node at (0,1){$\bullet$};
\node at (-0.2,1){$1$};
\node at (-0.2,1.5){$3/2$};
\node at (0,2){$\bullet$};
\node at (-0.2,2){$2$};
\node at (-0.2,2.5){$5/2$};
\node at (0,3){$\bullet$};
\node at (-0.2,3){$3$};
\node at (-0.2,0.5){$1/2$};
\node at (-0.2,3.5){$7/2$};

\node at (-0.2,-0.2){$0$};
\node at (1,-0.2){$1$};
\node at (2,-0.2){$2$};
\node at (3,-0.2){$3$};
\node at (-1,-0.2){$-1$};
\node at (-2,-0.2){$-2$};
\node at (-3,-0.2){$-3$};

\node at (-3,0){$\bullet$};
\node at (-2,0){$\bullet$};
\node at (-1,0){$\bullet$};
\node at (0,0){$\bullet$};
\node at (3,0){$\bullet$};
\node at (2,0){$\bullet$};
\node at (1,0){$\bullet$};

\node at (-3,1){$\bullet$};
\node at (-2,1){$\bullet$};
\node at (-1,1){$\bullet$};
\node at (0,1){$\bullet$};
\node at (3,1){$\bullet$};
\node at (2,1){$\bullet$};
\node at (1,1){$\bullet$};

\node at (-3,2){$\bullet$};
\node at (-2,2){$\bullet$};
\node at (-1,2){$\bullet$};
\node at (0,2){$\bullet$};
\node at (3,2){$\bullet$};
\node at (2,2){$\bullet$};
\node at (1,2){$\bullet$};

\node at (-3,2){$\bullet$};
\node at (-2,2){$\bullet$};
\node at (-1,2){$\bullet$};
\node at (0,2){$\bullet$};
\node at (3,2){$\bullet$};
\node at (2,2){$\bullet$};
\node at (1,2){$\bullet$};

\node at (-3,3){$\bullet$};
\node at (-2,3){$\bullet$};
\node at (-1,3){$\bullet$};
\node at (0,3){$\bullet$};
\node at (3,3){$\bullet$};
\node at (2,3){$\bullet$};
\node at (1,3){$\bullet$};

\node at (-3,-1){$\bullet$};
\node at (-2,-1){$\bullet$};
\node at (-1,-1){$\bullet$};
\node at (0,-1){$\bullet$};
\node at (3,-1){$\bullet$};
\node at (2,-1){$\bullet$};
\node at (1,-1){$\bullet$};

\node at (0.5,1.5){$\blue{\bullet}$};
\node at (-0.5,1.5){$\blue{\bullet}$};
\node at (0.5,2.5){$\blue{\bullet}$};
\node at (-0.5,2.5){$\blue{\bullet}$};
\node at (1.5,2.5){$\blue{\bullet}$};
\node at (-1.5,2.5){$\blue{\bullet}$};

\node at (2.5,1.5){$\blue{\bullet}$};
\node at (1.5,1.5){$\blue{\bullet}$};
\node at (0.5,1.5){$\blue{\bullet}$};
\node at (-0.5,1.5){$\blue{\bullet}$};
\node at (-1.5,1.5){$\blue{\bullet}$};
\node at (-2.5,1.5){$\blue{\bullet}$};

\node at (2.5,2.5){$\blue{\bullet}$};
\node at (1.5,2.5){$\blue{\bullet}$};
\node at (0.5,2.5){$\blue{\bullet}$};
\node at (-0.5,2.5){$\blue{\bullet}$};
\node at (-1.5,2.5){$\blue{\bullet}$};
\node at (-2.5,2.5){$\blue{\bullet}$};

\node at (2.5,3.5){$\blue{\bullet}$};
\node at (1.5,3.5){$\blue{\bullet}$};
\node at (0.5,3.5){$\blue{\bullet}$};
\node at (-0.5,3.5){$\blue{\bullet}$};
\node at (-1.5,3.5){$\blue{\bullet}$};
\node at (-2.5,3.5){$\blue{\bullet}$};

\end{tikzpicture}
\end{center}
	\caption{\emph{The generators of $w_{1+\infty}^{-\text{ spins}}$, $w_{1+\infty}$, and $w_\wedge$ correspond to all the black dots, the black dots above the dashed line, and the black dots in the red wedge region respectively. $\mathfrak{ham}(\mathbb{C}^2)$ corresponds to all dots in the red region, where the blue dots represent the half-integer spin generators that are not present in $w_\wedge$. In the full gravitational celestial chiral algebra $\lham$, infinitely many copies of all the dots in the wedge region are present.}} \label{fig:Wedge}
\end{figure}
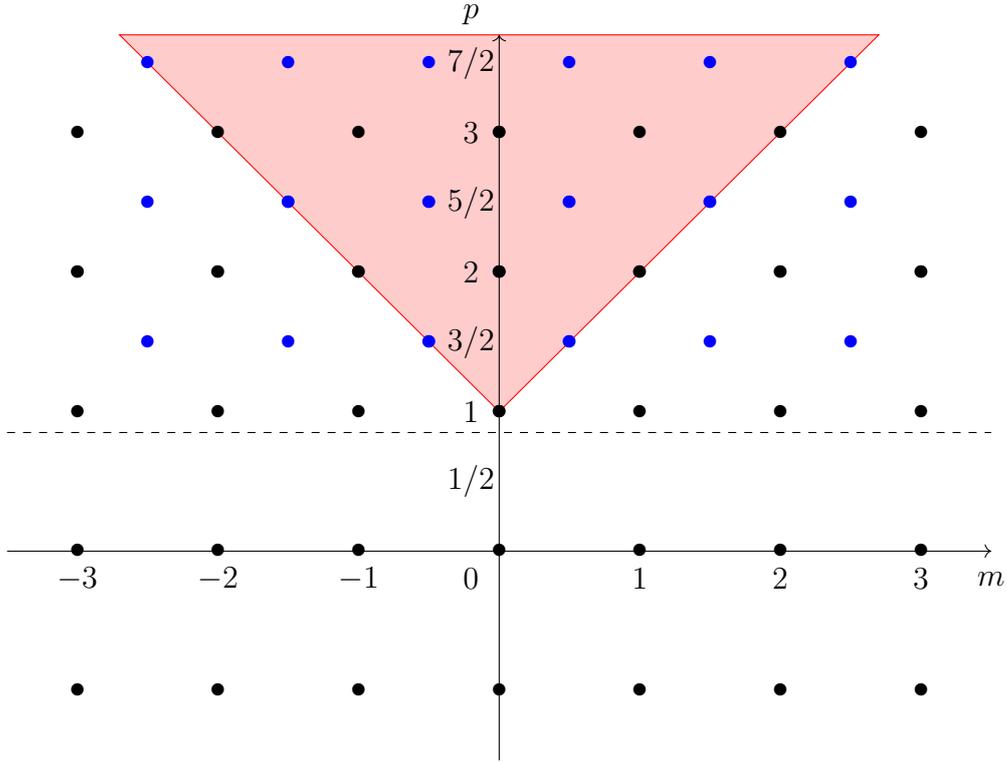

So, the Lie-algebras $ w_{1+\infty}, w_\wedge,$ $\mathfrak{ham}(\mathbb{C}^2)$, $w_{1+\infty}^{- \text{spins}}$,  and their loop algebras all have the same structure constants but the index-range on their generators differ. We have displayed this in figure \ref{fig:Wedge}.
The element $w^1_0$ is always central as can be seen immediately from \eqref{eq:wstructureconstants}.

These algebras arise in many different contexts. Of particular relevance to this thesis is the fact~\cite{Bakas:1989mz,Bakas:1989xu,Hoppe:1988gk} that $\mathfrak{ham}(\mathbb{C}^2)$ can be represented as the space of diffeomorphisms on the plane, preserving the standard Poisson bracket. Let $(\mu^{\dot{0}},\mu^{\dot{1}})\in \mathbb{C}^2$ be coordinates on this plane equipped with the standard holomorphic Poisson bracket
\begin{equation}
\label{Poisson}
\left\{f,g\right\} = \frac{\partial f}{\partial \mu^{\dot{0}}}\frac{\partial g}{\partial \mu^{\dot{1}}} - \frac{\partial f}{\partial \mu^{\dot{1}}}\frac{\partial g}{\partial \mu^{\dot{0}}}\,,
\end{equation}
for any pair of (smooth) functions $f,g$. Diffeomorphisms that preserve this Poisson bracket are generated by Hamiltonian vector fields so that $V = \{h, \ \}$ for some function $h$ on $\mathbb{C}^2$. These Hamiltonians are hence generated by polynomials. In particular, one recovers~\eqref{eq:wstructureconstants} by considering the following basis of monomials
\begin{equation}
\label{w_generator}
    w^p_m = \big(\mu^{\dot{0}}\big)^{p+m-1}\,\big(\mu^{\dot{1}}\big)^{p-m-1}\,,
\end{equation}
and using the Poisson bracket~\eqref{Poisson} as Lie bracket. 

\subsubsection*{The twistor basis}

From the Hamiltonian perspective \eqref{Poisson}, a much more natural basis is given by
\be
\label{eq:TwistorGenerators}
w[a,b]=\big(\mu^{\dot{0}}\big)^{a}\,\big(\mu^{\dot{1}}\big)^{b}=w^{(a+b)/2+1}_{(a-b)/2}\,,
\ee
where the labels $a,b$ take ranges
\begin{itemize}
    \item $a,b \in \mathbb{N}_0$ for $\mathfrak{ham}(\mathbb{C}^2)$\,,
    \item $a,b \in \mathbb{N}_0$ with $a+b\equiv 0\,\,(2)$ for $w_\wedge$\,,
    \item $a,b \in \mathbb{Z}$ with $a+b\geq 0$ and $a+b\equiv 0\,\,(2)$ for $w_{1+\infty}$\,.
       \item $a,b \in \mathbb{Z}$ and $a+b\equiv 0\,\,(2)$ for  $w^{- \text{ spins}}_{1+\infty}$.
\end{itemize}
This basis makes it manifest that $w_\wedge$ is generated by even-degree polynomials, which are invariant under the $\mathbb{Z}_2$-action
\be
\binom{\mu^{\dot{0}}}{\mu^{\dot{1}}}\mapsto \binom{-\mu^{\dot{0}}}{-\mu^{\dot{1}}}\,,
\ee
which means
\be
w_\wedge\cong\mathfrak{ham}(\mathbb{C}^2)^{\mathbb{Z}_2}=\mathfrak{ham}(\mathbb{C}^2/\mathbb{Z}_2)\,.
\ee
This insight is one motivation to consider the orbifold $\mathbb{C}^2/\mathbb{Z}_2$ in the celestial holography context as will be done below in chapter \ref{chapter4}. For further reasons that will become apparent in chapter \ref{chapter4}, we will refer to the generators \eqref{eq:TwistorGenerators} as the \emph{twistor basis}. Both generators $w^p_m$ and $w[a,b]$ are used in different parts of the celestial holography literature \cite{Costello:2022wso, Strominger:2021lvk}. Pictorially, the twistor basis arises when we rotate the axes in figure \ref{fig:Wedge} by $45$ degrees to the right before labelling the generators.

\subsubsection*{The S-algebra}

We discussed the Lie-algebra $\mathcal{L}\mathfrak{ham}(\mathbb{C}^2)$ which arises from gravitational amplitudes in quite a lot of detail. Although, in this thesis, we will put more emphasis on the gravitational case, in complete analogy such an algebra can also be derived from gluon amplitudes \cite{Guevara:2021abz} in a gauge theory with semisimple gauge algebra $\fg$ that admits an invariant bilinear form $\tr$. This celestial chiral algebra is often referred to as the \emph{S-algebra} and it is given by $\mathcal{L}\mathfrak{g}[\mathbb{C}^2]$, the loop algebra of the Lie algebra of polynomial maps from the complex 2-plane to $\mathfrak{g}$. Let $t_a$ with $a=1,\dots,\dim({\mathfrak{g}})$ be a basis of the Lie-algebra $\mathfrak{g}$, whose Lie-brackets are determined by the structure constants
\be
[t_a,t_b]=f_{ab}^c t_c\,.
\ee
In the original basis \cite{Strominger:2021lvk}, $\mathfrak{g}[\mathbb{C}^2]$ then consists of generators
\be
S^{p}_{m, a}=  \big(\mu^{\dot{0}}\big)^{p+m-1}\,\big(\mu^{\dot{1}}\big)^{p-m-1}t_a \,,
\ee
with index ranges
\be
p\in\{1,\tfrac{3}{2}, 2, \tfrac{5}{2}, \dots \}, \quad m\in \{1-p,2-p,\dots, p-2,p-1\}\,,
\ee
and a Lie-bracket simply given by the Lie-bracket of $\mathfrak{g}$. This leads to the commutation relations\footnote{In the original basis, there is an additional factor of $-\im$ in the structure constants which can be absorbed into the generators.} 
\be
[S^{p}_{m,a}, S^{q}_{n,b}]= f_{ab}^c\,S^{p+q-1}_{m+n,c}\,.
\ee
Once again, from the perspective of this thesis, a slightly more natural basis is the twistor basis of \cite{Costello:2022wso} given by 
\be
I_a[m,n]=\big(\mu^{\dot{0}}\big)^{m}\,\big(\mu^{\dot{1}}\big)^{n}t_a=S^{(m+n)/2+1}_{(m-n)/2,a}
\ee
with index ranges 
\be
m,n \in \mathbb{N}_0
\ee
and commutation relations
\be
[I_a[p,q],I_b[r,s]]=f_{ab}^c\,I_c[p+r,q+s]\,.
\ee

\paragraph{The Virasoro subalgebra} Returning to the gravitational case, the generators $\ell_m = -\tfrac{1}{2}w^2_m$ of~\eqref{w_generator} with $p=2$ fixed, form a subalgebra of $w_{1+\infty}$ that we recognise as the Witt algebra
\begin{equation}
\label{Wittalgebra}
\left[\ell_m,\ell_n\right] = (m-n)\,\ell_{m+n}\,,
\end{equation}
It is well known that the Witt algebra is the classical limit of the Virasoro algebra
\begin{equation}
\label{Virasoro}
\left[L_m,L_n\right] = (m-n)\,L_{m+n} + cm(m^2-1)\,\delta_{m+n,0}\,.
\end{equation}
that characterises a 2d CFT of central charge $c$. The Virasoro generators $L_m$ are the Laurent modes of the (holomorphic) stress tensor $T(z)$ of the CFT, with the algebra~\eqref{Virasoro} appearing from the $TT$ OPE. The appearance of the central charge $c$ is a quantum effect in the CFT. From this perspective, it is natural to expect that $w_{1+\infty}$ also admits a quantization. 

\subsection*{$W_{1+\infty}, W_\infty, \difff$ and the $W(\mu)$-family}

A quantization of $w_{1+\infty}$ was discovered~\cite{Fateev:1987zh,FATEEV1987644,Zamolodchikov:1985wn} by studying 2d CFTs with higher spin symmetry, and is known as $W_{1+\infty}$. Their generators $W^p_m$ are the Laurent modes of higher spin conserved currents. There is a large literature on $W_{1+\infty}$ (and other $W_N$-algebras) and their CFT realisations, and we refer the reader to {\it e.g.}~\cite{Pope:1991ig} for a review. $W_{1+\infty}$ extends to a Lie-algebra deformation of  $w^{-\text{ spins}}_{1+\infty}$ and it can be viewed as a member of a 1-parameter family of infinite dimensional Lie algebra deformations $W^{-\text{ spins}}(\mu)$ of $w^{- \text{ spins}}_{1+\infty}$ for $\mu \in \mathbb{R}$. For a generic $\mu$, $W^{-\text{ spins}}(\mu)$ will not restrict to a deformation of $w_{1+\infty}$, however it will always restrict to a deformation of $w_{\wedge}$ that we will refer to as $W(\mu)$.

So the $W(\mu)$ algebras are best thought of as wedge-subalgebras of larger algebras $W^{-\text{ spins}}(\mu)$ with generators $W^p_m$ for $p,m\in \mathbb{Z}$. The commutation relations of these generators are given by\footnote{Note that we have shifted the upper index $p$ and $q$ of the generators by 2 compared to the labels in~\cite{Pope:1991ig}.} 
\begin{equation}\label{capitalW_algebra}
    \left[W_m^p,W_n^q\right]=\sum_{l\geq 0}{\qq}^{2l} f^{pq}_{2l}(m,n;\sigma)\, W_{m+n}^{p+q-2l-2} + c_p(m)\,\qq^{2(p-2)}\,\delta^{p,q}\,\delta_{m+n,0}\,.
\end{equation}

Here, $\qq$ and $\sigma$ are parameters, with $\sigma$ related to $\mu$ by $\mu=\sigma(\sigma+1)$. The functions $f_{2l}^{pq}(m,n;\sigma) = -f_{2l}^{qp}(n,m;\sigma)$ are structure constants (depending on the parameter $\sigma$) and $c_p(m)$ are central charges. By demanding that the bracket in~\eqref{capitalW_algebra} obeys a Jacobi identity, Pope {\it et al.}~\cite{Pope:1989sr} found a solution for which the structure constants take the form
\begin{subequations}
\begin{equation}
\label{W_coefficients}
    f^{pq}_{2l}(m,n;\sigma)=\frac{1}{2(2l+1)!}\,\phi^{pq}_{2l}(\sigma)\,N^{pq}_{2l}(m,n)\,,
\end{equation}
where
\begin{equation}
\label{hypergeometric}
    \phi_{2l}^{pq}(\sigma)= \pFq[4]{4}{3}{-\frac{1}{2}-2\sigma,\frac{3}{2}+2\sigma,-l-\frac{1}{2},-l}{\frac{3}{2}-p,\frac{3}{2}-q,p+q-\frac{3}{2}-2l}{\, 1\,}
\end{equation}    
in terms of the generalized hypergeometric function ${}_4F_3$, and where
\begin{equation}\label{Symplecton_coefficient}
  N^{pq}_{2l}(m,n)= \sum_{i= 0}^{2l+1}(-1)^i\binom{2l\!+\!1}{i}[p\!-\!1\!+\!m]_{2l+1-i}\,[p\!-\!1\!-\!m]_{i}\,[q\!-\!1\!-\!n]_{2l+1-i}\,[q\!-1\!+n]_{i}
 \end{equation}
\end{subequations}
in terms of the descending Pochhammer symbol, defined by 
\be
[a]_b= \frac{\Gamma(a+1)}{\Gamma(a-b+1)}\,.
\ee
 Similarly, the central charges are constrained to be
\begin{equation}
 c_p(m) = c\,\frac{2^{2p-7}\,p!\,(p-2)!}{(2p-3)!!\,(2p-1)!!}\, \prod_{k=1-p}^{p-1}(m-k)\,.
\label{W_centre}
\end{equation}
In particular, all central charges are fixed in terms of the Virasoro central charge $c$. The algebra  \eqref{capitalW_algebra}, i.e. $W^{-\text{ spins}}(\mu)$  does generally not admit a consistent truncation to positive spins $p\geq 0$. However, it admits a consistent truncation to $|m|\leq p-1$ because the Pochhammer symbols in $N^{pq}_{2l}(m,n)$ ensure that the structure constants vanish if $|m+n|> p+q-2l-3$. 
This leads to the wedge subalgebra $W(\mu) \subset W^{-\text{ spins}}(\mu) $ which is a $1$-parameter  deformation of $w_\wedge$.

Importantly, $c_p(m)$ vanishes inside the wedge, where the condition
\be
m\in\{1-p,2-p,\dots,p-2,p-1\}
\ee
is imposed, so that
\be
 \prod_{k=1-p}^{p-1}(m-k)=0\,.
\ee
The latter means that although $c_p(m)$ is a central extension of $W^{-\text{ spins}}(\mu)$, it does not lead to a non-trivial central extension of $W(\mu)$. Hence, this $c_p(m)$ is irrelevant to our celestial holography context where we will restrict all generators to lie in the wedge $|m|\leq p-1$.  This was briefly explained above in section \ref{section1.5}.
\medskip 

Let us make some further remarks. Firstly, the parameter $\qq$ controls the deformation away from $w_\wedge$, in the sense that~\eqref{capitalW_algebra} reduces to~\eqref{eq:wstructureconstants} when $\qq\to0$. However, if $\qq\neq0$ it can be removed from~\eqref{capitalW_algebra} by rescaling $W^p_m \to \qq^{p-2}\,W^p_m$, 
so that the actual value of $\qq$ has no meaning. This justifies denoting the algebra by $W(\mu)$. In chapter \ref{chapter4}, we will occasionally denote $W(\mu)$ by $W(\mu;\mathfrak{q})$ to make its dependence on $\mathfrak{q}$ manifest and consider certain scaling limits.  Secondly, because the hypergeometric function in~\eqref{hypergeometric} is invariant under $\sigma\rightarrow -\sigma-1$, the algebras are more properly labelled by $\mu=\sigma(\sigma+1)$. In particular, the wedge-subalgebra of $W_{1+\infty}$ corresponds to setting $\mu=-\frac{1}{4}$ (and so $\sigma=-\frac{1}{2}$),  while $W(0)$ is related to the wedge subalgebra of $W_\infty$ \cite{Pope:1989sr}. 
Thirdly, we would like to point out that when negative spins are included, all the different algebras $W^{-\text{ spins}}(\mu)$ are in fact isomorphic to each other and an explicit isomorphism from an arbitrary $W^{-\text{ spins}}(\mu)$ to $W^{-\text{ spins}}(-3/16)$ is provided in \cite{Fairlie:1990wv}. The isomorphism is induced by a change of variables
\be
W'^p_m=\sum_{l\geq 0}\chi(l,p,m;\mu)W^{p-l}_m
\ee
for some complicated function $\chi$ \cite{Fairlie:1990wv}. This isomorphism does not generically restrict to an isomorphism of the wedge subalgebras $W(\mu)$ and $W(-3/16)$.

\medskip

Unlike $w_{1+\infty}, w_\wedge,$ and $\mathfrak{ham}(\mathbb{C}^2)$ with their relation to Poisson diffeomorphisms of the plane, the geometric interpretation of the generic $W(\mu)$ algebras is not straightforward in terms of the $\mathbb{C}^2$-plane\footnote{in fact, it is related to Eguchi-Hanson space as will be discussed below in chapter \ref{chapter4}.}. However, 
there is a particular member of the $W(\mu)$ family, occurring  when $\mu=-3/16$, for which such an interpretation is known. Fixing $\mu=-\frac{3}{16}$ implies either $2\sigma+\frac{1}{2}=0$ or $2\sigma+\frac{3}{2}=0$, so that one of the arguments in the top line of the hypergeometric function in~\eqref{hypergeometric} vanishes. In either of these cases, the hypergeometric function \eqref{hypergeometric} reduces to 1, so that the structure constants in this algebra simplify and the relations~\eqref{capitalW_algebra} become
\begin{equation}
\label{W_algebra}
    \left[W_m^p,W_n^q\right]=\sum_{l\geq 0}\frac{{\qq}^{2l}}{2(2l+1)!} \,N^{pq}_{2l}(m,n)\, W_{m+n}^{p+q-2l-2}
\end{equation}
within the wedge where the central charge vanishes.

The algebra~\eqref{W_algebra} can now be realised geometrically by equipping the $(\mu^{\dot{0}}, \mu^{\dot{1}})$-plane with a Moyal bracket, deforming the earlier Poisson bracket~\cite{Pope:1989sr}. That is, we define the Moyal bracket of a pair of functions $f,g$ by~\cite{moyal_1949}
\begin{subequations} 
\begin{equation}
\label{Moyal_bracket}
    \{f,g\}_{\qq} = \qq^{-1}(f\star g - g\star f)\,,
\end{equation}
where the Moyal star product is given by
\begin{equation}
\label{Moyal_star}
    f\star g = f\, \exp\left[ \qq \left(\overleftarrow{\partial_{\dot{0}}}\,\overrightarrow{\partial_{\dot{1}}} - \overleftarrow{\partial_{\dot{1}}}\,\overrightarrow{\partial_{\dot{0}}}\right)\right]\,g\,.
\end{equation}
\end{subequations}
The Moyal bracket is a deformation of the Poisson bracket, in the sense that 
\begin{equation}
    \{f,g\}_{\qq} = \{f,g\} + \mathcal{O}(\qq)\ .
\end{equation} 
In fact, it is the unique deformation constructed purely from the Poisson bracket, such that the deformed bracket still obeys a Jacobi identity~\cite{cmp/1103922592, Fletcher:1990ib}. Just like $\mathfrak{ham}(\mathbb{C}^2)$ and $w_\wedge$, the algebra~\eqref{W_algebra} can be realised by acting with the Moyal bracket~\eqref{Moyal_bracket} on the same ($\mathbb{Z}_2$-invariant) generators $W^p_m = \big(\mu^{\dot{0}}\big)^{p+m-1}\,\big(\mu^{\dot{1}}\big)^{p-m-1}$. Deforming the Poisson bracket to the Moyal bracket thus corresponds to deforming $w_\wedge$ to the $W(\mu)$ algebra at $\mu=-\frac{3}{16}$. For this reason, $W(-3/16)$ plays a special role and it is sometimes called the {\it symplecton}  \cite{Pope:1989sr}.

\medskip

The structure constants \eqref{W_coefficients} of the $W(\mu)$-algebras are importantly only defined for integer spin generators, i.e. $w^p_m$ with $p\in\mathbb{Z}$ displayed as the black dots in figure \ref{fig:Wedge}. The hypergeometric function in equation \eqref{hypergeometric} diverges for such half-integer values of $p$ for generic $\mu$. However, for the algebra $\mathfrak{ham}(\mathbb{C}^2)$ that arises in the celestial context $p$ importantly takes half-integer values\footnote{E.g. $p=3/2$ in the case of superrotations \cite{Guevara:2021abz}.} and the blue dots of figure \ref{fig:Wedge} have to be included.  Remarkably, the only case in which it is possible to augment $W(\mu)$ to positive half-integer values of $p$ is $W(-3/16)$\footnote{We thank Roland Bittleston for clarifying this point to us.}. 
This half-integer augmentation of the symplecton $W(-3/16)$ is the so-called \emph{Weyl algebra} $\difff$. $\difff$ is the quotient of the free algebra on two generators $\mu^{\dot{0}}, \mu^{\dot{1}}$ over $\bbC\llbracket\fq\rrbracket$ by the ideal
\be
\mathrm{span}\{ \mu^{\dot{0}}\mu^{\dot{1}}-\mu^{\dot{1}}\mu^{\dot{0}} = \fq\}\,.
\ee
$\difff$ is the only such half-integer augmentation since, for generic $\mu$, the hypergeometric function in~\eqref{hypergeometric} diverges if any of the arguments on its bottom line is a negative integer. The exception is for $\sigma(\sigma+1)=-\frac{3}{16}$, where the hypergeometric function reduces to $1$. Thus, since including all the conformally soft gravitons in the celestial holography context requires the augmentation to half-integer $p,m$, the only possible extension of $\mathfrak{ham}(\mathbb{C}^2)$ as a Lie algebra is $\difff$.

If we were to consider $w_\wedge=\mathfrak{ham}(\mathbb{C}^2/\mathbb{Z}_2)$, by working on the orbifold $\mathbb{C}^2/\mathbb{Z}_2$ then the blue dots in figure \ref{fig:Wedge} are not present anymore and all of the $W(\mu)$-algebras are consistent deformations. This will be the content of chapter \ref{chapter4}. Note, that there are even further consistent Lie-algebra deformations if we consider the full loop algebra $\mathcal{L}\mathfrak{ham}(\mathbb{C}^2)$\footnote{We thank Andrew Strominger for suggesting this out to us.}. Such deformations are not expected to arise from a Ricci-flat self-dual bulk spacetime\footnote{Such Ricci-flat self-dual spacetimes have a twistor space that fibres holomorphically over $\mathbb{CP}^1$ which will always lead to the loop algebra of some smaller algebra.} but we will see in chapter \ref{chapter5}, that the inclusion of a cosmological constant can lead to such deformations. 

\subsubsection*{$W(\mu)$ in the twistor basis}

Let us briefly express $W(\mu)$ in the twistor basis \eqref{eq:TwistorGenerators}. Once again, we introduce the generators
\be
W[a,b]=W^{(a+b)/2+1}_{(a-b)/2}\,,
\ee
where the wedge condition $a,b \in \mathbb{N}$ with $a+b\equiv 0\,\,(2)$ is imposed. Then, the algebra $W(\mu)$ reads
\bea \label{eq:W(mu)-commutator}
&[ W[p,q], W[r,s]] \\
&= \sum_{\ell\geq0}\fq^{2\ell}R_{2\ell+1}(p,q,r,s)\Psi_{2\ell+1}\bigg(\frac{p\!+\!q}{2},\frac{r\!+\!s}{2};\sigma\bigg) W[p\!+\!r-\!2\ell\!-\!1,q\!+\!s\!-\!2\ell\!-\!1]\,,
\eea
where
\be R_\ell(p,q,r,s) = \frac{1}{\ell!}\sum_{k=0}^\ell(-)^k\binom{\ell}{k}[p]_{\ell-k}[q]_k[r]_k[s]_{\ell-k}\,, \ee
$\Psi$ is the hypergeometric function
\be 
\label{eq:Psi-ell} 
\Psi_\ell(m,n;\sigma) = 
\pFq{4}{3}{-1/2\!-\!2\sigma,3/2\!+\!2\sigma,-\!\ell/2,(1\!-\!\ell)/2}{1/2\!-\!m,1/2\!-\!n,m\!+\!n\!+\!3/2\!-\!\ell}{1}
\ee
and $\sigma$ is the real parameter that gives $\mu=\sigma(\sigma+1)$.

\section{Self-dual gravity and its Moyal deformation}
\label{Action_section}

In this section, we briefly review actions for self-dual gravity in spacetime. We show how the conformally soft modes of the (positive helicity) graviton correspond to generators $w^p_m(z)$ of $\mathcal{L}\mathfrak{ham}(\mathbb{C}^2)$, with $2p\in\mathbb{Z}_{\geq 2}$. We then consider the \emph{Moyal deformation} of self-dual gravity by turning on a non-commutative $\mathbb{R}^4$-background and compute its 3-pt tree-level $\overline{\rm MHV}$ amplitude.

\subsection*{The Chalmers-Siegel action for self-dual gravity}
\label{sec:Chalmers-Siegel}

In the absence of a cosmological constant, self-dual gravity can be described by the Chalmers-Siegel action~\cite{Chalmers:1996rq}
\begin{equation}
\label{Chalmers-Siegel}
    S[\tilde\phi,\phi] = \int \tilde\phi\left(\Box \phi + \frac{\kappa}{2}\left\{\partial^{\dal}\phi,\partial_{\dal}\phi\right\}\right)\,d^4x\,.
\end{equation}
Here $\phi$ and $\tilde\phi$ are scalar fields representing the positive and negative helicity states of the graviton, respectively, while $\kappa = \sqrt{32\pi G_{\rm N}}$ is the coupling. To write the interaction, we have defined 
$\partial_{\dal}= \alpha^\alpha(\partial/\partial x^{\alpha\dal})$ for some choice of spinor $|\alpha\rangle$, and also introduced 
\begin{equation}
\label{R4Poisson}
    \left\{f,g\right\} = (\partial^{\dal}\!f)\,(\partial_{\dal} g) 
    =\epsilon^{\dal\dot{\beta}}\,(\partial_{\dot{\beta}}f)\,(\partial_{\dal}g)
\end{equation}
as a Poisson bracket on $\R^4$. The presence of this Poisson bracket is the origin of the fact that amplitudes in self-dual gravity possess $\mathcal{L}\mathfrak{ham}(\mathbb{C}^2)$  symmetry.  Notice that $\{\partial^{\dal}\phi,\partial_{\dal}\phi\}= \partial^{\dal}\partial^{\dot{\beta}}\phi\,\partial_{\dal}\partial_{\dot{\beta}}\phi$ so that the interaction involves four derivatives in total.

This action may be understood as follows (see also {\it e.g.}~\cite{Siegel:1992wd,Adamo:2021bej}). Any self-dual Ricci-flat  $M$ is hyperk{\"a}hler\footnote{Or pseudo-hyperk{\"a}hler in (2,2) signature.} and so possesses an $S^2$ family of complex structures, labelled by the spinor $|\lambda\rangle$ up to scale.  A hyperk{\"a}hler manifold also has an $S^2$'s worth of symplectic structures, which for our 4-manifold $M$ are given up to scale by 
\begin{equation}
    \Sigma(\lambda) = \lambda^\alpha\nabla^{\dal}_{\ \alpha}\lrcorner\,(\lambda^\beta \nabla_{\dot{\beta}\beta}\lrcorner\,({\rm vol}(M)))= \frac{1}{2}e^{\dal\beta}\wedge e_{\dot{\beta}}^{\ \alpha}\,\lambda_\al\lambda_\beta\,.
\end{equation}  
Here $\nabla_{\dal\al}$ is the connection on the tangent bundle, ${\rm vol}(M) = \frac{1}{4!}e^{\dal\al}\wedge e^{\ \beta}_{\dal}\wedge e^{\dot{\beta}}_{\ \al}\wedge e_{\dot{\beta}\beta}$ is the volume form on $M$ and $e^{\dal\al}$ the vierbein 1-forms dual to $\nabla_{\dal\al}$. $\Sigma(\lambda)$ is the so-called \emph{Gindikin $2$-form} \cite{gindikin1986construction}.

The hyperk{\"a}hler condition is equivalent to the triple $\Sigma^{\al\beta}=\e^{\dal(\al}\wedge e_{\dal}^{\ \beta)}$ of 2-forms being closed. In particular, we can identify an open patch $U\subset M$ with a patch of $\C^2$, by picking a basis $(|\alpha\ra,|\hat\al\ra)$ for our spinors (it will be convenient to choose $\la\al\hat\al\ra=1$) and letting $(u,v) = (x|\hat\alpha\rangle)^{\dal}$ be holomorphic coordinates in the complex structure defined by $|\lambda\rangle=|\hat\alpha\rangle$. In these coordinates, the vierbeins can be chosen to have components
\begin{equation}
\label{vierbeins}
 e^{\dal\al}\hat{\al}_\al = \d x^{\dal\al}\,\hat{\al}_{\al}\qquad\text{and}\qquad
 e^{\dal\al}{\al}_\al = \d x^{\dal\al}\al_{\al} - 
 \kappa\,\p^{\dal}\p_{\dot{\beta}}\phi\ \d x^{\dot\beta \beta}\hat\al_{\beta}
\end{equation}
for some scalar $\phi(x)$. The constant $\kappa$ controls the deformation away from flat space. With these vierbeins, closure of $\Sigma^{\al\beta}\hat{\alpha}_\al\hat{\alpha}_\beta$ and $\Sigma^{\al\beta}\hat{\al}_\al\al_{\beta}$ are automatic, while closure of the remaining $\Sigma(\al)$ requires that $\phi$ obeys
\begin{equation}
\label{2nd_Plebanski}
\Box\phi +\frac{\kappa}{2}\left\{\p^{\dal}\phi,\p_{\dal}\phi\right\}=0\,.
\end{equation}
This is known as the second Pleba\'{n}ski equation and arises as the field equation by varying $\tilde\phi$ in of~\eqref{Chalmers-Siegel}.
Notice that the Poisson bracket in~\eqref{Chalmers-Siegel} \&~\eqref{2nd_Plebanski} is the inverse of the symplectic form $\Sigma(\hat\alpha)$ that has type $(2,0)$ in our chosen complex structure. 

\medskip

The choice of $|\alpha\rangle$ means the action~\eqref{Chalmers-Siegel} respects only a subgroup\footnote{The subgroup is $SU(2)\times B$ where $B$ is the Borel subgroup of $SU(2)$ represented by unimodular upper triangular matrices; {\it i.e.} the subgroup of $SU(2)$ that preserves the spinor $|\alpha\rangle$ up to scale.} of $SO(4)$. However, provided the external states of momentum\footnote{In Euclidean signature, the linearised on-shell condition $p^2=0$ requires that the external momenta are complex.} $p=|\lambda\ra[\tilde{\lambda}|$ are normalized\footnote{The normalization is fixed by little group scaling.} as 
\begin{equation}
    \label{plane_wave_normalisation}
    \phi_p(x) = \langle\alpha\lambda\rangle^{-4}\,e^{{\rm i}p\cdot x}
     \qquad\qquad
    \tilde\phi_p(x) = \langle \alpha\lambda\rangle^4\,e^{{\rm i}p\cdot x}\,,
\end{equation}
the amplitudes it gives rise to are invariant under the full $SO(4)$, as we would expect from its origin as self-dual gravity. In fact, the only potentially non-vanishing amplitudes of self-dual gravity are the tree-level amplitude with one negative helicity and $n-1$ positive helicity gravitons, and the $n$-particle, all $+$ amplitude at 1-loop\footnote{A simple graph theoretic argument shows that the only Feynman diagrams that can be constructed from \eqref{Chalmers-Siegel} contribute to these amplitudes}. The tree-level amplitude vanishes unless all external particles are (holomorphically) collinear, but the 1-loop amplitude exists for generic $p_i$, subject only to $p_i^2=0$ and $\sum_{i=1}^n p_i=0$. Furthermore, these tree-level $-+\cdots +$ and 1-loop $+\cdots+$ amplitudes computed from~\eqref{Chalmers-Siegel} agree, at the same order of perturbation theory, with the corresponding amplitudes computed from the full Einstein-Hilbert action. However, in full gravity, these amplitudes receive further loop corrections.

\medskip

The generators of $\mathcal{L}\mathfrak{ham}(\mathbb{C}^2)$ are usually described as coming from scattering conformally soft gravitons, rather than plane waves~\cite{Guevara:2021abz}. For the positive helicity outgoing graviton, these are obtained by parametrizing 
\begin{equation}
    \label{spinor_parametrize}
    \lambda_\alpha= \sqrt{\omega}\,(1,z) = \sqrt{\omega}\,z_\alpha\,,
    \qquad\qquad
    \tilde\lambda_{\dal} = \sqrt{\omega}\,(1,\tilde z) = \sqrt{\omega}\,\tilde{z}_{\dal} 
\end{equation}
and then taking the residue of the Mellin-transformed momentum eigenstate
\begin{equation}
\label{Mellin}
G^\Delta_{z,\tilde{z}}(x) = \int_0^\infty \frac{d\omega}{\omega}\,\omega^{\Delta}\,\phi_p(x) = \frac{\im^{\Delta-2}}{\la\alpha z\ra^4}\frac{\Gamma(\Delta-2)}{(x^{\alpha\dal}z_\alpha \tilde{z}_{\dal})^{\Delta-2}}
\end{equation}
at integer values of $\Delta$. The normalisation factors in~\eqref{plane_wave_normalisation}, which ensure that $\phi$ and $\tilde\phi$ represent states of helicity +2 and $-2$ respectively, mean that the residue is non-zero only for $\Delta = k\in\{ 2,1,0,-1,-2,\ldots\}$.  The residues have conformal weights $(\frac{k+2}{2},\frac{k-2}{2})$, and in particular admit a (binomial) mode expansion
\begin{equation}
\label{tilde_z_mode_expansion}
    {\rm Res}_{\Delta=k}\,\left(G^\Delta_{z,\tilde{z}}(x)\right)= \frac{(-\im)^{2-k}}{\la\alpha z\ra^4}\frac{(x^{\alpha\dal}z_\alpha\tilde{z}_{\dal})^{2-k}}{(2-k)!}= \sum_{m=1-p}^{p-1} \frac{\tilde{z}^{p-m-1}\,w^p_{m}(z)}{(p-m-1)!\,(p+m-1)!}
\end{equation}
in $\tilde{z}$. Following~\cite{Adamo:2021lrv}, in the final equality we have relabelled $k= 4-2p$ to agree with the conventions in~\eqref{eq:wstructureconstants}, and defined
the conformally soft modes
\begin{equation}
\label{conformal_modes}    
w^p_m(z) = \frac{(-1)^{p-1}}{\la\alpha z\ra^{4}}\, (x^{\alpha\dot{0}}z_\alpha)^{p+m-1}(x^{\alpha\dot{1}}z_\alpha)^{p-m-1}\,.
\end{equation}
These $w^p_m(z)$ are the generators of $\mathcal{L}\mathfrak{ham}(\mathbb{C}^2)$, with modes $w^p_{m,r}$ coming from further expanding in $z$.  The fact that the residues of $G^\Delta_{z,\tilde{z}}(x)$ involved only positive powers of $x^{\alpha\dal}z_\alpha \tilde{z}_{\dal}$ is the origin of the restriction to the wedge subalgebra. Note again that the indices $p,m$ can each be (simultaneously) either an integer or half-integer which justifies the existence of the blue dots in figure \ref{fig:Wedge}. The structure of the algebra itself will come from the interactions between these modes and can be seen in the corresponding amplitudes. 

\medskip

Let us also point out that, at the classical level, self-dual gravity and its relation to $\mathcal{L}\mathfrak{ham}(\mathbb{C}^2)$ has long been known to be closely related to twistor theory. See {\it e.g.}~\cite{Penrose:1976js} for an original reference. As in self-dual Yang-Mills~\cite{Costello:2021bah,Costello:2022upu,Costello:2022wso}, the situation at the quantum level is more subtle, see~\cite{Bittleston:2022nfr}. We will introduce twistors below in chapter \ref{chapter3} to make this point more explicit.

\subsection*{Moyal deformed self-dual gravity}
\label{sec:Moyal-gravity}

The origin of $\lham$ in self-dual gravity amplitudes is ultimately the presence of the Poisson bracket~\eqref{R4Poisson} on $\R^4$. The fact that the Weyl algebra arises as the Lie algebra of functions on the plane under the Moyal bracket strongly suggests that, to obtain a theory whose amplitudes respect the loop algebra of the Weyl algebra  $\mathcal{L}\mathfrak{diff_{\mathfrak{q}}(\mathbb{C})}$, we should deform the action by changing~\eqref{R4Poisson} to a Moyal bracket. That is, we consider Moyal deformed self-dual gravity, by which we mean the theory with action
\begin{equation}
    \label{Moyal_action}
    S_{\qq}[\tilde\phi,\phi]= \int \tilde\phi\left(\Box\phi + \frac{\kappa}{2}\left\{\partial^{\dal}\phi,\partial_{\dal}\phi\right\}_{\qq}\right)\,d^4x \,,
\end{equation}
where $\{\ , \ \}_{\qq}$ is the Moyal bracket defined via
$\{f,g\}_{\qq} = \qq^{-1}\left( f\star g - g\star f\right)$ with 
\begin{equation}
    \label{R4_Moyal_star}
    f\star g = f\, \exp\left[ \qq \left(\epsilon^{\dal\dot{\beta}}\,\overleftarrow{\partial_{\dal}}\,\overrightarrow{\partial_{\dot{\beta}}}\right)\right]\,g\,
\end{equation}
the Moyal star on $\R^4\cong \C^2$, just as in~\eqref{Moyal_bracket}-\eqref{Moyal_star}. We emphasise that this Moyal bracket acts on spacetime itself, rather than on phase space as is common in applications to deformation quantization~\cite{moyal_1949,Kontsevich:1997vb,Groenewold1946OnTP,Dito:1990rj,Fairlie:1998rf}. The deformed theory \eqref{Moyal_action} can hence best be thought of as self-dual gravity on a non-commutative $\mathbb{R}^4$-background. Moyal deformed self-dual gravity has been considered previously in~\cite{Strachan:1992em,Strachan_1995}, where the non-commutative version of the Pleba\'{n}ski equations and associated non-linear graviton construction were considered from the perspective of integrable systems. 
We note that this Moyal star product breaks the  Lorentz group $\text{SU}(2)\times \text{SU}(2)$ to $\text{SU}(2)\times B$ where $B\subset SU(2)$ is a Borel subgroup that fixes the spinor $|\alpha\ra$. To introduce a Moyal bracket in a way compatible with full $\text{SU}(2)\times \text{SU}(2)$ invariance requires moving to a higher spin theory, see {\it e.g.}~\cite{Krasnov:2021nsq,Tran:2022tft}.

\medskip

Since the kinetic term in~\eqref{Moyal_action} is undeformed, at the linearised level we scatter the same states as in the Chalmers-Siegel theory \eqref{Chalmers-Siegel}. In particular, momentum eigenstates are again normalised as in~\eqref{plane_wave_normalisation}, so the Moyal deformed theory possesses the same set of conformally soft gravitons as usual.  This corresponds to the fact that, as we saw in section~\ref{sec:algebra}, $\difff$ has the same set of generators as $\ham(\mathbb{C}^2)$, with only the structure of the algebra itself being deformed. This deformation of course arises from the deformed interaction. While the Moyal bracket is a complicated, non-local operator on spacetime, its action on momentum eigenstates is remarkably simple. We have
\begin{equation}
    \label{Moyal-momentum}
    \left\{\phi_{p_1},\phi_{p_2}\right\}_{\qq} = \frac{1}{\qq}\sinh\left(\qq \,\langle \alpha|p_1p_2|\alpha\rangle\right)\,\phi_{p_1}\phi_{p_2}
\end{equation}
for any pair of 4-momenta $p_{1,2}$ that may be off-shell.

In particular, the 3-particle $\overline{\rm MHV}$ tree amplitude that follows from~\eqref{Moyal_action} is given by $M^{0,3}_\qq= \delta^4(\sum_i p_i)\,\cM^{0,3}_\qq$, with
\begin{equation}
    \label{Moyal_MHV_bar}
\begin{aligned}
    \cM^{0,3}_{\qq}(p_1^-,p_2^+,p_3^+) &= \kappa\left(\frac{\langle\alpha 1\rangle}{\langle\alpha 2\rangle\langle\alpha 3\rangle}\right)^4\frac{[23]\langle\alpha 2\rangle\langle\alpha3\rangle}{\qq}\,\sinh\left(\qq\,[23]\langle\alpha2\rangle\langle\alpha3\rangle\right)\\
     &= \kappa\, \frac{[23]^7}{([12][23][31])^2}\,[23]_{\qq}\ .
\end{aligned}
\end{equation}
The second expression here holds on the support of momentum conservation and involves the deformed symplectic product of pairs of dotted spinors, defined as
\begin{equation}
\label{Moyal_deformed_square}
    [ij]_{\qq}\,=\, \frac{\sinh\left(\qq\,[ij]\la\alpha i\ra\la\alpha j\ra\right)}{\qq\,\la\alpha i\ra\la\alpha j\ra} \,.
\end{equation}
Notice that, just like the usual spinor product $[ij]=\epsilon^{\dal\dot\beta}\tilde\lambda_{i\dot\beta}\tilde\lambda_{j\dal}$, this deformed product obeys $[ij]_\qq = -[ji]_{\qq}$, behaves as $[ij]_\qq \mapsto (r_ir_j)^{-1} [ij]_\qq$ under the scaling $(\lambda_i,\tilde\lambda_i)\mapsto (r_i\lambda_i,r_i^{-1}\tilde\lambda_i)$, and is 
invariant under $SL(2)$ transformations acting on dotted spinor indices. However, the fact that $[ij]_\qq$ depends on a choice of undotted spinor $|\alpha\ra$ shows that amplitudes in the Moyal deformed theory are not fully Lorentz invariant. Finally, we notice that $\lim_{\qq\to 0}\, [ij]_\qq = [ij]$, so that the Moyal deformed amplitude~\eqref{Moyal_MHV_bar} reduces to the usual three-point $\overline{\rm MHV}$ tree amplitude as $\qq\to0$.

\section{Celestial OPEs and $\ldifff$}
\label{sec:OPEs}

In this section, we will see explicitly that the $\lham$ symmetry of self-dual gravity gets deformed to $\ldifff$ in the Moyal theory. This $\ldifff$ symmetry is perturbatively exact in analogy to \cite{Ball:2021tmb} in the undeformed case.

\subsection*{The holomorphic collinear limit of $\cM^{1,n}_\qq$}
\label{sec:Collinear}

In the celestial holography program,  collinear limits of amplitudes are interpreted as providing information about the structure of OPEs in a 2d theory living on the celestial sphere as depicted in figure \ref{fig:celestial}. In particular, in self-dual gravity, these limits reveal that any such celestial dual theory must contain operators that generate $\lham$. 

While the Moyal deformed theory~\eqref{Moyal_action} will also generate 1-loop all plus amplitudes, at present we do not understand their explicit form.  Fortunately, in any quantum field theory, the behaviour of amplitudes in the \emph{true} collinear limit $p_i\propto p_j$ is fixed on general grounds, with the $\ell$-loop, $n$ particle amplitude factorizing into a sum of $k\leq\ell$-loop, $n-1$ particle amplitudes and an $(\ell-k)$-loop splitting function that describes how the two collinear particles connect to the remainder of the amplitude~\cite{Bern:1998sv,Bern:1998xc}. In both self-dual gravity and the Moyal theory, the only non-trivial amplitudes (for $n>3$) are the 1-loop all plus $\cM^{1,n}_\qq$, so only the tree-level splitting function is relevant within the self-dual theory on a flat background\footnote{However, notice that the $1$-loop deformation of the splitting amplitude is still of relevance \cite{Bittleston:2022jeq, Costello:2022upu} when perturbing away from self-dual gravity on $\mathbb{R}^4$. It is the basis of the \emph{celestial chiral algebra bootstrap}, which made it possible to bootstrap certain $2$-loop QCD amplitudes \cite{Costello:2023vyy, Dixon:2024mzh, Dixon:2024tsb}}. In particular, we must have
\begin{equation}
    \cM^{1,n}_\qq \xrightarrow{1\parallel 2} \text{Split}_{\qq}(1^+,2^+)\,\cM^{1,n-1}_\qq
\end{equation}
in the true collinear limit, where $\text{Split}_\qq$ is the tree-level splitting function associated to~\eqref{Moyal_action}. Explicitly, this splitting function is
\begin{equation}
     \text{Split}_{\qq}(1^+,2^+)= -\kappa \, \cM^{0,3}_\qq (1^+,2^+,-p^-)\times \frac{1}{p^2}=\frac{-\kappa}{2} \frac{[12]^4}{[2p]^2[p1]^2}\frac{[12]_{\qq}}{\la 12\ra} \,,
\end{equation}
where $p=p_1+p_2$ is the momenta in the propagator and we have used the $\overline{\text{MHV}}$ 3-point amplitude \eqref{Moyal_MHV_bar} and the undeformed propagator $1/p^2$.

Importantly, because $\sinh(z)$ that appears in~\eqref{Moyal_deformed_square} is an entire function, $[12]_\qq$ introduces no new singularities.  However, in the true collinear limit, $[12]\to0$ as well as $\langle 12\rangle\to0$, so that $[12]_\qq\to[12]$ and we have the same collinear limit as at $\qq=0$.

For celestial holography, what is actually needed is the \emph{holomorphic} collinear limit, where $\langle 12\rangle\to0$ with $[12]$ unchanged. Fortunately, it was argued in~\cite{Ball:2021tmb} that for two positive helicity gravitons, the structure of the holomorphic collinear limit is the same as that of the true collinear limit. In the Moyal case, this is still true provided we take the limit in a way such that $\langle 1\alpha\rangle$ remains non-zero  to avoid generating singularities from $[12]_\qq$. Parametrizing the momenta in the holomorphic collinear limit as usual by
\begin{equation}
 |p\rangle = \frac{1}{\sqrt{t}}|1\ra = \frac{1}{\sqrt{1-t}}|2\ra \,,
\end{equation} 
the Moyal deformed amplitude $\cM^{1,n}_{\qq}$ must behave in the holomorphic collinear limit $\langle12\rangle\to0$ as
\begin{equation}
\label{holomorphic_collinear_Mq}
    \cM^{1,n}_\qq(1^+,2^+,\dots,n^+) \xrightarrow{1\parallel 2}\frac{-\kappa/2}{t(1-t)}\, \frac{[12]_{\qq}}{\la 12\ra}\  \cM^{1,n-1}_\qq(p^+,3^+,\dots,n^+)\,.
\end{equation}
The crucial difference compared to self-dual gravity is that the overall factor of $[12]$ is deformed to $[12]_\qq$, arising from the fact that the splitting function involves the Moyal deformed vertex between the two collinear particles.

\subsection*{$\ldifff$ algebra from the holomorphic collinear limit}
\label{sec:collinear}

As mentioned in section \ref{section1.5}, celestial amplitudes are defined to be the Mellin transform of massless momentum space amplitudes, whose external null momentum $p_i$ can be parametrized by an energy scale $\omega_i$ for each particle and local coordinates $z_i$, $\tilde{z}_i$ on the celestial sphere
\cite{Pasterski:2016qvg}
\begin{equation}
   p_{\mu}=\omega\, z_\alpha \tilde{z}_{\dot{\alpha}} = \frac{\omega}{\sqrt{2}}\left(1+|z|^2,\,z+\tilde{z},\,-\im (z-\tilde{z}),\,1-|z|^2\right)\,.  
\end{equation}
Then, a Mellin transform in the energy scale $\omega_i$ gives the celestial amplitude
\begin{equation}
    \widetilde{\mathcal{M}}^{1,n}_\qq(\Delta_i,z_i,\tilde{z}_i)= \left[\prod_{i=1}^n\int_{0}^\infty \frac{d\omega_i}{\omega_i} \,\omega_i^{\Delta_i}\right]\cM_{\qq}^{1,n}(p_i).
\end{equation}
And each individual external particle is taken from the momentum eigenstate $\phi$ from equation \eqref{plane_wave_normalisation} to a \emph{boost eigenstate}, 
which we shall label by $G_{\Delta}(z,\tilde{z})$ in the following discussions. We now would like to examine the holomorphic collinear limit of our $\qq$-deformed 1-loop amplitude \eqref{holomorphic_collinear_Mq} in the conformal primary basis. After relabeling energy scales of particles 1 and 2 in the collinear regime as $\omega_1=t\omega_p$ and $ \omega_2=(1-t)\omega_p$, performing the $\omega_p$ integral gives
\begin{equation}
\begin{aligned}
  &\widetilde{\cM}^{1,n}_\qq(\Delta_i,z_i,\tilde{z}_i)\ \overset{z_{12}\rightarrow0}{\longrightarrow} \,  -\frac{\kappa}{2}\, \sum_{l=0}^{\infty} \frac{(\qq\, z_{\alpha 1}z_{\alpha 2})^{2l}}{(2l+1)!}\,   \frac{\tilde{z}_{12}^{2l+1}}{z_{12}}\ \times\\
    &\int_0^1  dt\, t^{\Delta_1+2l-2}(1-t)^{\Delta_2+2l-2}\, 
      \widetilde{\cM}^{1,n-1}_\qq(\Delta_1\!+\!\Delta_2\!+\!4l,z_2,\tilde{z}_2\!+\!t\tilde{z}_{12};\cdots)\ +\  \cO(z_{12}^0)\,,
\end{aligned}
\end{equation}
where $z_{ij}=z_i-z_j \, , \tilde{z}_{ij}= \tilde{z}_i-\tilde{z}_j$ and we have used spinor parametrization \eqref{spinor_parametrize} to capture collinear singularity. The additional sum over $l$ comes from expanding the $\sinh$ in $\qq$.
Taylor expanding the celestial amplitude on the right-hand side in $\tilde{z}_{12}$ and performing the $t$-integral leads to
\begin{equation}
\begin{aligned}
    &\widetilde{\cM}^{1,n}_\qq(\Delta_i,z_i,\tilde{z}_i)\ \overset{z_{12}\rightarrow0}{\longrightarrow}-\frac{\kappa}{2z_{12}}\sum_{l=0}^{\infty}\frac{(-1)^l(\qq\, z_{\alpha 1}z_{\alpha 2})^{2l}}{(2l+1)!}\ \times\\
    &\left[\sum_{n=0}^{\infty} \frac{
     \tilde{z}_{12}^{n+2l+1}}{n!}\, B(\Delta_1\!-\!1\!+\!2l\!+\!n,\Delta_2\!-\!1\!+2l)\ \tilde{\partial}^n_2\widetilde{\cM}^{1,n\!-\!1}_\qq(\Delta_1\!+\!\Delta_2\!+\!4l,z_2,\tilde{z}_2;\cdots)\right]\\
     &\qquad +\  \mathcal{O}(z_{12}^0)\,,
\end{aligned}
\end{equation}
where $\tilde\partial_2 = \partial/\partial\tilde{z}_2$. From this, we can read off the momentum space splitting function in boost eigenstate as two conformal primary gravitons approaching each other on the celestial sphere producing the celestial OPE. Here $B(x,y)=\frac{\Gamma(x+y)}{\Gamma(x)\Gamma(y)}$ is the Euler Beta function. Recalling that  conformally soft gravitons are defined by $H^k(z,\tilde{z})=\text{Res}_{\Delta=k}\, G_{\Delta}(z,\tilde{z})$ for $k=2,1,0,-1,\dots$, we obtain the celestial OPE of $H^k$ and $H^j$ as:
\begin{equation}
\label{eq:OPEH}
\begin{aligned}
    &H^k(z_1,\tilde{z}_1)\,H^j(z_2,\tilde{z}_2)\\ 
         &\sim-\frac{\kappa}{2z_{12}}\sum_{l=0}^{\infty}\sum_{n=0}^{1-k-2l}\frac{(-1)^l}{(2l+1)!} \binom{2-k-j-4l-n}{1-2l-j}(\qq z_{\alpha 1}z_{\alpha 2})^{2l}  \frac{\tilde{z}_{12}^{n+2l+1}}{n!} \tilde{\partial}^n H^{k+j+4l}(z_2,\tilde{z}_2).
\end{aligned}
\end{equation}
As a consistency check, we see that taking the $\qq\rightarrow 0$ limit reduces~\eqref{eq:OPEH} to the OPE found in~\cite{Guevara:2021abz, Ball:2021tmb}.\\

In the case of self-dual gravity, one extracts the OPE between conformally soft modes 
\begin{equation}
    H^k_n(z)=\oint_{|\tilde{z}|<\epsilon} \frac{d\tilde{z}}{2\pi i}\,\tilde{z}^{n+\frac{k-4}{2}}\,H^k(z,\tilde{z})
\end{equation} 
as a contour integral of the $\qq=0$ limit of~\eqref{eq:OPEH}. To obtain the $\mathcal{L}w_{\wedge}$-algebra, one must then use the Vandermonde identity $\sum_{t=0}^r \binom{n}{t}\binom{m}{r-t}=\binom{m+n}{r}\,$ to perform the residual sum \cite{Ball:2021tmb}. We can perform the same contour integrals and residual sum in the $\qq$-deformed theory where we need the following generalised Vandermonde identity
\begin{equation} 
\label{eq:GenVand}
      \sum_{t=0}^r\  [t]_l\,\binom{n}{t}\binom{m}{r-t}=\frac{[r]_l\,[m]_l}{[m+n]_l}\binom{m+n}{r}\,.
\end{equation}

In order to perform the contour integrals over equation \eqref{eq:OPEH} in $\tilde{z}_1$ and $\tilde{z}_2$, we use
\begin{equation}
\label{eq:contouridentity}
    \oint_{|\tilde{z}_1|<\epsilon} d\tilde{z}_1 \frac{\tilde{z}_1^{m+\frac{k-4}{2}}}{2\pi i} \tilde{z}_{12}^{n+2l+1}=\frac{(n+2l+1)!}{(\frac{2-k}{2}-m)!\,(n+2l+m+\frac{k}{2})!}(-\tilde{z}_2)^{m+n+2l+\frac{k}{2}}
 \end{equation} 
if $-\frac{k}{2}-m\leq n+2l$, and zero otherwise.
This shows that
\begin{align}
    &H^k_m(z_1)\,H^j_{n}(z_2) \nonumber\\ 
    \quad&\sim  -\frac{\kappa}{2z_{12}} \sum_{l=0}^{C(m,n,k,j)} \sum_{t=-k/2-m-2l}^{1-k-2l} \left\{\frac{(-1)^{l+m+t+2l+k/2}}{(2l+1)!\,t!} (qz_{\alpha 1}z_{\alpha 2})^{2l}\binom{2-k-j-4l-t}{1-2l-k-t}\right. \nonumber \\ 
    \quad&\qquad\qquad\times\  \left.\frac{(t+2l+1)!}{(\frac{2-k}{2}-m)!\,(t+2l+m+\frac{k}{2})!}  \left[\frac{2-k-j}{2}-2l-m-n\right]_t H^{k+j+4l}_{m+n}(z_2)\right\}
\end{align}
where $C(m,n,k,j)=\lfloor\tfrac{1}{4}(|m+n|+\frac{2-k-j}{2}) \rfloor$ is the upper bound for $l$, for which the inequality in the second case of equation \eqref{eq:contouridentity} holds. Note that when $l=0$, this reduces to the undeformed self-dual gravity calculation where the sum over $t$ can be performed straight away. In the given context this is more subtle but after shifting $t\rightarrow t-k/2-m-2l$ and rearranging the terms to collect terms that involve $t$, the right-hand side becomes
\begin{align}
\label{eq:lasttsum}
 &H^k_m(z_1)H^j_{n}(z_2)\sim -\frac{\kappa}{2z_{12}}\sum_{l=0}^{C(m,n,k,j)}\frac{(-1)^{2l+1-k/2+m}}{(2l+1)!}\\ \nonumber
 &\frac{((2-k)/2-m+(2-j)/2-n-2l-1)!}{(\frac{2-k}{2}-m)!\,(\frac{2-j}{2}-n)!} (qz_{\alpha 1}z_{\alpha 2})^{2l} F(k,j,l,m,n)\,H^{k+j+4l}_{m+n}(z_2)\,.
\end{align}
where $F(k,j,l,m,n)=\sum_{t=0}^{(2-k)/2+m} 
[t-k/2-m+1]_{2l+1}\binom{j+2l-2}{1-k/2+m-t}\binom{\frac{2-j}{2}-n}{t} $ is the sum over $t$. The first term $[t-k/2-m+1]_{2l+1}$ can be expressed as
\begin{equation}
    \left[t+\frac{2-k}{2}-m\right]_{2l+1}=\sum_{i=0}^{2l+1}\binom{2l+1}{i}[t]_{i}\left[\frac{2-k}{2}-m\right]_{2l+1-i}\,.
\end{equation}
Then, the generalized Vandermonde-identity \eqref{eq:GenVand} can be used to perform the sum over $t$ which leads to 
\begin{align}
F(k,j,l,m,n)& =(-1)^{2l+(2-k)/2+m} \\
   & \sum_{i=0}^{2l+1}\binom{2l+1}{i}\left[\frac{2-k}{2}-m\right]_{2l+1-i}\frac{[\frac{2-k}{2}+m]_i[\frac{2-j}{2}-n]_i}{[\frac{j-2}{2}-n+2l]_i}\binom{\frac{j-2}{2}-n+2l}{\frac{2-k}{2}+m}\,.\nonumber
\end{align}   
Absorbing the $(-1)^{2l+(2-k)/2+m}$ into the remaining binomial coefficients using $(-1)^k \binom{n}{k}=\binom{k-n-1}{k}$ leaves us with
\begin{align}
    F(k,&j,l,m,n)=\frac{(\frac{2-k}{2}+m+\frac{2-j}{2}+n-2l-1)!}{(\frac{2-k}{2}+m)!\,(\frac{2-j}{2}+n)!}\\\nonumber
    & \sum_{i=0}^{2l+1}\binom{2l+1}{i}\left[\frac{2-k}{2}-m\right]_{2l+1-i}\left[\frac{2-k}{2}+m\right]_i\left[\frac{2-j}{2}-n\right]_i\left[\frac{2-j}{2}+n\right]_{2l+1-i}.
\end{align}

\noindent After substituting this back into equation \eqref{eq:lasttsum} one finds the desired result that
the OPE of soft graviton modes in the $\qq$-deformed theory is
\begin{align}
    &H^k_m(z_1)\,H^j_{n}(z_2)\sim \\
    \nonumber&\frac{-\kappa}{2z_{12}}\sum_{l=0}^{C(k,j,m,n)}\frac{(-1)^l(\qq\, z_{\alpha 1}z_{\alpha 2})^{2l}}{(2l+1)!} \bigg(\frac{(\frac{2-k}{2}\!-\!m\!+\!\frac{2-j}{2}\!-\!n\!-\!2l\!-\!1)!\,(\frac{2-k}{2}\!+\!m\!+\!\frac{2-j}{2}\!+\!n\!-\!2l\!-\!1)!}{(\frac{2-k}{2}-m)!(\frac{2-j}{2}-n)!(\frac{2-k}{2}+m)!(\frac{2-j}{2}+n)!}\bigg)\times\\
    \nonumber&\ \sum_{i=0}^{2l+1}(-1)^i\binom{2l\!+\!1}{i}\left[\frac{2\!-\!k}{2}\!-\!m\right]_{2l+1-i}\left[\frac{2\!-\!k}{2}\!+\!m\right]_i\left[\frac{2\!-\!j}{2}\!-\!n\right]_i\left[\frac{2\!-\!j}{2}\!+\!n\right]_{2l+1-i}H^{k+j+4l}_{m+n}(z_2)\,,
\end{align}
where the upper limit of the sum 
\[
C(k,j,m,n)=\left\lfloor\frac{|m+n|}{4}+\frac{2-k-j}{8} \right\rfloor\, .
\]
Using the relabelling $k=4-2p$, we rewrite the modes to absorb some of the factorials in the coefficient 
\begin{equation}
W^p_m(z)=\frac{1}{\kappa} (p-m-1)!\,(p+m-1)!\,H^{4-2p}_m(z)
\end{equation}
as discussed in section \ref{sec:Chalmers-Siegel} to obtain
\begin{equation}
\begin{aligned}\label{OPE_result}
    & W^p_m(z_1)\,W^q_n(z_2)\sim \frac{1}{2z_{12}} \sum_{l=0}^{C(p,q,m,n)} \frac{(-1)^l (\qq\,z_{\alpha 1}z_{\alpha 2})^{2l}}{(2l+1)!}\sum_{i=0}^{2l+1}(-1)^i\binom{2l+1}{i}\\
    &[p-1+m]_{2l+1-i}[p-1-m]_i[q-1+n]_i[q-1-n]_{2l+1-i}W_{m+n}^{p+q-2-2l}(z_2)
\end{aligned}
\end{equation}
with $C(p,q,m,n)=\lfloor\tfrac{1}{4}(|m+n|+p+q-3) \rfloor$, which enforces the wedge condition.
Comparing \eqref{OPE_result} and \eqref{W_algebra}, we see that we have arrived at $\ldifff$ on the nose. Recall that although the sum over $l$ does not have an explicit upper limit in \eqref{W_algebra}, an implicit cut-off is in place by restricting to the wedge subalgebra. 

%% file: Chapter2/chapter2.tex

\chapter{Background on twistor theory}
\label{chapter3}

The gravitational celestial chiral algebras discussed in the previous chapter \ref{chapter2}, particularly in the light of their Hamiltonian interpretation \eqref{Poisson}, have a natural interpretation in \emph{twistor space}. Moreover, twistor space methods will become important in order to derive further deformations of celestial chiral algebras on curved spacetimes in chapters \ref{chapter4}, \ref{chapter5} and \ref{chapter6}. 

In the present chapter, we will review the relevant basic twistor theory and fix conventions that will be used throughout the rest of this thesis. For a more detailed treatment, we refer to the many existing review articles and books which include \cite{Adamo:2017qyl, Mason:1991rf, Huggett_Tod_1994, Penrose:1986ca}. 

This chapter is organized as follows: In section \ref{section2.1}, we will define the flat twistor space $\mathbb{PT}$ and discuss its relation to flat complexified Minkowski space via the \emph{incidence relations} and a certain \emph{double fibration}. Taking real slices in different signatures is discussed after that. Section \ref{sec:TwistorActions} then discusses twistor actions for various self-dual theories. Section \ref{sectionLinearPenrose} discusses the linear Penrose transform and how in this simple example the spacetime action for a massless scalar in Euclidean signature can be obtained from gauge fixing a certain twistor action. Section \ref{section2.2} discusses the classical Penrose-Ward correspondence and how it leads to a twistor action for self-dual Yang-Mills theory in Euclidean signature. Section \ref{section2.3} discusses the classical non-linear graviton construction and how it leads to a twistor action for self-dual Einstein gravity.

\section{Geometry of twistor space} 
\label{section2.1}

Throughout this chapter and the rest of this thesis we make free use of spinor conventions $\la\lambda\kappa\ra = \lambda^\al\kappa_\al=\epsilon^{\al\beta}\lambda_\beta\kappa_\al$ and similarly $[\tilde\lambda\tilde\kappa]=\tilde\lambda^\da\tilde\kappa_\da=\epsilon^{\da \db}\tilde\lambda_\db\tilde\kappa_\da$. The \emph{twistor space of flat } $\bbR^4$ is given by
\be
\mathbb{PT}=\mathcal{O}(1)\oplus \mathcal{O}(1)\rightarrow \mathbb{CP}^1\,.
\ee
We will often use homogeneous coordinates $\lambda_\alpha$ on the $\CP^1$ base, and $\mu^\da$ on the fibres, collectively denoting these by $Z^a=(\mu^\da,\lambda_\al)$ with $a\in \{1,2,3,4\}$. $Z^a$ can be viewed as a homogenous coordinate on $\mathbb{CP}^3$ with the standard equivalence relation $Z^a\sim r Z^a$ for any $r\in \bbC^*$. From this and the fact that $\lambda^\al\neq 0$, we can also view twistor space as
\be
\mathbb{PT}=\mathbb{CP}^3\setminus \mathbb{CP}^1\,.
\ee
$\mathbb{PT}$ is equipped with an action of the (complexified) conformal group $\textrm{SL}(4,\mathbb{C})$ acting on $Z^a$ in the natural way that respects the standard volume form induced by $\epsilon_{a b c d}$. The Lorentz group is the subgroup  $\textrm{SL}(2,\mathbb{C})\times \textrm{SL}(2,\mathbb{C}) \subset \textrm{SL}(4,\mathbb{C})$ that acts on $\mu^\da$ and $\lambda_\al$. Under this Lorentz group, $\mu^\da$ and $\lambda_\al$ transform as left- and right-handed Weyl-spinors respectively. 

The complexified flat spacetime $\mathbb{C}^4=\mathbb{R}^4 \otimes\mathbb{C}$ can be reconstructed from $\mathbb{PT}$ as the space of degree one holomorphic sections. Any such section is described by a $x^{\al \da}\in \mathbb{C}^4$ which in our conventions is given by
\be
\label{eq:Matrixform}
x^{\alpha \da}=\frac{1}{\sqrt{2}} \sigma^{\alpha \da}_\mu x^\mu= \frac{1}{\sqrt{2}}\begin{pmatrix}
x^0+x^3 & x^1 -\im x^2 \\
x^1+\im x^2 & x^0-x^3
\end{pmatrix}\,,
\ee with $x^0,x^1,x^2,x^3 \in \mathbb{C}$. Each such  $x^{\al \da}\in \mathbb{C}^4$ leads to an incidence relation
\be
\label{eq:incidencerel}
\mu^\da=x^{\al \da}\lambda_\al\,.
\ee
This relation between (complexified) spacetime and twistor space (in the above case, $\mathbb{C}^4$ and $\mathbb{PT}$) is generally referred to as the \emph{twistor correspondence} and the degree one holomorphic sections (here given by equation \eqref{eq:incidencerel}) are referred to as \emph{twistor lines}. 

This twistor correspondence is non-local in both directions. 
It is most clear when presented through the following \emph{double fibration}, which can be extended to much more general homogeneous spaces \cite{baston2016penrose}
\begin{center}
\label{tikz:doublefib}
    \begin{tikzpicture}
    	\node (PS) at (0,1.5) {$ \mathbb{PS} $};
    	\node (PT) at (-1.5,0) {$ \mathbb{PT} $};
    	\node (C4) at (1.5,0) {$ \mathbb{C}^4 $};
    	\draw[->] (PS) edge node[above, right] {$ \pi_1$} (C4);
		\draw[->] (PS) edge node[above, left] {$ \pi_2$} (PT);
	\end{tikzpicture}\,.
    \end{center}
Here, $\mathbb{PS}$ is called the correspondence space and is given by 
\be
\mathbb{PS}=\{(x^{\al \da}, \lambda_\al)\in \mathbb{C}^4\times \mathbb{CP}^1\}\,.
\ee
The maps $\pi_1$ and $\pi_2$ are defined in the natural way
\bea
\pi_1(x^{\al \da},\lambda_\beta)&=x^{\al \da}\\
\pi_2(x^{\al \da},\lambda_\beta)&=(x^{\beta \da}\lambda_\beta,\lambda_\alpha)\,.
\eea From this double fibration we can explicitly see that a spacetime point $x^{\alpha \da}\in \mathbb{C}^4$ corresponds to all points $(\mu^\da,\lambda_\al)$ obeying the incidence relations \eqref{eq:incidencerel} for $x^{\alpha \da}$, i.e. a holomorphic linearly embedded $\mathbb{CP}^1$, a \emph{twistor line}. In fact, it can be seen that every holomorphic linearly embedded $\mathbb{CP}^1$ arises as the solution to \eqref{eq:incidencerel} for some point in $\mathbb{C}^4$ as discussed in \cite{Adamo:2017qyl}. We will denote the twistor line over a spacetime point $x$ by $\mathbb{CP}^1_x$. Going in the other direction, a point $Z^a\in \mathbb{PT}$ is generically intersected by infinitely many twistor lines corresponding to infinitely many spacetime points. These points are pairwise null-separated and form a so-called \emph{$\alpha$-plane} \cite{Huggett_Tod_1994}. To see this, consider two twistor lines $X$ and $Y$ corresponding to spacetime points $x^{\alpha \da}$ and $y^{\alpha \da}$. Since they both intersect the point $Z^a=(\mu^\da,\lambda_\al)$ they solve the equations
\bea
\mu^\da&=x^{\al \da}\lambda_\al\\
\mu^\da&=y^{\al \da}\lambda_\al\,,
\eea
which imply $(x-y)^{\al \da}\lambda_\al=0$. Since $\lambda_\al\neq 0$, we know that the $2\times 2$ matrix $(x-y)^{\al \da}$ has vanishing determinant and hence $(x-y)^2=0$. 

\subsection*{Signatures}

So far, we always considered a complexified spacetime even though our universe of course seems to have dimension $4$ over $\bbR$ rather than over $\bbC$. We would ideally like to talk about Minkowski space, i.e. $\mathbb{R}^4$ with a flat metric of signature $(3,1)$. However, we will see below that self-dual theories generally do not admit real solutions in signature $(1,3)$, whereas they do in signatures $(4,0)$, \emph{Euclidean signature,} or $(2,2)$, \emph{Kleinian signature}. 

For most of this thesis we will be working in Euclidean signature since beyond the double fibration, there is a further (non-holomorphic) fibration of twistor space over $\mathbb{R}^4$ which we will discuss now.

\subsubsection*{Euclidean signature} 

To pick out the real Euclidean slice of our complexified spacetime, we choose the following anti-holomorphic involution on twistor space
\bea
\sigma:\mathbb{PT}&\rightarrow\mathbb{PT}\\
(\mu^\da,\lambda_\alpha)&\mapsto (\hat{\mu}^\da,\hat{\lambda}_\alpha)\,,
\eea
where this hat-operation acts equally on dotted as well as undotted spinors by
\bea
(\hat{\mu}^{\dot{0}},\hat{\mu}^{\dot{1}})&=(-\bar{\mu}^{\dot{1}},\bar{\mu}^{\dot{0}})\\
(\hat{\lambda}_{0},\hat{\lambda}_{1})&=(-\bar{\lambda}_{1},\bar{\lambda}_{0})\,.
\eea
When acting on $x^{\al \da}$ from equation \eqref{eq:Matrixform}, the hat-operation is then given by
\be
\label{eq:HatoponMatrix}
\hat{x}^{\alpha \da}= \frac{1}{\sqrt{2}}\begin{pmatrix}
\bar{x}^0-\bar{x}^3 & -\bar{x}^1 +\im \bar{x}^2 \\
-\bar{x}^1-\im \bar{x}^2 & \bar{x}^0+\bar{x}^3
\end{pmatrix}\,.
\ee
We can explicitly see that the positive definite metric 
\be
x^2=(x^0)^2+(y^1)^2+(y^2)^2+(y^3)^2
\ee
is obtained from demanding that $x^{\al \da}$ is preserved by this hat-operation \eqref{eq:HatoponMatrix} leading to real points
\be
x^{\alpha \da}\rvert_{x=\hat{x}} = \frac{1}{\sqrt{2}}\begin{pmatrix}
x^0+\im y^3 & \im y^1 + y^2 \\
\im y^1 + y^2 & x^0-\im y^3 
\end{pmatrix}\,,
\ee
with $x^0, y^1, y^2, y^3 \in \mathbb{R}$. The fact that the hat-operation does not leave any spacetime points of the form $x^{\alpha \da}=\kappa^\alpha \sigma^{\da}$ invariant corresponds to the fact that there are obviously  no real null-vectors in $\mathbb{R}^{4,0}$. This in turn means that any point in twistor space corresponds to a unique spacetime point, leading to an important simplification of the twistor correspondence in Euclidean signature. Beyond the double fibration, there exists the following non-holomorphic fibration 
\bea
\label{eq:TwistorPiEucl}
\pi:\,\,\mathbb{PT}&\rightarrow \mathbb{R}^{4}\\
(\mu^\da, \lambda_\al)&\mapsto x^{\al \da}=\frac{\lambda^\al\hat{\mu}^{\da}- \lambda^\al\hat{\mu}^{\da}}{\langle \lambda \hat{\lambda} \rangle}\,.
\eea
This map \eqref{eq:TwistorPiEucl} shows that twistor space as a smooth manifold is simply the product
\be
\label{eq:DecompositionPT}
\mathbb{PT}\cong S^2 \times\mathbb{R}^4\,,
\ee
which allows us to view the twistor correspondence as a special kind of Kaluza-Klein reduction along the $\mathbb{CP}^1$ factor as emphasised in \cite{Costello:2021bah}. Note that the decomposition \eqref{eq:DecompositionPT} does not hold in the category of complex manifolds in which we have $\mathbb{PT}\cong\mathbb{CP}^3\setminus \mathbb{CP}^1\not\cong  \mathbb{CP}^1 \times\mathbb{C}^2$.

\subsubsection*{Lorentzian signature} 

If we want to single out a real subspace $\mathbb{R}^{3,1}\subset \mathbb{C}^4$ of signature $(3,1)$ the correct reality condition is given by
\be
\label{eq:DagoponMatrix}
\big(x^{\alpha \da}\big)^\dag= \frac{1}{\sqrt{2}}\begin{pmatrix}
\bar{x}^0+\bar{x}^3 & \bar{x}^1 -\im \bar{x}^2 \\
\bar{x}^1+\im \bar{x}^2 & \bar{x}^0-\bar{x}^3
\end{pmatrix}\,.
\ee
When acting on spinors, $\dag$ exchanges dotted with undotted spinors and for $\kappa^\al=(\kappa^0, \kappa^1)$ or $\tilde{\kappa}^\da=(\tilde{\kappa}^{\dot{0}}), \tilde{\kappa}^{\dot{0}}$, it is given by
\bea
\big(\kappa^\al\big)^\dag&=\bar{\kappa}^\da=(\bar{\kappa}^{\dot{0}}, \bar{\kappa}^{\dot{1}})\,,\\
\big(\tilde{\kappa}^\da\big)^\dag&=\bar{\tilde{\kappa}}^\al=(\bar{\tilde{\kappa}}^0, \bar{\tilde{\kappa}}^1)\,.
\eea
In contrast to Euclidean signature, there are real null-vectors in Lorentzian signature and it can be seen that all of them are of the form $v^{\al \da}=\kappa^\al\tilde{\kappa}^\da$. 

To extend this conjugation map to twistor space, we need to introduce the notion of the \emph{dual twistor space} $\mathbb{PT}^\vee$, which is defined through the same open subset $\mathbb{CP}^3\setminus \mathbb{CP}^1\subset \mathbb{CP}^3$ but now with homogenous coordinates $W_a=(\tilde{\lambda}_\da, \tilde{\mu}^\al)$. 
A twistor gets now mapped to a \emph{dual twistor} under the map
\bea
\bar{\cdot}:&\mathbb{PT}\longrightarrow\mathbb{PT}^\vee\\
&Z^a=(\mu^\da, \lambda_\al) \mapsto \bar{Z}_a=(\bar{\lambda}_\da, \bar{\mu}^\al)\,.
\eea
There exists a natural pairing of $\mathbb{PT}$ with $\mathbb{PT}^\vee$ given by contracting the twistor index
\bea
\cdot:&\mathbb{PT} \times \mathbb{PT}^\vee\longrightarrow \mathbb{C}\\
&\big((\mu^\da, \lambda_\al),(\tilde{\lambda}_\da, \tilde{\mu}^\al)\big)\mapsto [\mu \tilde{\lambda}]+\langle\tilde{\mu} \lambda\rangle\,,
\eea
and we can identify all the spacetime points obeying $\big(x^{\al \da}\big)^\dag= x^{\al \da}$ as 
\be
\label{eq:definingPN}
\mathbb{PN}=\{Z\in \mathbb{PT}\vert Z\cdot \bar{Z}=0\}\,.
\ee
$\mathbb{PN}$ is often referred to as the \emph{space of null twistors} and an element $Z\in \mathbb{PN}$ indeed obeys
\be
0=\im x^{\al \da} \lambda_\alpha\bar{\lambda}_\da-\im \big(x^{\al \da}\big)^\dag \lambda_\alpha\bar{\lambda}_\da=\im \big(x-x^\dag\big)^{\al \da}\lambda_\alpha\bar{\lambda}_\da\,,
\ee
for some $x\in \mathbb{C}^4$ with $x=x^\dag$.

Note that when working in Lorentzian signature, 
we include a conventional $\im$ in the incidence relations. We will do so in chapter \ref{chapter5} and then, the incidence relations read
\be
\label{eq:LorentzianIncidence}
\mu^\da=\im x^{\al \da} \lambda_\al\,,
\ee
for an element $Z^a=(\mu^\da, \lambda_\al) \in \mathbb{PT}$ and respectively
\be
\tilde{\mu}^\al=-\im x^{\al \da} \tilde{\lambda}_\da\,,
\ee
for $W_a=(\tilde{\lambda}_\da, \tilde{\mu}^\al) \in \mathbb{PT}^\vee$.

\subsubsection*{Kleinian signature}

If we wish to single out a real subspace $\mathbb{R}^{2,2}\subset \mathbb{C}^4$ of signature $(2,2)$ the correct reality condition is
\be
\label{eq:KleinBaronMatrix}
\bar{x}^{\alpha \da}= \frac{1}{\sqrt{2}}\begin{pmatrix}
\bar{x}^0+\bar{x}^3 & \bar{x}^1 +\im \bar{x}^2 \\
\bar{x}^1-\im \bar{x}^2 & \bar{x}^0-\bar{x}^3
\end{pmatrix}\,,
\ee
with real points given by
\be
x^{\alpha \da}\rvert_{x=\bar{x}} = \frac{1}{\sqrt{2}}\begin{pmatrix}
x^0+ x^3 & x^1 + y^2 \\
x^1 + y^2 & x^0- x^3 
\end{pmatrix}\,,
\ee
for $x^0,x^1,y^2,x^3\in \mathbb{R}$.
These real points can be easily seen to give rise to the Kleinian inner product
\be
x^2=(x^0)^2+(y^2)^2-(x^1)^2-(x^3)^2
\ee
The action of \eqref{eq:KleinBaronMatrix} on $x^{\al \da}$ is given by component-wise complex conjugation (as opposed to Hermitian conjugation in \eqref{eq:DagoponMatrix} in the Lorentzian case) so that spinors will also just be component-wise complex conjugated without exchanging their respective spinor representation. More explicitly, we have
\bea
\label{eq:KleinianSpinors}
&\kappa^\alpha=(\kappa^0,\kappa^1)\mapsto \bar{\kappa}^\alpha=(\bar{\kappa}^0,\bar{\kappa}^1)\,,\\
&\tilde{\kappa}^\da=(\tilde{\kappa}^{\dot{0}},\tilde{\kappa}^1)\mapsto \bar{\tilde{\kappa}}^\da=(\bar{\tilde{\kappa}}^{\dot{0}},\bar{\tilde{\kappa}}^{\dot{1}})\,.
\eea
Equation \eqref{eq:KleinianSpinors} means that we have fewer subtleties in defining a complex conjugation on twistor space. The map reads
\be
Z^a=(\mu^\da, \lambda_\al)\mapsto \bar{Z}^a=(\bar{\mu}^\da, \bar{\lambda}_\al)
\ee
so that the twistor space of $\mathbb{R}^{2,2}$ is given by the subspace of twistor space whose homogenous coordinates are all real
\be
\mathbb{PT}_{\mathbb{R}}=\mathbb{PT}\cap \mathbb{RP}^3\,.
\ee

\section{Twistor actions for self-dual theories}
\label{sec:TwistorActions}

In this section, we will briefly review standard results of twistor theory and their formulation in a modern QFT language. Historically, twistor theory was used to generate on-shell data on spacetime, such as instanton solutions to the equations of self-dual Yang-Mills \cite{Ward:1977ta, Atiyah:1978ri} or self-dual gravity \cite{Penrose:1976jq, Atiyah:1978wi}. However, if we are working with the aim to eventually quantize theories on twistor space \cite{Costello:2021bah}, we need to consider off-shell data as well. In particular, it will be desirable to write down actions on twistor space that are equivalent (off-shell) to the relevant spacetime actions of SDYM or SDGR. We will do so by working in the Dolbeaut framework of sheaf cohomology rather than the Čech framework which is used in most of the older twistor literature and will also be commented on below in section \ref{sec:graviton_symmetries}.

\subsection{The linear Penrose transform}
\label{sectionLinearPenrose}

The usefulness of twistor space stems from the slogan 
\begin{center}
    \emph{Holomorphic data on twistor space is equivalent to conformal data on spacetime.}
\end{center}
There are many manifestations of this slogan which will be discussed below and all of them, in some way, translate the task of solving some PDE on spacetime into choosing some data on twistor space which is only constrained by holomorphicity. PDEs which have such a description in twistor space are 'solvable' through twistor methods which turn the respective field theories into \emph{integrable field theories} \cite{Mason:1991rf}. Examples that we will discuss below include self-dual Yang-Mills theory, self-dual conformal gravity, self-dual Einstein gravity and \emph{Mabuchi gravity}\footnote{For the latter two we need to break conformal invariance in a controlled way so that the above slogan needs to be slightly modified by including extra data, such as the \emph{infinity twistor}, on twistor space.}. All these more involved examples aside, let us begin with the simplest non-trivial example: a free massless scalar. 

It is described by the spacetime-action
\be
\label{eq:phiaction}
S[\phi]=\int_{\mathbb{R}^4} \dif^4x \, \partial_\mu \phi\, \partial^\mu \phi\,,
\ee 
 whose linear equation of motion
\be
\label{eq:Laplace}
\Delta \phi=0\,,
\ee
with $\Delta=\partial_\mu\partial^\mu$\,, turns out to be related to twistor space. The \emph{linear Penrose transform} states that there is a $1:1$ correspondence
\be
\label{eq:linearPenrose}
\{\text{Solutions to \eqref{eq:Laplace} on }\mathbb{C}^4\}\,{\overset{\,\,1:1\,\,}\longleftrightarrow}\, H^{0,1}(\mathbb{PT},\mathcal{O}(-2))\,.
\ee
For a detailed discussion of this theorem, in particular, including a definition of cohomology classes and the bundles $\mathcal{O}(n)$ in the given context, see \cite{Adamo:2017qyl} and for a detailed proof see \cite{Eastwood:1981jy}, the content of which we will sketch very briefly.

Given an element $f\in  H^{0,1}(\mathbb{PT},\mathcal{O}(-2))$, the incidence relations can be used to check that
\be
\phi(x)=\int_{\mathbb{CP}^1_x} \,\langle\lambda\dif\lambda \rangle \wedge f|_{\mathbb{CP}^1_x}\,
\ee
is indeed a solution to \eqref{eq:Laplace} that is obtained  simply by integrating $f$ over the twistor lines.
Note that this statement can be performed in any signature and after obtaining $\phi$ on complexified Minkowski space $\mathbb{C}^4$, we can restrict to any real slice thereof. Proving the other direction of the $1:1$ correspondence \eqref{eq:linearPenrose} is much more subtle and in general there is no natural way to construct an explicit representative of the cohomology class corresponding to a fixed $\phi$ that solves \eqref{eq:Laplace}. However, Woodhouse  showed that \cite{Woodhouse:1985id} given a solution $\phi$ on Euclidean $\mathbb{R}^4$, there is a canonical way to construct such a representative\footnote{We follow the conventions of  \cite{Adamo:2017qyl} concerning  $(0,1)$-forms on twistor space.}
\be
f=\bar{e}^0f_0+\bar{e}^\da f_\da
\ee
which obeys the additional gauge condition
\be
\label{eq:Woodhouse}
\bar{\partial}^*|_{\mathbb{CP}^1_x}f_0=0
\ee
on a fibre $\mathbb{CP}^1_x$ over any arbitrary point $x\in \mathbb{R}^4$.
\eqref{eq:Woodhouse} is commonly referred to as \emph{Woodhouse gauge}.

In Euclidean signature something else happens which is rather special and not expected in other signatures: The action \eqref{eq:phiaction} admits an equivalent description on twistor space with an action that can be reduced to \eqref{eq:phiaction} for arbitrary off-shell spacetime-fields that might not solve \eqref{eq:Laplace}. In contrast, the original statement \eqref{eq:linearPenrose} of the linear Penrose transform only makes a statement about on-shell fields in arbitrary signature \cite{Penrose:1969ae, Adamo:2017qyl}. 

Let us consider a theory on $\mathbb{PT}$ whose field content is given by
\be
\Phi\in \Omega^{0,1}(\mathbb{PT},\mathcal{O}(-2))
\ee
together with the action
\be
\label{eq:TwistorActionlinear}
S[\Phi]=\int_{\mathbb{PT}} \text{D}^3 Z \,\Phi\wedge\bar{\partial} \Phi\,,
\ee
where the holomorphic volume form of weight $4$ is given by
\be
\label{eq:holomorphicvolumeform}
\text{D}^3Z=\frac{1}{4!}\epsilon_{abcd} Z^a\dif Z^b\wedge \dif Z^c\wedge \dif Z^d \in \Omega^{3,0}(\mathbb{PT},\mathcal{O}(4))\,.
\ee
The action \eqref{eq:TwistorActionlinear} leads to the equation of motion
\be
\bar{\partial}\Phi=0\,,
\ee
and has the gauge symmetry
\be
\label{eq:gaugetrafoscalar}
\Phi\mapsto \Phi+\bar{\partial}\chi\,.
\ee

Using \eqref{eq:TwistorPiEucl} gives a diffeomorphism
\be
\label{twistorspacetimeDiffeo}
\mathbb{PT}\cong S^2\times \mathbb{R}^4\,, 
\ee
and we can apply a standard Kaluza-Klein reduction to rewrite the theory \eqref{eq:TwistorActionlinear} as a theory on $\mathbb{R}^4$ with a priori infinitely many massive fields. 
Let us perform this reduction explicitly by following the argument of \cite{Costello:2021bah}. $\Phi$ can be decomposed into a singlet $\Phi_{\bar{z}}$ and a doublet $\Phi_{\dot{\alpha}}$ whose $(0,1)$-form part points in the $\mathbb{CP}^1$ or $\mathbb{C}^2$ directions respectively. $\Phi_{\bar{z}}$ and  $\Phi_{\dot{\alpha}}$ can then be expanded in fourier modes along $\mathbb{CP}^1$, where the components $\Phi_{\bar{z}}^{(2j)}$ transform in the spin $2j$ representation of SU$(2)$ with $j>0$ and $\Phi_{\dot{\alpha}}^{(2j+1/2)}$  transform in the  $2j+1/2$ representation of SU$(2)$ with $j>0$. The gauge variation $\chi$ can equally be expanded in Fourier modes $\chi^{(2j)}$ which importantly transform in integer representations with $j\geq1$ rather than $j\geq 0$ because the form degree is different to that of $\Phi$. 

The twistor action \eqref{eq:TwistorActionlinear} can then be written in terms of these Fourier modes and schematically reads
\be
\label{eq:SchematicLinearAction}
S[\Phi]=\sum_{j\geq 0} \int_{\mathbb{PT}} \Phi_{\dot{0}}^{(2j+1/2)} \Phi_{\dot{1}}^{(2j+1/2)}+ \Phi_{\dot{\alpha}}^{(2j+1/2)} D^{\dot{\alpha}} \Phi_{\bar{z}}^{(2j)}
\ee
where $D^{\dot{\alpha}}$ is a certain derivative on spacetime-coordinates (and not on $\mathbb{CP}^1$ coordinates). From \eqref{eq:SchematicLinearAction} we can immediately see that all the fields $\Phi_{\dot{\alpha}}^{(2j+1/2)}$ are auxilliary and can be integrated out leaving us with the tower of fields $\Phi^{(2j)}_{\bar{z}}$.

On $\Phi^{(2j)}_{\bar{z}}$, the gauge transformation \eqref{eq:gaugetrafoscalar} turns out to read 
\be
\delta \Phi_{\bar{z}}^{(2j)}=\chi^{(2j)}\,,
\ee
with $j\geq 1$, which means that all non-zero modes $\Phi_{\bar{z}}^{(2j)}$ with $j\geq 1$ are unphysical and can be set to $0$ by a gauge transformation. The only remaining physical field is the zero mode $\Phi_{\bar{z}}^{(0)}$ which can be viewed as the scalar field $\phi$ on spacetime. The remaining action \eqref{eq:SchematicLinearAction} then  turns out to read \cite{Costello:2021bah}
\be
S[\phi]=\int_{\mathbb{R}^4} \dif^4x \, \phi \Delta \phi\,,
\ee
which after integration by parts and choosing appropriate boundary conditions is equivalent to \eqref{eq:phiaction}.

This is the magic of local holomorphic theories on twistor space: If we were to consider a generic local field theory involving massless fields on $\mathbb{R}^4\times \mathbb{CP}^1$ and we KK-reduce it along the $\mathbb{CP}^1$ factor, then we would find a theory on $\mathbb{R}^4$ with a massless zero-mode as well as infinitely many massive KK-modes. In the IR, all of these massive fields get suppressed so that the massless $4$-dimensional theory is a low-energy effective description of the $6$-dimensional theory. On the other hand, if we begin with a holomorphic theory on twistor space such as \eqref{eq:TwistorActionlinear} with an infinite-dimensional gauge group such as \eqref{eq:gaugetrafoscalar}, then this KK-reduction is more robust and holds beyond the low-energy limit to all orders.

This perspective on the Penrose transform was discussed in the beautiful paper \cite{Costello:2021bah} where a mathematically rigorous proof of the above results is given using the language of \emph{factorization algebras} \cite{Costello:2021jvx, Costello:2023knl}. In this language, the $1:1$ correspondence \eqref{eq:linearPenrose}\footnote{and its restriction to arbitrary open subsets $U\subset \mathbb{R}^4$} can be viewed as an isomorphism of factorization algebras \cite{Costello:2021bah}.

Equation \eqref{eq:linearPenrose} can be extended from scalars to zero rest mass (z.r.m.) fields of arbitrary helicity $h$, where $h=0$ reduces to the scalar case. The only difference is that for arbitrary helicities the weight of the cohomology class is given by $2h-2$ as discussed in \cite{Adamo:2017qyl} 
\be
\label{eq:linearPenroseh}
\{\text{z.r.m. fields on }\mathbb{C}^4\text{ of helicity $h$}\}\,{\overset{\,\,1:1\,\,}\longleftrightarrow}\, H^{0,1}(\mathbb{PT},\mathcal{O}(2h-2))\,.
\ee
For helicities $h\in{\pm 1}$ and $h\in{\pm 2}$, this correspondence leads to photons/gluons and gravitons of positive or negative helicity, respectively. However, it turns out that it is possible to do better and obtain solutions to the equations of gauge theory and gravity beyond the linearised level. This will be the content of the rest of this chapter.

\subsection{Self-dual Yang-Mills theory} 
\label{section2.2}

Perhaps the most prominent example of a $4$-dimenional integrable system is the self-dual Yang-Mills equation \cite{belavin1978yang}
\be
\label{eq:SDYMequ}
F^-=0\,,
\ee
where $F=F^++F^-$ is the standard decomposition of the field strength two form into self-dual and anti-self-dual part
\be
\Omega^2(\mathbb{R}^4)=\Omega^2(\mathbb{R}^4)^+\oplus \Omega^2(\mathbb{R}^4)^-\,.
\ee

\subsection*{The Penrose-Ward correspondence}

The Penrose-Ward correspondence \cite{Ward:1977ta} gives a way to construct solutions to the equation \eqref{eq:SDYMequ} equation from twistor data. More precisely, the Penrose-Ward correspondence states a one-to-one correspondence between
\begin{enumerate}
    \item Solutions to the self-dual Yang-Mills equation \eqref{eq:SDYMequ} with gauge group GL$(n)$ considered up to gauge equivalence, and
    \item holomorphic vector bundles $E\rightarrow \mathbb{PT}$ of rank $n$ that are topologically trivial on each twistor line $\mathbb{CP}^1_x$. 
\end{enumerate}
This means that solutions to \eqref{eq:SDYMequ} correspond to free holomorphic data on twistor space which is the underlying reason for the integrability of \eqref{eq:SDYMequ}.

\subsection*{Twistor space is not Calabi-Yau}

Such holomorphic vector bundles on a general Calabi-Yau $3$-fold $X$ are known to arise from connections, whose partial connection $(0,1)$-forms $A\in \Omega^{0,1}(X, \mathfrak{g})$ obeys the condition
\be
0=F^{0,2}=\bar{\partial}A+[A\wedge A]\,.
\ee
Such connections arise as stationary points of the \emph{holomorphic Chern-Simons} action \cite{Witten:1992fb}
\be
S_{\textrm{hCS}}[A]=\int_X \Omega\wedge \tr \big(A\wedge \bar{\partial} A+ \tfrac{2}{3}A\wedge A\wedge A\big)\,,
\ee
where $\Omega\in \Omega^{3,0}(X)$ is the holomorphic Calabi-Yau volume form and $A\in \Omega^{0,1}(X, \mathfrak{g})$ is the partial connection $(0,1)$-form. As mentioned in chapter \ref{chapter1}, $\mathbb{PT}$ is not a Calabi-Yau manifold and there does not exist a globally defined $\Omega\in \Omega^{3,0}(\mathbb{PT})$. Using the holomorphic volume form of weight $4$
\be
\textrm{D}^3Z=\frac{1}{4!}\epsilon_{abcd}Z^a\dif Z^b\wedge \dif Z^c \wedge \dif Z^d\in \Omega^{3,0}(\mathbb{PT},\mathcal{O}(4))\,,
\ee
there are several ways to circumvent this problem. The first way was considered by Witten in the seminal paper \cite{Witten:2003nn} where holomorphic Chern-Simons theory was defined on the \emph{Calabi-Yau super manifold} $\mathbb{PT}^{3\vert 4}$ on which the opposite scaling of fermionic variables can be used to define a weightless holomorphic volume form
\be
\Omega=\textrm{D}^3Z \, \dif \psi_1 \, \dif \psi_2 \, \dif \psi_3 \, \dif \psi_4 \in \Omega^{3,0}(\mathbb{PT}^{3\vert 4})
\ee
where $\psi_i$ are fermionic coordinates. In the resulting holomorphic Chern-Simons action on super twistor space
\be
\label{eq:SupertwistorHCS}
S_{\textrm{hCS}}[A]=\int_{\mathbb{PT}^{3\vert 4}}\Omega \wedge  \tr \big(A\wedge \bar{\partial} A+ \tfrac{2}{3}A\wedge A\wedge A\big)\,,
\ee
the fermionic integrals can be performed which results in an $\mathcal{N}=4$ supersymmetric theory on twistor space which leads to an $\mathcal{N}=4$ completion of self-dual Yang-Mills theory in spacetime \cite{Boels:2006ir, Boels:2007qn}. The self-dual Yang-Mills equations occur from the first and last term in the fermionic expansion of $A$ which schematically looks like
\be
A(Z^a,\psi_i)=a(Z^a)+\dots+B(Z^a)\, \epsilon_{ijkl}\psi^i\psi^j\psi^k\psi^l\in\Omega^{0,1}(\mathbb{PT}^{3\vert 4},\mathfrak{g})\,.
\ee
Indeed, after integrating out the fermions, the action \eqref{eq:SupertwistorHCS} contains the term
\be
\label{eq:InterestingTermHCS}
S_{\textrm{hCS}}[A]=\int_{\mathbb{PT}}\textrm{D}^3Z  \, \tr \big(B\wedge F^{0,2}(a)+\dots \big)\,,
\ee
 which imposes $F^{0,2}(a)=0$ through the \emph{Lagrange multiplier} $B\in \Omega^{0,1}(\mathbb{PT},\mathcal{O}(4))$. To clean up the notation, we define a new Lagrange multiplier field
\be
b=\textrm{D}^3Z  \, \tr(B, \,-\,) \in \Omega^{3,1}(\mathbb{PT},\mathfrak{g}^\vee). 
\ee
in terms of which the action \eqref{eq:InterestingTermHCS} simply reads
\be
\label{eq:firstterminSUSYHCS}
S_{\textrm{hCS}}[A]=\int_{\mathbb{PT}}b\wedge F^{0,2}(a) +\dots\,
\ee

There are alternatives to get around the problem that $\mathbb{PT}$ is not Calabi-Yau without supersymmetry which include choosing some divisor in $\mathbb{PT}$ on which $\Omega$ is allowed to have poles such as in the Burns holography context \cite{Bittleston:2022cmz, Bittleston:2020hfv, Costello:2021bah, Cole:2023umd, Cole:2024sje}  we briefly discussed in section \ref{section1.4}
\be
\Omega=\frac{\textrm{D}^3Z}{\langle\alpha\lambda \rangle^2 \langle\lambda \beta \rangle^2}  \in \Omega^{3,0}\big(\mathbb{PT}\setminus(\mathbb{C}^2_\alpha\cup \mathbb{C}^2_\beta)\big) \,.
\ee
$\mathbb{C}^2_\alpha$ and $\mathbb{C}^2_\beta$ hereby denote the fibers of $\mathcal{O}(1)\oplus\mathcal{O}(1)\rightarrow \mathbb{CP}^1$ over the points $\alpha,\beta\in \mathbb{CP}^1$.
It is also possible to consider some bundle over $\mathbb{PT}$ whose total space is a Calabi-Yau manifold. For instance, 
\be
\mathcal{O}(-1)\oplus\mathcal{O}(-3)\rightarrow \mathbb{PT}
\ee
is a Calabi-Yau $5$-manifold which was considered as the target space of a B-model and leads to many new top-down examples of holography in asymptotically flat spacetimes \cite{Bittleston:2024rqe}.

\subsection*{Holomorphic BF theory}

In this thesis, we will not be working with the supersymmetric theory \eqref{eq:SupertwistorHCS} but rather with its truncation that only involves the first term of \eqref{eq:firstterminSUSYHCS}. This theory is known as \emph{holomorphic BF theory} and one can obtain it formally from \eqref{eq:SupertwistorHCS} via quotienting by the discrete $\mathbb{Z}_4$-action on $\mathbb{PT}^{3\vert 4}$ given by
\be
\Psi_i\mapsto e^{\tfrac{\pi \im}{2}}\Psi_i\,.
\ee
We will view this 'derivation' of holomorphic BF theory from the string field theory \eqref{eq:SupertwistorHCS} as a vague motivation but from now on, we will consider the holomorphic BF theory in its own right.

The field content of holomorphic BF theory is given by the two fields 
\bea
a\in \Omega^{0,1}(\mathbb{PT},\mathfrak{g})\,,\qquad 
b\in \Omega^{3,1}(\mathbb{PT},\mathfrak{g}^\vee)\,,
\eea
and its action is defined by 
\be
\label{eq:HoloBFaction}
S_{\textrm{hBF}}[a,b]=\int_{\mathbb{PT}}\tr\big( b\wedge F^{0,2}(a)\big)\,.
\ee
The action \eqref{eq:HoloBFaction} is invariant under two gauge transformations. The first one is the usual one under which $F^{0,2}(a)$ transforms in the adjoint
\begin{subequations}
\label{subeq:GaugeTrafosYM}
\bea
\delta a= \bar{\partial} c+ [a,c]\,,\qquad
\delta b=  [b,c]\,,
\eea
where $c\in \Omega^{0,0}(\mathbb{PT},\mathfrak{g})$ is the gauge parameter. The second set of gauge transformations only acts on $b$ and depends on a gauge parameter  $d\in \Omega^{3,0}(\mathbb{PT},\mathfrak{g}^\vee)$
\bea
\label{eq:secondGaugeTrafo}
\delta a= 0\,,\qquad
\delta b=\bar{\partial}d+  [a,d]\,.
\eea
\end{subequations}
\eqref{eq:secondGaugeTrafo} leaves the action \eqref{eq:HoloBFaction} invariant due to the Bianchi identity.

The action \eqref{eq:HoloBFaction} leads to solutions of the SDYM equation \eqref{eq:SDYMequ} since on-shell we have
\be
(\bar{\partial}+a)^2=F^{0,2}(a)=0
\ee
so that $a$ defines a holomorphic vector bundle on $\mathbb{PT}$ and hence a solution to \eqref{eq:SDYMequ} by the Penrose-Ward correspondence. This \emph{on-shell} statement holds in any signature. 

Moreover, showing that the action \eqref{eq:HoloBFaction} (on-shell as well as off-shell) is equivalent to a spacetime action describing SDYM is possible in Euclidean signature where twistor space fibres over spacetime by the map \eqref{eq:TwistorPiEucl}  \cite{Boels:2006ir, Mason:2005zm}. This is analogous to the off-shell reduction of the free scalar we discussed in section  \ref{sectionLinearPenrose}.

\medskip

The twistor action \eqref{eq:HoloBFaction} is chiral so that it is expected to suffer from a gauge anomaly which is indeed the case \cite{Costello:2021bah}. This anomaly can be cancelled through the inclusion of additional fields in which case the twistor description \eqref{eq:HoloBFaction} also holds at the quantum level \cite{Costello:2021bah}. While the twistorial anomaly is a fundamental feature in the quantization of holomorphic field theories on twistor space, this thesis will mostly be concerned with tree-level results.

\subsection{Self-dual gravity} 
\label{section2.3}

\subsection*{The non-linear graviton construction}

Even though up until this point we only described the flat twistor space $\mathbb{PT}$, twistor theory can be extended to self-dual curved spacetimes through the non-linear graviton construction \cite{Penrose:1976jq, Penrose:1976js, Atiyah:1978wi}. Similarly to the Penrose-Ward correspondence we discussed in section \ref{section2.2} and the linear Penrose transform we discussed in section \ref{sectionLinearPenrose}, the non-linear graviton construction relates holomorphic data on twistor space to conformal data on spacetime. The big difference is that now, the twistor data does not consist of some holomorphic data in a fixed complex structure. Rather, the holomorphic data on twistor space consists of the complex structure itself. Deforming the complex structure on $\mathbb{PT}$ in a controlled way will correspond to deforming the (conformal class of the) flat metric on $\mathbb{R}^4$. More precisely, the non-linear graviton construction \cite{Penrose:1976jq, Penrose:1976js, Atiyah:1978wi} consists of the $1:1$ correspondence between

\begin{enumerate}
    \item Four-dimensional manifolds $M$ together with a conformal class $[g]$ of Riemannian metrics with self-dual Weyl curvature
    \be
    \mathcal{W}^-=0\,,
    \ee
    and
    \item Complex $3$-manifolds $\mathcal{PT}$ that possess at least one rational curve $\mathcal{L}_x\cong \mathbb{CP}^1$ with normal bundle $N=\mathcal{O}(1)\oplus \mathcal{O}(1)$, together with an antiholomorphic involution $\sigma:\mathcal{PT}\rightarrow \mathcal{PT}$ that acts as the antipodal map on $\mathcal{L}_x$.
\end{enumerate}
A more detailed discussion and a proof can be found in \cite{Penrose:1976jq, Penrose:1976js, Atiyah:1978wi}. Briefly summarizing this, Kodaira theory \cite{Kodaira:1962, Kodaira:1963} ensures that once there is a single such rational curve  $\mathcal{L}_x$, then there is a four-parameter family of them, and their moduli space $M_\mathbb{C}$ is the complexification
of the curved spacetime $M$. To recover $M$ from $M_\mathbb{C}$, one has to consider only those twistor lines that are fixed by $\sigma$. 

Since in perturbative QFT, we would like to consider small perturbations around a background $\mathbb{R}^4$, we would also like to consider self-dual spacetimes $M$ whose twistor space arises as a perturbation of $\mathbb{PT}$. This is formalized by \emph{variation of complex structure}. The complex structure of a complex manifold is determined in some local coordinates $\{z^i\}$  by its Dolbeaut operator \be
\bar{\partial}=\dif \bar{z}^{\bar{i}}\frac{\partial}{\partial \bar{z}^{\bar{i}}}
\ee
which can be deformed by a so-called \emph{Beltrami differential}
\be
V=V^j_{\bar{i}}\dif \bar{z}^{\bar{i}}\frac{\partial}{\partial z^{j}}\,,
\ee
via
\be
\label{eq:DeformbyBeltrami}
\bar{\partial}\mapsto \bar{\nabla}= \bar{\partial}+\mathcal{L}_V=\dif \bar{z}^{\bar{i}}\Big(\frac{\partial}{\partial \bar{z}^{\bar{i}}}+V^j_{\bar{i}} \frac{\partial}{\partial z^{j}}\Big) \,,
\ee
where $\mathcal{L}_V$ denotes the Lie-derivative along $V$. For a generic $V$ this is merely an almost complex structure since $\bar{\nabla}$ might not be integral, i.e. $\bar{\nabla}^2\neq 0$. $\bar{\nabla}$ is integrable if and only if its Nijenhuis tensor 
\be
N=\frac{1}{2}[\bar{\nabla},\bar{\nabla}]=\bar{\partial}V+\frac{1}{2}[V,V]
\ee
vanishes. As we discussed above in section \ref{sectionLinearPenrose}, the twistor correspondence is generally conformally invariant which is reflected in the non-linear graviton construction by the fact that we only obtain a conformal class of metrics.

\subsection*{Holomorphic Poisson BF theory}

If we wish to obtain an actual Einstein metric with a self-dual Weyl tensor rather than a whole conformal class of metrics with a self-dual Weyl tensor we need to break the conformal invariance on twistor space somehow. If the curved twistor space admits a fibration
\be
\mathcal{PT}\rightarrow\mathbb{CP}^1
\ee
with an $\mathcal{O}(2)$-valued holomorphic symplectic form on the fibres,  then we obtain a unique Ricci-flat representative $g\in[g]$. This further data can also be seen as an $\mathcal{O}(-2)$-valued holomorphic Poisson bracket
\be
\label{eq:PoissonGeneralInfty}
\{\,\,,\,\,\}= I^{ab}\frac{\partial}{\partial Z^a}\frac{\partial}{\partial Z^b}\,,
\ee
where $I^{ab}=I^{[ab]}$ is the \emph{infinity twistor}. It changes the output of the non-linear graviton construction from solutions of the self-dual \emph{conformal gravity} equations to solutions of the self-dual \emph{Einstein gravity} equations. In the case of flat space, $\mathbb{PT}$ fibres over $\mathbb{CP}^1$ by definition and the $\mathcal{O}(2)$-valued holomorphic symplectic form on the fibres is simply given by
\be
\omega=\tfrac{1}{2}\dif \mu^\da \wedge\dif \mu_\da\,.
\ee
Generic deformations of the form \eqref{eq:DeformbyBeltrami} will not preserve $\omega$. However, such a deformation will preserve $\omega$, if $V$ is Hamiltonian in the sense that there exists some $h\in \Omega^{0,1}(\mathbb{PT},\mathcal{O}(2))$ such that
\be
V=\{h,\,\,\}=\epsilon^{\da \db}\big( \mathcal{L}_{\partial_\da} h\big) \partial_\db\,.
\ee
Here, $\{-,-\}=\omega^{-1}$ is the holomorphic Poisson bracket of weight $-2$ on the fibres, explicitly defined by
\be
\label{eq:PoissonFlatspace}
\{f,g\}=\epsilon^{\da \db}\mathcal{L}_{\partial_\da}f\wedge\mathcal{L}_{\partial_\da}g
\ee
 for general $(p,q)$-forms $f,g$ on $\mathbb{PT}$ and we defined 
\be
\partial_\da=\frac{\partial}{\partial \mu^\da}\,.
\ee
The Poisson bracket \eqref{eq:PoissonFlatspace} is of the form \eqref{eq:PoissonGeneralInfty}, with the infinity twistor given by\footnote{In chapter \ref{chapter5}, we will see a generalization of this including a non-zero cosmological constant.}
\begin{equation}
I^{ab}=\left(\begin{array}{cc}
     \varepsilon^{\dot\alpha\dot\beta}&0  \\
     0&0 
\end{array}\right)\,.
\end{equation}

In the case of a Beltrami differential, we will usually suppress the Lie-derivatives and write
\be
\label{eq:Beltramifromh}
V=\{h,\,\,\}=\epsilon^{\da \db}\big( \partial_\da h\big) \partial_\db\,,
\ee
for $h\in \Omega^{0,1}(\mathbb{PT},\mathcal{O}(2))$, which leads to a weightless $V$ since $\{-,-\}$ has weight $-2$. For a Beltrami differential of this form \eqref{eq:Beltramifromh}, the Nijenhuis tensor takes the form
\be
N^{0,2}=(\bar{\partial}+\{h,\,\,\})^2=\{T^{0,2}(h),\,\,\}\,,
\ee
where
\be
T^{0,2}(h)=\bar{\partial}h+\frac{1}{2}\{h,h\}\in \Omega^{0,2}(\mathbb{PT},\mathcal{O}(2))\,.
\ee

\subsection*{Holomorphic Poisson BF theory}

The condition that a Hamiltonian complex structure deformation is integrable, i.e.
\be
\label{eq:IntegrabilityofHamBeltrami}
0=T^{0,2}(h)=\bar{\partial}h+\frac{1}{2}\{h,h\}\in \Omega^{0,2}(\mathbb{PT},\mathcal{O}(2))\,,
\ee
can be imposed by a BF-type action on twistor space similar to the case of Yang-Mills theory in section \ref{section2.2}. Such an action was first identified in \cite{Mason:2007ct} and in analogy to the Yang-Mills case it can be viewed as a non-supersymmetric truncation of a twistor theory describing $\mathcal{N}=8$ supergravity \cite{Skinner:2013xp}.  The corresponding theory is a twistor description of $4$-dimensional self-dual gravity and we will refer to it as \emph{holomorphic Poisson BF theory}. Its action reads
\begin{equation}
\label{eq:twistor-sd-gravity-action}
S[g,h] = \int_\PT\Dif^3Z\wedge g\wedge \left(\bar\p h + \frac{1}{2}\{h,h\}\right),
\end{equation}
for fields $h\in\Omega^{0,1}(\PT,\cO(2))$ and $g\in \Omega^{0,1}(\PT,\cO(-6))$. 

The action~\eqref{eq:twistor-sd-gravity-action} is invariant under Hamiltonian diffeomorphisms of twistor space, acting as
\begin{subequations}
\label{eq:gravityGauge}
\begin{equation}
    \label{eq:Ham-diffs-twistor}
    \delta h = \bar\p\chi + \{h,\chi\}\,,\qquad\qquad \delta g = \{g,\chi\}
\end{equation}
for a smooth function $\chi\in\Omega^0(\PT,\cO(2))$, as well as the transformations
\begin{equation}
\label{eq:g-field-trans-twistor}
\delta g = \bar\p\xi +\{h,\xi\}\,,\qquad\qquad\delta h=0\,,
\end{equation}
\end{subequations}
 with $\xi\in\Omega^{0}(\PT,\cO(-6))$. These gauge transformations are analogous to equations \eqref{subeq:GaugeTrafosYM} in the gauge theory case. 

The field equations that follow from~\eqref{eq:twistor-sd-gravity-action} are the integrability condition \eqref{eq:IntegrabilityofHamBeltrami} and 
\be
\label{eq:HoloPoissonBFgeq}
\bar\p g + \{h,g\}=0\,.
\ee
Using Penrose's non-linear graviton construction, solutions to the first equations lead to solutions of the self-dual Einstein equations with vanishing cosmological constant. The second equation \eqref{eq:HoloPoissonBFgeq} means that the
field $g$ then represents a massless field of helicity\footnote{In our conventions, the self-dual background can be viewed as a coherent state of gravitons of helicity $+2$.} $-2$ propagating on this self-dual background via the linear Penrose transform of section \ref{sectionLinearPenrose}.

Including a non-vanishing cosmological constant leads to a variation of the non-linear graviton construction \cite{LeBrun:1982vjh, Salamon:1982,  Hitchin:1982vry, Ward:1980am}  and holomorphic Poisson BF theory \cite{Mason:2007ct}. We will discuss these and their application to celestial chiral algebras below in chapter \ref{chapter5}.

In similarity to the gauge theory case \eqref{eq:HoloBFaction}, the self-dual gravity action \eqref{eq:twistor-sd-gravity-action} is also chiral so that it suffers from a similar anomaly. Once again, this anomaly can be cancelled through the inclusion of additional fields \cite{Bittleston:2022nfr}.

In \cite{Bittleston:2022nfr, Sharma:2021pkl}, it was explicitly shown that the action \eqref{eq:twistor-sd-gravity-action} is classically equivalent to 
\be
S[\Psi, \Sigma, \Gamma]=\int_{\mathbb{R}^4} \Sigma^{\alpha \beta}\wedge \dif \Gamma_{\alpha \beta}+\frac{1}{2}\Psi_{\alpha \beta \gamma \delta}\Sigma^{\alpha\beta}\wedge\Sigma^{\gamma\delta}
\ee
which describes self-dual Einstein gravity with $\Lambda=0$ in $\mathbb{R}^4$. $\Psi$ can be viewed as a Lagrange multiplier that imposes the simplicity constraint
\be
\Sigma^{(\alpha \beta}\Sigma^{\gamma \delta)}=0\,,
\ee
which is equivalent to the existence of some vierbein $1$-forms $\{e^{\da\alpha}\}$ such that 
\be
\Sigma^{\alpha \beta}=e^{\da\alpha}\wedge e_\da^{\,\beta}\,.
\ee
Integrating this Lagrange multiplier out then results in the action 
\be
\label{eq:AlmostChalmersSiegel}
S[e,\Gamma]=\int_{\mathbb{R}^4} e^{\da\alpha}\wedge e_\da^{\,\beta} \wedge \dif \Gamma_{\alpha \beta}\,,
\ee
which can be seen as the $\kappa \rightarrow 0$ limit of the Pleba\'{n}ski action for full Einstein gravity \cite{Smolin:1992wj, Capovilla:1991qb, Ashtekar:1987qx, Plebanski:1977zz}. The action \eqref{eq:AlmostChalmersSiegel} is in fact equivalent to the Chalmers-Siegel action \eqref{Chalmers-Siegel}, that we used in chapter \ref{chapter2}. 

%% file: Chapter5/chapter5.tex
\chapter{Celestial chiral algebras on Eguchi-Hanson space}
\label{chapter4}

In chapter \ref{chapter1} we discussed that Burns holography has the unphysical feature of involving a version of self-dual conformal gravity, Mabuchi gravity, in the bulk. This chapter will provide a first step towards finding a similar holographic duality for self-dual Einstein gravity with $\Lambda=0$ in the bulk as suggestively displayed in figure \ref{fig:EHholography}. An essential step in Burns holography was played by the backreaction of D$1$-branes. We will formally repeat this backreaction in the twistor description of self-dual Einstein gravity. There, however, we are not necessarily working with a string theory so it is not sensible to talk about D$1$-branes. We will instead view the sources of our backreaction as a \emph{defect operator}. 

\medskip

This chapter is based on the paper \cite{Bittleston:2023bzp} and it is organized as follows: In section~\ref{sec:deforming-twistor-space}, as in twisted holography~\cite{Costello:2018zrm} and the top-down celestial holographic model of~\cite{Costello:2022jpg}, we couple a defect wrapping a complex curve $\CP^1\subset\PT$ to the holomorphic Poisson BF theory discussed in the previous section \ref{section2.3}. We show that the backreaction sourced by this defect deforms $\PT$ to the twistor space $\cPT$ of Eguchi-Hanson space. In an effort to keep this chapter self-contained, we review the important features of $\cPT$, including its projection to $\CP^1$, the weight 2 symplectic structure on the fibres of this projection, and the 4-parameter family of holomorphic sections, from which the space-time and Eguchi-Hanson metric may be recovered. In section~\ref{sec:twistor-algebra} we use the twistor space to construct (at the classical level) the celestial chiral algebra (CCA) of self-dual gravity on Eguchi-Hanson space, identifying this with the loop algebra of the wedge subalgebra of a scaling limit of $W(\mu)$, as discussed above. We also consider the CCA of self-dual Yang-Mills for a complex semisimple Lie algebra $\fg$. On flat space its CCA is $\cL\fg[\bbC^2]$, the loop algebra of the Lie algebra of polynomial maps from $\bbC^2$ into $\fg$. We show that on Eguchi-Hanson this is deformed to an algebra we denote $S_\wedge(\infty)$, the detailed structure of which is given in section \ref{subsec:twistor-CCA-SDYM}.

The results of section~\ref{sec:twistor-algebra} essentially follow just from the ring of holomorphic functions on the fibres of $\cPT\to\CP^1$. In section~\ref{sec:space-time-algebra} we recover the same algebras by calculating, entirely on space-time, the gravitational and Yang-Mills splitting functions for scattering on the Eguchi-Hanson background. In section~\ref{sec:non-commutativity}, we allow the twistor space to become non-commutative, showing that the CCA now corresponds to the loop algebra of the wedge subalgebra of a generic $W(\mu)$ algebra, with $\fq$ governing the non-commutativity and $\fq^2\mu$ determined by the curvature scale set by the Eguchi-Hanson space bulk.  We conclude in section~\ref{sec:discussion} with a brief discussion of some open directions.

\section{Deforming twistor space with a defect operator}
\label{sec:deforming-twistor-space}

Inspired by the twisted holography of Costello \& Gaiotto~\cite{Costello:2018zrm}, we consider the effect of introducing a defect into the twistor space description of self-dual gravity \eqref{eq:twistor-sd-gravity-action}. We choose to consider a defect that couples electrically to $g$, so take the action to be
\begin{equation}
    \label{eq:twistor-defect-action}
    S[g,h] =  \int_\PT\Dif^3Z\wedge g\wedge \left(\bar\p h + \frac{1}{2}\{h,h\}\right) - \frac{\pi^2c^2}{2} \int_{\CP^1}\la\lambda\,\dif\lambda\ra\wedge(\la\al\lambda\ra\la\lambda{\beta}\ra)^2\,  g \,,
\end{equation}
where the final term describes a defect wrapping the zero section $\mu^{\dot\alpha}=0$ of $\PT\to\CP^1$.  In this term, $c^2$ is a real coupling constant that measures the strength of the coupling to the defect. Since $g$ has homogeneity $-6$, the electrical coupling requires that we pick a holomorphic function of homogeneity $4$ on the $\CP^1$ defect; we chose this to be the square of,\footnote{By Pontecorvo's theorem~\cite{Pontecorvo:1992twi} twistor spaces admitting such a holomorphic function correspond to space-times with a preferred scalar-flat K{\"a}hler metric if $\beta=\hat{\alpha}$.} $\la\al\lambda\ra\la\lambda{\beta}\ra$ where $\{|\al\ra,\,|\beta\ra\}$ are an arbitrary dyad normalized so that $\la\al\beta\ra=1$.  We will use the notation
\begin{equation}
    \label{eq:p(lambda)-def}
    c(\lambda) = c\,\la\alpha\lambda\ra\la\lambda\beta\ra\,,
\end{equation} 
for later convenience.

In the presence of this defect, the equation of motion for $h$ becomes
\begin{equation}
\label{eq:deformed}
\bar\p h + \frac{1}{2}\{h,h\} = 2\pi^2 c^2(\lambda)\,\bar\delta^2(\mu)\,,
\end{equation}
where $\int \dif \mu^{\dot0}\wedge\dif\mu^{\dot1}\wedge\bar\delta^2(\mu)=1$. The equation of motion for $g$ itself is unaffected. This sourced equation is solved by
\begin{equation}
\label{eq:h-in-terms-of-phi}
 h = \frac{c^2(\lambda)}{2} \frac{[\hat\mu\,\dif\hat\mu]}{\,[\mu\,\hat\mu]^2} \,.
\end{equation}
To see this, first notice that
\begin{equation}
\label{eq:Poisson-phi}
\{h, \ \,\} = -c^2(\lambda) \,\frac{[\hat\mu\,\dif\hat\mu]}{\,[\mu\,\hat\mu]^3} \,\hat{\mu}^\da\cL_{\dot\alpha}
\end{equation}
so that $\{h,h\}=0$, both because $[\hat\mu \,\dif\hat\mu]\wedge[\hat\mu\,\dif\hat\mu]=0$ and because $\hat\mu^{\da}\cL_\da h= 0$. Now let
 \begin{equation}
\label{eq:phiEH}
\phi =  \frac{1}{4\pi^2}\frac{[\hat\mu\,\dif\hat\mu]}{\,[\mu\,\hat\mu]^2}\in\Omega^{0,1}(\PT,\cO(-2))
\end{equation}  
so $h=2\pi^2c^2(\lambda)\phi$. Then provided $[\mu\,\hat\mu]\neq0$ we have
\begin{equation}
\begin{aligned}
\bar\p\phi &= \frac{1}{4\pi^2}\left(\frac{1}{[\mu\,\hat\mu]^2}\,[\dif\hat\mu\wedge\dif\hat\mu] - 2\frac{[\mu\,\dif\hat\mu]\wedge [\hat\mu\,\dif\hat\mu]}{[\mu\,\hat\mu]^3}\right)\,\\
&= \frac{1}{4\pi^2}\left(\frac{1}{[\mu\,\hat\mu]^2}\,[\dif\hat\mu\wedge\dif\hat\mu] - \frac{[\mu\,\hat\mu]\,[\dif\hat\mu\wedge\dif\hat\mu]}{[\mu\,\hat\mu]^3}\right) =0\,.
\end{aligned}
\end{equation}
Since $[\mu\,\hat\mu]=|\mu^{\dot{0}}|^2 + |\mu^{\dot1}|^2$,
we see that $\bar\p h + \frac{1}{2}\{h,h\}$ vanishes away from the zero section $\mu^\da=0$. Furthermore, since $\phi$ is a $(0,1)$-form of homogeneity $-2$, we must have $\bar\p\phi \propto \bar\delta^2(\mu)$. We can fix the normalization by integrating $\phi$ over the $S^3$ given by $[\mu\,\hat\mu]=1$ at constant $\lambda$. On this sphere we have
\begin{equation}
\begin{aligned}
  \frac{1}{2}\int_{S^3}[\dif\mu\wedge\dif\mu]\wedge\phi 
  &= \frac{1}{8\pi^2}\int_{S^3}[\dif\mu\wedge\dif\mu]\wedge [\hat\mu\,\dif\hat\mu] 
  \\
  &= \frac{1}{8\pi^2}\int_{B}[\dif\mu\wedge\dif\mu]\wedge[\dif\hat\mu\wedge\dif\hat\mu]
  = \frac{2}{\pi^2}\,\mathrm{Vol}(B) = 1\,.
\end{aligned} 
\end{equation}
where $B$ is a unit 4-ball in the $\mathbb{C}^2$ fibre. Thus $\phi$  is correctly normalized to obey $\bar\p\phi=\bar\delta^2(\mu)$, while the corresponding $h$ obeys~\eqref{eq:deformed}.

\medskip

Let us remark that the field~\eqref{eq:phiEH} is essentially identical to the one used in section 4 of~\cite{Costello:2018zrm} in the context of twisted holography for the topological string. More precisely, the relation is
\begin{equation}
    \eta = \frac{1}{2}\phi\wedge[\dif\mu\wedge\dif\mu]\in \Omega^{2,1}_\mathrm{cl}(\PT)\,,
\end{equation}
where $\eta$ is the closed string field of the topological B model, and $\phi$ is as above. In that context, $\eta$ is related to a Beltrami differential by $\eta = V\ip \Omega$, where $\Omega$ is the holomorphic (3,0)-form on $\bbC^3$. There, the backreacted geometry sourced by the defect deforms $\bbC^3$ to $\gSL(2,\bbC)\cong\mathrm{AdS}_3\times S^3$ as we saw in chapter \ref{chapter1}. Similarly, the same $\phi$ plays an important role in the asymptotically flat holography of~\cite{Costello:2022jpg}, where it deforms $\PT$ to the twistor space of Burns space~\cite{Burns:1986,LeBrun:1991}, a particular scalar-flat K{\"a}hler manifold. However, in our case the relation between the Beltrami differential and the field $\phi$ is different: instead of the B-model relation $\eta = V\ip \Omega$, we have $V = \{h,\ \} = 2\pi^2c^2(\lambda)\,\{\phi,\ \}$ using the Poisson structure. In addition, here the defect acts as a source for $\bar\p h+\frac{1}{2}\{h,h\}$ rather than $\p^{-1}\eta$. This means that the deformed twistor space we obtain will be very different, in particular corresponding to the twistor space of a self-dual Ricci-flat manifold.

\medskip

In~\cite{Costello:2018zrm,Costello:2022jpg}, the defect was interpreted as a D1-brane of the topological B model, wrapping a holomorphic curve. This D1-brane supports a chiral algebra that is holographically dual to the bulk theory. It would clearly be very interesting to find a string theory realisation of the bulk theory and defect of the present work.


\subsection*{The Eguchi-Hanson twistor space}
\label{sec:EH-twistor}

We'll now show that the solution~\eqref{eq:h-in-terms-of-phi}-\eqref{eq:phiEH} implies that the effect of the backreaction of the defect causes $\PT$ to be deformed to the twistor space of Eguchi-Hanson space. 

\medskip

We seek coordinates that are holomorphic with respect to the deformed Dolbeault operator 
\be \label{eq:nabla-bar}
\bar\nabla = \bar\p + \{h,\ \} = \bar\p - c^2(\lambda)\,\frac{[\hat\mu\,\dif\hat\mu]}{\,[\mu\,\hat\mu]^3}\,\hat{\mu}^{\dot\gamma} \cL_{\dot\gamma}\,.
\ee
By construction the $\lambda_\al$ are still holomorphic, so as mentioned above our curved twistor space $\mathcal{PT}$ admits a holomorphic fibration over $\CP^1$. However, the coordinates $\mu^{\da}$ are no longer holomorphic. 

Instead, consider the three functions 
\be \label{eq:newholomorphic}
X^{\da\dot{\beta}} = X^{(\da\dot\beta)} = \mu^{\da}\mu^{\dot\beta} - c^2(\lambda)\,\frac{\hat{\mu}^{\da}\hat{\mu}^{\dot\beta}}{[\mu\,\hat\mu]^2} \,,
\ee
each of homogeneity $+2$. We find
\bea
\bar\nabla X^{\da\dot\beta} &= -c^2(\lambda)\,\bar\p\left(\frac{\hat{\mu}^{\da}\hat{\mu}^{\dot\beta}}{[\mu\,\hat\mu]^2}\right) - c^2(\lambda)\,\frac{[\hat\mu\,\dif\hat\mu]}{\,[\mu\,\hat\mu]^3}\left(\mu^\da\hat\mu^{\dot\beta}+\hat{\mu}^{\da}\mu^{\dot\beta}\right) \\
&= - c^2(\lambda)\,[\mu\,\hat\mu]^3\left[ \left([\mu\,\hat\mu]\,\dif\hat{\mu}^\da -[\mu\,\dif\hat{\mu}] \,\hat{\mu}^\da +[\hat\mu\,\dif\hat\mu]\,\mu^\da\,\right)\hat{\mu}^{\dot\beta} + (\da\leftrightarrow\db)\,\right] \\
&=0
\eea
as a consequence of the Schouten identity. Thus we can take $(X^{\da\dot\beta},\lambda_\al)$ as holomorphic coordinates on the deformed twistor space. These coordinates are not all independent, but are subject to the scaling relations $(r^2 X,r\lambda)\sim(X,\lambda)$ for $r\in\mathbb{C}^*$, as well as the constraints
\begin{equation}
\label{eq:def-twistor-X-coords}
    X^{\da\dot\beta}X_{\da\dot\beta} = -2c^2(\lambda)\,.
\end{equation}
Equivalently, setting $X^{\dot{0}\dot{0}} =X$, $X^{\dot{1}\dot{1}}=Y$ and $X^{\dot{0}\dot{1}} = X^{\dot{1}\dot{0}} = Z$, this may be written as 
\begin{equation}
\label{eq:twistor-constraint}
    XY = \left(Z-c(\lambda)\right)\left(Z+c(\lambda)\right)\,.
\end{equation}
In other words, the backreaction deforms the twistor space to the subvariety of the total space of 
\[
\cO(2)\oplus\cO(2)\oplus\cO(2)\to\CP^1
\] 
defined by these equations. This is essentially\footnote{$c(\lambda)$ vanishes when $\lambda=\al$ or $\lambda=\beta$, so these two fibres remain singular. A more precise description of the twistor space of Eguchi-Hanson space involves resolving these remaining singularities by blowing up a $\CP^1$ in each of these two fibres. See~\cite{Hitchin:1979rts} for more details.}  the twistor space of Eguchi-Hanson space~\cite{Eguchi:1978xp}. The novelty here is that we have obtained it as the backreaction of the action~\eqref{eq:twistor-sd-gravity-action} from the presence of a defect inserted in flat twistor space $\mathbb{PT}$. Strictly speaking, we obtain the twistor space of Eguchi-Hanson space from a backreaction on $\mathbb{PT}/\mathbb{Z}_2$, the twistor space of the flat orbifold $\mathbb{R}^4/\mathbb{Z}_2$. This will be discussed in detail below and the backreaction is schematically depicted in figure \ref{fig:EHspacePic}. Without this $\mathbb{Z}_2$-quotient, the backreacted twistor space would be a singular double cover of the twistor space of Eguchi-Hanson space. The same double cover of Eguchi-Hanson space appears in \cite{Bittleston:2024efo}.


\subsection*{The space-time metric}
\label{subsec:EH-space-time}

Here we briefly review how the twistor space above corresponds to Eguchi-Hanson space-time; see \emph{e.g.}~\cite{Hitchin:1979rts,Burnett:1979st,Sparling:1981nk,Tod:1982mmp} for further details.

Recall that we saw in section \ref{section2.1} how complexified space-time $\cM_\bbC$ arises as the space of holomorphic sections of $\cPT\to\CP^1$, \emph{i.e.}, holomorphic sections of $\cO(2)\oplus\cO(2)\oplus\cO(2)\to\CP^1$ obeying the  constraint~\eqref{eq:twistor-constraint}. For example, we may describe a section by the incidence relations
\be
\label{eq:incidence}
X^{\dot\al\dot\beta} = x^{\dot\alpha\alpha}x^{\dot\beta\beta}\left(\lambda_\alpha\lambda_\beta - \frac{4c^2\la\al\lambda\ra^2}{x^4} {\beta}_\alpha{\beta}_\beta\right)
\ee
where $x^{\da\al}$ will turn out to be Kerr-Schild coordinates on Eguchi-Hanson space\footnote{The incidence relation \eqref{eq:incidence} breaks the symmetry $\al\leftrightarrow{\beta}$. (We could of course have chosen to break it in the other direction.) 
It is possible to write incidence relations
\[
X^{\dot\al\dot\beta} = 2\im c\frac{x^{\dot\alpha\alpha}x^{\dot\beta\beta}}{x^2} \left(\al_\al\al_\beta\la\beta\lambda\ra^2 + \beta_\al\beta_\beta\la\al\lambda\ra^2\right)
\]
that obey the hypersurface constraint and preserve this symmetry, but the flat space limit is less obvious here. For this reason, we prefer to use the symmetry-breaking case~\eqref{eq:incidence}. One consequence is that, even if we choose $|\beta\ra=|\hat\al\ra$ as is natural in Euclidean signature, the real Euclidean structure of the metric is not manifest in Kerr-Schild form. Instead, we can give it a natural ultrahyperbolic structure if we take the spinors $\alpha,\beta$ to each be real.}. The incidence relations~\eqref{eq:incidence} are easily verified to obey $X^{\dot\al\dot\beta}X_{\dot\al\dot\beta} = -2c^2(\lambda)$. In addition, we see that as either the defect coupling $c\to0$ or as $|x|\to\infty$ they reduce to the flat space incidence relations  $\mu^\da\mu^\db = x^{\da\al}x^{\db\beta}\lambda_\al\lambda_\beta$ for the orbifold $\bbR^4/\bbZ_2$.  It will sometimes be useful to write~\eqref{eq:incidence} as $X^{\da\db} = M^{(\da}_+(\lambda)\,M_-^{\db)}(\lambda)$ in terms of 
\be
M^{\da}_\pm(\lambda) = x^{\da\al}\lambda_\al \pm \frac{2c\la\al\lambda\ra}{x^2}\,x^{\da\al}\beta_\al\,.
\ee
The incidence relation describes a holomorphic curve $\CP^1\subset\cPT$ and the conformal structure of the corresponding space-time is given by declaring that two points $x,y\in\cM_{\bbC}$ are null separated iff the corresponding curves $\CP^1_x$ \& $\CP^1_y$ intersect. To fix the scale, we also need the $\cO(2)$-valued symplectic structure on the fibres of $\cPT\to\CP^1$. In terms of the coordinates $(X,Y,Z)$ on the fibres of $\cPT\to\CP^1$ introduced in~\eqref{eq:def-twistor-X-coords}, this is given by
\bea \label{eq:symplectic1}
\omega=\frac{1}{2\pi\im}\oint \frac{\dif X\wedge\dif Y\wedge\dif Z}{2(XY-Z^2 + c^2(\lambda))} \ =\  
\begin{cases}
 \displaystyle{\frac{\dif X\wedge\dif Z}{2X}}\qquad &\text{if $X\
\neq0$}\\[2ex]
 \displaystyle{-\frac{\dif Y\wedge \dif Z}{2Y}} \qquad &\text{if $Y\neq0$\,,}
\end{cases}
\eea
where the normalization is chosen to agree with the corresponding form $\dif\mu^{\dot0}\wedge\dif\mu^{\dot1}$ on $\PT/\bbZ_2$. More globally, this 2-form may be written  as
\be
    \label{eq:symplectic2}
    \omega = \frac{1}{8c^2(\lambda)}X^{\da\db}\,\dif X_{\dc\da}\wedge\dif X^\dc_{~\,\db}
\ee
where the $X^{\da\db}$ obey~\eqref{eq:twistor-constraint}. It is worth emphasising that this $\omega$ is still just $\dif \mu^{\dot0}\wedge \dif\mu^{\dot1}$; indeed, this follows by construction from the fact that $\cPT$ was obtained via Hamiltonian deformation. However, the $\mu^\da$ are no longer holomorphic on $\cPT$, so it is more useful to write $\omega$ in terms of the coordinates $X^{\da\db}$. The Gindikin 2-form $\Sigma(\lambda)$ is defined to be the pullback of $\omega$ to (the conormal bundle of) a $\CP^1$~\eqref{eq:incidence}.  A short calculation shows that this is
\bea
    &\Sigma(\lambda) = \frac{[M_-\dif M_-]\wedge[M_+\dif M_+]}{4c(\lambda)}\\
    &= \frac{\la\lambda\beta\ra^2}{2}[\dif u\wedge\dif u] + \la\al\lambda\ra\la\lambda\beta\ra[\dif u\wedge\dif\tilde  u] + \frac{\la\alpha\lambda\ra^2}{2}[\dif\tilde u\wedge\dif\tilde u] + \frac{c^2\la\al\lambda\ra^2}{[u\,\ut]^3}[\tilde u\,\dif u]\wedge[\tilde u\,\dif\tilde u]\,,
\eea
where we have introduced $u^{\da}=x^{\da\al}\al_\al$ and $\tilde{u}^{\da}=x^{\da\al}{\beta}_\al$. The Gindikin 2-form can be written as $\Sigma(\lambda) = e^{\da\al}\wedge e_{\da}^{\ \beta}\lambda_\al\lambda_\beta/2$, where 
\be
\label{eq:viebein-def}
    e^{\da\al} = \dif x^{\da\al} - \frac{c^2\ut^\da\al^\alpha[\ut\,\dif u]}{[u\,\ut]^3} = \bigg(\dif\ut^\da - \frac{c^2\ut^\da[\ut\,\dif u]}{[u\,\ut]^3}\bigg)\alpha^\al - \dif u^\da\beta^\alpha
\ee
defines a space-time vierbein. The corresponding metric is
\be \label{eq:EH-metric}
\dif s^2 = e^{\da\al}\odot e_{\da\al} = 2[\dif u\odot\dif\ut] + \frac{2c^2}{[u\,\ut]^3}[\ut\,\dif u]\odot[\ut\,\dif u] = \delta + \frac{16c^2}{x^6}[\tilde u\,\dif u]^{\odot 2}\,,
\ee
where $\delta= 2[\dif u\odot\dif\ut] = \dif x^{\da\al}\odot\dif x_{\da\al}$ is the flat metric. This is the Eguchi-Hanson metric, in the Kerr-Schild coordinates\footnote{Formally, we should view this as a complexified metric on the complexified space-time with complex coordinates $u^\da,\ut^\da$. The metric is real in ultrahyperbolic signature if the dyad $\{\alpha,\beta\}$ are real spinors. In this case (or in the complexified setting) we can take the limit $\alpha\to\beta$ to obtain a metric
\[ 
\dif s^2 = \delta + \frac{16c^2}{x^6}[u\,\dif u]^{\odot 2} \]
with $2^{\rm nd}$ Pleba{\'n}ski scalar $\Theta(x) = 2c^2/x^2$.} first found by Sparling \& Tod~\cite{Sparling:1981nk,Burnett:1979st,Berman:2018hwd}. (Other coordinates would be obtained from other choices of incidence relations.) It can also be written as
\be
    \dif s^2 = \delta + \tilde\p_\da \tilde\p_\db\Theta\,\dif u^\da\odot\dif u^\db
\ee
where $\tilde\p_\da = \al^{\al}\p_{\da\al}$ and the scalar
\be
\label{eq:2nd-Plebanski-scalar}
\Theta(x) = 2c^2\left(\frac{\ut^{\dot 1}}{u^{\dot 1}}\right)^2 \frac{1}{x^2}
\ee
obeys the $2^\mathrm{nd}$ Pleba{\'n}ski equation $\Delta\Theta - \frac{1}{2}\tilde\p^\da\tilde\p^\db\Theta\,\tilde\p_\da\tilde\p_\db\Theta=0$, where $\Delta = \p^{\da\al}\p_{\da\al}$ is the flat space Laplacian.

As is well known, the Eguchi-Hanson metric as written in equation \eqref{eq:EH-metric} has a singularity at the origin, which can be rectified by identifying $u^\da\sim-u^\da$. The cost of this is that space-time is no longer globally asymptotically Euclidean, but rather only locally so. Indeed, it is the canonical example of an ALE (asymptotically locally Euclidean) spacetime. This $\bbZ_2$ quotient plays a crucial role in our narrative: We saw in section \ref{sec:UniqueDeformation} that the unique non-trivial Lie algebra deformation of $\ham(\bbC^2)$ is the Weyl algebra related to space-time non-commutativity. However, its fixed point subalgebra under the $\bbZ_2$ action is the wedge subalgebra of $w_{1+\infty}$,
\be
\ham(\mathbb{C}^2)^{\mathbb{Z}_2}\cong \mathfrak{ham}(\mathbb{C}^2/\mathbb{Z}_2)\cong w_\wedge
\ee
admitting the larger family of deformations $W(\mu)$. We shall see that a certain scaling limit of $W(\mu)$ corresponds to deforming $\bbR^4/\bbZ_2$ to Eguchi-Hanson space.

\section{Celestial chiral algebras from twistor space} \label{sec:twistor-algebra} 

In~\cite{Adamo:2021lrv,Costello:2022wso}, it was shown that the celestial chiral algebra of self-dual Einstein gravity is simply the loop algebra of the Poisson algebra of holomorphic functions on the fibres of twistor space over $\CP^1$. For the twistor space $\PT$ of flat space, this is simply $\cL\ham(\bbC^2)$. On the deformed twistor space above, we will see that it is instead isomorphic to a $\mu\to\infty$ scaling limit of the deformed family of $W(\mu)$ algebras~\cite{Pope:1989sr,Pope:1991ig}. This coincides with the complexification of $\mathrm{sdiff}(S^2)$, the Lie algebra of area-preserving vector fields on the sphere.

As explained in section \ref{sec:UniqueDeformation}, the unique member of the $W(\mu)$ family that arises as a deformation of $\ham(\bbC^2)$ is the Weyl algebra. In~\cite{Bu:2022iak} this Weyl algebra was shown to arise by making $\PT$ non-commutative as explained in chapter \ref{chapter2}. The reason our deformed $\cPT$ yields a different algebra is that it is really a deformation of the orbifold $\PT/\bbZ_2$, where the non-trivial element of $\bbZ_2$ acts on the coordinates of $\PT$ by
\be
    \label{eq:Z2-action-on-twistors}
    (\mu^\da,\lambda_\al) \mapsto (-\mu^\da,+\lambda_\al)\,.
\ee
Notice that this action is independent of the overall scalings $(\mu^\da,\lambda_\al)\to r(\mu^\da,\lambda_\al)$ of the homogeneous coordinates. The source term $c^2(\lambda)\,\bar\delta^2(\mu)$, the corresponding twistor field $h = (c^2/2) [\hat\mu\,\dif\hat\mu]/[\mu\,\hat\mu]^2$ and the new holomorphic coordinates $X^{\da\db}$ are all invariant under this $\bbZ_2$ action. Indeed, when the coupling to the defect is turned off, these coordinates reduce to $X_0 = (\mu^{\dot0})^2$, $Y_0=(\mu^{\dot1})^2$ and $Z_0=\mu^{\dot0}\mu^{\dot1}$, and the relation $X_0Y_0=Z_0^2$ defines the twistor space $\PT/\bbZ_2$ corresponding to the space-time orbifold $\bbR^4/\bbZ_2$.

Only the generators of $\ham(\bbC^2)$ that are invariant under this action descend to holomorphic functions on $\PT/\bbZ_2$: these are
\bea
\label{eq:orbifold-generators}
    w[2r,2s] &= (\mu^{\dot0})^{2r}(\mu^{\dot1})^{2s} = X_0^r\,Y_0^s\\
    w[2r+1,2s+1] &= (\mu^{\dot0})^{2r+1}(\mu^{\dot1})^{2s+1} = X_0^r\,Y_0^s\, Z_0
\eea
with $r,s\in\mathbb{N}_0$. These generators transform in integer spin representations of the $\fsl_2(\bbC)$ acting on dotted spinors. Their Poisson algebra (with respect to the usual Poisson bracket $\{f,g\}=\p^\da \!f\,\p_\da g$) is exactly the wedge subalgebra of $w_{1+\infty}$, or equivalently $\ham(\bbC^2)^{\bbZ_2}=\ham(\bbC^2/\bbZ_2)$. It is thus the fact that only half the generators of the celestial chiral algebra of self-dual gravity on flat space descend to the orbifold which permits us to obtain new deformations of this algebra.


\subsection*{CCA for self-dual gravity}
\label{subsec:twistor-CCA-SDG}

In this section, we compute the structure constants of the Poisson algebra (\emph{i.e.} the Lie algebra of polynomial functions under the Poisson-bracket) on a fibre $\mathcal{M}_{\lambda}$ of twistor space of the Eguchi-Hanson space for a generic $\lambda \in \mathbb{CP}^1$. We will then prove that it is isomorphic to a certain $\mu \rightarrow \infty$ scaling limit of the deformed family of $W(\mu)$-algebras \cite{Pope:1989sr}. Following the arguments of~\cite{Adamo:2021lrv}, the corresponding loop algebra of it is to be identified as the celestial chiral algebra of self-dual Einstein gravity on an Eguchi-Hanson background. We will verify this by an explicit space-time calculation in section~\ref{sec:space-time-algebra}.

\medskip

The weight $-2$ Poisson bracket associated to the $\cO(2)$-valued (2,0)-form~\eqref{eq:symplectic1} is
\be\label{eq:deformed-Poisson}
\{f,g\} = 2X\left(\frac{\p f}{\p X}\frac{\p g}{\p Z} - \frac{\p f}{\p Z}\frac{\p g}{\p X}\right)
\ee
Acting on the basic coordinates, this Poisson bracket gives 
\be \label{eq:Poisson-bracket}   
 \{X,Y\} = 4Z\,,\qquad \{X,Z\} = 2X\,,\qquad \{Y,Z\} = -2Y\,,
\ee
which are just the defining relations of $\fsl_2$. Then the constraint $XY - Z^2 =- c^2(\lambda)$ is just the statement that the quadratic Casimir of this $\fsl_2$ takes the value $c^2(\lambda)$. In particular, these relations imply that the ideal 
\be
\label{eq:twistor-ideal}
\scrI = \mathrm{span}\{XY - Z^2 + c^2(\lambda)\}
\ee
is a Poisson ideal, in the sense that $\{\cO,\scrI\}\subset\scrI$.  Because of this, the Poisson bracket can straightforwardly be extended to act on the full coordinate ring
\be 
\label{eq:EHCoordinateRing}
\cO_{\cM_{\lambda}} = \bbC[X,Y,Z]\,\big/\, \mathscr{I}\,.
\ee
A natural choice of basis for this ring is
\be 
\label{eq:EH-twistor-basis}
V[2p,2q] =  X^pY^q\,,\qquad\qquad V[2p+1,2q+1] = X^pY^qZ 
\ee
with $p,q\in\mathbb{N}_0$, because any polynomial involving higher powers of $Z$ can be traded for these generators using the ideal. We emphasise the states $V[m,n]$ are defined only when $m+n\equiv0\Mod{2}$. This basis is a  deformation of the basis~\eqref{eq:orbifold-generators} to the deformed twistor space. 

 In the basis~\eqref{eq:EH-twistor-basis} we find the algebra 
\begin{subequations} \label{eqs:algebra-in-twistor-basis}
\begin{align}
\begin{split} \label{eq:alg-in-twistor-basis-1}
    &\big\{V[2p,2q]\,,\,V[2r,2s]\big\} = 4(ps-qr)\,V[2(p\!+\!r\!-\!1)\!+\!1,2(q\!+\!s\!-\!1)\!+\!1]\,,
\end{split} \\[2ex]
\begin{split} \label{eq:alg-in-twistor-basis-2}
&\big\{V[2p,2q]\,,\,V[2r\!+\!1,2s\!+\!1]\big\} = 2(p(2s+1)-q(2r+1))\,V[2(p\!+\!r),2(q\!+\!s)] \\
    &\hspace{5cm} + 4c^2(\lambda)\,(ps-qr)\,V[2(p\!+\!r\!-\!1),2(q\!+\!s\!-\!1)]\,,
\end{split} \\[2ex]
\begin{split} \label{eq:alg-in-twistor-basis-3}
 &\big\{V[2p\!+\!1,2q\!+\!1]\,,\,V[2r\!+\!1,2s\!+\!1]\big\}  = ((2p\!+\!1)(2s\!+\!1)-(2q\!+\!1)(2r\!+\!1))\,V[2(p\!+\!r)\!+\!1,2(q\!+\!s)\!+\!1] \\
    &\hspace{6cm} + 4c^2(\lambda)\,(ps-qr)\,V[2(p\!+\!r)\!-\!1,2(q\!+\!s)\!-\!1]\,.
\end{split}
\end{align}
\end{subequations}
The terms proportional to $c^2(\lambda)$ in~\eqref{eq:alg-in-twistor-basis-2} \&~\eqref{eq:alg-in-twistor-basis-3} are not present in $w_{1+\infty}$. These terms are easily seen to represent a non-trivial element in Lie algebra cohomology: the deformation preserves the $\fsl_2$ subalgebra, so any redefinition undoing it must act as an intertwiner. Since  $\ham(\bbC^2)^{\bbZ_2}$ decomposes into a direct sum over distinct integer spin representations, the only $\fsl_2$ equivariant redefinitions are rescalings of the generators. It's clear that such rescalings cannot trivialize the deformation \eqref{eqs:algebra-in-twistor-basis}. We identify this deformed algebra below.

\subsection*{CCA in the scattering basis} \label{subsec:states-propagators}

While~\eqref{eq:EH-twistor-basis} is probably the simplest basis of $\cO_{\cM_\lambda}$, it will be helpful to consider a different basis so as to make contact with ideas of celestial holography. There, one considers the generators as conformally soft modes of the graviton~\cite{Guevara:2021abz,Strominger:2021lvk}. In flat $\bbR^4$, these are Mellin modes of momentum eigenstates, representing the external states of a scattering process.

\medskip

In our context, we should look for scattering states defined on the background deformed twistor space $\cPT$, corresponding to linearized fluctuations of the graviton around the Eguchi-Hanson background. We seek solutions that look asymptotically like plane waves. These may be represented on twistor space by 
\begin{subequations}
\label{eq:twistor-scattering-states}
\begin{equation}
\label{twistorscatteringstate}
\delta h(X,\lambda) = \int \frac{\dif t}{t^3}  \,\bar\delta^2(t|\lambda\ra - |\kappa\ra) \,\cos\left(t\sqrt{ -[\tilde\kappa|X|\tilde\kappa]}\right)
\end{equation}
for the helicity $+2$ graviton and 
\begin{equation}
\delta g(X,\lambda) = \int\frac{\dif t}{t}\, t^6 \,\bar\delta^2(t|\lambda\ra-|\kappa\ra)\,\cos\left(t\sqrt{ -[\tilde\kappa|X|\tilde\kappa]}\right)
\end{equation}
\end{subequations}
for the helicity $-2$ graviton \cite{Adamo:2022mev}, where $|\kappa\ra$ and $|\tilde\kappa]$ are fixed spinors and $[\tilde\kappa|X|\tilde\kappa] = -X^{\da\db}\tilde\kappa_{\da}\tilde\kappa_\db$. \eqref{eq:twistor-scattering-states} are manifestly holomorphic for $\bar{\nabla}$ since they only depend on $\lambda_\alpha$ and the deformed holomorphic coordinates $X^{\da \db}$. Taking their weight into account means that they represent elements in the appropriate cohomology classes
\be
\delta h\in H^{0,1}(\cPT, \mathcal{O}(2))\,,\quad \delta g\in H^{0,1}(\cPT, \mathcal{O}(-6))\,.
\ee

In space-time, the fluctuation in the $2^{\rm nd}$ Pleba{\'n}ski scalar may be obtained from $\delta h$ via the linear Penrose transform we reviewed in section \ref{sectionLinearPenrose}\footnote{Extracting the fluctuation in the Pleba{\'n}ski scalar from $\delta h\in H^{0,1}(\cPT,\cO(2))$ makes use of the twistor recursion operator for the hyperk{\"a}hler hierarchy. See~\cite{Dunajski:2000iq} for details.}
\bea
\label{eq:twistor-scatttering-state-scalar}
    \delta\Theta(x) &= \int_{\CP^1_x} \frac{\la\lambda\,\dif\lambda\ra}{\la\al\lambda\ra^4} \wedge \delta h\\
    &=\frac{1}{\la\al\kappa\ra^4}\cos\left(\sqrt{(k\cdot x)^2 - \frac{4c^2\la\al| kx|{\beta}\ra^2}{x^4}}\right)
\eea
where $k=|\kappa\ra[\tilde\kappa|$ and we have used the incidence relations~\eqref{eq:incidence}. Although it follows from the twistor construction, one can easily verify directly that this state indeed obeys the $2^{\rm nd}$ Pleba{\'n}ski equation, linearized around the Eguchi-Hanson background. Note that at asymptotically large distances  $|x|\to\infty$ in space-time, it approaches $\cos(k\cdot x)$, which is a `momentum eigenstate' on the orbifold $\bbR^4/\bbZ_2$. 

\medskip

The celestial chiral algebra is usually written in terms of the soft modes of the scattering states of the graviton. Concretely, these are the coefficients $W[p,q]$ of $(-)^{(p+q)/2}\lt_\dzero^p\lt_\done^q/p!q!$ in the expansion 
\be \label{eq:cos-Taylor-expand}
\cos\sqrt{-t[\lt|X|\lt]} = \sum_{m=0}^\infty\frac{t^{2m}[\lt|X|\lt]^m}{(2m)!}
\ee
in the twistor space scattering state. Trinomially expanding $[\lt|X|\lt]^m = (-)^m(X\lt_{\dot0}^2+Y\lt_{\dot1}^2+2Z\lt_{\dot0}\lt_{\dot1})^m$ gives
\bea
\label{eq:softbasis}
&\frac{[\lt|X|\lt]^m}{(2m)!}=  \frac{(-)^m}{(2m)!} \sum_{i+j+l=m} \binom{m}{i,j,l} X^iY^j (2Z)^l\, \Tilde{\lambda}_{\dot{0}}^{2i+l}\Tilde{\lambda}_{\dot{1}}^{2j+l}\\
&= \sum_{p+q=m}\left(\frac{(-)^{p+q}}{(2p+2q)!}\sum_{\ell=0}^{\min(p,q)}\binom{p+q}{p-\ell,q-\ell,2\ell}X^{p-\ell}Y^{q-\ell}(2Z)^{2\ell}\right)\Tilde{\lambda}_{\dot{0}}^{2p}\Tilde{\lambda}_{\dot{1}}^{2q}\\
&+ \sum_{p+q=m-1}\left(\frac{(-)^{p+q+1}}{(2p\!+\!2q\!+\!2)!}\sum_{\ell=0}^{\min(p,q)}\binom{p\!+\!q\!+\!1}{p\!-\!\ell,q\!-\!\ell,2\ell\!+\!1}X^{p-\ell}Y^{q-\ell}(2Z)^{2\ell+1}\right)\Tilde{\lambda}_{\dot{0}}^{2p+1}\Tilde{\lambda}_{\dot{1}}^{2q+1}\\
&= \sum_{p+q=m}W[2p,2q] \frac{(-)^{p+q}\Tilde{\lambda}_{\dot{0}}^{2p}\Tilde{\lambda}_{\dot{1}}^{2q}}{(2p)!(2q)!}\ \ +\sum_{p+q=m-1} W[2p+1,2q+1] \frac{(-)^{p+q+1}\Tilde{\lambda}_{\dot{0}}^{2p+1}\Tilde{\lambda}_{\dot{1}}^{2q+1}}{(2p+1)!(2q+1)!}\,,
\eea
where we have defined the generators in the scattering basis
\bea \label{eq:W-even-def}
    W[2p,2q]  &=\frac{(2p)!\,(2q)!}{(2p+2q)!}\sum_{\ell=0}^{\min(p,q)}\binom{p+q}{p-\ell,q-\ell,2\ell}X^{p-\ell}Y^{q-\ell}(2Z)^{2\ell} \\
    W[2p\!+\!1,2q\!+\!1] & = \frac{(2p+1)!\,(2q+1)!}{(2p+2q+2)!}\sum_{\ell=0}^{\min(p,q)}\binom{p\!+\!q\!+\!1}{p\!-\!\ell,q\!-\!\ell,2\ell\!+\!1}X^{p-\ell}Y^{q-\ell}(2Z)^{2\ell+1}\,.
\eea
The terms with $\ell>0$ involve higher powers of $Z$, so may be traded for powers of $X,Y$ and $c^2(\lambda)$ using~\eqref{eq:twistor-ideal}. Doing so, we find that this scattering basis is related to the basis~\eqref{eq:EH-twistor-basis} by an upper triangular transformation of the form
\begin{subequations}
\label{eq:basis-transformation}
\bea
W[2p,2q] &= \sum_{\ell=0}^{\min(p,q)}(2c(\lambda))^{2\ell} \,C_0(p,q,\ell) \,V[2p\!-\!2\ell,2q\!-\!2\ell]\,, \\
W[2p\!+\!1,2q\!+\!1] &= \sum_{\ell=0}^{\min(p,q)}(2c(\lambda))^{2\ell}\,C_1(p,q,\ell)\, V[2p\!-\!2\ell\!+\!1,2q\!-\!2\ell\!+\!1]\,,
\eea
with coefficients
\be
\label{eq:c0c1}
C_0(p,q,\ell)=\frac{[p]_\ell\,[q]_\ell\,[p+q]_\ell}{\ell!\,[2(p+q)]_{2\ell}}\,,\qquad\qquad
C_1(p,q,\ell)= \frac{[p]_{\ell}
\,[q]_{\ell}\,[p+q+1]_{\ell}}{\ell!\,[2(p+q+1)]_{2\ell}}\,,
\ee
\end{subequations}
where $[p]_\ell = p!/(p-\ell)!$ is the descending Pochhammer symbol. Note that the two bases coincide when the coupling $c$ to the defect is sent to zero.   

In terms of the scattering basis, the algebra takes the form
\bea
\label{eq:sl-infinity} 
&\big\{W[p,q],W[r,s]\big\} \\
&= \sum_{\ell\geq0} (2c(\lambda))^{2\ell}\,R_{2\ell+1}(p,q,r,s)\,\psi_{2\ell+1}\bigg(\frac{p\!+\!q}{2},\frac{r\!+\!s}{2}\bigg)\,W[p\!+\!r\!-\!2\ell\!-\!1,q\!+\!s\!-\!2\ell\!-\!1]\,, 
\eea
where 
\be
\psi_{2\ell+1}(m,n) = (-)^\ell\frac{[\ell+1/2]_\ell}{4^{2\ell}\,[m-1/2]_\ell\,[n-1/2]_\ell\,[m+n-1/2-\ell]_\ell}\,.
\ee
Comparing this to the commutation relations~\eqref{eq:W(mu)-commutator} of the $W(\mu)$-algebras, we see that the function $\Psi_{2\ell+1}(m,n;\sigma)$ in~\eqref{eq:Psi-ell} has been replaced by $\psi_{2\ell+1}(m,n)$. It's instructive to compare these functions in the case $\ell=1$. We have
\be \label{eq:Psi-3} 
\Psi_3(m,n;\sigma) = 1 - \frac{3(4\sigma+1)(4\sigma+3)}{(2m-1)(2n-1)(2(m+n)-3)}\,. \ee
It's then clear that in the scaling limit $\sigma\to\infty$, $\fq\to0$ with $4\fq\sigma = c(\lambda)$ held fixed
\be 
\lim_{{\sigma\to\infty}\atop{\fq\to0}}\fq^2\Psi_3(m,n;\sigma) = - \frac{3c^2(\lambda)}{(2m-1)(2n-1)(2(m+n)-3)} = 4c^2(\lambda)\,\psi_3(m,n)\,. 
\ee
Similarly, 
\be
\lim_{{\sigma\to\infty}\atop{\fq\to0}}\fq^{2\ell}\Psi_{2\ell+1}(m,n;\sigma) =  (2c(\lambda))^{2\ell}\,\psi_{2\ell+1}(m,n)\,.
\ee
It follows from the fact that the scattering and twistor bases are related by the (invertible) upper triangular transformation~\eqref{eq:basis-transformation} that the algebras~\eqref{eqs:algebra-in-twistor-basis} \&~\eqref{eq:sl-infinity} are isomorphic. (In appendix \ref{app:isomorphism} we check explicitly that this transformation respects the Lie brackets, see also~\cite{Pope:1989sr,Bergshoeff:1989ns}). The algebras on different generic,\footnote{In the twistor space defined by~\eqref{eq:twistor-constraint} the fibres $\lambda=\al$ and $\lambda=\beta$ remain singular. In the true twistor space of Eguchi-Hanson space-time these singularities are resolved by blowing up $T^*\CP^1\to\bbC^2/\bbZ_2$ \cite{Hitchin:1979rts}. It's nonetheless easy to verify that the Poisson algebra of global regular functions on $T^*\CP^1$ (in its standard complex structure) remains isomorphic to $w_\wedge$.} fibres $\cM_\lambda$ are all isomorphic, as may be seen by rescaling the generators by an appropriate power of $c(\lambda)$. We call this algebra
\be
    W(\infty) = \lim_{{\fq\to0}\atop{\mu\to\infty}} W(\mu)\,,\qquad\fq\sqrt{\mu} \quad\text{fixed}\,.
\ee

\medskip

The CCA of self-dual gravity on the Eguchi-Hanson background therefore has OPEs
\bea
&W[p,q](\lambda_1)\,W[r,s](\lambda_2) \\
&\sim - \frac{\tau^{p+q-3}}{2\la12\ra}\sum_{\ell\geq0} (2c(\lambda_2))^{2\ell}\,R_{2\ell+1}(p,q,r,s)\,\psi_{2\ell+1}\bigg(\frac{p\!+\!q}{2},\frac{r\!+\!s}{2}\bigg)\,W[p\!+\!r\!-\!2\ell\!-\!1,q\!+\!s\!-\!2\ell\!-\!1](\lambda_2)\,,
\eea
where $\tau = \la\al1\ra/\la\al2\ra$ has been introduced to give both sides the appropriate weight. (Note that $\tau$ is independent of $\alpha$ working modulo $\la12\ra$.) In the inhomogeneous coordinates $z_i = \la\al i\ra/\la i\beta\ra$ this reads\footnote{The soft modes $W[p,q](\lambda)$ should be viewed as sections of $\cO(p+q-4)$ over the celestial sphere. Working inhomogeneously in terms of $z = \la\alpha\kappa\ra/\la\kappa\beta\ra$ these coincide with the modes obtained via Mellin transform.}
\bea
&W[p,q](z_1)\,W[r,s](z_2) \\
&\sim - \frac{1}{2z_{12}}\sum_{\ell\geq0} (2cz_2)^{2\ell}\,R_{2\ell+1}(p,q,r,s)\,\psi_{2\ell+1}\bigg(\frac{p\!+\!q}{2},\frac{r\!+\!s}{2}\bigg)\,W[p\!+\!r\!-\!2\ell\!-\!1,q\!+\!s\!-\!2\ell\!-\!1](z_2)\,. \eea
Decomposing $W[p,q](z)$ into Laurent modes in $z$ we can see the CCA is isomorphic to $\cL W(\infty)$\footnote{It's intriguing that this isomorphism requires rescaling generators in a $z$ dependent way. We expect this to modify the modules at $z=0,\infty$ determining the vertex algebra vacua.}.


\subsection*{CCA for self-dual Yang-Mills}
\label{subsec:twistor-CCA-SDYM}

It is straightforward to extend these considerations to the celestial chiral algebra of self-dual Yang-Mills (for the semisimple gauge algebra $\fg$ with invariant bilinear form $\tr$), the S-algebra we discussed in section \ref{section2.1}. Again, this algebra is deformed when describing self-dual Yang-Mills on an Eguchi-Hanson background. Classically, this may described by the holomorphic BF action \eqref{eq:HoloBFaction} on the curved twistor space
\be \label{eq:sdYM-twistor-action}
S[b,a]= \int_\cPT\Omega\wedge\tr(b\wedge f)\,
\ee
for the fields $b\in\Omega^{0,1}(\cPT,\cO(-4)\otimes\fg)$  and $a\in\Omega^{0,1}(\cPT,\fg)$, where $f =\bar\nabla a+\frac{1}{2}[a,a]$. The main difference compared to self-dual gravity is that the vertex now involves the Lie bracket on $\fg$ rather than the Poisson bracket.\footnote{The Dolbeault operator $\bar\nabla$ appearing in the curvature $(0,2)$-form is just the usual $\bar\p$ operator when written in terms of holomorphic coordinates $(X,Y,Z,\lambda)$ on $\cPT$.}  The deformed $S$-algebra is thus simply the loop algebra of  $\fg\otimes\cO_{\cM_\lambda}$.

\medskip

As before, the most natural choice of basis for $\fg\otimes\cO_{\cM_\lambda}$ is 
\be 
\label{eq:GaugePolynomials} I_\sfa[2p,2q] = t_\sfa X^pY^q\,,\qquad\qquad I_\sfa[2p+1,2q+1] = t_\sfa X^pY^qZ\,,
\ee
where the $t_\sfa$ form a basis of $\fg$. Note that again $I_\sfa[m,n]$ is only defined for $m+n\equiv0\Mod{2}$. The structure constants follow immediately from the coordinate ring $\cO_{\cM_\lambda}$ and are given by 
\bea \label{eq:twistor-S-structure-constants}
\big[I_\sfa[2p,2q],I_\sfb[2r,2s]\big] &= f_{\sfa\sfb}^{~~\sfc}\,I_\sfc[2p\!+\!2r,2q\!+\!2s]\,, \\
\big[I_\sfa[2p,2q],I_\sfb[2r\!+\!1,2s\!+\!1]\big] &= f_{\sfa\sfb}^{~~\sfc}\,I_\sfc[2p\!+\!2r\!+\!1,2q\!+\!2s\!+\!1]\,,\\
\big[I_\sfa[2p\!+\!1,2q\!+\!1],I_\sfb[2r\!+\!1,2s\!+\!1]\big] 
&= f_{\sfa\sfb}^{~~\sfc}\,\big(I_\sfc[2p\!+\!2r\!+\!1,2q\!+\!2s\!+\!1] + c^2(\lambda)\, I_\sfc[2p\!+\!2r,2q\!+\!2s]\big)\,.
\eea
Once again, we can change the basis to the soft modes appearing in the expansion of the scattering states, which now include a colour factor. Defining generators $J_\sfa[r,s]$ in the scattering basis via
\bea
\label{eq:twistor-wavefunction} 
    &t_\sfa\cos\sqrt{-[\lt|X|\lt]} = t_\sfa\sum_{m=0}^\infty\frac{[\lt|X|\lt]^m}{(2m)!}\\
    &=\sum_{p+q=m}J_\sfa[2p,2q] \frac{(-)^{p+q}\Tilde{\lambda}_{\dot{0}}^{2p}\Tilde{\lambda}_{\dot{1}}^{2q}}{(2p)!(2q)!}\ \ +\sum_{p+q=m-1} \!\!J_\sfa[2p\!+\!1,2q\!+\!1] \frac{(-)^{p+q}\Tilde{\lambda}_{\dot{0}}^{2p\!+\!1}\Tilde{\lambda}_{\dot{1}}^{2q+1}}{(2p\!+\!1)!\,(2q\!+\!1)!}\,,
\eea
the $I$ and $J$ bases are related by the same upper triangular transformation that we met in self-dual gravity:
\bea
J_\sfa[2p,2q]&=\sum_{\ell=0}^{\min(p,q)}(2c(\lambda))^{2\ell}\, C_0(p,q,\ell)\, I_\sfa[2(p-\ell),2(q-\ell)]\,,\\
J_\sfa[2p\!+\!1,2q\!+\!1]&= \sum_{\ell=0}^{\min(p,q)}(2c(\lambda))^{2\ell}\,C_1(p,q,\ell)\,I_\sfa[2(p-\ell)\!+\!1,2(q-\ell)\!+\!1]\,,
\eea
where $C_0$, $C_1$ were given in~\eqref{eq:c0c1}. In the scattering basis, the deformed $S$-algebra becomes
\bea
\label{eq:SDYM-CCA-twistor}
&[J_\sfa[p,q],J_\sfb[r,s]] = \\ &f_{\sfa\sfb}^{~~\sfc}\sum_{\ell=0}^{\infty}(2c(\lambda))^{2\ell}\, R_{2\ell}(p,q,r,s)\,\psi_{2\ell}\bigg(\frac{p+q}{2},\frac{r+s}{2}\bigg) 
J_\sfc[p+r-2\ell,q+s-2\ell]\,.
\eea
with 
\be
\psi_{2\ell}(m,n) = (-)^\ell\frac{[\ell-1/2]_\ell}{4^{2\ell}[m-1/2]_\ell[n-1/2]_\ell[m+n-1/2-\ell]_\ell}\,.
\ee
These structure constants arise as a scaling limit of a family $S_\wedge(\mu;\fq)$ of deformed $S$-algebras (in the case $\fg=\fgl(N)$). More precisely, in analogy to $W(\mu;\fq)$, we define $S_\wedge(\mu;\fq)$ by the relations
\bea \label{eq:S-sigma-algebra}
&\left[\widetilde J_\sfa[p,q],\widetilde J_\sfb[r,s]\right] = f_{\sfa\sfb}^{~~\sfc}\sum_{\ell=0}^\infty\fq^{2\ell}R_{2\ell}(p,q,r,s)\Psi_{2\ell}\bigg(\frac{p+q}{2},\frac{r+s}{2};\sigma\bigg)\widetilde J_\sfc[p+r-2\ell,q+s-2\ell] \\
&+ d_{\sfa\sfb}^{~~\sfc}\sum_{\ell=0}^\infty\fq^{2\ell+1}R_{2\ell+1}(p,q,r,s)\Psi_{2\ell+1}\bigg(\frac{p+q}{2},\frac{r+s}{2};\sigma\bigg)\widetilde J_\sfc[p+r-2\ell,q+s-2\ell]
\eea
where $\Psi_{\ell}(m,n;\sigma)$ is given in \eqref{eq:Psi-ell} 
Here $d_{\sfa\sfb\sfc} = \tr(t_\sfa\{t_\sfb,t_\sfc\})$, which arises in the non-commutative setting but drops out in the scaling limit. See \cite{Monteiro:2022xwq} for further details. Sending $\sigma\to\infty$, $\fq\to0$ with $4\sigma\fq=c(\lambda)$ fixed gives
\bea
&\lim_{\sigma\to\infty}\fq^{2\ell}\Psi_{2\ell}(m,n;\sigma) = (2c(\lambda))^{2\ell}\psi_{2\ell}(m,n) \\
&= (2c(\lambda))^{2\ell}(-)^\ell\frac{[\ell-1/2]_\ell}{4^{2\ell}[m-1/2]_\ell[n-1/2]_\ell[m+n+1/2-\ell]_\ell}\,,
\eea
agreeing with \eqref{eq:SDYM-CCA-twistor} and defining a family of Lie algebras $S_\wedge(\infty;c)$. For $\lambda\neq\alpha,\beta$ and $c\neq0$ these are all isomorphic to $S_\wedge(\infty)\cong S_\wedge(\infty;1)$, so we obtain the same algebra on generic twistor fibres.

Therefore, the defining OPEs of the self-dual Yang-Mills CCA on Eguchi-Hanson are
\bea
&J_\sfa[p,q](\lambda_1)J_\sfb[r,s](\lambda_2) \\
&\sim - \frac{\tau^{p+q-1}f_{\sfa\sfb}^{~~\sfc}}{2\la12\ra}\sum_{\ell=0}^{\infty}(2c(\lambda))^{2\ell}\, R_{2\ell}(p,q,r,s)\,\psi_{2\ell}\bigg(\frac{p+q}{2},\frac{r+s}{2}\bigg) 
J_\sfc[p+r-2\ell,q+s-2\ell](\lambda_2)\,.
\eea
Or, in inhomogeneous coordinates
\bea
&J_\sfa[p,q](z_1)J_\sfb[r,s](z_2) \\
&\sim - \frac{f_{\sfa\sfb}^{~~\sfc}}{2z_{12}}\sum_{\ell=0}^{\infty}(2cz_2)^{2\ell}\, R_{2\ell}(p,q,r,s)\,\psi_{2\ell}\bigg(\frac{p+q}{2},\frac{r+s}{2}\bigg) 
J_\sfc[p+r-2\ell,q+s-2\ell](z_2)\,.
\eea


\section{Splitting functions on the Eguchi-Hanson background} \label{sec:space-time-algebra}

As we saw in section \ref{Action_section}, in space-time, self-dual gravity may be described perturbatively around a self-dual background by the Chalmers-Siegel action~\cite{Siegel:1992wd,Chalmers:1996rq} 
\begin{equation}
\label{eq:Chalmers-Siegel-SDGR-action}
    S[\Tilde{\Theta},\Theta] = \int \dif^4x\,\bigg(\p^{\da\al}\Tilde{\Theta}\,\p_{\da\al}\Theta + \frac{1}{2}\Tilde{\Theta}\, \tilde\p^\da\tilde\p^\db\Theta\,\tilde\p_\da\tilde\p_\db\Theta\bigg)\,, 
\end{equation}
where, as before, $\tilde\p_{\da}=\al^\al\p_{\da\al}$. This action is equivalent to the twistor action~\eqref{eq:twistor-sd-gravity-action} at the classical level as discussed in section \ref{section2.3}. Varying $\Tilde{\Theta}$ leads to Pleba{\'n}ski's second heavenly equation
\begin{equation}
    \label{eq:2nd-Plebanski}
    \Delta\Theta - \frac{1}{2}\tilde\p^\da\tilde\p^\db\Theta\,\tilde\p_\da\tilde\p_\db\Theta=0\,,
\end{equation}
with $\Delta=\p^{\da\al}\p_{\da\al}$. The Laplacian $\Delta_g$ on the self-dual background defined by a solution of~\eqref{eq:2nd-Plebanski} can be written as
\be
\Delta_g = \Delta - (\tilde\p^\da\tilde\p^\db\Theta)\tilde\p_\da\tilde\p_\db\,,
\ee
so that the remaining field equation reads
\be
0=\Delta\Tilde{\Theta} - \{\tilde\p^\da\Theta,\tilde\p_\da\Tilde\Theta\} = \Delta_g\Tilde\Theta\,.
\ee
We will often write $\tilde\p^\da\tilde\p^\db\Theta\,\tilde\p_\da\tilde\p_\db\Theta= \{\tilde\p^\db\Theta,\tilde\p_\db\Theta\}$, where $\{\ , \ \}$ is a Poisson bracket on space-time which coincides with the twistor bracket on the fibre over $\lambda=\beta$.

\begin{figure}[t!]
    \centering
    \includegraphics[scale=0.4]{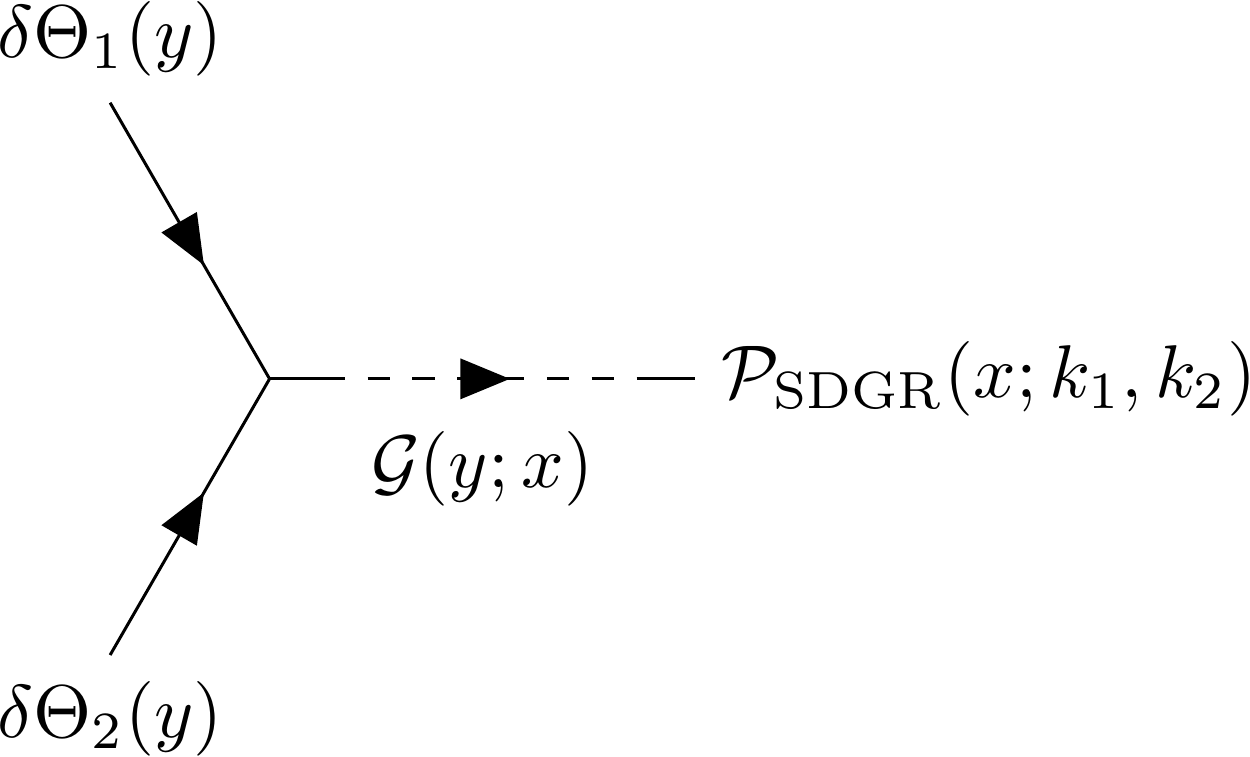}
    \caption{\emph{Tree contribution to the perturbiner in self-dual gravity. This is the position space counterpart of figure \ref{fig:splitting}, which determines the tree splitting function.}} \label{fig:perturbiner}
\end{figure}

\medskip

In this section, we will recover the celestial chiral algebras obtained in section~\ref{sec:twistor-algebra}  by considering the splitting functions for positive helicity fluctuations of self-dual gravity and self-dual Yang-Mills around the Eguchi-Hanson background $\Theta=2c^2(\tilde u^{\dot1}/u^{\dot1})^2(1/x^2)$. These calculations provide an independent check of the results obtained by twistor theory above. In particular, in the gravitational case, we  compute the residue of the holomorphic collinear singularity in 
\be \label{eq:SDGR-perturbiner-int} 
\cP_\mathrm{SDGR}(x;k_1,k_2) = \int_{\widetilde{\bbR^4/\bbZ_2}}\dif^4y\,\cG(x;y)\,\{\tilde\p^\da\delta\Theta_1(y),\tilde\p_\da\delta\Theta_2(y)\}\,,
\ee
where $\delta\Theta_i(y)$ are fluctuations in the Pleba{\'n}ski scalar and $\cG(x,y)$ is the Green's function for the background Laplacian. 
$\cP_\mathrm{SDGR}$ is known as the \emph{perturbiner} of self-dual gravity~\cite{Berends:1987me,Rosly:1996vr,Rosly:1997ap}. The integral \eqref{eq:SDGR-perturbiner-int} can be obtained from the partially off-shell Feynman diagram illustrated in figure \ref{fig:perturbiner}. This is simply the position space analogue of the diagram leading to the gravitational splitting function, as depicted in figure \ref{fig:splitting}. It encodes the same information and can be used to extract the celestial OPE.

Differentiating under the integral shows that the perturbiner obeys
\be \label{eq:perturbiner-def}
    \Delta_g \cP_\mathrm{SDGR} = \{\tilde\p^\da\delta\Theta_1,\tilde\p_\da\delta\Theta_2\}\,,
\ee
where $\Delta_g$ is the Eguchi-Hanson Laplacian. As above, we consider the fluctuations 
\be 
\label{eq:Plebanski-fluctuations}
    \delta\Theta_i(x) = \frac{1}{\la\al i\ra^4} \cos\sqrt{(k_i\cdot x)^2 - \frac{4c^2\la\alpha|k_i x|\beta\ra^2}{x^4}}
\ee
around Eguchi-Hanson space, while the scalar Green's function is~\cite{Page:1979ga,Atiyah:1981ey}
\be
\cG(x;y) = -\frac{x^2 + y^2}{2\pi^2}\Bigg((x^2+y^2)^2 - 4[u\,\vt]^2w(y) - 4[v\,\ut]^2w(x) - 8[u\,\vt][v\,\ut]\bigg(1+\frac{4c^2}{x^2y^2}\bigg)\Bigg)^{-1}\,, \ee
in the Kerr-Schild coordinates, where $u = x|\al\ra$, $\ut=x|\beta\ra$, $v=y|\al\ra$ and $\vt=y|\beta\ra$, and where we have set  $w(x) = 1 - 4c^2/x^4$.

The space-time calculation turns out to be considerably more involved than the twistor space arguments. We will content ourselves with expanding both sides of equation \eqref{eq:SDGR-perturbiner-int} in powers of $c^2$ as
\bea
&\cP_\mathrm{SDGR}(x;k_1,k_2) = \sum_{n=0}^\infty c^{2n}\,\cP_\mathrm{SDGR}^{(n)}(x;k_1,k_2)\,, \\
&\cG(x;y) = \sum_{n=0}^\infty c^{2n}\,\cG^{(n)}(x;y)\,,\qquad \delta\Theta_i(x) = \sum_{n=0}^\infty c^{2n}\,\delta\Theta_i^{(n)}(x)\,,
\eea
and working just to the first non-trivial order in $c^2$.


\subsection*{CCA for self-dual gravity on the orbifold} 
\label{subsec:CCA-orbifold}

At zeroth order in $c^2$ we expect to recover the fixed point subalgebra of $\cL\ham(\bbC^2)$ under $\bbZ_2$, \emph{i.e.}, the loop algebra of the wedge subalgebra of $w_{1+\infty}$. To check this, we use the zeroth order states $\delta\Theta_i^{(0)}(x) = \cos(k_i\cdot x)/\la\al i\ra^4$ and
propagator
\be
\cG^{(0)}(x;y) = - \frac{x^2+y^2}{2\pi^2(x^2-y^2)^2} = - \bigg(\frac{1}{4\pi^2(x-y)^2} + \frac{1}{4\pi^2(x+y)^2}\bigg)\,.
\ee
Strictly, in this section we're working in ultrahyperbolic signature and our propagator should be defined using an $\im\eps$ prescription. To evaluate our integrals, we Wick rotate to Euclidean signature and complexify the momenta. We leave these steps implicit for the remainder of the section. Plugging the above propagator and states into the zeroth order part of equation~\eqref{eq:SDGR-perturbiner-int} gives
\bea \label{eq:SDGR-orbifold-perturbiner}
\cP_\mathrm{SDGR}^{(0)}(x;k_1,k_2) &= - \frac{[12]^2}{\la\alpha1\ra^2\la\alpha2\ra^2}\int_{\bbR^4}\dif^4y\,\frac{\cos(y\cdot k_1)\cos(y\cdot k_2)}{4\pi^2(x-y)^2}\\
&=-\frac{[12]}{\la12\ra} \frac{\cos(x\cdot k_-)-\cos(x\cdot k_+)}{4\la\al1\ra^2\la\al2\ra^2}\,,
\eea
where $k_\pm = k_1\pm k_2$.  We see that $\cP^{(0)}_{\rm SDGR}$ is the usual gravitational splitting function $[12]/\la12\ra$ times a wavefunction on the orbifold that combines the two momentum of the original states.

In the holomorphic collinear limit, modulo non-singular terms, we can take $k_\pm = (\tau|1]\pm|2])\la2|$ and set $\tau = \la\al1\ra/\la\al2\ra$. The familiar flat space correspondence between null momentum eigenstates and hard graviton generating functions $\delta\Theta_k(x)\leftrightarrow w(\kt,\kappa)$ goes through largely unchanged, although we now require that $w(\kt,\kappa) = w(-\kt,\kappa)$. Making this identification in equation \eqref{eq:SDGR-orbifold-perturbiner} we recover the celestial OPE
\be w(\kt_1,\kappa_1)\,w(\kt_2,\kappa_2)\sim\frac{[12]}{4\la12\ra}\tau^{-2}\big(w(\tau\kt_1-\kt_2,\kappa_2) - w(\tau\kt_1+\kt_2,\kappa_2)\big)\,.
\ee
Note that the difference on the right-hand side simply projects onto the $\bbZ_2$-invariant terms in both $\kt_1$ and $\kt_2$. Extracting the soft modes via Mellin transforms
\be \mathrm{Res}_{\Delta=-2m}\int_0^\infty\dif\omega\,\omega^{\Delta-1}\delta\Theta_k(x) = (-)^m\sum_{p+q=2m}\frac{\tilde z^q}{p!q!}w[p,q](z) \ee
we find that
\be \label{eq:CCA-orbifold} w[p,q](z_1)w[r,s](z_2)\sim - \frac{ps-qr}{2z_{12}}w[p\!+\!r\!-\!1,q\!+\!s\!-\!1](z_2)\,. \ee
Here we've expressed the OPE in terms of the inhomogeneous coordinates $z_i = \la\alpha\kappa_i\ra/\la\kappa_i\beta\ra$ chosen so that $\kappa_i = \beta$ lies at $z_i=\infty$. As expected, these are the defining relations of $\cL\ham(\bbC^2)^{\bbZ_2}$.


\subsection*{The correction at order \texorpdfstring{$c^2$}{c2}} \label{subsec:simplify-first-order}
Now let's consider the first-order correction to the perturbiner in $c^2$. The Laplacian on Eguchi-Hanson space is 
\be 
\label{eq:4.73}
\Delta_g = \Delta^{(0)} + c^2\Delta^{(1)} = \Delta_\delta - \frac{16c^2\ut^\da\ut^\db}{x^6}\dt_\da\dt_\db\,, \ee
so the first-order part of the perturbiner obeys 
\be \label{eq:SDGR-perturbiner-dif-1} \Delta^{(0)}\cP_\mathrm{SDGR}^{(1)}(x;k_1,k_2) = \{\tilde\p^\da\delta\Theta_1^{(0)}(x),\tilde\p_\da\delta\Theta_2^{(1)}(x)\} + (1\leftrightarrow2) - \Delta^{(1)}\cP^{(0)}_\mathrm{SDGR}(x;k_1,k_2)\,. 
\ee
We've already seen that in the holomorphic collinear singularity in $\cP^{(0)}_\mathrm{SDGR}(x;k_1,k_2)$ is a linear combination of null momentum eigenstates on flat space. The effect of the third term on the right-hand side is simply to shift these null momentum eigenstates to their curved space counterparts at first-order in $c^2$. Hard graviton generating functions in the CCA are identified with null momentum eigenstates on the Eguchi-Hanson background; therefore, this term does not modify the singular part of the celestial OPE.\footnote{This argument is a little too slick. It could be that the non-singular part of the flat space perturbiner $\cP^{(0)}_\mathrm{SDGR}(x;k_1,k_2)$ involves terms of the form $\la12\ra\log\la12\ra$. These can generate holomorphic collinear singularities when we differentiate under the integral sign to perform the Fourier transform. Terms of this type are in fact present, but it's not hard to show that they don't contribute to the celestial OPE.}

It's therefore sufficient to compute the holomorphic collinear singularity in
\be \label{eq:SDGR-singularity-integral} \int_{\bbR^4/\bbZ_2}\dif^4y\,\cG^{(0)}(x;y)\{\tilde\p^\da\delta\Theta_1^{(0)},\tilde\p_\da\delta\Theta_2^{(1)}\}
= - \frac{1}{4\pi^2}\int_{\bbR^4}\frac{\dif^4y}{(x-y)^2}\,\{\tilde\p^\da\delta\Theta_1^{(0)},\tilde\p_\da\delta\Theta_2^{(1)}\}\,. 
\ee
The order $c^2$ piece of the null momentum eigenstate is
\be \label{eq:first-order-eigenstate} 
\delta\Theta_i^{(1)}(y) = \frac{2}{\la\alpha i\ra^4}\bigg(\frac{\la\alpha|k_i y|\beta\ra}{y^2}\bigg)^2\frac{\sin(k_i\cdot y)}{k_i\cdot y} = \frac{2}{\la\alpha i\ra^4}\bigg(\frac{\la\al|k_i y|\beta\ra}{y^2}\bigg)^2\int_0^1\dif s\,\cos(s\,y\cdot k_i)\,, 
\ee
and so
\bea \label{eq:SDGR-vertex}
&\{\tilde\p^\da\delta\Theta_1^{(0)}(y),\tilde\p_\da\delta\Theta_2^{(1)}(y)\} \\
&= \frac{4\cos(y\cdot k_1)}{\la\alpha 1\ra^2\la\alpha 2\ra^2}\Bigg(\frac{2[\vt 1][v 2]}{y^6}\bigg([12] - \frac{6[v1][\vt 2]}{y^2}\bigg)\int_0^1\dif s\,\cos(s\,y\cdot k_2) \\
&+ \frac{[\vt 2]\la\alpha 2\ra[12]}{y^4}\bigg([12] - \frac{2([\vt 1][v2]+[v1][\vt 2])}{y^2}\bigg)\int_0^1\dif s\,s\sin(s\,y\cdot k_2) \\
&+ \frac{[\vt 2]^2\la\alpha2\ra^2[12]^2}{2y^4}\int_0^1\dif s\,s^2\cos(s\,y\cdot k_2)\Bigg)\,,
\eea
where again $|v]=y|\alpha\ra$ and $|\vt] = y|\beta\ra$.

\medskip

Given the number of terms present, evaluating the collinear singularity in equation \eqref{eq:SDGR-singularity-integral} is somewhat tedious. As such, we relegate the detailed computation to appendix \ref{app:SDGR-calcs}, and simply sketch the calculation for the final term; that is, we wish to compute the holomorphic collinear singularity in
\be -\frac{[12]^2}{2\pi^2\la\alpha1\ra^2}\int_0^1\dif s\,s^2\int_{\bbR^4}\frac{\dif^4y}{(x-y)^2y^4}\,[\vt2]^2\cos(y\cdot k_1)\cos(s\,y\cdot k_2)\,. 
\ee
The coefficient of the integrals depending only on spinor-helicity variables can be ignored for the moment. The first step is to combine the cosines using the double angle formula and replace the $[\vt2]^2$ factor in the integrand with derivatives to get
\be \label{eq:SDGR-simplified-integral} \frac{[12]^2}{4\pi^2\la\alpha1\ra^2}\la\beta\p_{\lambda_2}\ra^2\int_0^1\dif s\,\int_{\bbR^4}\frac{\dif^4y}{(x-y)^2y^4}\,\big(\cos(y\cdot k_-(s)) + \cos(y\cdot k_+(s))\big)\,, \ee
where $k_\pm(s) = k_1\pm sk_2$. Let's consider the integral 
\be \label{eq:Im} \cI_m(x;k) = \int_{\bbR^4}\frac{\dif^4y}{(x-y)^2y^{2(m+1)}}\,\cos(y\cdot k) \ee
for $m\in\bbZ$, which when $m=1$ appears twice in the inner integral of \eqref{eq:SDGR-simplified-integral}. It suffers from a divergence of order $2(m-1)$ as $y\to0$ for $m\in\bbZ_{\geq1}$. However, after taking the derivatives with respect to $\lambda_2$ in equation \eqref{eq:SDGR-simplified-integral} will obtain a finite answer.\footnote{This step is not strictly necessary: we can retain factors of $[vi],[\vt i]$ in the integrand, which are ultimately integrated against a Gaussian in a straightforward way. This would keep our integrals finite throughout the calculation. However, it's more computationally convenient to absorb these factors into derivatives with respect to spinor helicity variables.} We can rewrite \eqref{eq:Im} using standard tricks. First, Feynman parametrisation gives
\be \label{eq:Im-simplified} \cI_m(x;k) = (m+1)\int_0^1\dif t\,(1-t)^m\int_{\bbR^4}\frac{\dif^4y}{((y-tx)^2 + t(1-t)x^2)^{m+2}}\,\cos(y\cdot k)\,. \ee
Shifting $y\mapsto \tilde y = y+tx$
\bea &\cI_m(x;k) = (m+1)\int_0^1\dif t\,(1-t)^m\cos(t\,x\cdot k)\int_{\bbR^4}\frac{\dif^4\tilde y}{(\tilde y^2 + t(1-t)x^2)^{m+2}}\,\cos(\tilde y\cdot k) \\
&= \frac{1}{m!}\int_0^1\dif t\,(1-t)^m\cos(t\,x\cdot k)\int_0^\infty\dif r\,r^{m+1}e^{-rt(1-t)x^2}\int_{\bbR^4}\dif^4\tilde y\,\cos(\tilde y\cdot k)e^{-r\tilde y^2}\,.
\eea
The space-time integral is now a straightforward Fourier transform of a Gaussian, giving
\bea \label{eq:Im-fully-simplified}
&\cI_m(x;k) = \frac{\pi^2}{m!}\int_0^1\dif t\,(1-t)^m\cos(t\,x\cdot k)\int_0^\infty\dif r\,r^{m-1}e^{-rt(1-t)x^2 - k^2/4r} \\
&= \frac{\pi^2}{m!}\int_0^1\dif t\,(1-t)^m\cos(t\,x\cdot k)\int_0^\infty\dif r\,r^{m-1}e^{-rt(1-t)x^2-k^2/4r}\,. \eea
The $r$ integral can be performed directly by making the substitution $\nu = -r t(1-t)x^2 - k^2/4r$. We have
\be \label{eq:BesselK} \int_0^\infty\dif r\,r^{m-1}e^{-rt(1-t)x^2 - k^2/4r} = 2\bigg(\frac{k^2}{4t(1-t)x^2}\bigg)^{m/2}K_m(\sqrt{t(1-t)x^2k^2})\,, \ee
where $K_m(z)$ denotes a modified Bessel function of the second kind. For $m\in\bbZ_{\geq1}$ this has a pole of order $m$ in $t(1-t)x^2$, leading to a divergence in the outer integral of \eqref{eq:Im-fully-simplified} as $t\to0$. This reflects the divergence in the original expression \eqref{eq:Im} as $y\to0$. For $m\in\bbZ_{\leq-1}$ equation \eqref{eq:BesselK} has a pole order $-m$ in $k^2$, and at $m=0$ it has logarithmic singularity in both $t(1-t)x^2$ and $k^2$.

In particular, we find that $\cI_1(x,k_\pm(s))$ is non-singular in the holomorphic collinear limit (in which $k_\pm(s)^2 = \pm2s\la12\ra[12]\to0$). However, we have yet to take the derivatives with respect to $\lambda_2$ in equation \eqref{eq:SDGR-simplified-integral}. Doing so gives
\bea
&\la\beta\p_{\lambda_2}\ra^2\cI_1(x;k_\pm(s)) \\
&= \pi^2s^2\Bigg( - [\ut2]^2\int_0^1\dif t\,t^2(1-t)\cos(t\,x\cdot k_\pm(s))\int_0^\infty\dif r\,e^{-rt(1-t)x^2-k_\pm(s)^2/4r} \\
&- [\tilde u2]\la\beta1\ra[12]\int_0^1\dif t\,t(1-t)\sin(t\,x\cdot k_\pm(s))\int_0^\infty\frac{\dif r}{r}\,e^{-rt(1-t)x^2 - k_\pm(s)^2/4r} \\
&+ \frac{1}{4}\la\beta1\ra^2[12]^2\int_0^1\dif t\,(1-t)\cos(t\,x\cdot k_\pm(s))\int_0^\infty\frac{\dif r}{r^2}\,e^{-rt(1-t)x^2 - k_\pm(s)^2/4r}\Bigg)\,. \eea
The first of the above three terms is non-singular in the holomorphic collinear limit, but the second has a logarithmic divergence of the form $\log\la12\ra$. Logarithmic divergences of this type are expected to cancel: we demonstrate an analogous cancellation in the case of self-dual Yang-Mills explicitly in appendix \ref{app:logarithmic-cancellation}. This leaves the final term, which does contribute a first-order pole
\bea
&\int_0^\infty\frac{\dif r}{r^2}\,e^{-rt(1-t)x^2-k_\pm(s)^2/4r} = 2\bigg(\frac{4t(1-t)x^2}{k_\pm(s)^2}\bigg)^{1/2}K_1(\sqrt{t(1-t)x^2k_\pm(s)^2})\\ 
&\sim \frac{4}{k_\pm(s)^2} + \cO(\log(k_\pm(s)^2)) \sim \pm\frac{2}{s\la12\ra[12]} + \cO(\log\la12\ra)\,. \eea
Therefore
\be \label{eq:SDGR-singularity-term3} \la\beta\p_{\lambda_2}\ra^2\cI_1(x;k_\pm(s))\sim \pm\frac{s\pi^2\la1\beta\ra^2[12]}{2\la12\ra}\int_0^1\dif t\,(1-t)\cos(t\,x\cdot k_{\pm}(s)) + \cO(\log\la12\ra)\,. \ee
In sum, the holomorphic collinear singularity in equation \eqref{eq:SDGR-simplified-integral} takes the form
\be - \frac{\la1\beta\ra^2[12]^3}{4\la\alpha1\ra^2\la12\ra}\int_0^1\dif s\,s\int_0^1\dif t\,(1-t)\sin(t\,x\cdot k_1)\sin(st\,x\cdot k_2) + \cO(\log\la12\ra)\,. \ee
By exploiting the identity $\la\alpha1\ra\la2\beta\ra - \la\alpha2\ra\la1\beta\ra + \la12\ra = 0$, and rescaling $s$ by a factor of $1/t$ so that it now takes values in the range $[0,t]$, we can rewrite this in a slightly more symmetric form as
\be \label{eq:SDGR-final-term} - \frac{\la1\beta\ra\la2\beta\ra[12]^3}{4\la\alpha1\ra\la\alpha2\ra\la12\ra}\int_{0\leq s\leq t\leq 1}\dif s\,\dif t\,\frac{s(1-t)}{t^2}\sin(t\,x\cdot k_1)\sin(s\,x\cdot k_2) + \cO(\log\la12\ra)\,. \ee
We show in appendix \ref{app:SDGR-calcs} that the first and second terms in equation \eqref{eq:SDGR-vertex} contribute first-order poles of the form
\bea \label{eq:SDGR-remaining-terms}
- \frac{\la1\beta\ra\la2\beta\ra[12]^3}{4\la\alpha1\ra\la\alpha2\ra\la12\ra}&\int_{0\leq s\leq t\leq 1}\dif s\,\dif t\,\frac{s(1-t)^3}{t^2}\sin(t\,x\cdot k_1)\sin(s\,x\cdot k_2) + \cO(\log\la12\ra)\,, \\
\frac{2\la1\beta\ra\la2\beta\ra[12]^3}{4\la\alpha1\ra\la\alpha2\ra\la12\ra}&\int_{0\leq s\leq t\leq 1}\dif s\,\dif t\,\frac{s(1-t)^2}{t^2}\sin(t\,x\cdot k_1)\sin(s\,x\cdot k_2) + \cO(\log\la12\ra)
\eea
respectively. Putting together equations \eqref{eq:SDGR-final-term} and \eqref{eq:SDGR-remaining-terms}, and then symmetrising under the exchange $1\leftrightarrow2$ as indicated in equation \eqref{eq:SDGR-perturbiner-dif-1}, gives the simple pole
\be \label{eq:SDGR-perturbiner-singularity} - \frac{\la1\beta\ra\la2\beta\ra[12]^3}{4\la\alpha1\ra\la\alpha2\ra\la12\ra}\int_0^1\dif s\,\int_0^1\dif t\,\min(s,t)(1-\max(s,t))\sin(t\,x\cdot k_1)\sin(s\,x\cdot k_2)\,. \ee
This encodes the relevant first-order correction to the collinear singularity in the perturbiner $\cP_{\mathrm{SDGR}}(x;k_1,k_2)$\footnote{As mentioned above, there is a further contribution in equation \eqref{eq:4.73} which cannot be attributed to the deformation of the zeroth order orbifold perturbiner to its curved counterpart.}.
It therefore determines the order $c^2$ correction to the celestial OPE on Eguchi-Hanson.


\subsection*{CCA for self-dual gravity on Eguchi-Hanson}

We now recast the collinear singularity in the perturbiner of self-dual gravity on the Eguchi-Hanson background, evaluated in equation~\eqref{eq:SDGR-perturbiner-singularity}, using the soft expansion. We then identify the resulting deformation with that computed using twistor methods in section \ref{sec:twistor-algebra}.

\medskip

Let's proceed by expanding \eqref{eq:SDGR-perturbiner-singularity} in powers of $\kt_1,\kt_2$. We have
\bea &\int_0^1\dif s\,\int_0^1\dif t\,\min(s,t)(1-\max(s,t))\sin(t\,x\cdot k_1)\sin(s\,x\cdot k_2) \\
&= \sum_{m,n=0}^\infty\frac{(-)^{m+n}(x\cdot k_1)^{2m+1}(x\cdot k_2)^{2n+1}}{(2m+1)!(2n+1)!}\int_0^1\dif s\,\int_0^1\dif t\,\min(s,t)(1-\max(s,t))t^{2m+1}s^{2n+1} \\
&= \sum_{m,n=0}^\infty\frac{(-)^{m+n}(x\cdot k_1)^{2m+1}(x\cdot k_2)^{2n+1}}{(2m+1)!(2n+1)!(2m+3)(2n+3)(2m+2n+5)}\,.
\eea
Writing $\mu_i^\da = x^{\da\alpha}\kappa_{i\alpha}$, in the holomorphic collinear limit $x\cdot k_1\to\tau\mu_2^\da\kt_{1\da}$, $x\cdot k_2\to\mu_2^\da\kt_{2\da}$. The above is therefore equivalent to
\be \sum_{m,n=0}^\infty\sum_{p=0}^{2m+1}\sum_{r=0}^{2n+1}\frac{(-)^{m+n}\tau^{2m+1}(\kt_{1\dzero})^p(\kt_{1\done})^{2m+1-p}(\kt_{2\dzero})^r(\kt_{2\done})^{2n+1-r}(\mu_2^\dzero)^{p+r}(\mu_2^\done)^{2(m+n+1)-p-r}}{p!(2m+1-p)!r!(2n+1-r)!(2m+3)(2n+3)(2m+2n+5)}\,. \ee
 Changing dummy variables to $q = 2m + 1 - p$, $s = 2n + 1 - r$ gives
\be \label{eq:SDGR-singularity-soft}
- \sum_{\substack{p,q=0 \\ p+q\equiv1\,(2)}}^\infty\sum_{\substack{r,s=0 \\ r+s\equiv1\,(2)}}^\infty(-)^{(p+q+r+s)/2}\frac{\tau^{p+q}(\kt_{1\dzero})^p(\kt_{1\done})^q(\kt_{2\dzero})^r(\kt_{2\done})^s(\mu_2^\dzero)^{p+r}(\mu_2^\done)^{q+s}}{p!q!r!s!(p+q+2)(r+s+2)(p+q+r+s+3)}\,.
\ee
On Eguchi-Hanson, we identify null momentum eigenstates $\delta\Theta_k(x)$ with hard generating functions in the CCA (which we denote by $W(\kt,\kappa)$ on the curved background). As discussed in subsection \ref{subsec:simplify-first-order}, we've already accounted for the change in the zeroth order OPE \eqref{eq:CCA-orbifold} from this redefinition of the states by discarding the corresponding contribution to the collinear singularity in the self-dual gravity perturbiner. Since we're working to first-order in $c^2$, we can therefore continue to use the flat space identification $W(\kt,\kappa)\leftrightarrow\cos(x\cdot k)/\la\alpha\kappa\ra^4 + \cO(c^2)$. Decomposing into soft modes
\be W(\kt,\kappa) = \sum_{\substack{p,q=0 \\ p+q\equiv0\,(2)}}\frac{(-)^{(p+q)/2}(\kt_\dzero)^p(\kt_\done)^q}{p!q!}W[p,q](\kappa) \ee
this becomes $W[p,q](\kappa)\leftrightarrow (\mu^\dzero)^p(\mu^\done)^q + \cO(c^2)$. Under this identification, equation \eqref{eq:SDGR-singularity-soft} reads
\be - \la\alpha2\ra^4\sum_{\substack{p,q=0 \\ p+q\equiv1\,(2)}}^\infty\sum_{\substack{r,s=0 \\ r+s\equiv1\,(2)}}^\infty\frac{(-)^{(p+q+r+s)/2}\tau^{p+q}(\kt_{1\dzero})^p(\kt_{1\done})^q(\kt_{2\dzero})^r(\kt_{2\done})^s}{p!q!r!s!(p+q+2)(r+s+2)(p+q+r+s+3)}\,W[p+r,q+s](\kappa_2)\,. \ee
Reintroducing the coefficient from equation \eqref{eq:SDGR-perturbiner-singularity}, and absorbing the factor of $[12]^3$ into the sum through a shift of the dummy variables gives
\bea
&- \frac{2\la\al2\ra^3\la1\beta\ra\la2\beta\ra}{\la\al1\ra\la12\ra}\sum_{\substack{p,q=0 \\ p+q\equiv0\,(2)}}^\infty\sum_{\substack{r,s=0 \\ r+s\equiv0\,(2)}}^\infty\frac{(-)^{(p+q+r+s)/2}\tau^{p+q-3}(\kt_{1\dzero})^p(\kt_{1\done})^q(\kt_{2\dzero})^r(\kt_{2\done})^s}{p!q!r!s!} \\
&\qquad R_3(p,q,r,s)\,\phi_3\bigg(\frac{p+q}{2},\frac{r+s}{2}\bigg)\,W[p+r-3,q+s-3](\kappa_2)\,.
\eea
The indices in this sum are restricted to the range $p+q,r+s\geq4$, $p+r,q+s\geq 3$, and we have defined
\be \phi_3(m,n) = - \frac{3}{4(2m-1)(2n-1)(2(m+n)-3)}\,. \ee
The order $c^2$ correction to the OPE of soft modes can read off as
\bea \label{eq:SDGR-OPE-first}
&W[p,q](\kappa_1)\,W[r,s](\kappa_2) \\
&\sim - \frac{2c(\kappa_1)c(\kappa_2)}{\la12\ra}\tau^{p+q-5}R_3(p,q,r,s)
\,\phi_3\bigg(\frac{p+q}{2},\frac{r+s}{2}\bigg)\,W[p+r-3,q+s-3](\kappa_2) \\
&\sim - \frac{2c^2(\kappa_2)}{\la12\ra}\tau^{p+q-3}R_3(p,q,r,s)
\,\phi_3\bigg(\frac{p+q}{2},\frac{r+s}{2}\bigg)\,W[p+r-3,q+s-3](\kappa_2)\,, \eea
where in the second equality we've employed a Schouten identity. Equivalently in inhomogeneous coordinates
\bea \label{eq:SDGR-OPE-first-inhom}
&W[p,q](z_1)\,W[r,s](z_2) \\
&\sim - \frac{2z_2^2c^2}{z_{12}}R_3(p,q,r,s)\,\phi_3\bigg(\frac{p+q}{2},\frac{r+s}{2}\bigg)\,W[p+r-3,q+s-3](z_2)\,.
\eea
This is precisely the order $c^2$ contribution to the algebra we found working directly on twistor space in section \ref{sec:twistor-algebra}.

\medskip

Under the assumption that a deformed celestial chiral algebra exists and that it remains the loop algebra of some Lie algebra deformation of $w_\wedge$, this first-order computation is enough to determine it uniquely. This is because the $W(\mu)$ algebras (including the scaling limit $\mu\to\infty$) are the most general filtered deformations of $w_\wedge$ \cite{Post:1996gl}, the grading on $w_\wedge$ coinciding with the one induced by space-time dilations. Our explicit computations in this section show that we get the scaling limit.


\subsection*{Self-dual Yang-Mills} \label{subsec:SDYM-space-time}

Here we provide space-time calculation of the CCA of self-dual Yang-Mills on the Eguchi-Hanson background,  again working to first-order in $c^2$. The calculation is parallel to that in self-dual gravity and we leave many of the details to appendix~\ref{app:SDYM-perturbiner}. Again we show that this self-dual Yang-Mills perturbiner calculation recovers the algebra $\cL S_\wedge(\infty)$, to first-order in $c^2$.

\medskip

On a background Eguchi-Hanson space, self-dual Yang-Mills may be described perturbatively by a Chalmers-Siegel action~\cite{Siegel:1992wd,Chalmers:1996rq} which is analogous to the gravitational action we discussed in section \ref{Action_section}
\begin{equation}
\label{eq:Chalmers-Siegel-SDYM-action}
    S[\Tilde{\Phi},\Phi] = -\int \dif^4x\,\tr\left(\Tilde{\Phi}\bigg(\Delta_g\Phi - \frac{1}{2} \left[\tilde\p^\da\Phi,\tilde\p_\da\Phi\right]\bigg)\right)\,, 
\end{equation}
where again $\Delta_g$ is the Eguchi-Hanson Laplacian and the fields $\Phi,\Tilde\Phi\in \Omega^0(\cM,\fg)$. The vertex now involves the Lie bracket $[\ ,\ ]$ on $\fg$. This action is equivalent to the twistor action~\eqref{eq:sdYM-twistor-action} at the classical level. The SDYM perturbiner $\cP_{\rm SDYM}$ satisfies
\be \label{eq:SDYM-perturbiner-dif} \Delta_g\cP_\mathrm{SDYM}(x;k_1,k_2) = [\tilde\p^\da \delta\Phi_1(x),\tilde\p_\da \delta\Phi_2(x)]\,,
\ee
where the wavefunctions $\delta\Phi_i$ for fluctuations of the positive helicity gluon may be obtained by dressing the graviton states~\eqref{eq:Plebanski-fluctuations} with Lie algebra generators $t_\sfa$ and shifting their normalizations
\be 
\delta\Phi_{i\sfa}(x) = \frac{t_\sfa}{\la\al i\ra^2} \cos\sqrt{(k_i\cdot x)^2 - \frac{4c^2\la\alpha|k_i x|\beta\ra^2}{x^4}}\,. \ee

\medskip

At zeroth order in $c^2$ it's easy to compute the holomorphic collinear singularity in the perturbiner to find the celestial OPE
\be j_\sfa(\kt_1,\kappa_1)\,j_\sfb(\kt_2,\kappa_2) \sim \frac{1}{4\la12\ra}\tau^{-1}f_{\sfa\sfb}^{~~\sfc}\big(j_\sfc(\tau\kt_1 - \kt_2,\kappa_2) + j_\sfc(\tau\kt_1 + \kt_2,\kappa_2)\big)\,. \ee
Equivalently, in terms of soft modes
\be j_\sfa[p,q](\kappa_1)\,j_\sfb[r,s](\kappa_2)\sim - \frac{f_{\sfa\sfb}^{~~\sfc}}{2\la12\ra}\tau^{p+q-1}j_\sfc[p+r,q+s](\kappa_2)\,, \ee
where the indices are restricted by $p+q\equiv r+s\equiv0\Mod{2}$. Working in inhomogeneous coordinates
\be j_\sfa[p,q](z_1)\,j_\sfb[r,s](z_2)\sim - \frac{f_{\sfa\sfb}^{~~\sfc}}{2z_{12}}j_\sfc[p+r,q+s](z_2) \ee
Unsurprisingly, the CCA on the orbifold is isomorphic to $\cL\fg[\bbC^2]^{\bbZ_2}$, where we recall that $\cL\fg[\bbC^2]$ is the S-algebra of section \ref{sec:UniqueDeformation}.

\medskip

The  order $c^2$ part of equation \eqref{eq:SDYM-perturbiner-dif} is
\be \label{eq:SDYM-perturbiner-dif-1} \Delta^{(0)}\cP^{(1)}_\mathrm{SDYM}(x;k_1,k_2) = [\tilde\p^\da\delta\Phi^{(0)}_1(x),\tilde\p_\da\delta\Phi^{(1)}_2(x)] - (1\leftrightarrow2) - \Delta^{(1)}\cP^{(0)}_\mathrm{SDYM}(x;k_1,k_2)\,. 
\ee
As we saw in section \ref{subsec:simplify-first-order}, the final term is responsible for shifting the states in the zeroth order OPE to their curved counterparts. It is therefore sufficient to find the holomorphic collinear singularity induced by the first two terms. Employing the arguments sketched above for gravity, we find that the leading simple pole takes the form
\be \label{eq:SDYM-perturbiner-singularity}
\frac{\la1\beta\ra\la2\beta\ra[12]^2}{4\la12\ra}\int_0^1\dif s\,\int_0^1\dif t\,(1-\max(s,t))\cos(t\,x\cdot k_1)\cos(s\,x\cdot k_2)\,. \ee
For completeness we have included the full calculation in appendix \ref{app:SDYM-perturbiner}. Furthermore, in appendix \ref{app:logarithmic-cancellation} we show explicitly that there's no subleading logarithmic singularity of the form $\log\la12\ra$. Expanding in terms of soft modes (which we denote by $J_\sfa[m,n]$ on the curved background) the first-order correction to the celestial OPE is
\bea
&J_\sfa[p,q](\kappa_1)\,J_\sfb[r,s](\kappa_2) \\
&\sim - \frac{2c(\kappa_1)c(\kappa_2)}{\la12\ra}\tau^{p+q-3}R_2(p,q,r,s)\,\psi_2\bigg(\frac{p+q}{2},\frac{r+s}{2}\bigg)\,f_{\sfa\sfb}^{~~\sfc}\,J_\sfc[p\!+\!r\!-\!2,q\!+\!s\!-\!2](\kappa_2)\,, \\
&\sim - \frac{2c^2(\kappa_2)}{\la12\ra}\tau^{p+q-1}R_2(p,q,r,s)\,\psi_2\bigg(\frac{p+q}{2},\frac{r+s}{2}\bigg)\,f_{\sfa\sfb}^{~~\sfc}\,J_\sfc[p\!+\!r\!-\!2,q\!+\!s\!-\!2](\kappa_2)\,,
\eea
or equivalently
\bea
&J_\sfa[p,q](z_1)J_\sfb[r,s](z_2) \\
&\sim - \frac{2z_2^2c^2}{z_{12}}\psi_2\bigg(\frac{p+q}{2},\frac{r+s}{2}\bigg)R_2(p,q,r,s)f_{\sfa\sfb}^{~~\sfc}J_\sfc[p+r-2,q+s-2](z_2)
\eea
in inhomogeneous coordinates. Here
\be \label{eq:psi-2} 
\psi_2(m,n) = - \frac{1}{4(2m-1)(2n-1)(2(m+n)-1)}\,,
\ee
so this agrees with the order $c^2$ correction to the $S$-algebra we found working directly on twistor space in section \ref{sec:twistor-algebra}.


\section{Switching on non-commutativity} \label{sec:non-commutativity}

So far we've seen how to obtain $\cL W(\infty)$ as the CCA of self-dual gravity on the Eguchi-Hanson background. On the other hand, the loop algebra of the symplecton $\cL W(-3/16)$, is the CCA of Moyal-deformed self-dual gravity on the orbifold $\bbR^4/\bbZ_2$ \cite{Bu:2022iak} as discussed in section \ref{sec:UniqueDeformation}. It's then natural to ask whether $\cL W(\mu)$ for generic $\mu$ can arise as the CCA of a gravitational theory. In this section, we argue that this is indeed the case, by considering both, non-commutativity and an Eguchi-Hanson background. In this section, we will consider this Moyal deformed self-dual gravity on the Eguchi-Hanson background.


\subsection*{The CCA for Moyal deformed self-dual gravity} \label{subsec:Moyal}

Let's begin by reviewing the Moyal deformation of self-dual gravity on flat space \cite{Strachan:1992em} from a twistor perspective. Classically this theory is described by the previously considered space-time action \eqref{Moyal_action}, however, it can also be described from a twistor action which will simplify the identification of the corresponding celestial chiral algebra.
See also {\it e.g.}~\cite{Kapustin:2000ek,Bridgeland:2020zjh,Gilson:2008hp,Hannabuss:2001xj,Monteiro:2022lwm,Guevara:2022qnm} for other treatments of self-dual gravity and self-dual Yang-Mills on non-commutative twistor spaces. Moyal deformed self-dual gravity involves switching on non-commutativity associated to the Poisson bracket $\p^\da\vee\p_\da/2$ on the fibres of $\PT\to\CP^1$. However, there's a catch: the bracket is twisted by $\cO(-2)$. In the twistor uplift of self-dual gravity as Poisson-BF theory this is compensated by also twisting $h$, but to turn on non-commutativity we must instead specify an unweighted bracket, which can be achieved by fixing a holomorphic section of $\cO(2)$.\footnote{Another way to achieve a consistent action is to consider infinitely many fields of increasing weight leading to chiral higher spin theories \cite{Tran:2022tft, Krasnov:2021nsq, Monteiro:2022xwq,Adamo:2022lah}.}  It's natural to make the same choice as for the defect in section \ref{sec:deforming-twistor-space}, so that the weightless Poisson structure reads
\be \pi_0 = \frac{1}{2}\la\al\lambda\ra\la\lambda\beta\ra\,\p^\da\vee\p_\da\,. \ee
We can then switch on a Moyal product associated to this Poisson structure, with formal parameter $\fq$. This parameter has weight 2 under scaling the twistor fibres. We can then write down a non-commutative analogue of Poisson BF-theory on twistor space in which the Poisson bracket in \eqref{eq:twistor-sd-gravity-action} is replaced by the Moyal bracket
\bea \label{eq:Moyal-bracket}
\{f,g\} \mapsto [f,g]_\fq &= \frac{2}{\fq(\lambda)} m\circ\sin(\fq\pi_0)(f\otimes g) \\
&= \sum_{k=0}^\infty\frac{\fq^{2k}(\lambda)}{2^{2k}\,(2k+1)!}\p_{\da_1}\dots\p_{\da_{2k+1}}f\ \p^{\da_1}\dots\p^{\da_{2k+1}}g\,. 
\eea
Here $m$ is the product map $f\otimes g\mapsto fg$, and for convenience we've defined $\fq(\lambda) = \fq\la\al\lambda\ra\la\lambda\beta\ra$. Since the Moyal bracket is defined using holomorphic bidifferential operators, it extends to $(0,q)$-forms in a straightforward way. It's also the commutator of an associative star product
\be [f,g]_\fq = \frac{1}{\fq(\lambda)}(f\star_\fq g - g\star_\fq f)\,, \ee
where
\be f\star_\fq g = m\circ\exp(\fq\pi_0)(f\otimes g)\,. \ee
We can similarly introduce a non-commutative analogue of holomorphic BF theory for the Lie algebra $\fg$ by replacing the Lie bracket appearing in \eqref{eq:sdYM-twistor-action} with its non-commutative counterpart. Since the Lie bracket is already antisymmetric, this depends on the star product through its anticommutator.

\medskip

In space-time, these correspond to Moyal deformed self-dual gravity and Yang-Mills on flat space, whose equations of motion describe non-commutative instantons. Their celestial chiral algebras were identified in \cite{Bu:2022iak} as discussed in chapter \ref{chapter2}, and can be straightforwardly recovered from the twistor description. In order to do so, let's first recall the definition of the Weyl algebra $\diff_\fq(\bbC)$. It's the quotient of the free algebra on two generators $u,v$ over $\bbC\llbracket\fq\rrbracket$ by the ideal $\mathrm{span}\{uv-vu = \fq\}$.

In the case of Moyal deformed self-dual gravity, the algebra of functions on the twistor fibre over $\lambda\neq\alpha,\beta$ inherits a Lie bracket from the interaction vertex of the non-commutative deformation of Poisson-BF theory. The resulting Lie algebra is isomorphic to $\diff_\fq(\bbC)$ equipped with the standard Moyal bracket. Forming the loop algebra gives the CCA.

For Moyal deformed self-dual Yang-Mills, the algebra of functions on the twistor fibre over $\lambda\neq\alpha,\beta$ inherits a Lie bracket from the interaction vertex of the non-commutative BF theory. The resulting Lie algebra structure on $\diff_\fq(\bbC)\otimes\fgl(N)$ is defined using the star product on the first factor and matrix multiplication on the second. The CCA is then obtained by taking the loop algebra.

Quotienting space-time by $\bbZ_2$, the Weyl algebra $\diff_\fq(\bbC)$ is restricted to its $\bbZ_2$-invariant subspace, the symplecton $W(-3/16;\fq)$ \cite{Pope:1989sr}. Making this replacement in the CCAs of Moyal deformed self-dual gravity and Yang-Mills on flat space gives their counterparts on the orbifold $\bbR^4/\bbZ_2$.


\subsection*{The CCA for Moyal deformed self-dual gravity on Eguchi-Hanson} 
\label{subsec:NCEH}

Now let's consider the result of coupling the non-commutative analogue of Poisson-BF theory with the Poisson bracket replaced by \eqref{eq:Moyal-bracket} to a holomorphic surface defect as in equation \eqref{eq:twistor-defect-action}. In the presence of this defect, the equation of motion for $h$ becomes
\be \label{eq:sourced-NCBF}
\bar\p h + \frac{1}{2}[h,h]_\fq = 4\pi^2 c^2(\lambda)\,\bar\delta^2(\mu)\,.
\ee
We can treat the parameter $c$ (which has weight 2 under scaling the twistor fibres) as a finite complex parameter, or as a formal parameter proportional to $\fq$ with constant of proportionality $\tilde c$. In both cases, the resulting CCA will be defined over $\bbC\llbracket\fq\rrbracket$. We solve for $h$ exactly as we did in section \ref{sec:deforming-twistor-space} to get
\be h = c^2(\lambda)\,\frac{[\hat\mu\,\dif\hat\mu]}{2[\mu\,\hat\mu]^2}\,. \ee
This induces a non-commutative analogue of a Dolbeault operator
\be \nbar_\fq = \dbar + [h,\ ]_\fq = \dbar - c^2(\lambda)[\hat\mu\,\dif\hat\mu]\sum_{k=0}^\infty\frac{(k+1)\fq^{2k}(\lambda)}{[\mu\,\hat\mu]^{2k+3}2^{2k}}\hat\mu^{\al_1}\dots\hat\mu^{\dot\al_{2k+1}}\p_{\da_1}\dots\p_{\dot\al_{2k+1}}\,. \ee
Functions, and more generally differential forms, in the kernel of this operator should be viewed as `holomorphic'. Since $\nbar_\fq$ distributes over the star product, star products of holomorphic functions in this non-commutative sense are themselves holomorphic. In particular, it's easy to check that the functions
\be X^{\da\db} = \mu^\da\mu^\db - c^2(\lambda)\frac{\hat\mu^\da\hat\mu^\db}{[\mu\,\hat\mu]^2} \ee
from equation \eqref{eq:newholomorphic} remain holomorphic. We can then determine
\bea \label{eq:deformed-star-product}
&X^{\da\db}\star_\fq X^{\dc\dd} = X^{\da\db}X^{\dc\dd} + \frac{\fq(\lambda)}{2}\big(\eps^{\da\dc}X^{\db\dd} + \eps^{\da\dd}X^{\db\dc} + \eps^{\db\dc}X^{\da\dd} + \eps^{\db\dd}X^{\da\dc}\big) \\
&+ \frac{\fq^2(\lambda)}{4}\bigg(\eps^{\da\dc}\eps^{\db\dd} + \eps^{\da\dd}\eps^{\db\dc} - \frac{12c^2(\lambda)\hat\mu^\da\hat\mu^\db\hat\mu^\dc\hat\mu^\dd}{[\mu\,\hat\mu]^4} \bigg)\,.
\eea
We remark that the explicit $\hat\mu$ dependence in the final term, which may seem surprising, is needed to compensate for the fact that the commutative product $X^{\da\db}X^{\dc\dd}$ is not in the kernel of $\nbar_\fq$. 

From the above, we infer that
\be \label{eq:commutator} [X^{\da\db},X^{\dc\dd}]_\fq = \eps^{\da\dc}X^{\db\dd} + \eps^{\da\dd}X^{\db\dc} + \eps^{\db\dc}X^{\da\dd} + \eps^{\db\dd}X^{\da\dc}\,. \ee
If there were no further constraints on the products of the $X^{\da\db}$, we'd learn that on each generic twistor fibre ($\lambda\neq\alpha,\beta$), under the star product $\star_\fq$ they generate the universal enveloping algebra (UEA) of\footnote{Here $\fg_{\fq(\lambda)}$ denotes the Lie algebra over $\bbC\llbracket\fq\rrbracket$ obtained by multiplying the structure constants of $\fg$ by $\fq(\lambda)$.} $\fsl_{2,\fq(\lambda)}$.  However, contracting indices in~\eqref{eq:deformed-star-product} gives
\be \label{eq:Casimir} 
X^{\da\db}\star_\fq X_{\da\db} = X^{\da\db}X_{\da\db} + \frac{3\fq^2(\lambda)}{2} = -2c^2(\lambda) + \frac{3\fq^2(\lambda)}{2}\,, 
\ee
where the second equality uses the constraint $X^{\da\db}X_{\da\db} = -2c^2(\lambda)$. At this point we notice that in the $\fq\to0$ limit we recover the commutative algebra generated by the  $X^{\da\db}$ subject to the constraint $X^{\da\db}X_{\da\db} = -2c^2(\lambda)$, equipped with the standard Poisson structure \eqref{eq:Poisson-bracket}. This is isomorphic to $W(\infty)$, as we saw in section \ref{sec:twistor-algebra}.

The standard normalization of the Casimir in the UEA of $\fsl_{2,\fq(\lambda)}$ is $C = -X^{\da\db}\star_\fq X_{\da\db}/8$, which has eigenvalues $\fq^2(\lambda)\mu = \fq^2(\lambda)\sigma(\sigma+1)$. Taking $c$ to be a formal parameter proportional to $\fq$, {\it i.e.}, $c = \fq\tilde c$, we have
\be C = -\frac{1}{8}X^{\da\db}\star_\fq X_{\da\db} = \frac{\fq^2(\lambda)}{16}(4\tilde c^2 - 3) = \fq^2(\lambda)\mu\,, \ee
so that $\mu = (4\tilde c^2 - 3)/16$. On each twistor fibre the $X^{\da\db}$ then generate $U(\fsl_{2,\fq(\lambda)})/\mathrm{span}\{C - \fq^2(\lambda)\mu\}$. The Lie algebra defined through the commutator is isomorphic to $W(\mu;\fq(\lambda))\cong W(\mu;\fq)$ \cite{Pope:1989sr}. Forming the loop algebra gives the CCA of Moyal deformed self-dual gravity on Eguchi-Hanson.

Setting the Eguchi-Hanson parameter $c=0$, the non-commutative algebra generated by the $X^{\da\db}$ equipped with the bracket \eqref{eq:commutator} is isomorphic to the symplecton $W(-3/16;\fq)$ \cite{Pope:1989sr}, consistent with the results of \cite{Bu:2022iak} and chapter \ref{chapter2}. Indeed, equation \eqref{eq:Casimir} is the unique consistent relation with the correct $\fq\to0$ limit and compatible with the grading induced by scaling the twistor fibres. Notice that the shift $3\fq^2(\lambda)/4$ to the Casimir for $X^{\da\db}\star_\fq X_{\da\db}$ compared to that for the commutative $X^{\da\db}X_{\da\db}$ is essential to find the symplecton algebra at $\mu=-3/16$. While this is the expected value~\cite{Pope:1991ig}, it is rather unusual -- 
for example, quantizing coadjoint orbits of $\fsl_2$ via the A-model or Duflo-Kirillov-Kontsevich map induces a shift to $\mu = -1/4$ \cite{Gukov:2008ve,Chervov:2004tw} -- and it is gratifying to obtain it so directly.

By varying $\tilde c$ we sweep out all possible choices of the parameter $\mu$. Two cases are of particular interest: setting $\tilde c^2 = -1/4$, we find that $W(\mu;\fq)$ is isomorphic to the wedge subalgebra of $W_{1+\infty}$, whilst at $\tilde c^2=3/4$ we have instead the wedge subalgebra of $W_\infty$. In particular, this gives a bulk interpretation for the deformation to $W_{1+\infty}$ as speculated in \cite{Strominger:2021lvk}. In Euclidean signature, it's natural to take $\tilde c^2\geq0$ so that only those $W(\mu;\fq)$ algebras with $\mu\geq-3/16$ are attainable, excluding the wedge subalgebra of $W_{1+\infty}$. We do not see any restrictions on the sign of $\tilde c^2$ in ultrahyperbolic signature.

\section{Discussion}
\label{sec:discussion}

One motivation for this chapter was to clarify the relationship between the vertex algebras arising in the celestial holography literature, and the infinite-dimensional Lie algebras arising as wedge subalgebras of infinite $W$-algebras. CCAs for the class of self-dual theories considered here are loop algebras of infinite-dimensional Lie algebras. Let us briefly summarise those which feature in this chapter.

For self-dual gravity on the Eguchi-Hanson background, we've found that the appropriate infinite-dimensional Lie algebra is $W(\infty)$. Taking the limit of the Eguchi-Hanson parameter $c\to0$, the geometry degenerates to the orbifold $\bbR^4/\bbZ_2$. This is reflected in the CCA: the $W(\infty)$ algebra contracts to $w_\wedge$, the wedge subalgebra of $w_{1+\infty}$. $w_\wedge$ is also a $\bbZ_2$ quotient of $\ham(\bbC^2)$, corresponding to the CCA of self-dual gravity on flat space-time.

Allowing the twistor space to become non-commutative with formal parameter $\fq$, corresponds to considering the Moyal deformation of self-dual gravity. Working on an Eguchi-Hanson background with formal parameter $c = \tilde c\fq$, the appropriate infinite-dimensional Lie algebra is likewise deformed to $W(\mu;\fq)$, where
\be \label{eq:mu-again} \mu = \frac{4\tilde c^2 - 3}{16}\,. \ee
In particular, when $\tilde c=0$ we recover the symplecton, itself the $\bbZ_2$ quotient of the Weyl algebra. This is the infinite-dimensional algebra determining the CCA of Moyal deformed self-dual gravity on flat space-time as discussed in chapter \ref{chapter2} \cite{Bu:2022iak}. 

All of the above statements carry over to self-dual Yang-Mills on Eguchi-Hanson, and, taking appropriate care to track powers of $\fq$, to its Moyal deformation.

\medskip

We conclude with a brief discussion of some future directions suggested by the results of this chapter.

Firstly, as mentioned above, it is a natural question whether the tree-level dictionary \eqref{eq:mu-again} can be uplifted to a fully fledged holographic duality. We hope that in the future it will be possible to engineer such a duality from some topological string theory on twistor space or some closely related space.

Secondly, it is clear that many of the considerations of this chapter can be extended to so-called \emph{ALE spaces}, and (self-dual) gravitational instantons more generally as will be commented on in chapter \ref{chapter6} and further discussed in future work. Particularly, these ALE spaces are given by hyperkähler metrics on $\widetilde{\mathbb{C}^2/\Gamma}$ \cite{Kronheimer:1989zs}, where $\Gamma \subset \text{SU}(2)$ is a finite subgroup. Such finite groups $\Gamma$ admit an \emph{ALE-classification} and Eguchi-Hanson space corresponds to the simplest case A$_1$ where $\Gamma=\mathbb{Z}_2$.
For a more general finite subgroup $\Gamma \subset \text{SU}(2)$, it turns out that $\mathfrak{ham}(\mathbb{C}^2)^{\Gamma}=\mathfrak{ham}(\mathbb{C}^2/\Gamma)$ admits a $(\text{rk}(\Gamma)+1)$-parameter deformation corresponding to non-commutativity and the radii of various $\mathbb{CP}^1$s that are present in the bulk geometry of the corresponding ALE space. Obtaining these ALE spaces from a (multi-centred) backreaction will be discussed in future work\footnote{See figure \ref{fig:moreDefects}.}. In the A$_k$-case with $\Gamma=\mathbb{Z}_{k+1}$, the metrics are the well-known multi-centred \emph{Gibbons-Hawking metrics} \cite{Gibbons:1978tef} which will be briefly discussed below in chapter \ref{chapter6}.

\medskip

In \cite{Mago:2021wje,Ren:2022sws} the authors explore certain Lie algebra deformations of tree-level CCAs for theories of self-dual gravity and Yang-Mills non-minimally coupled to matter. The Jacobi identity constrains the deformed OPEs, and hence the possible coupled theories admitting CCAs. The Lie algebras obtained there are not directly related to wedge subalgebras of $W_{1+\infty}$ and $W_{\infty}$, which only arise as deformations of the $\bbZ_2$ fixed point subalgebra of $\ham(\bbC^2)$ as discussed above. They're more closely related to higher spin analogues of the symplecton algebra restricted to $|s|\leq2$ \cite{Monteiro:2022xwq}.

The CCAs in this note are all loop algebras of infinite-dimensional Lie algebras. It's natural to ask whether still more complicated celestial chiral algebras can arise geometrically; say, with higher order poles or non-linear OPEs? A potentially related issue is that self-dual Einstein gravity has trivial (or more precisely distributional) tree amplitudes, a fact that remains true on any on-shell background.\footnote{As suggested to us by Kevin Costello: since the twistor space of any self-dual Ricci-flat metric fibres over $\CP^1$, scattering states lift to twistor space with support in an arbitrarily small neighbourhood of their left-handed spinor helicity variable. Any tree diagram must include a vertex with two external legs; therefore, beyond $n=3$ it necessarily vanishes for generic kinematics, just as in flat space.}


One concrete way to obtain higher-order poles and non-linearities  is by incorporating loop corrections \cite{Costello:2022upu,Bittleston:2022jeq}, though this requires introducing states in the chiral algebra corresponding to negative helicity fields. Classically these transform in the adjoint representation of the algebra generated by the positive helicity states. At 1-loop the perturbiner gets corrected by the diagram illustrated in figure \ref{fig:loopsplitting}. Since all the external legs are incoming, this describes a correction to the OPE of positive helicity states which is proportional to the negative helicity states.

Deforming self-dual gravity at first-order, {\it e.g.}, to full Einstein gravity, we get non-vanishing loop amplitudes whose collinear singularities receive 1-loop corrections from this diagram. However, the collinear singularities of these amplitudes will not be universal unless the 1-loop all-plus amplitudes vanish. This is reflected in the chiral algebra, which does not have an associative operator product unless certain anomalies on twistor space, which can be identified with the space-time 1-loop all-plus amplitudes, vanish. 

In \cite{Bittleston:2022nfr} a number of methods of cancelling the twistorial anomaly/1-loop all-plus amplitudes in self-dual gravity were presented, inspired by analogous methods for self-dual Yang-Mills \cite{Costello:2021bah}. The simplest anomaly-free variant is $\cN=1$ self-dual supergravity. Another possibility is to couple to a gravitational axion with 4\textsuperscript{th}-order kinetic term, which cancels the 1-loop amplitudes through tree-level exchange. Putting either of these theories on twistor space will lead to consistent quantum-deformed CCAs.

%% file: Chapter6/chapter6.tex
\chapter{Switching on a cosmological constant}
\label{chapter5}

We saw that one of the successes of celestial holography \cite{Pasterski:2021raf} has been the identification of new  symmetry algebras of perturbative amplitudes in flat space~\cite{Guevara:2021abz, Strominger:2021lvk}, the celestial chiral algebras discussed previously.  It is important to understand how widely such new celestial chiral algebras apply, both in terms of their appearance in different theories on flat space and in terms of whether they exist beyond flat space\footnote{Particularly, beyond asymptotically (locally) flat spaces, such as Burns space or Eguchi-Hanson space which were discussed previously.}. In particular, if celestial holography is to be thought of as some limit of conventional AdS/CFT, could these symmetries have some precursor there and, if so, what role might they play in that context?  An answer to this question has been provided by Taylor \& Zhu in~\cite{Taylor:2023ajd}, at least as a first-order deformation in the cosmological constant $\Lambda$. In this work, we show that 
their answer naturally extends   to all orders in $\Lambda$ in a  twistorial representation of these symmetries as local holomorphic Hamiltonian diffeomorphisms of twistor space; these have a natural action  on the self-dual sector of Einstein gravity.

\medskip

In flat space, the celestial chiral algebra was first introduced~\cite{Guevara:2021abz} by examining collinear limits and splitting functions of amplitudes in gravity and Yang-Mills  at null infinity $\scri$ as discussed in chapter \ref{chapter1} and chapter \ref{chapter2}. This analysis revealed that the celestial chiral algebra associated with positive helicity gravitons is $\lham$, the loop algebra of the Lie-algebra of Hamiltonian vector fields on $\C^2$~\cite{Strominger:2021lvk}. In \cite{Adamo:2021lrv, Mason:2022hly}, it was explained how this essentially inverts the Penrose transform~\cite{Penrose:1969ae, Eastwood:1981jy} to realize positive helicity gravitons as deformations of twistor space.

The role of $\lham$ on twistor space was first identified by Penrose \cite{Penrose:1976js, Penrose:1976jq} in his non-linear graviton construction that was discussed in section \ref{section2.3}. Locally,  curved twistor spaces $\mathcal{PT}$ are deformations of a region in flat twistor space. These deformations are not arbitrary but are required to preserve a degenerate Poisson structure\footnote{Taking values in $\cO(-2)$, the square root of the canonical bundle, \emph{i.e.}, of homogeneity degree $-2$.} $\{\,\,,\,\,\}$, defined by a choice of skew bi-twistor $I^{ab}$ known as the \emph{infinity twistor}. (The reason for the name will become apparent below.) The Poisson algebra of Hamiltonians\footnote{Taking values in $\cO(2)$, the square root of the anti-canonical bundle.} preserving $\{\,\,,\,\,\}$ can be readily identified with $\lham$ as we saw in chapter \ref{chapter2} and the algebra has a two-fold role in this correspondence \cite{Adamo:2021lrv,Mason:2022hly}.  Firstly, it arises as those local holomorphic diffeomorphisms of twistor space that preserve the global geometric structure. Such local symmetries have a second interpretation as defining infinitesimal deformations of the complex structure on $\mathcal{PT}$; the non-linear graviton construction then realizes these as self-dual gravitons in space-time. 

\medskip
  
The non-linear graviton construction was extended to incorporate a cosmological constant by Ward \cite{Ward:1980am} - see also \cite{LeBrun:1982vjh, Salamon:1982}. 
When the cosmological constant is non-vanishing, the non-linear graviton simply relaxes the degeneracy requirement on the infinity twistor and the Poisson structure $\{\,\,,\,\,\}_\Lambda$ becomes non-degenerate\footnote{Strictly speaking, it becomes a non-degenerate Poisson structure on non-projective twistor space. On projective twistor space, it defines a so-called (holomorphic, twisted) \emph{Jacobi structure}.} with an additional term that can be chosen to be proportional to $\Lambda$. 

In this paper, our first aim is to show that the deformation of $\lham$ found by~\cite{Taylor:2023ajd} is indeed the algebra of Hamiltonians for $\{\,\,,\,\,\}$ on a region $\C^2\times \C^*\subset \PT$, which we will refer to as  $\LHL$. Specifically, $\LHL$ is given by
\begin{equation}
	\{w^p_{m,a},w^{q}_{n,b}\}_\Lambda=(m(q-1)-n(p-1))w^{p+q-2}_{m+n,a+b}-\Lambda(a(q-2)-b(p-2))w^{p+q-1}_{m+n,a+b}\,,
\end{equation}  
as we derive in equation~\eqref{eq:alg-neq0} below. Similarly to the $\Lambda=0$ case, the algebra may be interpreted as both describing infinitesimal diffeomorphisms of twistor space which preserve the Poisson structure, and as the Penrose transform of linearized gravitons. We also point out that different (non-isomorphic) algebras are possible depending on the model of Euclidean AdS$_4$ one considers since this changes the subset $\C^2\times \C^*\subset\PT$. The above algebra is adapted to hyperbolic space being presented as a ball. If one instead uses that upper-half space presentation, a different version of the algebra is obtained. 

The Hamiltonians considered above generate symmetries of the \emph{Mason-Wolf action}, which is the twistor action for self-dual gravity with cosmological constant, first constructed in~\cite{Mason:2007ct}. We will obtain the corresponding Noether charges directly in twistor space, showing that they reduce on-shell to pure boundary terms. This places our work as part of a long tradition of the study of hidden symmetries of the self-duality equations, together with associated conserved quantities, hierarchies and their Hamiltonian origins; see \emph{e.g.}~\cite{Boyer:1985aj,Park:1989fz,Park:1989vq,Mason:1990,Dunajski:2000iq,Dunajski:2003gp} for self-dual gravity and~\cite{Forgacs:1980su,Chau:1981gi,Mason:1991rf} for self-dual Yang Mills.  

\medskip

This chapter is loosely based on the paper \cite{Bittleston:2024rqe} and organized as follows: in section~\ref{sec:TwistorsforAdS} we review the basic construction of the twistor space of AdS$_4$, highlighting the role of the infinity twistor. In Section~\ref{sec:graviton_symmetries}, we explain how one can think of the celestial chiral algebra as the algebra of holomorphic symmetries of the complex structure on twistor space. We show explicitly how the cosmological constant deforms the standard $\lham$ algebra. In section~\ref{sec:symm} we explain how the symmetries arise as symmetries of the twistor action of \cite{Mason:2007ct}  for the self-dual Einstein sector  and identify the associated charges on twistor space.

\section{Twistors for (A)dS\texorpdfstring{$_4$}{4}}
\label{sec:TwistorsforAdS}

 As reviewed in chapter \ref{section2.1}, the construction of twistor space $\PT$ is conformally invariant. Indeed, $\CP^3$ naturally carries an action of $\SL(4,\C)$, the spin group of the complexification of the conformal group $\SO(2,4)$ in four dimensions, acting linearly on the homogeneous coordinates $Z^a$. We can describe the incidence relations \eqref{eq:LorentzianIncidence} more invariantly using embedding space coordinates: we first notice that we can pick a rational line by choosing any pair of distinct twistors $Z_1,Z_2 \in \mathbb{PT}$. Thus we can coordinatize space-time points in terms of skew bi-twistors of the form $X^{ab}= Z_1^{[a}Z_2^{b]}$. Conversely, any bi-twistor $X^{ab}$ is of this form for some $Z_1,Z_2$ if it satisfies the simplicity condition
\begin{equation}
    X\cdot X= \epsilon_{abcd}X^{ab}X^{cd}=0\,.
\end{equation}
Skew bi-twistors are natural homogeneous coordinates on $\CP^5$, so this construction identifies complexified, conformally compactified Minkowski space as the quadric $\{X\cdot X=0\}\subset\CP^5$. In this sense, the skew bi-twistors $X^{ab}$ are embedding coordinates. In terms of these, the incidence relations read
\begin{equation}
    \epsilon_{abcd}Z^aX^{bc}=0\,,
\end{equation}
which ensures that $Z$ lies on the line defined by the simple $X$.

\subsection*{The infinity twistor and Poisson structure}

A conformal scale is encoded by a choice of skew bi-twistor $I^{ab}_\Lambda$ known as the \emph{infinity twistor}.  As our notation emphasises, this bi-twistor depends on the cosmological constant $\Lambda$, and is normalized to obey
\begin{equation}
			I^{ab}_{\Lambda}I^\Lambda_{cb}=\Lambda\,\delta^a_c\,,
\label{eq:non-degenerate}
\end{equation}
where $I_{ab}^\Lambda=\frac{1}{2}\epsilon_{abcd}I_\Lambda^{cd}$ is the dual of $I^{ab}_\Lambda$. In particular, this implies (and in fact is implied by $I^{ab}_\Lambda I^{\Lambda}_{ab} = 4\Lambda$) that only when the cosmological constant is zero does the infinity twistor define a line $L_I\subset\PT$, or a point in space-time. A standard representation is
\begin{subequations}
\begin{equation}
I^{ab}_\Lambda=\left(\begin{array}{cc}
     \varepsilon^{\dot\alpha\dot\beta}&0  \\
     0&\Lambda\,\varepsilon_{\alpha\beta} 
\end{array}\right)
\end{equation}
for the infinity twistor, or equivalently 
\begin{equation}
I^\Lambda_{ab}=\left(\begin{array}{cc}
     \Lambda\,\varepsilon_{\dot\alpha\dot\beta}&0  \\
     0&\varepsilon^{\alpha\beta} 
\end{array}\right)
\end{equation}
\label{eq:infinity_twistors}
\end{subequations} for its dual. Using the infinity twistor, we may define an $\cO(-2)$-valued holomorphic Poisson structure\footnote{Formally, this is only a Poisson structure when considered on the $8$-dimensional non-projective twistor space. On the 6-dimensional twistor space, it is merely a so-called \emph{Jacobi-structure} which is not required to obey the Leibniz rule. We will discuss the distinction below.}
\begin{equation}
\label{eq:Lambdabracket}
    \{f,g\}_\Lambda =I_\Lambda^{ab}\frac{\p f}{\p Z^a}\frac{\p g}{\p Z^b}=\varepsilon^{\dot\alpha\dot\beta}\frac{\p f}{\p\mu^{\dot\alpha}}\frac{\p g}{\p\mu^{\dot\beta}}+\Lambda\,\varepsilon_{\alpha\beta}\frac{\p f}{\p\lambda_\alpha}\frac{\p g}{\p\lambda_\beta}\,,
\end{equation}
where the second equalities follow from using the representation
\eqref{eq:infinity_twistors}. When $\Lambda=0$, this simply is the degenerate Poisson bracket that we used in section \ref{section2.3} to define holomorphic Poisson BF theory.
Dually, we can use $I^\Lambda_{ab}$ to define the $\cO(2)$-valued 1-form
\begin{equation}
    \tau_\Lambda= I^{\Lambda}_{ab}\,Z^a\d Z^b=\langle\lambda\,\d\lambda\rangle+\Lambda[\mu\,\d\mu]\,.
\label{eq:taudef}
\end{equation}
We can state the non-degeneracy condition~\eqref{eq:non-degenerate} more geometrically by using D$^3Z$, the $\cO(4)$-valued holomorphic volume form of equation \eqref{eq:holomorphicvolumeform}.
Then~\eqref{eq:non-degenerate} can also be written as  
\begin{equation}
\tau_\Lambda\wedge\d\tau_\Lambda=2\Lambda\, \text{D}^3Z\,.
\end{equation}
The Poisson structure is then defined as the bivector obtained by contracting $\tau_\Lambda$ with the inverse of $\Omega$.
In this context, $\tau_\Lambda$ is often known as a holomorphic $\mathcal{O}(2)$-valued \emph{contact form}. 

On $\mathbb{C}^2\times \mathbb{C}^*\subset \mathbb{PT}$, we can introduce the inhomogeneous coordinates $(v^{\dot{0}},v^{\dot{1}},z)=(\mu^{\dot{0}}/\lambda_1,\mu^{\dot{1}}/\lambda_1,\lambda_0/\lambda_1)$ in which \eqref{eq:Lambdabracket} takes the form
\be
\label{eq:LambdaBracketInhomogen}
\{f,g\}_\Lambda=\frac{\partial f}{\partial v^{\dot{\alpha}}}   \frac{\partial g}{\partial v_{\dot{\alpha}}} +\Lambda\bigg(v^{\dot{\alpha}} \frac{\partial f}{\partial v^{\dot{\alpha}}}\partial_z g-2f\partial_z g-v^{\dot{\alpha}} \frac{\partial g}{\partial v^{\dot{\alpha}}}\partial_z f+2g\partial_z f\bigg)\,,
\ee
where $f,g$ are sections of $\mathcal{O}(2)$ restricted to $\mathbb{C}^2\times \mathbb{C}^*\subset \mathbb{PT}$, on which the bundle becomes trivial. The two terms $\Lambda(2g\partial_z f-2f\partial_zg)$ resulting from this trivialisation make it manifest that \eqref{eq:LambdaBracketInhomogen} does not obey Leibniz's rule and hence is strictly speaking not a Poisson bracket but rather a so-called \emph{Jacobi bracket}.


\smallskip

Points at the conformal boundary $\scri_\Lambda$ of AdS$_4$ are characterised in embedding space coordinates by  
\begin{equation}
\scri_\Lambda=\{ X| X^{ab}I_{ab}=0\}\,.    \label{infinity}
\end{equation}
In twistor space, this is the condition that the restriction of  $\tau_\Lambda$ to the corresponding twistor line vanishes. Using the incidence relations~\eqref{eq:incidence} in~\eqref{eq:taudef}  gives the intersection of $\scri_\Lambda$ with the Euclidean patch as 
\begin{equation}
	\scri_\Lambda=\left\{x^{\alpha\dot\alpha}\in \R^4:  x^2=-2/\Lambda\right\}\, ,
\end{equation}
as expected from the standard form of the Euclidean (A)dS metric
\begin{equation}
    g=\frac{\d x^{\alpha\dot\alpha}\d x_{\alpha\dot\alpha}}{\left(1+\frac{1}{2}\Lambda x^2\right)^2}\,.\label{eq:ads_metric0}
\end{equation}
This fixing of the conformal boundary justifies calling $I_\Lambda$ the infinity twistor.

\begin{figure}[!t]
        \centering
            \begin{tikzpicture}[scale=2.5]
\draw [thick] (0,0) circle (1);
\draw [name path = A, thick, dashed] (-1,0)  .. controls (-1,0.4) and (1,0.4) ..  (1,0)  ;
\draw [name path = B, semithick] (-1,0)  .. controls (-1,-0.4) and (1,-0.4) ..  (1,0)  ;
\draw [name path = D, semithick, blue] (1,0) arc(0:-180:1);
 \begin{pgfonlayer}{bg}
\fill [blue!20,
          intersection segments={
            of=A and B,
            sequence={L2--R2}
          }];
          \fill [blue!40,
          intersection segments={
            of=B and D,
            sequence={L2--R2}
          }];
\end{pgfonlayer}
 \node[name = a] at (0,-1) {\blue{$\bullet$}};
  \node at (0.15,-1.15) {\blue{$0$}};
\node (b) at (0,1){$\bullet$}; 
\node at (0.15,1.15) {$\infty$};
\draw [thick, black] (3,0) circle (1);
\draw [name path = A2, thick, dashed] (3,-1)  .. controls (3.4,-1) and (3.4,1) ..  (3,1)  ;
\draw [name path = B2, semithick] (3,-1)  .. controls (2.6,-1) and (2.6,1) ..  (3,1)  ;
\draw [name path = D2, thick, blue] (3,1) arc(90:-90:1);
 \begin{pgfonlayer}{bg}
\fill [blue!20,
          intersection segments={
            of=A2 and B2,
            sequence={L2--R2}
          }];
          \fill [blue!40,
          intersection segments={
            of=B2 and D2,
            sequence={L2--R2}
          }];
\end{pgfonlayer}
\draw [thick, dashed] (2,0)  .. controls (2,0.4) and (4,0.4) ..  (4,0)  ;
\draw [semithick] (2,0)  .. controls (2,-0.4) and (4,-0.4) ..  (4,0)  ;

 \node[name = a] at (3,-1) {$\bullet$};
  \node at (3.15,-1.15) {$0$};
\node (b) at (3,1){$\bullet$}; 
\node at (3.15,1.15) {$\infty$};
\node at (-1,1) {$\blue{\mathbb{H}^4}\subset S^4$};
\end{tikzpicture}
\caption{\emph{The ball model of Euclidean AdS$_4$ is displayed on the left and the Poincare half space model is displayed on the right. Their twistor spaces are given by the total space of a (non-holomorphic) $\mathbb{CP}^1$-bundle over the blue region. Considering a blue region that covers $S^4\setminus\{\infty\}\cong\mathbb{R}^4$, leads to $\mathbb{PT}$. Considering a blue region that covers the entire $S^4$ leads to all of $\mathbb{CP}^3$, the twistor space of $S^4$.}}
\label{fig:AdS4Twistor}
    \end{figure}
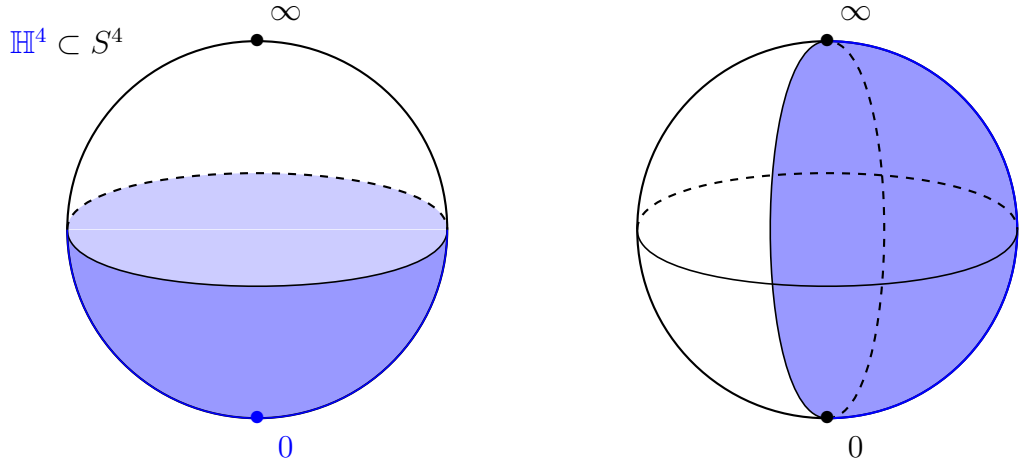

We will mostly focus on the AdS-case $\Lambda<0$ in which $\scri_\Lambda$ is not the empty set. However, most of the twistor methods discussed here also carry over to the case $\Lambda>0$ resulting in Euclidean dS$_4$, \emph{i.e.} the $4$-dimensional sphere $S^4$. The left of figure \ref{fig:AdS4Twistor} schematically displays the twistor space of this ball model of Euclidean AdS$_4$.

The form~\eqref{infinity} represents Euclidean AdS$_4$ as the interior of a ball of radius $\sqrt{-2/\Lambda}$.  We can instead present the infinity twistors so that Euclidean AdS$_4$ is represented as an upper half space in Poincaré coordinates. Introduce  a constant vector $z^{\alpha\dot\alpha}= o^\alpha\bar o^{\dot\alpha}-\iota^\alpha\bar\iota^{\dot\alpha}$ of length $\sqrt{2}$ so that the vertical coordinate of the upper half space representation of $\AdS_4$ can be defined to be $z= x^{\alpha\dot\alpha}z_{\alpha\dot\alpha}/\sqrt{2}$ on which the metric becomes
\begin{equation}
    g=\frac{\d x^{\alpha\dot\alpha}\d x_{\alpha\dot\alpha}}{\Lambda z^2}\, .
\end{equation}
 Eliminating dotted indices via  $\mu^{\alpha}= z^\alpha{}_{\dot\alpha}\mu^{\dot\alpha}$, the infinity twistors that reproduce the metric in Poincaré coordinates are
\begin{equation} I^\Lambda_{ab}=I^{ab}_{\Lambda}=\sqrt{\Lambda}\left(\begin{array}{cc}
         0&\varepsilon^{\alpha\beta}  \\
         -\varepsilon_{\alpha\beta}&0 
    \end{array}\right)\,\label{eq:infinity_twistor_poincare}\, .
\end{equation}
This gives 
the following (holomorphic, twisted) Jacobi bracket and contact form
\begin{equation}
\begin{aligned}
   \{\,\,,\,\,\}^{P}_\Lambda&=\sqrt{\Lambda}\,\frac{\p }{\p\lambda_\alpha}\wedge\frac{\p }{\p \mu^\alpha}\,,\\
   \tau_\Lambda&=\sqrt{\Lambda}\,\langle\mu\,\d\lambda\rangle-\sqrt{\Lambda}\,\langle\d\mu\,\lambda\rangle\,.
\end{aligned}
\end{equation}
For the most part, we will use the ball coordinates \eqref{eq:ads_metric0}, as they make the limit $\Lambda\to0$ to flat space straightforward. The half-space Poincare model arising from  $\{\,\,,\,\,\}^{P}_\Lambda$ is schematically depicted on the right of figure \ref{fig:AdS4Twistor}.

\subsection*{The non-linear graviton with cosmological constant}

Twistor theory extends beyond conformally flat space-times: As we discussed in section \ref{section2.3}, Penrose's non-linear graviton construction 
\cite{Penrose:1976jq,Penrose:1976js} establishes a correspondence between self-dual (SD) vacuum metrics and certain deformed twistor spaces.  Ward \cite{Ward:1980am} extended the non-linear graviton construction to include non-zero cosmological constant as follows\footnote{See also \cite{LeBrun:1982vjh, Salamon:1982,  Hitchin:1982vry} for extensions and variations of this construction.}

\begin{thm}[Ward '80]\label{th:non-linear_graviton_with_Lambda}
    There is a 1-to-1 correspondence between complex self-dual Einstein manifolds $(M,g)$ with cosmological constant $\Lambda$ and deformations $\mathcal{PT}$ of a neighbourhood of a line in $\PT$ preserving $\{\,\,,\,\,\}_\Lambda$. 

    A real slice of $M$ of signature $(2,2)$ or $(4,0)$ corresponds to an antiholomorphic involution $\sigma:\mathcal{PT}\rightarrow \mathcal{PT}$ that, for signature $(2,2)$, fixes a real slice $\mathcal{PT}_\R$. For Euclidean signature, the involution $\sigma$ has no fixed points. 
\end{thm}
Preservation of $\{\,\,,\,\,\}_\Lambda$ can be formulated dually in terms of the twisted, holomorphic contact structure $\tau_\Lambda$. $\tau_\Lambda$ can be defined as the 1-form with values in the square root of the  anti-canonical line bundle dual to $\{\,\,,\,\,\}_\Lambda$ (\emph{i.e.}, obtained by contracting  the Poisson structure into $\text{D}^3Z$ thought of as a 3-form with values in the anti-canonical bundle).

Briefly, the theorem is proved in the forward direction by constructing the curved twistor space as the space of totally null SD 2-surfaces in some small complexification; their existence follows from the vanishing of the SD Weyl curvature.  In Euclidean signature, these simply hit the real slice in a unique point with tangent plane defined by an SD spinor up to scale, so that the twistor space can be identified with the total space of the bundle of projective SD spinors~\cite{Atiyah:1978wi,Hitchin:1982vry,Salamon:1982}.  In the reverse direction, space-time is realized as the moduli space of degree-1 holomorphic curves in $\mathcal{PT}$; those at $\scri$ are those on which the contact form $\tau_\Lambda$ vanishes.  The real slice $M_\R\subset M$ is given by  those degree-1 holomorphic curves that are sent to themselves by the anti-holomorphic involution.


\section{Symmetries and gravitons}
\label{sec:graviton_symmetries}

The infinitesimal symmetries of AdS$_4$ are naturally defined by holomorphic functions of homogeneity degree 2 on twistor space, given as $h=\frac{1}{2}h_{AB}Z^AZ^B$ for a constant, symmetric $h_{AB}$. The associated Hamiltonian vector fields $X_h= \{h\,,\,\, \}_{\Lambda} =  Z^Ah_{AB}I^{BC}_\Lambda\p_C$  generate the corresponding motions of twistor space. By construction, flows along such Hamiltonian vector fields preserve $\{\,\,,\,\,\}_\Lambda$ and its dual contact 1-form. In space-time, this motion induces the standard isometries of AdS$_4$, arising as the spin group Sp$(2)$ of the more usual $\SO(3,2)$. The Lie algebra of these isometries form a subalgebra of the full celestial chiral algebra on AdS$_4$.

More generally, local holomorphic diffeomorphisms that are symmetries of $\{\,\,,\,\,\}_\Lambda$ allow singularities in the Hamiltonians, with  the algebra of such holomorphic symmetries being simply the Poisson structure between two generators. As discussed in chapters \ref{chapter1}-\ref{chapter4}, in flat space-time, it was shown in~\cite{Adamo:2021lrv} that this algebra can be identified with the celestial chiral algebra by using the regularity near $\mu^{\dot\alpha}=0$ to decompose such $h$ into modes of the form
\be \label{eq:basis}
    w^{p}_{m,a} = \frac{(\mu^{\dot 0})^{p+m-1}(\mu^{\dot 1})^{p-m-1}}{2\lambda_0^{p-a-2}\lambda_1^{p+a-2}}\,.
\ee
Here the parameters $p,m$ have been chosen to agree with their counterparts in \cite{Strominger:2021lvk} with ranges fixed by the requirement that $p\in\{1,3/2,2,5/2, \dots\}$ and $p\pm m-1\in \Z_{\geq 0}$. In this range,  negative powers of $\mu^{\dot\alpha}$ do not arise. The parameter $a\in\Z+p$ has been shifted relative to its analogue in \cite{Adamo:2021lrv} to match the conventions of \cite{Taylor:2023ajd}. It is simple to check that, using the Poisson bracket for $\Lambda=0$, the Poisson algebra of such generators yields the celestial chiral algebra
\be
    \{w^p_{m,a},w^q_{n,b}\}_{\Lambda=0}=(m(q-1)-n(p-1))w^{p+q-2}_{m+n,a+b}\,, \label{eq:alg-0}
\ee
of flat space-time. This algebra was originally derived via a bottom-up calculation of soft symmetry algebras at null infinity \cite{Guevara:2021abz}, followed by a light-ray transform \cite{Strominger:2021lvk} as we reviewed in chapters \ref{chapter1} and \ref{chapter2}.

Recall that we denote this algebra~\eqref{eq:alg-0} by  $\lham$ in this thesis because it is
the algebra of Hamiltonian functions on the $\C^2$ coordinatized by $\mu^{\dot\alpha}$. For $\Lambda=0$, the Jacobi structure becomes an honest Poisson structure defined by the degenerate infinity twistor \eqref{eq:infinity_twistors} which acts only on the $\mu^{\dot\alpha}\in \C^2$ variables and is global in this $\C^2$.
One gets the loop algebra because $\lambda_\alpha$ appear only as parameters that are only required to be holomorphic for $\lambda_0/\lambda_1 \in \C^*\supset S^1$  leading to the loop $\mathcal{L}$ in the notation $\lham$.  

\smallskip

The algebra could equivalently be denoted $\LHL|_{\Lambda=0}$ as the generators are holomorphic on  $(\mu^{\dot\alpha}/\lambda_1, \lambda_0/\lambda_1)\in \C^2\times \C^*$
and the $\Lambda=0$ Poisson structure does not see the $\lambda_\alpha$ variables.
The above considerations extend straightforwardly to the case of non-vanishing $\Lambda$. Explicitly, using the $\Lambda$-deformed Poisson bracket $\{\,\,,\,\,\}_\Lambda$, the flat space celestial chiral algebra is deformed to\footnote{Note that this Lie algebra deformation does \emph{not} arise as the loop algebra of a deformation of $\mathfrak{ham}(\C^2)$ itself. Indeed, the unique Lie algebra deformation of the latter is the Weyl algebra corresponding to a non-commutative self-dual gravity \cite{Bu:2022iak} as we discussed in section \ref{section2.1}.}
\be
	\{w^p_{m,a},w^q_{n,b}\}_\Lambda = (m(q-1)-n(p-1))w^{p+q-2}_{m+n,a+b}-\Lambda(a(q-2)-b(p-2))w^{p+q-1}_{m+n,a+b}\,, \label{eq:alg-neq0}
\ee
by the presence of a cosmological constant.   As above, for $2-p\leq a\leq p-2$ the $w^p_{m,a}$ are the quadratic Hamiltonians generating the standard AdS$_4$ isometries. 

The algebra~\eqref{eq:alg-neq0} agrees with that found by Taylor \& Zhu~\cite{Taylor:2023ajd}, who considered the holomorphic collinear limit of the Mellin transform of the leading order in $\Lambda$ correction to the 4-graviton tree level amplitude (sum of all tree-level Witten diagrams) in AdS$_4$. This Mellin amplitude was computed in~\cite{Alday:2022uxp, Alday:2022xwz, Alday:2023jdk, Alday:2023mvu}. Interestingly, \cite{Taylor:2023ajd} found that the true 4-pt amplitude does \emph{not} quite exhibit this algebra; the form of the $O(\Lambda)$ correction 
to be modified in order to ensure the Jacobi identity holds. For us, the fact that the algebra arises from a Jacobi bracket immediately ensures that the Jacobi identity is satisfied. The twistor construction suggests that the algebra~\eqref{eq:alg-neq0} is a true celestial symmetry of \emph{self-dual} gravity on AdS$_4$, at least at the classical level.


\subsection*{Elements of \texorpdfstring{$\LHL$}{ham(C2xC*)} as gravitons}

More general singular elements of the celestial chiral algebra correspond to allowing positive helicity gravitons as fluctuations on the background. In twistor space, the construction of Theorem \ref{th:non-linear_graviton_with_Lambda} identifies self-dual gravitons with infinitesimal deformations that are Hamiltonian. In the original \v Cech presentation~\cite{Penrose:1976jq,Penrose:1976js}, these are determined by a cohomology element $[h]\in H^{1}(\PT,\cO(2))$. Specifically, for a \v Cech cohomology description, one covers $\PT$ by two topologically trivial open sets $\PT=U_0\cup U_1$, where $U_0=\{ Z^A\in\PT:\lambda_0\neq 0\} $ and $U_1=\{Z^A\in\PT:\lambda_1\neq 0\}$. The class $[h]$ is then simply represented as a homogeneity $+2$ holomorphic function on the overlap $U_0\cap U_1$. Thus positive-helicity gravitons are equivalent to holomorphic symmetries of the contact structure and Poisson bracket on $U_0\cap U_1$ modulo gauge. 

The gauge modes with $a\leq p-2$ can be extended holomorphically over all of $U_1$, while those with $a\geq2-p$ extend holomorphically over all of $U_0$. In \v Cech cohomology, these modes are pure gauge, generating coordinate transformations (rather than genuine deformations) of $U_1$ and $U_0$ respectively.
Thus, as a symmetry algebra, it acts on triples consisting of the complex structure of $\mathcal{PT}$ together with  a coordinate charts on $U_0$ and $U_1$.

 If $2-p\leq a\leq p-2$, $w^p_{m,a}$ are the quadratic Hamiltonians whose associated vector fields generate the subgroup of the Poincar\'e group consisting of the translations and self-dual rotations.  More generally, for $p=3/2$ these generate certain holomorphic but singular supertranslations, while for $p=2$ they generate self-dual superrotations. (See \cite{Adamo:2021lrv} for further explanation.)


\subsection*{Celestial chiral algebras as vertex algebras} \label{subsec:VOA}

Lie algebras of local symmetries on twistor space are closely associated with vertex algebras supported on twistor lines \cite{Adamo:2021lrv,Costello:2022wso}. Mathematically, this vertex algebra is the Koszul dual to the algebra of local operators in the Mason-Wolf theory \eqref{eq:poisson-bf} which describes self-dual gravity with $\Lambda\neq 0$ and will be discussed below. For more details on the role of Koszul
duality in quantum field theory we refer to \cite{Paquette:2021cij}.

We can understand the vertex algebra corresponding to \eqref{eq:alg-neq0} in the following way. Suppose we couple the twistor uplift of self-dual gravity on AdS$_4$ to a 2d holomorphic theory living on a twistor line. In general such a coupling will take the  form
\be \label{eq:defect} \sum_{p\pm m\in\mathbb{N}}\frac{2}{(p+m-1)!\,(p-m-1)!}\int_{\CP^1_x}\la\lambda\,\d\lambda\ra\wedge\,w^p_m(\lambda)\,\p_{\mu^{\dot0}}^{p+m-1}\p_{\mu^{\dot 1}}^{p-m-1}H\,\,, 
\ee
for 2d operators $w^p_m(\lambda)$ depending meromorphically on $\lambda$ and labelled by $p,m$ with the same ranges as above. Here $H\in\Omega^{0,1}(\PT,\cO(2))$ is a Dolbeault representative corresponding to the \v Cech cocycle $[h]$. For the integrand to have vanishing homogeneity the operators $w^p_m(\lambda)$ must take values in $\cO(2p-6)$, \emph{i.e.}, they must have conformal spin $3-p$.

Our notation for the operators $w^p_m(\lambda)$ can be justified by choosing $H$ corresponding to the basis \eqref{eq:basis}. Explicitly, we fix $H = w^p_{m,a}\,\dbar B$ for $B$ a bump function on $\CP^1_x$ taking the value 1 in a neighbourhood of $\lambda_0=0$, 0 in a neighbourhood of $\lambda_1=1$ and non-constant on an annulus disconnecting $\lambda_0,\lambda_1$. Introducing inhomogeneous coordinates on $\CP^1_x$ such that $\lambda\sim(z,1)$, in the limit of an arbitrarily narrow annulus we may take $B = \Theta(|z|^2<1)$ for $\Theta$ the Heaviside step function. Substituting into \eqref{eq:defect} gives
\be \oint\frac{\la\lambda\,\d\lambda\ra}{\lambda_0^{p-a-2}\lambda_1^{p+a-2}}\,w^p_m(\lambda) = \oint\d z\,z^{a+2-p}w^p_m(z)\,. \ee
In this way the Hamiltonians $w^p_{m,a}$ are naturally identified with the modes of the operators $w^p_m(z)$.

BRST-invariance of the coupled bulk-defect system involving the coupling \eqref{eq:defect} under the local holomorphic diffeomorphism symmetry on twistor space necessitates the following operator products
\be
\begin{aligned} \label{eq:VOA-neq0}
&w^p_m(z_1)w^q_n(z_2)
\sim \frac{m(q-1)-n(p-1)}{z_{12}}w^{p+q-2}_{m+n}(z_2) \\
&- \frac{\Lambda}{z_{12}^2}\big((p+q-4)w^{p+q-1}_{m+n}(z_2) + z_{12}(p-2)\p_zw^{p+q-1}_{m+n}(z_2)\big)\,.
\end{aligned}
\ee
This can be seen explicitly by computing the BRST-variation of the defect coupling \eqref{eq:defect} and cancelling it with a certain bilocal term. For $\Lambda=0$ the details of this have been spelled out in \cite{Bittleston:2022jeq}, section 2.2 and the discussion can be immediately extended to $\Lambda\neq0$ by including the order $\Lambda$ correction in the interaction vertex. 

Equation \eqref{eq:VOA-neq0} is the (tree level) celestial chiral algebra of SD gravity on AdS$_4$ represented as a vertex algebra. We remark that the field $T(z) = w^1_0(z)/\Lambda$ plays the role of a stress tensor, with OPEs
\be \label{eq:Stresstensor} T(z_1)w^p_m(z_2) \sim \frac{1}{z_{12}^2}\big((3-p)w^p_m(z_2) + z_{12}\p_zw^p_m(z_2)\big)\,. \ee
The corresponding central charge vanishes. Furthermore, $T(z)$ is the field of conformal spin 2 in a vertex subalgebra generated by $w^p_0(z)/\Lambda$ for $p\in\Z_{\geq1}$. This resembles the $w_\infty$ vertex algebra which has the same defining operator products but for labels taking values in the range $p\in\Z_{\leq1}$. $w_\infty$ is generated by fields of all integer conformal spins $s\geq2$ rather than $s\leq2$. 

The fields $w^p_m(z)$ can be conveniently organised into hard generating functions depending on an auxiliary right-handed spinor $\tilde\lambda_{\dot\alpha}$, defined by
\be w(\tilde\lambda,z) = \sum_{p\pm m\in\Z_{\geq1}}\frac{2\tilde\lambda_{\dot0}^{p+m-1}\tilde\lambda_{\dot1}^{p-m-1}}{(p+m-1)!(p-m-1)!}w^p_m(z)\,. \ee
In terms of these hard generators the vertex algebra \eqref{eq:VOA-neq0} reads
\be
\begin{aligned}
&w(\tilde\lambda_1,z_1)w(\tilde\lambda_2,z_2) \\
&\sim \frac{[12]}{z_{12}}w(\tilde\lambda_1+\tilde\lambda_2,z_2) - \frac{\Lambda}{z_{12}^2}\Big(\big(\tilde\lambda_{\dot\alpha}\p_{\tilde\lambda_{\dot\alpha}} - 4\big)w(\tilde\lambda,z_2) + z_{12}\big(\tilde\lambda_{1\dot\alpha}\p_{\tilde\lambda_{\dot\alpha}} - 2\big)\p_zw(\tilde\lambda,z_2)\Big)\Big|_{\tilde\lambda = \tilde\lambda_1+\tilde\lambda_2}\,.
\end{aligned}
\ee
We recognise the coefficient $[12]/z_{12}$ as the tree graviton splitting function on flat space \cite{Bern:1998xc}. But the $\Lambda$ dependent coefficients are not simply functions of the spinor-helicity variables, instead taking the form of differential operators. This results from the loss of supertranslation invariance on AdS$_4$.


\subsection*{Variations and extensions}

Had we used the representation of the infinity twistor \eqref{eq:infinity_twistor_poincare} appropriate to the Poincar\'e patch, we'd have obtained the algebra
\be
\begin{aligned} \label{eq:Poincare-alg}
    &\{\hat w^p_{m,a},\hat w^q_{n,b}\}_\Lambda^P = \big((p+m-1)(q-b-2) - (q+n-1)(p-a-2)\big)\hat w^{p+q-3/2}_{m+n-1/2,a+b-1/2} \\
    &+ \big((p-m-1)(q+b-2) - (q-n-1)(p+a-2)\big)\hat w^{p+q-3/2}_{m+n+1/2,a+b+1/2}\,,
\end{aligned}
\ee
where the generators of the algebra have been redefined to 
\be
  \hat w^{p}_{m,a}  = \frac{2}{\sqrt{\Lambda}}w_{m,a}^{p},
\ee
Although these again provide an extension of the $\AdS_4$ symmetries, they are not suitable for expansion around $\Lambda=0$.  They are however well adapted to soft limits for momentum eigenstates based on translations of the Poincar\' e patch. We emphasise that the Lie algebra \eqref{eq:Poincare-alg} is \emph{not} isomorphic to \eqref{eq:alg-neq0}. The difference is  essentially the choice of the location of the line $\lambda_\alpha=0$: up to AdS$_4$ isometries there are two such choices, the first where the line is in the complement of the unit ball, and the second here where the line corresponds to a point of $\scri$. This difference can be observed by comparing the left and right sides of figure \ref{fig:AdS4Twistor}. These two choices provide the two algebras~\eqref{eq:alg-neq0} \&~\eqref{eq:Poincare-alg}, respectively. On the other hand, the choices of the sets $U_0$ and $U_1$ used to define our basis $\{w^p_{m,n}\}$ are more associated with the choice of cohomology representation. Indeed, these can be made canonically in split signature.

Following the steps outlined in section~\ref{subsec:VOA}, one may recover the vertex algebra associated with~\eqref{eq:Poincare-alg}, which reads
\be
\begin{aligned}
    &\hat w^p_m(z_1) \hat w^q_n(z_2) \sim \frac{(p+q+m+n-2)}{z_{12}^2}\hat w^{p+q-3/2}_{m+n-1/2}(z_2) + \frac{(p+m-1)}{z_{12}}\p_z\hat w^{p+q-3/2}_{m+n-1/2}(z_2) \\
    &- \frac{(p+q-m-n-2)}{z_{12}^2}z_2\hat w^{p+q-3/2}_{m+n+1/2}(z_2) - \frac{(p-m-1)}{z_{12}}\p_z(z_2\hat w^{p+q-3/2}_{m+n+1/2}(z_2)) \\
    &+ \frac{2\big((p-2)(n-1) - (q-2)(m-1)\big)}{z_{12}}\hat w^{p+q-3/2}_{m+n+1/2}(z_2)\,.
\end{aligned}
\ee
Moreover, it's straightforward to recast the above in terms of hard generators.

\smallskip

Both the Lie algebra adapted to the ball \eqref{eq:alg-neq0} and the Poincar\'{e} patch \eqref{eq:Poincare-alg} can easily be extended to incorporate free fields propagating on the gravitational background. By the linear Penrose transform reviewed in section \ref{sectionLinearPenrose} for the $\Lambda=0$ case, solutions to free field equations of spin $s$ on AdS$_4$ are in bijection with cohomology classes $[\varphi]\in H^1(\PT,\cO(2s-2))$. Fluctuations of such fields can be represented in the \v Cech language by holomorphic functions on $U_0\cap U_1$, with homogeneity $2s-2$. Regularity near $\mu^{\dot\alpha}=0$ allows us to decompose $\varphi$ into modes
\be x^p_{m,a} = \frac{(\mu^{\dot0})^{p+m-1}(\mu^{\dot1})^{p-m-1}}{2\lambda_0^{p-a-s}\lambda_1^{p+a-s}} \ee
where as above $p\pm m-1\in\Z_{\geq0}$. The Hamiltonians $w^p_{m,a}$ naturally act on these modes via the Poisson bracket, furnishing us with modules for the Lie algebras $\eqref{eq:alg-neq0}$ and \eqref{eq:Poincare-alg}. For example, in the case of the ball model, this action reads
\be \label{eq:spin-s-module}
	\{w^p_{m,a},x^q_{n,b}\}_\Lambda = (m(q-1)-n(p-1))x^{p+q-2}_{m+n,a+b}-\Lambda(a(q-s)-b(p-2))x^{p+q-1}_{m+n,a+b}\,.
\ee
Extending by these modules gives symmetry algebras for the coupled systems. As we can see in equation \eqref{eq:spin-s-module}, for $\Lambda\neq0$ the structure constants of such extended algebras depend explicitly on $s$. This reflects the fact that for general fluctuations $h$ the resulting curved twistor space does not holomorphically fibre over $\CP^1$. The complex structure of the line bundles $\cO(2s-2)$ is deformed to $\mathcal{K}^{(1-s)/2}$ for $\mathcal{K}$ the canonical bundle of the curved twistor space. The $s$-dependent term in \eqref{eq:spin-s-module} is generated by this shift.

To incorporate self-dual Yang-Mills is no more difficult. Conformal invariance ensures that the $S$-algebra is undeformed on AdS$_4$, and the natural action of \eqref{eq:alg-neq0} and \eqref{eq:Poincare-alg} outlined above distributes over its commutators. Therefore we can extend in the same way.

It is also easy to extend these considerations to include supersymmetry. We adjoin  $\mathcal{N}$ fermionic  coordinates $\eta^I$ to give the homogeneous coordinates $\cZ^{\cI}= (Z^a,\eta^I)$ on $\CP^{3|\cN}$, acted on by the superconformal group $\SL(4|\mathcal{N})$.  As shown in~\cite{Wolf:2007tx, Mason:2007ct}, self-dual $\SO(\cN)$ gauged supergravity on AdS$_4$ may be described by the non-linear graviton construction, where the infinity twistor is extended to a non-degenerate graded skew supertwistor $I^{\cI\cJ}$ defining graded Poisson structure and contact form. This preserves an $\mathrm{OSP}(2|\cN)$ subgroup of the superconformal group. Ungauged supergravities can be obtained by considering infinity twistors that do not have maximal rank in the fermionic directions. In particular, the fully ungauged supergravity arises when we continue to use the non-supersymmetric Poisson structure. By augmenting the $w^p_{m,a}$ to a basis of functions of the homogeneous supertwistor coordinates $\cZ^{\cI}$ of homogeneity 2, we obtain supersymmetric extensions of the above algebras. These provide extensions of the relevant super-Lie algebras of symmetries for supersymmetric extensions of $\AdS_4$ depending the choice of infinity twistor. They also include twistor functions for all the supergravity modes.

In chapter \ref{chapter6} we will see that it is possible to deform AdS$_4$ to a certain gravitational instanton through a backreaction from including a defect operator wrapping a twistor line. This is analogous to the way we obtained Eguchi-Hanson space from a backreaction in chapter \ref{chapter4}. 


\section{Symmetries of the twistor action}\label{sec:symm}

These symmetries can be understood via Noether arguments applied to the twistor action. We give a brief sketch here. Self-dual gravity with $\Lambda\neq0$ may be described on twistor space by the \emph{Mason-Wolf action}
\begin{equation}
    S_{\Lambda}[G,H]=\int_\PT \text{D}^3Z\wedge G \wedge \left(\dbar H+\frac{1}{2}\{H,H\}_\Lambda\right)\,,\label{eq:poisson-bf}
\end{equation}
first introduced in~\cite{Mason:2007ct}. It is a $\Lambda\neq 0$ generalization of holomorphic Poisson BF theory which was introduced in section \ref{section2.3}. Here, $H\in\Omega^{0,1}(\PT,\mathcal{O}(2))$ is a Hamiltonian governing the complex structure deformation on twistor space via the deformed Dolbeault operator
\begin{equation}
    \bar\nabla:=\dbar+\{H,\,\,\}_\Lambda.
\end{equation} $G\in\Omega^{0,1}(\PT,\mathcal{O}(-6))$ is a Lagrange multiplier that, on-shell, may be interpreted as the Penrose transform of a linearized ASD Weyl spinor propagating on the SD background determined by $H$.  

\medskip 

The celestial chiral algebra above is related to  gauge transformations of this Lagrangian.  These are a combination of Poisson diffeomorphisms of twistor space, generated by the Hamiltonian vector field associated with a smooth function $\chi$ of homogeneity degree 2, and a further transformation generated by the Hamiltonian vector field of a smooth function $\tilde\chi $ of weight $-6$. These transformations act on the fields as
\begin{equation}
    \delta H= \bar\nabla \chi\, , \qquad \delta G= \bar\nabla \tilde\chi +\{\chi ,G\}_\Lambda\,.
\label{eq:gauge-transforms}
\end{equation}

To make contact with the celestial chiral algebra, we compute the Noether charges associated with these field transformations. The (pre-)symplectic form on the space of classical solutions of the twistor action~\eqref{eq:poisson-bf} can easily be seen to be
\begin{equation}
    \omega(\delta H,\delta G )=\int_{\Sigma} \delta H\wedge \delta G \wedge \text{D}^3Z \,,
\end{equation}
where $\Sigma$ is a real co-dimension 1 slice of $\PT$. The Noether charges corresponding to~\eqref{eq:gauge-transforms} are then given by the integrals 
\begin{equation}
\begin{aligned}
\mathcal{H}_{\chi} &= \int_\Sigma \bar\nabla\chi \wedge G \wedge \text{D}^3Z\\
\mathcal{H}_{\tilde\chi} &= \int_\Sigma \bar\nabla \tilde \chi\wedge H\wedge\text{D}^3Z\,,
\end{aligned}
\end{equation}
On shell, after integrating by parts these reduce to boundary terms
\begin{equation}
\begin{aligned}
\mathcal{H}_{\chi}&= \int_{\p\Sigma} \chi\,  G \wedge\text{D}^3Z\\
\mathcal{H}_{\tilde\chi} &=  \int_{\p\Sigma} \tilde \chi\, H\wedge\text{D}^3Z\, .
\end{aligned}
\label{eq:boundaryNoether}
\end{equation}
as expected for Noether charges of gauge transformations. The real four-manifold  $\p\Sigma$ can for example be taken to be the 4-surface swept out by Riemann spheres $L_x$ as $x$ varies over some choice of 2-surface in space-time. In this case, the Penrose transform may be used to express the integrals~\eqref{eq:boundaryNoether} in terms of fields on space-time.

The Noether charge integrals~\eqref{eq:boundaryNoether} are not themselves gauge invariant unless $\chi, \tilde \chi$ are holomorphic.  Such global holomorphic $\chi$ of homogeneity degree two are, as described above, the Hamiltonians that generate the global isometries of AdS$_4$. Because $\tilde \chi $ has weight $-6$, there are no such global $\tilde \chi$. If we wish to allow singularities so as to extend the $\chi$ to be the generators \eqref{eq:basis} of our extended algebra, we must  extend our on-shell  phase space by imposing boundary conditions so that $H=G=0$ near $\lambda_0=0$ and $\lambda_1=0$.  This is equivalent to choosing holomorphic Darboux coordinates on small neighbourhoods of $\lambda_0=0$ and $\lambda_1=0$ as discussed earlier. The Noether charges~\eqref{eq:boundaryNoether} make sense on this extended phase space and generate the celestial chiral algebra for $\Lambda\neq0$ described earlier.  

\section{Discussion}

We have seen that the first-order deformation to the flat space celestial chiral algebra found in \cite{Taylor:2023ajd} naturally arises from local Poisson diffeomorphisms of twistor space, once the flat-space Poisson structure is replaced by the holomorphic contact structure $\tau_\Lambda$, or dually Jacobi structure $\{\,\,,\,\,\}_\Lambda$, defined by a non-degenerate infinity twistor. In \cite{Taylor:2023ajd}, the algebra was constructed via the AdS amplitudes first found in \cite{Alday:2022uxp, Alday:2022xwz, Alday:2023jdk, Alday:2023mvu}, although in a rather \emph{ad hoc} manner: the algebra that arises na\"ively from the AdS amplitudes fails to satisfy the Jacobi identity, so a suitable modification of the $\Lambda$-deformed graviton OPE is needed to restore associativity. In contrast, here we have constructed the algebra from first principles directly from the Poisson bracket on twistor space, so the Jacobi identity is automatically satisfied. We thus see that this algebra is the celestial chiral algebra of self-dual gravity of AdS$_4$, at least at the classical level.  We have seen further that the algebra can be understood as symmetries of an extension of the twistor action for the SD Einstein equations leading to Noether charges on an extended phase space.

\medskip

Perhaps the most interesting question is the extent to which these symmetries can yield useful insights beyond the self-dual sector. However, many more modest  directions deserve to be explored in further work:


\paragraph{Perturbiner calculations.} In flat space, the celestial chiral algebra can be seen in the splitting function of gravity amplitudes as discussed in chapter \ref{chapter1}. This is essentially a pertubiner: two on-shell positive helicity states joined to a propagator at a trivalent vertex, taken in the limit that the external momenta become (holomorphically) collinear. It would be interesting to perform these calculations in AdS$_4$, with bulk-to-boundary propagators as external states. This would bring together the perspective of~\cite{Taylor:2023ajd} with the current work, albeit in the context of self-dual gravity. There are various spacetime descriptions of self-dual gravity in the presence of a cosmological constant  \cite{Krasnov:2024qkh, Lipstein:2023pih, Przanowski:1984qq, Neiman:2023bkq, Hoegner:2012sq, Krasnov:2016emc, Adamo:2021bej} and it would be a useful starting point to understand their precise relations to each other and to the Mason-Wolf action \cite{Mason:2007ct} we used in this chapter.

\paragraph{Space-time realizations and AdS/CFT.}  Here we have focused on formulations in twistor space.  There are by now several space-time formulations of many of these ideas at $\Lambda=0$ such as \cite{Freidel:2021ytz,Donnay:2024qwq} and it would be interesting to extend these to $\Lambda\neq 0$, perhaps following on from the frameworks developed by \cite{Hoegner:2012sq,Lipstein:2023pih, Neiman:2023bkq}.  In a different direction, it will be interesting to identify the self-dual sectors of conventional AdS$_4$/CFT$_3$ correspondences such as, for example, those described by ABJM \cite{Aharony:2008ug}, so as to see how these structures arise there. It would be particularly interesting to  understand the role of the Virasoro subalgebra discussed above  \eqref{eq:Stresstensor}.  There are also a number of other $\Lambda$-BMS proposals to be compared to from a space-time perspective such as \cite{Compere:2019bua,Compere:2020lrt,Fiorucci:2020xto}.

\paragraph{Further deformations with $\Lambda \neq 0$.}
As discussed in the section \ref{section1.5} there is by now a fairly large number of works on classical deformations of celestial chiral algebras (or the absence thereof) in asymptotically (locally) flat space-times. 
All of these are loop algebras of deformations of $\mathfrak{ham}(\mathbb{C}^2)$ or $\mathfrak{ham}(\mathbb{C}^2/\Gamma)$ for some finite subgroup $\Gamma \subset \SU(2)$ which are very restricted. The Lie algebra \eqref{eq:alg-neq0} in turn is not the loop algebra of a deformation of $\mathfrak{ham}(\mathbb{C}^2)$. It is natural to ask whether there are further deformations of this form and what their geometric bulk-interpretations are. There are many examples of gravitational instantons in the presence of a cosmological constant and one such example is known to lead to a further deformation of  \eqref{eq:alg-neq0} as will be discussed in chapter \ref{chapter6} based on \cite{Bogna:2024gnt}. 

\paragraph{Quantum corrections.} For $\Lambda=0$ holomorphic Poisson-BF theory on twistor space is anomalous \cite{Bittleston:2022nfr}, signalling a loss of integrability in SD gravity at the quantum level. Certainly, we should expect a similar anomaly in the Mason-Wolf action with $\Lambda\neq0$. It would be interesting to compute this, particularly to see whether there exist alternative methods of anomaly cancellation. Successful cancellation of the anomaly would presumably lead to consistent quantum counterparts of the tree-level celestial chiral algebra \eqref{eq:VOA-neq0} along the lines of \cite{Costello:2022upu,Bittleston:2022jeq}.

%% file: Chapter7/chapter7.tex
\chapter{Towards self-dual black holes}
\label{chapter6}

Astrophysical black holes described by the Kerr metric are known to have non-vanishing components of the self-dual as well as the anti-self-dual halves of their Weyl tensor. In particular, they are \emph{not} self-dual spacetimes so that the twistor methods of this thesis can not directly apply.
However, introducing a NUT-charge leads to the possibility of \emph{self-dual black holes} \cite{Crawley:2021auj, Crawley:2023brz} which can teach us lessons about Kerr black holes by perturbing around the self-dual sector \cite{Guevara:2023wlr, Guevara:2024edh}.

In the simplest case, a self-dual black hole is described by the famous \emph{self-dual Taub-NUT metric} \cite{Crawley:2021auj, Crawley:2023brz}. In Euclidean signature, this metric does not have a horizon\footnote{In fact, the 'horizon' is simply a point.} and is not commonly thought of as a black hole. However, after Wick-rotating the metric to Kleinian signature it does have a genuine horizon and it is possible to continue the metric past this horizon where the maximal continuation encounters a curvature singularity \cite{Crawley:2021auj}. This justifies referring to such self-dual Taub-NUT geometries as \emph{self-dual black holes} \cite{Crawley:2023brz}.

We will see that the twistor space of self-dual Taub-NUT arises through a backreaction that is similar to the backreaction resulting in Eguchi-Hanson space we discussed in chapter \ref{chapter4}. The main difference is that we apply a conformal inversion that formally results in a defect operator wrapping the twistor line $\mathbb{CP}^1_{\lambda_\alpha=0}$ of the 'point at $\infty$' rather than the twistor line $\mathbb{CP}^1_{\mu^{\da}=0}$ of the point at $0$. Since twistor spaces are a priori signature agnostic, it is possible to obtain the self-dual black hole of \cite{Crawley:2021auj, Crawley:2023brz} from this backreaction.

A much more general class of (non-self dual) black hole metrics with a NUT charge was constructed by Pleba\'{n}ski-Demia\'{n}ski in \cite{Plebanski:1976gy}. We will discuss how these metrics can be made self-dual by imposing a certain relation between their parameters given by mass, NUT-charge, angular momentum and cosmological constant\footnote{Including electromagnetic charges and a so-called acceleration-parameter is also possible but we will not discuss this here.}. In the absence of angular momentum and in Euclidean signature, we will explicitly see that the resulting self-dual Einstein metric with negative cosmological constant is isometric to the so-called \emph{Pedersen metric}. The Pedersen metric has been studied a long time ago in the differential geometry literature \cite{Pedersen:1986vup,pedersen1985geometry} and in particular, Pedersen constructed its twistor space. This twistor space fibres over $\mathbb{CP}^1\times\mathbb{CP}^1$ viewed as the \emph{minitwistor space} of the $3$-sphere\footnote{Or Euclidean AdS$_3$ after removing the antiholomorphic diagonal}. We conjecture that Pedersen's twistor space arises from a backreaction in self-dual gravity with $\Lambda\neq 0$ on twistor space, described by the Mason-Wolf action \eqref{eq:poisson-bf}, and we will discuss strong  evidence in favour of this conjecture.

 The Pedersen metric depends on two parameters, a mass parameter $M$ and a cosmological constant $\Lambda$. Further relations can be imposed between these in which it reduces to previously studied self-dual geometries such as self-dual Taub-NUT,  a singular double cover of Eguchi-Hanson space, Euclidean AdS$_4$, and non-compact $\mathbb{CP}^2$, which is conformally equivalent to Burns space. From its $2$-parameter twistor space we derive a $2$-parameter deformation of $\mathcal{L}w_\wedge=\mathcal{L}\mathfrak{ham}(\mathbb{C}^2/\mathbb{Z}_2)$ which reduces to the expected deformations in various limits.  In this sense, the $2$-parameter family of algebras derived in this chapter interpolates between the two  celestial chiral algebras discussed in chapters \ref{chapter4} and \ref{chapter5}: $\mathcal{L}W(\infty)$ obtained from Eguchi-Hanson space in chapter \ref{chapter4} and $\LHL$ obtained from AdS$_4$ in chapter \ref{chapter5}\footnote{Strictly speaking, a $\mathbb{Z}_2$-invariant subalgebra thereof $\LHL^{\mathbb{Z}_2}\subset \LHL$.}.

The Pedersen metric is conformally equivalent to a $2$-parameter family of scalar flat Kähler manifolds i.e. solutions to the classical field equations of Mabuchi gravity \cite{Costello:2022jpg, Costello:2021bah} which interpolates between Burns space and a singular double cover of Eguchi-Hanson space. This means that it might be possible to obtain Pedersen's twistor space from the top-down constructions of Costello and collaborators \cite{Costello:2023hmi, Costello:2022jpg, Bittleston:2024efo}.


\medskip

This chapter is partly based on \cite{Bogna:2024gnt} and organized as follows: In section \ref{sec:SDPlebDem}, we will discuss how the  Pleba\'{n}ski-Demia\'{n}ski black holes can be made self-dual by imposing a relation between their (complexified) parameters. In the non-spinning case, this leads to the metric of interest in this chapter: the Pedersen metric. Different representations of the Pedersen metric, its relation to Mabuchi gravity and its various limits are discussed in the rest of section \ref{sec:SDPlebDem}. In section \ref{sec:sec3}, we will derive the twistor space of self-dual Taub-NUT from a backreaction in analogy to the results of section \ref{sec:deforming-twistor-space} in the case of Eguchi-Hanson space. 
Then we will generalise this result to $\Lambda< 0$ and give evidence for our conjecture that the twistor space of the Pederson metric arises from a backreaction with $\Lambda< 0$. 
In section \ref{subsec:CelSymmTwistor} we will then derive a celestial chiral algebra of self-dual gravity on the Pedersen background from Pedersen's twistor space. It is a consistent $2$-parameter deformation of $\mathcal{L}w_\wedge$ that respects the expected symmetries of the Pedersen metric and reproduces the correct limiting cases.

\section{Self-dual Pleba\'{n}ski-Demia\'{n}ski spacetimes and their limits}
\label{sec:SDPlebDem}

\subsection*{Self-dual Pleba\'{n}ski-Demia\'{n}ski Spacetimes
}
\label{subsec:sdPlebanski}

A well-known $7$-parameter generalization of the Kerr metric was constructed by Pleba\'{n}ski-Demia\'{n}ski in \cite{Plebanski:1976gy}. We will not be interested in acceleration or electromagnetic charges, so we set the corresponding parameters to $0$. The resulting $4$-parameter family of rotating Taub-NUT-AdS black holes reads \cite{Griffiths:2005qp, Rodriguez:2021hks}
\bea
\label{eq:TNAdS4}
\dif s^2=& -\frac{\Delta}{\Sigma}\left(\dif t+(2n\cos\theta-a\sin^2\theta)\frac{\dif\phi}{\Xi}\right)^2\\
&+\frac{\Delta_\theta}{\Sigma}\left(a\,\dif t-(r^2+a^2+n^2)\frac{\dif \phi}{\Xi}\right)^2\\
&+\frac{\Sigma}{\Delta}\,\dif r^2+\frac{\Sigma}{\Delta_\theta}\sin^2\theta\,\dif \theta^2\,,
\eea
where
\label{eq:Parameters}
\bea
\Sigma&=r^2+(n+a\,\cos \theta)^2\,,\\
\frac{\Delta_\theta}{\sin^2\theta}&=1-\frac{4an \cos \theta}{l^2}-\frac{a^2\cos^2 \theta}{l^2}\,,\\
\Delta&=r^2+a^2-2mr-n^2\\
&\quad+\frac{3(a^2-n^2)\,n^2+(a^2+6n^2)\,r^2+r^4}{l^2}\,,\\
\Xi&=1-\frac{a^2}{l^2}\,.
\eea
The metric \eqref{eq:TNAdS4} is known to solve Einstein's equations in the presence of a cosmological constant $\Lambda=-3/l^2$. The remaining three parameters $(m,n,a)$ are related to the mass, the NUT charge and the angular momentum of the black hole. 
\medskip

In the absence of $\Lambda$, i.e. in the limit $l\rightarrow \infty$, the metric \eqref{eq:TNAdS4} has a self-dual limit $n=-\im M$, $m=M$, in which it can be analytically continued to give a Euclidean self-dual metric commonly referred to as self-dual Taub-NUT \cite{Hawking:1976jb}. Alternatively, it can be continued to Kleinian signature \cite{Crawley:2021auj, Easson:2023ytf, Giribet:2025ihk}, where the metric has a genuine horizon beyond which it can be extended with the maximal extension having a singularity at $r=-M$. This justifies the previously used terminology \emph{self-dual black hole}. 

We will now argue that for a non-spinning black hole, $a=0$, and in the presence of a non-vanishing  cosmological constant $\Lambda\neq0$ there exists a similar self-dual limit given by\footnote{And, correspondingly, an anti-self-dual limit given by \be
n=\im M\qquad\qquad m=M\left(1-\frac{4M^2}{l^2}\right)\,.\label{eq:asd-point}
\ee
If $a\neq 0$, the (A)SD points are shifted to
\be
n=\pm\im M\,,\qquad\qquad m=M\left(1-\frac{a^2+4M^2}{l^2}\right)\,.
\ee
}
\be
n=-\im M\qquad \qquad  m=M\left(1-\frac{4M^2}{l^2}\right)\,.\label{eq:sd-point}
\ee
To see this, let us introduce the null tetrad
\begin{equation}
    \begin{aligned}
    \ell&=-\frac{1}{\sqrt{2(r^2+n^2)}}\left(\frac{r^2+n^2}{\sin\theta}\p_t+\sin\theta\,\p_r\right)\,,\\ n&=\frac{1}{\sqrt{2(r^2+n^2)}}\left(-\frac{r^2+n^2}{\sin\theta}\p_t+\sin\theta\,\p_r\right)\,,\\m&=\frac{1}{\sqrt{2(r^2+n^2)}}\left(2n\cot\theta\,\p_t-\im\sin\theta\,\p_\theta-\frac{1}{\sin\theta}\p_\phi\right)\,,\\\bar m&=\frac{1}{\sqrt{2(r^2+n^2)}}\left(2n\cot\theta\,\p_t+\im\sin\theta\,\p_\theta-\frac{1}{\sin\theta}\p_\phi\right)\,.
    \end{aligned}
\end{equation}
In terms of this tetrad, the only non-vanishing component of the anti self-dual Weyl tensor in the NP formalism is
\begin{equation}
    \psi_2=\frac{4n^3+l^2(-\im m+n)}{l^2(n-\im r)^3}\,,
\end{equation}
together with its complex conjugate $\tilde\psi_2$. So we find a self-dual (complex) space-time precisely for the choice \eqref{eq:sd-point}. The same limit has been previously found in \cite{Chamblin:1998pz}\footnote{To our knowledge, the limit \eqref{eq:sd-point} first appeared in \cite{Gibbons:1978zy}.}, where it was argued that a so-called \emph{regular nut singularity} exists only when \eqref{eq:sd-point} holds. In particular, this means that the Misner string \cite{Misner:1963fr} is unobservable and there are no conical singularities when \eqref{eq:sd-point} holds. The spacetime is diffeomorphic to $\mathbb{R}^4$.

In the self-dual limit \eqref{eq:sd-point}, $\Delta$ from equation \eqref{eq:Parameters} simplifies to\footnote{In the spinning case with $a=\im A\neq0$, $\Delta$ simplifies to
\be
\Delta= (r-M-A)(r-M+A)\bigg(1+\frac{(r-M)(r+3M)}{l^2}\bigg)\,.
\ee
}
\be
\Delta= (r-M)^2\bigg(1+\frac{(r-M)(r+3M)}{l^2}\bigg)\,,
\ee
so that after a Wick-rotation $t\mapsto \im t$, the metric \eqref{eq:TNAdS4} becomes the Euclidean metric
\be
\label{eq:SdTNADS}
\dif s^2=\frac{\dif r^2}{U(r)}+ U(r)(\dif t-2M\cos \theta\, \dif \phi)^2+(r^2-M^2)\,\d\Omega_2^2
\ee
where
\be
U(r)=\frac{r-M}{r+M}\left(1+\frac{(r-M)(r+3M)}{l^2}\right)\,,
\ee
and $\d\Omega_2^2=\d\theta^2+\sin^2\theta\d\phi^2$ denotes the canonical metric on $S^2$. The Euclidean metric in equation \eqref{eq:SdTNADS} is the most commonly used form of self-dual Euclidean AdS-Taub-NUT \cite{Hawking:1979pi, Zoubos:2002cw}. The coordinate range is $r\in(M,\infty)$, $t\in \R$, $(\theta,\phi)\in S^2$. The singularity at $r=M$ is only apparent, as the Kretschmann scalar is
\begin{equation}
R^{\mu\nu\rho\sigma}R_{\mu\nu\rho\sigma}= 24\frac{4M^2(l^2-4M^2)^2+(r+M)^6}{l^4(r+M)^6}\,.
\end{equation}
There is no horizon either, because the radius of the transverse 2-spheres vanish at $r=M$ as $\sqrt{r-M}$, so the locus $r=M$ is just a point in Euclidean signature \cite{Adamo:2023fbj}. If we impose $t$ to be periodic as $t\sim t+8\pi M$, we thus end with a space-time which is topologically $\R^4$.

We will now show that the metric \eqref{eq:SdTNADS} can be brought into a triaxial form with a suitable diffeomorphism. This form of the metric was first constructed by Pedersen \cite{Pedersen:1986vup, pedersen1985geometry} so that we will refer to \eqref{eq:SdTNADS} as the \emph{Pedersen metric}.

\subsection*{The Pedersen metric}
\label{sec:Pedersenmetric}

Performing the diffeomorphism
\begin{equation}
    \rho=\sqrt{2Ml^2\frac{r-M}{l^2+2M(r-M)}}\,,\qquad \psi=\frac{t}{2M}\,,
\end{equation}
and introducing the new parameter
\begin{equation}
    \nu^2=\frac{1}{4M^2}-\frac{1}{l^2}\,,
\end{equation}
 brings the metric into the form\footnote{After applying the diffeomorphism, we relabel $\rho\mapsto r$.}
\be
\label{eq:PedMet}
\d s^2=f^2(r)\left(h_r(r)\d r^2+h_{12}(r)(\sigma_1^2+\sigma_2^2)+h_{3}(r)\sigma_3^2\right)\,.
\ee
where
\bea
    f(r)&=\frac{2}{1-r^2/l^2}\,,\qquad h_r(r)=\frac{1+\nu^2r^2}{1+\nu^2r^4/l^2}\,,\\ h_{12}(r)&=r^2(1+\nu^2r^2)\,,\qquad h_3(r)=\frac{r^2}{h_r(r)}\,,
\eea
and $0\leq r< l$ and $0\leq \psi<4\pi$. Here, we introduced the standard SU$(2)$-invariant $1$-forms
\bea
    \sigma_1&=\frac{1}{2}(\cos\psi\,\d\theta+\sin\psi\,\sin\theta\,\d\phi)\,,\\
    \sigma_2&=\frac{1}{2}(-\sin\psi\,\d\theta+\cos\psi\,\sin\theta\,\d\phi)\,,\\
     \sigma_3&=\frac{1}{2}(\d\psi-\cos\theta\,\d\phi)\,.
\eea

We can see that the conformal boundary of the spacetime is a squashed 3-sphere known as the \emph{Berger sphere}: Setting $r=l$ in \eqref{eq:PedMet}, we find the boundary metric
\begin{equation}
\label{eq:oblatesquashedS3}
    \d s^2_{3}=\sigma_1^2+\sigma_2^2+\frac{1}{1+\nu^2l^2}\sigma_3^2\,,
\end{equation}
up to an overall constant. Since $(1+\nu^2l^2)^{-1}\leq 1$, the metric \eqref{eq:oblatesquashedS3} is an oblate squashing of the $3$-sphere. For this reason, we will also refer to the metric in \eqref{eq:PedMet} as the \emph{oblate Pedersen metric}. The spacetime metric represents an explicit realization of a theorem by LeBrun \cite{LeBrun:1982vjh}. In \cite{LeBrun:1982vjh}, it is shown that any 3-manifold with a Riemannian conformal structure is the conformal boundary at infinity of a self-dual Riemannian 4-manifold satisfying Einstein's equations with a negative cosmological constant. The theorem ensures the existence of the bulk 4-manifold only in a collared neighbourhood of its conformal boundary, but the Pedersen metric is a complete metric inside the entire $4$-ball with radius $l$. In the language of Lebrun \cite{lebrun1991complete}, this means that the Berger sphere is \emph{of positive frequency} (with the appropriate orientation). More generally, Hitchin showed that that every left-invariant conformal structure on $S^3$, in particular a generalisation of the Berger sphere that is squashed along all three axes, has positive frequency leading to a generalisation of the Pedersen metric \cite{Hitchin:1995hxv}\footnote{We conjecture that Hitchin's metric is related to turning on further parameters of the  Pleba\'{n}ski-Demia\'{n}ski metric in a way preserving self-duality  \cite{Martelli:2011fw, Cvetic:2002kj}}.

\subsection*{Relation to Mabuchi gravity}

Applying a further change of coordinates of the radial direction
\begin{equation}
    r=\sqrt{\frac{l}{\nu}\tan\chi}\,,
\end{equation}
it can be seen that the Pedersen metric is diffeomorphic to
\begin{equation}
    \d s^2=\frac{\nu l^3}{(\sin\chi-\nu l\cos\chi)^2}\left(V^{-1}_S(\d\tau+A_S)^2+V_S\d \Omega^2_3\right)\,,\label{eq:s3_gibbons-hawking}
\end{equation}
where we introduced a so-called \emph{Higgs field} and \emph{gauge potential} on the 3-sphere
\begin{equation}
    V_S=\nu l+\cot\chi\,,\qquad A_S=-\cos\theta\,\d\phi\,.
\end{equation}
$\d \Omega_3^2 =\d\chi^2+\sin^2\chi\,\d\Omega_2^2$ denotes the canonical metric on the 3-sphere. It is straightforward to show that the pair $(V,A)$ satisfies the Bogomolny equation $\star_3\,\d V=\d A$, where $\star_3$ denotes the Hodge star operator on $S^3$ with respect to $\d\Omega_3^2$.

It is also possible to analytically continue the metric \eqref{eq:PedMet}  $\nu \mapsto \im \nu$ to a region with $\nu^2<0$. Provided that $\nu l < 1$, the metric turns out to still be complete on the open ball $r<l$ \cite{Zoubos:2002cw}. We will refer to this as the \emph{prolate Pedersen metric}, because its boundary conformal structure is a prolate squashing of $S^3$ \eqref{eq:oblatesquashedS3} in contrast to the oblate Pedersen metric \eqref{eq:PedMet}. A similar diffeomorphism
\begin{equation}
    r=\sqrt{\frac{l}{\nu}\tanh\chi}\,.
\end{equation}
can be applied to the prolate Pedersen metric. There, it brings the metric into the form
\be
\label{eq:eq:h3_gibbons-hawking}
\d s^2=\frac{\nu l^3}{(\sinh\chi-\nu l\cosh\chi)^2}(V_H^{-1}(\d\tau+A_H)^2+V_H\d s_{\mathbb{H}^3}^2)\,,
\ee
where now the pair $(V_H,A_H)$ is
\begin{equation}
    V_H=-\nu l+\coth\chi\,,\qquad A_H=-\cos\theta\,\d\phi\,,
\end{equation}
and $\d s_{\mathbb{H}^3}^2=\d\chi^2+\sinh^2\chi\,\d\Omega_2^2$ denotes the standard metric on $\mathbb{H}^3$.
The Pedersen metric can thus be understood as arising from a generalized Gibbons-Hawking ansatz \cite{gibbons1978gravitational} where one replaces the $\R^3$ with $S^3$ or $\mathbb{H}^3$, respectively in the oblate and prolate case \cite{pedersen1985geometry, Pedersen:1986vup}\footnote{This leads to the twistor space of the Pedersen metric fibring over the \emph{minitwistor space} of $S^3$ or $\mathbb{H}^3$ respectively in the oblate and prolate case.}.

A conformal factor can be included to turn \eqref{eq:eq:h3_gibbons-hawking} into
\be
\label{eq:ProlPedMetconf}
\d s^2=\nu l^3q(\chi,\theta, \phi)^2(V_H^{-1}(\d\tau+A_H)^2+V_H\d s_{\mathbb{H}^3}^2)\,,
\ee
which has been shown to be scalar flat Kähler by Lebrun \cite{lebrun1991explicit}. $q(\chi,\theta, \phi)$ is a horospherical height function \cite{lebrun1991complete}, which is explicitly given by the coordinate $q$ after transforming to half-plane coordinates 
\be
\dif s_{\mathbb{H}^3}^2=q^{-2}(\dif x^2+\dif y^2 +\dif q^2)
\ee
on the $\mathbb{H}^3$ factor \cite{costa2001description}. Since twistor space is a priori agnostic about conformal factors, we see that the Pedersen twistor space also leads to a two-parameter family of solutions to the equations of motion of \emph{Mabuchi gravity} \cite{Costello:2023hmi, Costello:2021bah} which we discussed in chapter \ref{chapter1} to play an important role in Burns holography and more recent top-down constructions. In particular, we will see below that it is a class of scalar-flat Kähler metrics that interpolates between Burns space and a double-cover of Eguchi-Hanson space which both appear in \cite{Bittleston:2024efo}  from giving a VEV to certain two operators.

\subsection*{Various limits}

\begin{figure}[t!]
	\centering
 	\scalebox{.7}{\begin{tikzpicture}[scale=1.7]
\draw[black] (0.5,0) -- (5,0);
\draw[->] (5,0) -- (10,0);
\draw[->] (5,0) -- (5,4);

\fill[yellow!30] (5,0)--(5,4)--(0.5,4);
\fill[yellow!30] (5,0)--(5,4)--(10,4)--(10,0);

\draw[green, ultra thick] (5,0) -- (5,4);
\draw[violet, ultra thick] (5,0) -- (10,0);
\draw[blue, ultra thick] (5,0) -- (0.5,4); 
\begin{scope}
    \clip (4,3) rectangle (5,4);
    \draw[orange, very thick] (5,4) circle(0.5);
\end{scope}

\draw (0.5,4) -- (5,4);
\draw (5,4) -- (10,4);
\draw (10,4) -- (10,0);

\node at (10.3,0.2){$\nu^2$};
\node at (4.5,4.5){$-\Lambda/3$};
\node at (10,0){$\bullet$};
\node at (10,-0.3){$\infty$};
\node at (5,4){$\bullet$};
\node at (5.3,4.2){$\infty$};
\node at (0.5,0){$\bullet$};
\node at (0.5,-0.3){$-\infty$};

\filldraw[red] (5,0) circle (5pt);
\filldraw[orange] (5,4) circle (5pt);

\end{tikzpicture}}
\caption{\emph{The horizontal axis represents $\nu^2$ and the vertical axis represents $-\Lambda/3$. The Pedersen metric corresponds to points in the yellow region and has various limits: a singular double cover of Eguchi-Hanson space (orange), self-dual Taub-NUT (violet), $\widetilde{\mathbb{CP}}^2$ (blue), Euclidean AdS$_4$ (green), and $\mathbb{R}^4$ (red). $\mu^2<0$ makes up the prolate case while $\mu^2>0$ makes up the oblate case.}} 
\label{Pedersen}
\label{fig:Ped}
\end{figure}
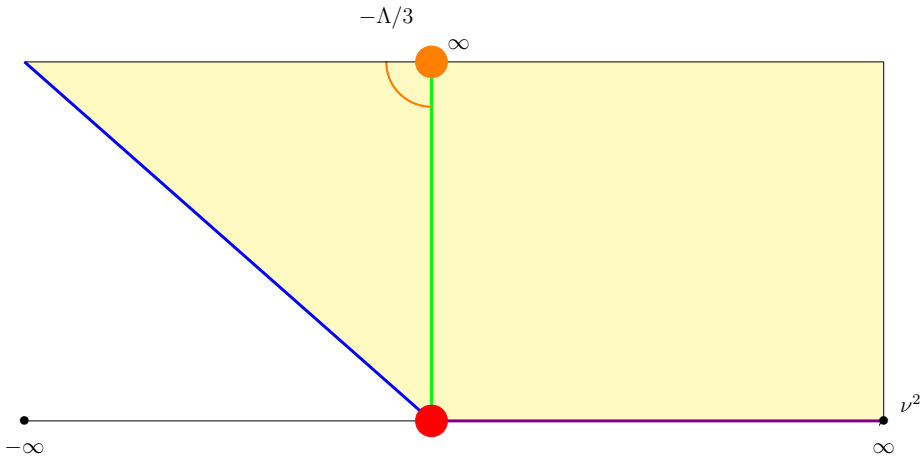

The oblate Pedersen metric \eqref{eq:PedMet} and its continuation to the prolate Pedersen metric \eqref{eq:eq:h3_gibbons-hawking} depend on two parameters $\nu^2$ and $\Lambda=-3/l^2$. Note that $\nu$ in \eqref{eq:PedMet} is either real with any  $l>0$ (oblate case) or purely imaginary with $|\nu| l<1$ (prolate case) to ensure completeness. The resulting region is displayed in yellow in figure \ref{fig:Ped}. There are various limiting cases in which the Pedersen metric turns into spacetimes that were previously studied in the context of celestial and twisted holography \cite{Crawley:2021auj, Crawley:2023brz, Bittleston:2023bzp, Bittleston:2024rqe, Costello:2023hmi, Costello:2022jpg, Costellotalk}. These limits are displayed in figure \ref{fig:Ped}. Let us discuss them case by case. 

\paragraph{Self-dual Taub-NUT}
\label{para:SDTNmetric}
We first consider the limit in which the cosmological constant is vanishing, $l\to\infty$. It's easier to understand this limit from the form \eqref{eq:SdTNADS} of the Pedersen metric, which reduces to
\bea
\label{eq:GibHawk}
\d s^2&= V^{-1}(\d t+2M\,\cos\theta\,\d\phi)^2+V(\d r^2+r^2\d\Omega_2^2)\,,
\eea
after taking $l\to\infty$ and shifting $r\mapsto r+M$. Here, $V$ is the Higgs field defined on $\mathbb{R}^3$ by
\be
V(r,\theta,\phi)=1+\frac{2M}{r}
\ee
This is precisely the self-dual Taub-NUT metric in the Gibbons-Hawking form used in \cite{Adamo:2023fbj}. For a more general discussion of Gibbons-Hawking metrics see \cite{Gibbons:1978tef}.

\paragraph{AdS$_4$}
\label{para:AdS4met}
The second natural limit is the limit $\nu\to 0$. In this case, the (prolate or oblate) Berger sphere \eqref{eq:oblatesquashedS3} reduces to the round $S^3$. Correspondingly in the bulk, the  (prolate or oblate)  Pedersen metric in the triaxial form \eqref{eq:PedMet} reduces to
\begin{equation}
    \d s^2=\frac{4}{(1-r^2/l^2)^2}(\d r^2+r^2(\sigma_1^2+\sigma_2^2+\sigma_3^2))\,.
\end{equation}
This is Euclidean AdS$_4$ which has a round $3$-sphere as its conformal boundary at infinity and the boundary metric is the canonical metric on the 3-sphere parametrized by $(\psi,\theta,\phi)$\footnote{Be careful not to confuse this 3-sphere with the 3-sphere parametrized by $(\chi,\theta,\phi)$.}.

\paragraph{A singular double cover of Eguchi-Hanson space}
\label{para:EH2met}
In this and the following limit, we consider the prolate Pedersen metric \eqref{eq:eq:h3_gibbons-hawking}. Consider the limit $\nu,l\to 0$ with $\alpha^2\equiv l/\nu$ held constant and set 
\begin{equation}
\label{eq:conf}
    r=\frac{\alpha^2}{\varrho}\,.
\end{equation}
The metric in the triaxial form then becomes
\begin{equation}
\label{eq:EH}
    \d s^2=\frac{\d \varrho^2}{1-\alpha^4/\varrho^4}+\varrho^2(\sigma_1^2+\sigma_2^2)+\varrho^2\left(1-\frac{\alpha^4}{\varrho^4}\right)\sigma_3^2\,,
\end{equation}
up to an overall factor of $l^4/4\alpha^2$. The space-time metric can be extended to the region $\varrho>\alpha$ and is seen to locally be described be the Eguchi-Hanson metric \cite{Eguchi:1978xp} that was the content of chapter \ref{chapter4}. More precisely, the singularity at $\varrho=\alpha$ is not removable without taking a $\Z_2$ quotient and here we did not take this $\Z_2$ quotient. So the spacetime is in fact a double cover of Eguchi-Hanson space similar to the one described in \cite{Bittleston:2024efo}.

\paragraph{$\widetilde{\mathbb{CP}}^2$ and Burns space}
\label{para:Burns}
Setting $\nu l=1$ in the prolate Pedersen metric \eqref{eq:eq:h3_gibbons-hawking}, we recover the Burns space metric as described in equation (3.63) of  \cite{Costello:2023hmi} up to a conformal prefactor. Performing the change of coordinates 
\begin{equation}
    r^2=\frac{R^2}{2-R^2/l^2}\,,
\end{equation}
in the triaxial form of the metric, we find 
\begin{equation}
    \d s^2=\frac{2}{(1-R^2/l^2)^2}(\d R^2+R^2\sigma_3^2)+\frac{2}{1-R^2/l^2}(\sigma_1^2+\sigma_2^2)\,,
\end{equation}
which is the Fubini-Study metric on an open subset of a non-compact version of $\CP^2$ commonly referred to as Bergmann space and denoted by $\widetilde{\mathbb{CP}}^2$. As a homogeneous space it is described as $\widetilde{\mathbb{CP}}^2=SU(2,1)/U(2)$ and it is the non-compact dual of $\mathbb{CP}^2=SU(3)/U(2)$ in the sense of \cite{Helgason:1968bvw}. $\widetilde{\mathbb{CP}}^2$ is conformally equivalent to (an open subset of)  $\mathbb{CP}^2$ with a point removed 
which is conformally equivalent to an open subset of Burns space \cite{Costello:2023hmi}. 


\section{More backreactions in self-dual Einstein gravity}
\label{sec:sec3}

The Pedersen metric arises from a twistor space, that we will refer to as the \emph{Pedersen twistor space}. In this section, we will first see that the twistor space of its $\Lambda\rightarrow0$ limit, \emph{i.e.} that of self-dual Taub-NUT, arises  from a backreaction similar to the one in chapter \ref{chapter4}. A slight variation thereof gives rise to the so-called  \emph{multi-centred Taub-NUT space} with $2$ centres \cite{Hawking:1976jb} sometimes also referred to as the \emph{A$_1$-ALF gravitational instanton}.
Then we will discuss how the full Pedersen twistor space conjecturally arises from a $\Lambda\leq 0$ generalisation of the self-dual Taub-NUT backreaction. There is an analogous backreaction with $\Lambda>0$ that we conjecture to lead to the singular \emph{Taub-NUT-de Sitter} metric of \cite{Cvetic:2002kj}. We will not discuss this $\Lambda>0$ case in detail.

\medskip

We will consider the twistor uplift of self-dual gravity in the presence of a cosmological constant described by the Mason-Wolf action \cite{Mason:2007ct} 
\be
S_{\Lambda}[g,h]=\int_\PT \text{D}^3Z\wedge g \wedge \left(\dbar h+\frac{1}{2}\{h,h\}_\Lambda\right)\,,
\ee
where we are using the holomorphic $\mathcal{O}(-2)$-valued Jacobi bracket $\{\,\,,\,\,\}_\Lambda$ defined through the ball-model infinity twistor \eqref{eq:infinity_twistors}
\bea
\label{eq:LambdaBrack}
&\{f,g\}_\Lambda=\varepsilon^{\dot\alpha\dot\beta}\frac{\p f}{\p\mu^{\dot\alpha}}\frac{\p g}{\p\mu^{\dot\beta}}+\Lambda\,\varepsilon_{\alpha\beta}\frac{\p f}{\p\lambda_\alpha}\frac{\p g}{\p\lambda_\beta}\,.
\eea
A discussion of this theory was provided in chapter \ref{chapter5}. Note that considering homogeneous coordinates means that we are working on non-projective twistor space, where \eqref{eq:LambdaBrack} is a (twisted) Poisson bracket whereas it is a (twisted) Jacobi bracket on projective twistor space. Below, we will formally work on a generalized non-projective twistor space of a curved spacetime which is formalised by the total space of the Swann-bundle over the curved twistor space \cite{Adamo:2021bej, swann1990hyperkahler}.

\subsection*{Deforming twistor space with a defect operator}
\label{sec:Backreaction}

In chapter \ref{chapter4}, it was shown that the twistor space of Eguchi-Hanson space arises through a backreaction from including a defect operator wrapping $\mathbb{CP}^1_{\mu^{\dot{\alpha}}=0}$ in the flat twistor space of $\mathbb{C}^2/\mathbb{Z}_2$ \cite{Bittleston:2023bzp}. We will now formally include a defect operator wrapping the twistor line $\mathbb{CP}^1_{\lambda_\alpha=0}$ in $\mathbb{CP}^3\supset \mathbb{PT}$ that couples electrically to $g$. Doing this can equivalently be viewed as introducing a boundary condition for $h$.

\subsection*{Self-dual Taub-NUT}
\label{para:SDTN}
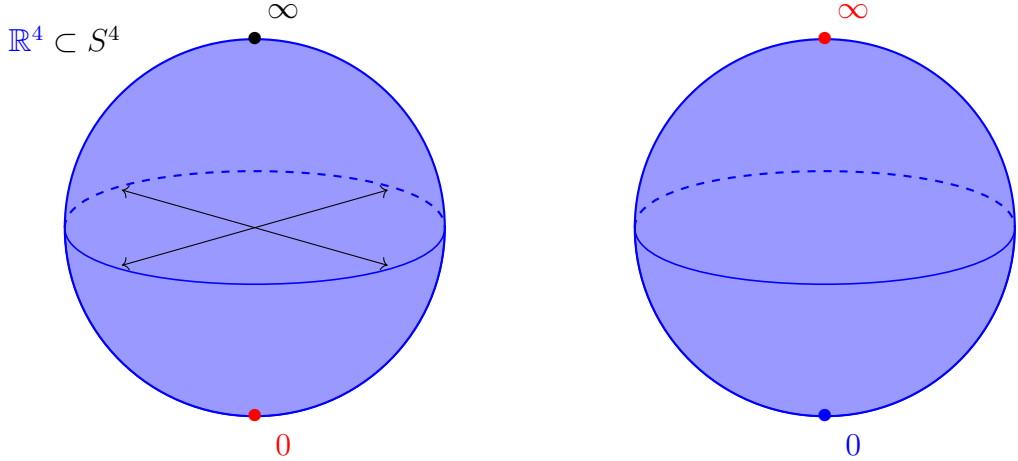
\begin{figure}[t!]
    \centering
\begin{tikzpicture}[scale=2.5]
\filldraw [color= blue, fill=blue!40, thick] (0,0) circle (1);
\draw [name path = A, thick, dashed, blue] (-1,0)  .. controls (-1,0.4) and (1,0.4) ..  (1,0)  ;
\draw [name path = B, semithick, blue] (-1,0)  .. controls (-1,-0.4) and (1,-0.4) ..  (1,0)  ;
\draw [name path = D, semithick, blue] (1,0) arc(0:-180:1);
 \begin{pgfonlayer}{bg}
\fill [blue!20,
          intersection segments={
            of=A and B,
            sequence={L2--R2}
          }];
          \fill [blue!40,
          intersection segments={
            of=B and D,
            sequence={L2--R2}
          }];
\end{pgfonlayer}
 \node[name = a] at (0,-1) {\red{$\bullet$}};
  \node at (0.15,-1.15) {\red{$0$}};
\node (b) at (0,1){$\bullet$}; 
\node at (0.15,1.15) {$\infty$};


\filldraw [color= blue, fill=blue!40, thick] (3,0) circle (1);
\draw [name path = A, thick, dashed, blue] (2,0)  .. controls (2,0.4) and (4,0.4) ..  (4,0)  ;
\draw [name path = B, semithick, blue] (2,0)  .. controls (2,-0.4) and (4,-0.4) ..  (4,0)  ;
\draw [name path = D, semithick, blue] (4,0) arc(0:-180:1);
 \begin{pgfonlayer}{bg}
\fill [blue!20,
          intersection segments={
            of=A and B,
            sequence={L2--R2}
          }];
          \fill [blue!40,
          intersection segments={
            of=B and D,
            sequence={L2--R2}
          }];
\end{pgfonlayer}

\draw[<->] (-0.7,-0.2)  -- (0.7,0.2);
\draw[<->] (-0.7,0.2)  -- (0.7,-0.2);
 \node[name = a] at (3,-1) {\blue{$\bullet$}};
  \node at (3.15,-1.15) {\blue{$0$}};
\node (b) at (3,1){$\red{\bullet}$}; 
\node at (3.15,1.15) {$\red{\infty}$};

\node at (-1,1) {$\blue{\mathbb{R}^4}\subset S^4$};

\end{tikzpicture}
\caption{\emph{Any point in the depicted $S^4$ represents a Euclidean twistor line inside $\mathbb{CP}^3\supset\mathbb{PT}$. Left: Coupling a defect operator wrapping the $\mathbb{CP}^1$ over $0$ (red) backreacts $\mathbb{PT}/\mathbb{Z}_2$ to the twistor space of Eguchi-Hanson space as discussed in chapter \ref{chapter4}. A so-called \emph{bolt singularity} \cite{Eguchi:1979yx} is present at $0$ which requires the $\mathbb{Z}_2$-quotient.
Right: Coupling a defect operator wrapping the $\mathbb{CP}^1$ over $\infty$ (red) backreacts $\mathbb{PT}$ to the twistor space of self-dual Taub-NUT. A \emph{regular NUT singularity} is present at $0$ so that no $\mathbb{Z}_2$-quotient is required.}}
    \label{fig:twoTwistorBranes}
\end{figure}

As a warm-up, let us consider the special case $\Lambda=0$ in which the backreacted twistor space is that of self-dual Taub-NUT. The deformed action reads
\bea
    \label{eq:twistor-defect-action1}
    S[g,h] &=  S_{0}[g,h]- \frac{\pi^2}{4M} \int_{\CP^1_{\lambda_\alpha=0}}\la\lambda\,\dif\lambda\ra\wedge\eta^2\,  g \,,
\eea
where  $\eta=\mu^{\dot{0}}\mu^{\dot{1}}=[\iota \mu][\mu\hat{\iota}]$ singles out a choice of dotted 
reference spinors $\iota, \hat{\iota}$ which breaks Lorentz invariance. Below, we will also use  $\mu^{+}=\mu^{\dot{0}}$ and $\mu^{-}=\mu^{\dot{1}}$ for notational convenience. The latter term in \eqref{eq:twistor-defect-action1} describes the coupling to the defect. $1/M$ is to be viewed as a coupling constant controlling the strength of the coupling. For $1/M\neq 0$, we can vary $g$ to obtain the deformed equation of motion for $h$
\begin{equation}
    \dbar h+\frac{1}{2}\{h,h\}_0=\frac{\pi^2\eta^2}{2M}\bar\delta^{(2)}(\lambda)\,.
\end{equation}
The equation for $g$ remains unchanged. Similar to the treatment of chapter \ref{chapter4} \cite{Bittleston:2023bzp}, we can solve the sourced equation by 
\begin{equation}
    h=\frac{\eta^2}{8M}\bar e^0\,,\qquad\bar e^0=\frac{\langle\hat\lambda\d\hat\lambda\ra}{\la\lambda\hat\lambda
    \ra^2}\,.\label{eq:sdtn_hamiltonian}
\end{equation}
since we have $\{h,h\}_0=0$, $\bar \partial h=0$ for $\la\lambda \hat{\lambda}\ra\neq 0$ and the correct normalization.
Note that up until this point, the backreaction is directly related to the one performed in chapter \ref{chapter4} \cite{Bittleston:2023bzp} by exchanging $\lambda_\alpha$ and $\mu^{\dot{\alpha}}$, which corresponds to a conformal inversion
\be
x^{\alpha\dal}\mapsto\frac{2 x^{\alpha\dal}}{x^2}
\ee
on spacetime, since
\be
\mu^\da =x^{\alpha \da}\lambda_\alpha \quad \iff \quad \frac{2 x_{\alpha \dal}}{x^2}\mu^\dal=\lambda_\alpha\,.
\ee
The location of the branes in the two different cases is depicted from a Euclidean spacetime point of view in figure \ref{fig:twoTwistorBranes}. $\mathbb{CP}^1_{\lambda_\alpha=0}$ importantly is not a twistor line in $\mathbb{PT}$ so that referring to our treatment as a backreaction is not quite obvious. Our treatment might better be interpreted as a boundary condition for $h$ or a non-trivial state in some dual theory similar to \cite{Crawley:2023brz}. However, switching on a positive cosmological constant means that twistor space becomes $\mathbb{CP}^3$ which genuinely contains $\mathbb{CP}^1_{\lambda_\alpha=0}$ so that the interpretation as a backreaction makes sense. The cosmological constant can be sent to zero in the end resulting in the above 'backreaction'. For this reason, we will keep using the term backreaction throughout this section.


Even though the Hamiltonian $h$ simply arises from  exchanging $\lambda_\alpha$ and $\mu^{\dot{\alpha}}$, the Beltrami differential will not do so. This is the case because the Poisson structure\footnote{or Jacobi structure when $\Lambda \neq 0$} $\{\,\,,\,\,\}_\Lambda$ in equation \eqref{eq:LambdaBrack} is not conformally invariant. We can see this explicitly by considering the deformed Dolbeaut operator which now reads
\begin{equation}
    \begin{aligned}
        \bar\nabla_0&=\dbar+\{h,-\}_0\\&=\dbar +\frac{\eta}{4M}\bar e^0\wedge(\mu^{\dot 1}\mathcal{L}_{\dot 1}-\mu^{\dot 0}\mathcal{L}_{\dot 0})\,,
    \end{aligned}
\end{equation}
where $\mathcal{L}_{\dot\alpha}=\mathcal{L}_{\p/\p\mu^{\dot\alpha}}$ and $\mathcal{L}^\alpha=\mathcal{L}_{\p/\p\lambda_\alpha}$ denote the Lie derivatives. The coordinates $\lambda_\alpha$ and $\eta$ are still holomorphic in the deformed complex structure, but $\mu^\da$ are not. We can construct patchwise  (over $\mathbb{CP}^1$) holomorphic coordinates
\bea
\label{eq:rhopm}
\rho^\pm = \mu^{\pm}\,\exp\bigg(\pm\frac{\eta f(\lambda)}{4M}\bigg)\,,
\eea
where we introduced the patchwise-defined function
\be
\label{eq:f(lambda)}
f(\lambda) = \begin{cases}
    \displaystyle \frac{1}{\la\lambda \hat{\lambda}\ra}\frac{\hat\lambda_0}{\lambda_0}\qquad &\lambda_0\neq0\\
    \vspace{-1.2em}\\
    \displaystyle\frac{1}{\la\lambda \hat{\lambda}\ra}\frac{\hat\lambda_1}{\lambda_1}\qquad &\lambda_1\neq0
\end{cases}\;.
\ee
$\rho^\pm$ are holomorphic as a consequence of $\dbar f=\bar e^0$. 

On the overlap, $\rho^\pm$ patch according to the transition function 
\bea
\label{eq:TransitionSDTN}
\rho^\pm&\mapsto\rho^\pm\,\exp\bigg(\mp \frac{\eta \hat{\lambda}_0}{4M\la\lambda \hat{\lambda}\ra \lambda_0} \bigg)\exp\bigg(\pm \frac{\eta \hat{\lambda}_1}{4M\la\lambda \hat{\lambda}\ra \lambda_1} \bigg)  \\
&=\rho^{\pm}\, \exp\bigg(\pm \frac{\eta}{4M\lambda_0 \lambda_1} \bigg)\,.
\eea
Moreover, they also satisfy $\rho^+\rho^-=\eta$ on each patch. This backreacted geometry precisely matches Hitchin's description of the twistor space of self-dual Taub-NUT \cite{Hitchin:1979rts, hitchin1992hyper, Besse:1987pua}.

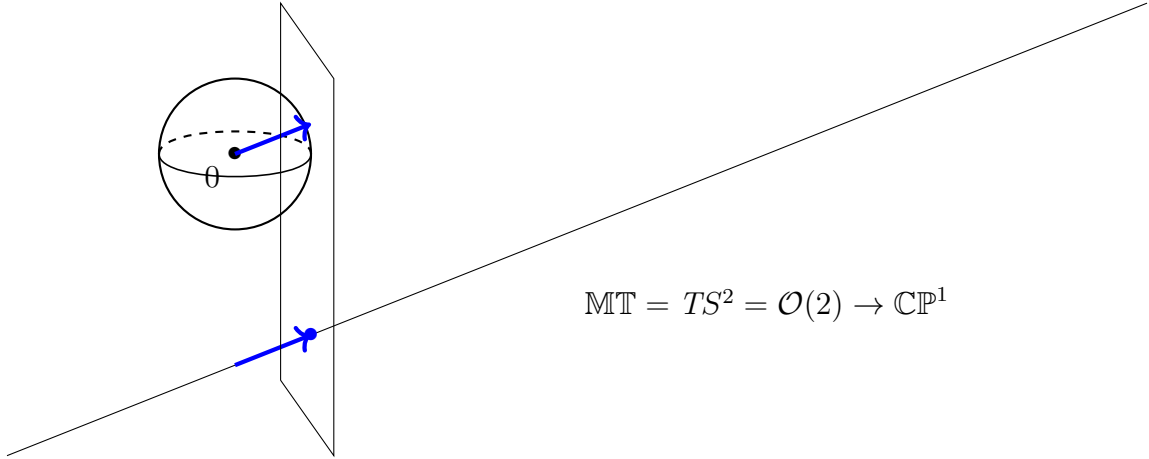
\begin{figure}[t!]
    \centering
\begin{tikzpicture}[scale=1]
    \draw (0,0)--(15,6);
    \node at (3,4) {$\bullet$};
  \node at (2.7,3.7) {$0$};
\draw[->, ultra thick, color= blue] (3,18/15)--(4,24/15);
  \draw [thick] (3,4) circle (1);
  \draw [ thick, dashed] (2,4)  .. controls (2,4.4) and (4,4.4) ..  (4,4)  ;
\draw [ semithick] (2,4)  .. controls (2,3.6) and (4,3.6) ..  (4,4)  ;
\draw[->, ultra thick, color= blue] (3,4)--(4,4+6/15);

\draw (3.6,1)--(3.6,6)--(4.3,5)--(4.3,0)--(3.6,1);
\node at (4,24/15) {$\blue{\bullet}$};
\node at (10,2) {$\mathbb{MT}=\textit{T}S^2=\mathcal{O}(2)\rightarrow\mathbb{CP}^1$};
\end{tikzpicture}
    \caption{\emph{An oriented geodesic in the Einstein space $\mathbb{R}^3$ is given by a straight line. A straight line is uniquely determined by a point on the unit sphere $S^2$ and a tangent vector to that point. Algebraically, the tangent bundle of $S^2$ is given by the total space of $\mathcal{O}(2)\rightarrow \mathbb{CP}^1$.}}
    \label{fig:MinitwistorR3}
\end{figure}

To define it, let us briefly discuss the \emph{minitwistor space} of the Einstein space $\mathbb{R}^3$. It is referred to as $\mathbb{MT}$ and given by the space of all oriented geodesics in $\mathbb{R}^3$ with the flat metric. This can be easily seen to be given by the total space
\be
\mathbb{MT}=\mathcal{O}(2)\rightarrow \mathbb{CP}^1\,,
\ee
as depicted in figure \ref{fig:MinitwistorR3}. Minitwistor spaces are defined for much more general so-called \emph{Einstein-Weyl geometries}. We will not discuss minitwistor theory here and simply refer to the literature for a review \cite{calderbank2001einstein}. 

Hitchin's twistor space is then defined by solutions of $\rho^+ \rho^-=\eta$ inside the total space of a sum of two line bundles $L^{1/4M}(1)=L^{1/4M}\otimes \mathcal{O}(1)$ and $L^{-1/4M}(1)= L^{-1/4M}\otimes \mathcal{O}(1)$, for the precise definition of which we refer to \cite{Hitchin:1979rts, hitchin1992hyper, Besse:1987pua}, over $\mathbb{MT}$
\bea
\label{eq:TwistorofSDTN}
&\{(\rho^+, \rho^-)\in L^{1/4M}(1)\oplus L^{-1/4M}(1): \rho^+ \rho^-=\eta\}\\
&\subset \text{Tot}(L^{1/4M}(1)\oplus L^{-1/4M}(1)\rightarrow \mathbb{MT})\,. 
\eea
The spacetime metric of self-dual Taub-NUT can be explicitly derived from this twistor space \cite{Besse:1987pua}. Note that the holomorphic coordinates \eqref{eq:rhopm} are formally equivalent to coordinates which previously appeared in the twisted holography context in \cite{Budzik:2023nnx}.

The fact that Hitchin's twistor space of equation \eqref{eq:TwistorofSDTN} fibres over $\mathbb{MT}$, the minitwistor space of $\mathbb{R}^3$, is reflected by the fact that the self-dual Taub-NUT metric itself fibres over $\mathbb{R}^3$ as is manifest from its description in Gibbons-Hawking form \eqref{eq:GibHawk}.

\subsection*{The A$_1$-ALF metric}

\begin{figure}[t!]
    \centering
\begin{tikzpicture}[scale=2.5]
\filldraw [color= blue, fill=blue!40, thick] (0,0) circle (1);
\draw [name path = A, thick, dashed, blue] (-1,0)  .. controls (-1,0.4) and (1,0.4) ..  (1,0)  ;
\draw [name path = B, semithick, blue] (-1,0)  .. controls (-1,-0.4) and (1,-0.4) ..  (1,0)  ;
\draw [name path = D, semithick, blue] (1,0) arc(0:-180:1);
 \begin{pgfonlayer}{bg}
\fill [blue!20,
          intersection segments={
            of=A and B,
            sequence={L2--R2}
          }];
          \fill [blue!40,
          intersection segments={
            of=B and D,
            sequence={L2--R2}
          }];
\end{pgfonlayer}
 \node[name = a] at (0,-1) {\red{$\bullet$}};
  \node at (0.15,-1.15) {\red{$0$}};
\node (b) at (0,1){\red{$\bullet$}}; 
\node at (0.15,1.15) {\red{$\infty$}};

\draw[<->] (-0.7,-0.2)  -- (0.7,0.2);
\draw[<->] (-0.7,0.2)  -- (0.7,-0.2);

\node at (-1,1) {$\blue{\mathbb{R}^4}\subset S^4$};

\end{tikzpicture}
\caption{\emph{Including defects at both twistor lines $\mathbb{CP}^1_{\lambda_\alpha=0}$ and $\mathbb{CP}^1_{\mu^{\da}=0}$ (red) leads to a backreaction in which the twistor space $\mathbb{PT}$ gets deformed to the twistor of the A$_1$-ALF space. Roughly speaking, this space is a 'taub-NUT version' of Eguchi-Hanson space.  Equivalently, it is a \emph{multi-centred Taub-NUT metric} with two centres \cite{Hawking:1976jb}.}}
    \label{fig:A1ALF}
\end{figure}
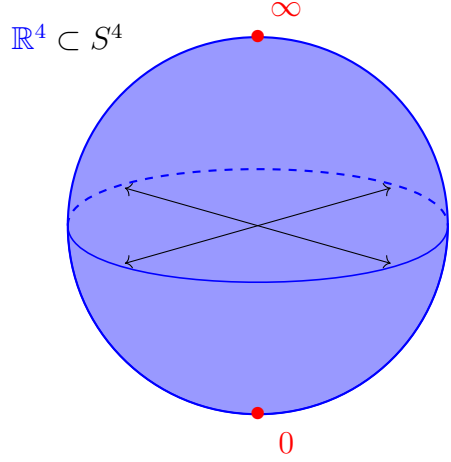

Let us briefly discuss a variation of the previous backreaction, in which we include two sources, at both twistor lines  $\mathbb{CP}^1_{\mu^{\da}=0}$, the twistor line over $0$, and $\mathbb{CP}^1_{\lambda_\alpha=0}$, the twistor line over $\infty$. The setup is depicted in figure \ref{fig:A1ALF} in analogy to figure \ref{fig:twoTwistorBranes}.

In equations, the backreaction is described by the sourced equation
\be
\label{eq:TN2source}
\dbar h + \frac{1}{2}\{h,h\} =2\pi^2 (M^2(\mu)\bar{\delta}^2(\lambda) +c^2(\lambda)\bar{\delta}^2(\mu))\,,
\ee
where we denote 
\be
M(\mu)= \frac{[\iota \mu][\mu\hat{\iota}]}{\sqrt{4M}}=  \frac{\eta}{\sqrt{4M}}\,,
\ee
in analogy to the $c(\lambda)= c\langle\alpha \lambda \rangle\langle \lambda \beta \rangle$ defined in chapter \ref{chapter4}. The solution for $h$ reads
\be
\label{eq:hTN2}
h=\frac{1}{2}c^2(\lambda)\frac{[\hat{\mu}d\hat{\mu}]}{[\mu \hat{\mu}]^2} +\frac{1}{2} M^2(\mu)\bar{e}^0-c^2(\lambda)M(\mu)\frac{M(\hat{\mu})}{[\mu \hat{\mu}]^2}\bar{e}^0 +\frac{1}{2}c^4(\lambda) \frac{M(\hat{\mu})^2}{[\mu \hat{\mu}]^4}\bar{e}^0
\ee
where the first two terms solve 
\be
\dbar \bigg(\frac{1}{2}c^2(\lambda)\frac{[\hat{\mu}d\hat{\mu}]}{[\mu \hat{\mu}]^2} + \frac{1}{2}M^2(\mu)\bar{e}^0\bigg) =2\pi^2 \big(M^2(\mu)\bar{\delta}^2(\lambda) +c^2(\lambda)\bar{\delta}^2(\mu)\big)\,,
\ee
as discussed above in the Taub-NUT and Eguchi-Hanson space cases. However, they also give rise to a non-linear term which is compensated by $\dbar$ of the third term in equation \eqref{eq:hTN2}:
\be
\begin{aligned}
\bigg\{c^2(\lambda) \frac{[\hat{\mu}d\hat{\mu}]}{[\mu \hat{\mu}]^2}, M^2(\mu)\bar{e}^0\bigg\}&=  \frac{-4 c^2(\lambda)M(\mu)}{\sqrt{2M}} \big([\iota \hat{\mu}][\mu \hat{\iota}]+[\iota \mu][\hat{\mu} \hat{\iota}]\big)\frac{[\hat{\mu}d\hat{\mu}]}{[\mu \hat{\mu}]^3}\wedge \bar{e}^0\\
&=\dbar\bigg(4c^2(\lambda)M(\mu)\bar{e}^0\frac{M(\hat{\mu})}{[\mu \hat{\mu}]^2}\bigg)\,.
\end{aligned}
\ee
In the first step, we used 
\be
\{[\mu \hat{\mu}],M(\mu)\}=\frac{1}{\sqrt{2M}}([\iota \hat{\mu}][\mu \hat{\iota}]+[\iota \mu][\hat{\mu} \hat{\iota}])\,,
\ee
and in the last step, we used the two identities
\bea
&\dbar\Big(\frac{\hat{\mu}^{\dal}}{[\mu \hat{\mu}]}\Big)=-\frac{[\hat{\mu}d\hat{\mu}]}{[\mu \hat{\mu}]^2}\mu^\dal\,,\\
&4c^2(\lambda)M(\mu)\frac{M(\hat{\mu})}{[\mu \hat{\mu}]^2}\dbar(\bar{e}^0)=0\,.
\eea
There is a further non-linear term that gets cancelled by $\dbar$ of the fourth term in \eqref{eq:hTN2}
\be
\begin{aligned}
&\bigg\{c^2(\lambda)\frac{[\hat{\mu}d\hat{\mu}]}{[\mu \hat{\mu}]^2},-c^2(\lambda)M(\mu)\bar{e}^0 \frac{M(\hat{\mu})} {[\mu \hat{\mu}]^2}\bigg\}=2c^4(\lambda) M(\bar{\mu})  \frac{[\hat{\mu}d\hat{\mu}]}{[\mu \hat{\mu}]^5}\bar{e}^0\bigg\{[\mu \hat{\mu}],M(\mu)\bigg\} \\
&=2c^4(\lambda) M(\hat{\mu})  \frac{[\hat{\mu}d\hat{\mu}]}{[\mu \hat{\mu}]^5}\bar{e}^0 \frac{1}{\sqrt{2M}}([\iota \hat{\mu}][\mu \hat{\iota}]+[\iota \mu][\hat{\mu} \hat{\iota}])\\
&=\dbar\Big(-c^4(\lambda)\bar{e}^0 \frac{M^2(\hat{\mu})}{[\mu \hat{\mu}]^4}\Big)\,.
\end{aligned}
\ee
All the other non-linear terms vanish either because they involve $\bar{e}^0\wedge \bar{e}^0=0$ or because they involve $\{[\mu \hat{\mu}],[\mu \hat{\mu}]\}=0$
so that equation \eqref{eq:TN2source} holds for \eqref{eq:hTN2}.

It can be explicitly checked that in a patch $\lambda_0\neq 0$ or $\lambda_1\neq 0$ the holomorphic coordinates for the Beltrami differential corresponding to the Hamiltonian  \eqref{eq:hTN2} are given by $\lambda_\alpha$ as well as
\be
\begin{aligned}
\label{eq:A1ALFcoords}
    X&=\Big([\iota \mu]^2-c^2(\lambda) \frac{[\iota \hat{\mu}]^2}{[\mu \hat{\mu}]^2}\Big) \exp(\tfrac{1}{4M}Zf(\lambda))\,,\\
    Y&=\Big([ \mu \hat{\iota}]^2-c^2(\lambda) \frac{[ \hat{\mu} \hat{\iota}]^2}{[\mu \hat{\mu}]^2}\Big) \exp(-\tfrac{1}{4M}Zf(\lambda))\,,\\
    Z&=[\iota \mu][\mu \hat{\iota}]-c^2(\lambda)\frac{[\iota \hat{\mu}][\hat{\mu}\hat{\iota}]}{[\mu \hat{\mu}]^2}\,.
\end{aligned}
\ee Again, $f(\lambda)$ which was defined in \eqref{eq:f(lambda)} depends on the chart. 

\medskip

When $\frac{1}{4M}\rightarrow0$, the coordinates \eqref{eq:A1ALFcoords} reduce to the coordinates \eqref{eq:newholomorphic} of the twistor space of Eguchi-Hanson space, which we recall to be given by the subvariety
\be
 XY = \left(Z-c(\lambda)\right)\left(Z+c(\lambda)\right)\,.
\ee
inside
\be
\mathcal{O}(2)\oplus\mathcal{O}(2)\oplus\mathcal{O}(2)\rightarrow \mathbb{CP}^1\,.
\ee
When  $\frac{1}{4M}\neq 0$, this Eguchi-Hanson twistor space gets deformed and the first as well as the second of these $\mathcal{O}(2)$s parametrized by $X$ and $Y$ get twisted by the bundles $L^{\pm 1/4M}$ that appeared in the case of self-dual Taub-NUT. The resulting twistor space is the subvariety 
\be
 XY = \left(Z-c(\lambda)\right)\left(Z+c(\lambda)\right)\,.
\ee
inside
\be
\label{eq:A1ALFMinitwistor}
 L^{ 1/4M}(2) \oplus L^{-1/4M}(2) \oplus\mathcal{O}(2)\rightarrow \mathbb{CP}^1\,,
\ee
where we defined
\be
L^{\pm1/4M}(2)=L^{\pm1/4M} \otimes\mathcal{O}(2)\,.
\ee

Note, that we can interpret the (untwisted) $Z$-coordinate over $\mathbb{CP}^1$ as defining $\mathbb{MT}$ so that equation \eqref{eq:A1ALFMinitwistor} can be viewed as
\be
\label{eq:A1ALFFiberover}
L^{ 1/4M}(2) \oplus L^{-1/4M}(2)\rightarrow \mathbb{MT}\,.
\ee

The fact that \eqref{eq:A1ALFFiberover} fibres over minitwistor space means that the metric is expected to be of Gibbons-Hawking form similar to \eqref{eq:GibHawk}. Indeed, the twistor space we have found above \eqref{eq:A1ALFMinitwistor} is the twistor space of the so-called A$_1$-ALF gravitational instanton \cite{Ionas:2016oxr, Hitchin:1979rts, Hawking:1976jb}. The spacetime metric that can be derived from it is given by a Gibbons-Hawking metric (similar to \eqref{eq:GibHawk}) with a Higgs field schematically of the form
\be
V(\Vec{x})=\frac{1}{2M}+\frac{1}{|\Vec{x}-\Vec{a}|}+ \frac{1}{|\Vec{x}+\Vec{a}|}\,,
\ee
where $\Vec{a}$ is related to the choice of $c^2(\lambda)$. For $\vec{a}\rightarrow0$, this reduces to the self-dual Taub-NUT spacetime\footnote{Up to a $\mathbb{Z}_2$-quotient.}  \eqref{eq:GibHawk} and for $\tfrac{1}{4M}\rightarrow0$, this reduces to the Eguchi-Hanson metric as expected from the twistor space.

\subsection*{The Pedersen twistor space}
\label{para:PedTwistor}
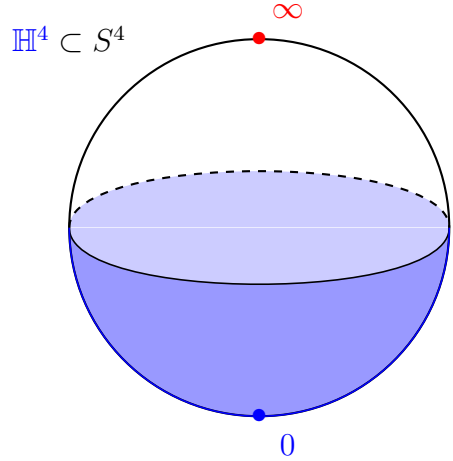
\begin{figure}
    \centering
\begin{tikzpicture}[scale=2.5]
\draw [thick] (0,0) circle (1);
\draw [name path = A, thick, dashed] (-1,0)  .. controls (-1,0.4) and (1,0.4) ..  (1,0)  ;
\draw [name path = B, semithick] (-1,0)  .. controls (-1,-0.4) and (1,-0.4) ..  (1,0)  ;
\draw [name path = D, semithick, blue] (1,0) arc(0:-180:1);
 \begin{pgfonlayer}{bg}
\fill [blue!20,
          intersection segments={
            of=A and B,
            sequence={L2--R2}
          }];
          \fill [blue!40,
          intersection segments={
            of=B and D,
            sequence={L2--R2}
          }];
\end{pgfonlayer}

 \node[name = a] at (0,-1) {\blue{$\bullet$}};
  \node at (0.15,-1.15) {\blue{$0$}};
\node (b) at (0,1){\red{$\bullet$}}; 
\node at (0.15,1.15) {\red{$\infty$}};
\node at (-1,1) {$\blue{\mathbb{H}^4}\subset S^4$};
\end{tikzpicture}
\caption{\emph{Coupling the Mason-Wolf action with $\Lambda<0$ to a defect operator wrapping the $\mathbb{CP}^1$ over $\infty$ (red) conjecturally backreacts the twistor space of AdS$_4$ (blue) to the twistor space of the Pedersen metric.}}
    \label{fig:BraneAdS}
\end{figure}

Let us now consider a backreaction using a 'source at $\infty$' in the presence of a cosmological constant $\Lambda<0$. The location of the defect is now displayed in figure \ref{fig:BraneAdS}.

\begin{figure}[t!]
    \centering
    \begin{tikzpicture}[scale=2.5]
\node at (-1,1) {$\mathbb{H}^3$};
    \filldraw[ fill=blue!10, thick] (0,0) circle (1);
    \node at (-1,0) {$\bullet$};
  \node at (0,1) {$\blue{\bullet}$}; 
   \draw[blue, ->] (-1,0)  .. controls (-0.3,0) and (0,0.3) ..  (0,1);
\end{tikzpicture}
    \caption{\emph{An oriented geodesic in the Einstein space $\mathbb{H}^3$ is uniquely determined by two points on its boundary $\mathbb{CP}^1$. The minitwistor space of $\mathbb{H}^3$ is hence given by  $\mathbb{CP}^1\times\overline{\mathbb{CP}}^1\setminus\text{diag}\cong  \mathbb{CP}^1\times\mathbb{CP}^1\setminus\overline{\text{diag}}$. Similarly, the minitwistor space of $S^3$ describes great circles inside $S^3$ which can be seen to be parametrized by $(S^3\times \mathbb{CP}^1)/S^1\cong \mathbb{CP}^1\times \mathbb{CP}^1$.}}
    \label{fig:minitwistoCurved}
\end{figure}
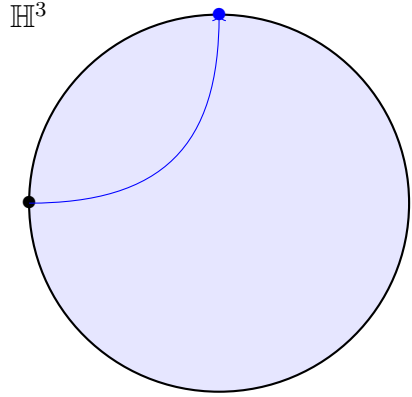

Consider the action 
\bea
    \label{eq:twistor-defect-action2}
    S[g,h] &=  S_{\Lambda}[g,h]- \frac{\pi^2\nu^2}{2} \int_{\CP^1_{\lambda_\alpha=0}}\la\lambda\,\dif\lambda\ra\wedge\eta^2\,  g \,,
\eea
with $\Lambda<0$. The solution to the equation of motion is given by the same Hamiltonian as in the self-dual Taub NUT case with $\Lambda=0$
\begin{equation}
    h=\frac{\nu^2\eta^2}{4}\bar e^0\,,
\end{equation}
 since $\{h,h\}_\Lambda=0$. However, the presence of the new term in $\{\,\,,\,\,\}_\Lambda$ deforms the Beltrami differential to
\begin{equation}
\label{eq:NablaQTN}
    \begin{aligned}
        \bar\nabla_\Lambda&=\dbar+\{h,-\}_{\Lambda}\\&=\dbar +\frac{\nu^2 \eta}{2}\bar e^0\wedge(\mu^{\dot 1}\mathcal{L}_{\dot 1}-\mu^{\dot 0}\mathcal{L}_{\dot 0})+\frac{{\Lambda}\nu^2\eta^2}{2}\frac{\hat\lambda_\alpha}{\la \lambda\,\hat\lambda\ra}\bar e^0\wedge\mathcal{L}^\alpha\,,
    \end{aligned}
\end{equation}
where we used the identity $\mathcal{L}^\alpha\bar e^0=2\hat\lambda^\alpha\bar e^0/\la\lambda\,\hat\lambda\ra$.
The coordinate $\eta=\mu^{\dot{0}}\mu^{\dot{1}}=[\iota \mu][\mu \hat{\iota}]$, which now breaks the Euclidean AdS$_4$ isometry group, is still manifestly holomorphic since $h$ only depends on $\mu^{\dot{\alpha}}$ through $\eta$ and $\{\eta, \eta\}=0$. However, now neither $\mu^{\dot \alpha}$ nor $\lambda_\alpha$ are holomorphic. Similar to equation \eqref{eq:newholomorphic}, the $\lambda_\alpha$ coordinates get deformed to 
\begin{equation}
\label{eq:XYZ}
Y^{\alpha\beta}=\lambda^\alpha\lambda^\beta-\frac{\Lambda\nu^2 \eta^2}{2}\frac{\hat\lambda^\alpha\hat\lambda^\beta}{\la\lambda\,\hat\lambda\ra^2}\,.
\end{equation} 
Indeed, it is straightforward to see that $Y^{\alpha \beta}$ is holomorphic in analogy to \eqref{eq:newholomorphic}
\begin{equation}
    \begin{aligned}
        \bar\nabla_\Lambda Y^{\alpha\beta}&=2\lambda^{(\alpha}\bar\nabla_\Lambda\lambda^{\beta)}-2\frac{\Lambda\nu^2 \eta^2}{2}\frac{\hat\lambda^{(\alpha}}{\la\lambda\,\hat\lambda\ra}\bar\nabla_\Lambda\frac{\hat\lambda^{\beta)}}{\la\lambda\,\hat\lambda\ra}\\&=2\frac{\Lambda\nu^2 \eta^2}{2}\bar e^0\frac{\lambda^{(\alpha}\hat\lambda^{\beta)}}{\la\lambda\,\hat\lambda\ra}-2\frac{\Lambda\nu^2 \eta^2}{2}\bar e^0 \frac{\hat\lambda^{(\alpha}\lambda^{\beta)}}{\la\lambda\,\hat\lambda\ra}\\&=0\,.
    \end{aligned}
\end{equation}
 Since the three holomorphic coordinates 
\begin{equation}
    X= Y^{11}\,,\quad Y= Y^{00}\,,\quad Z=-Y^{01}\,,  
\end{equation}
and $\eta$ all scale with the same weight, and obey the relation $Y^{\alpha \beta}Y_{\alpha \beta}=-\Lambda \nu^2\eta^2$, they form a quadric inside $\mathbb{CP}^3$
\be
Q_{\Lambda\nu^2}=\{(X,Y,Z,\eta)\in \mathbb{CP}^3: XY-Z^2+\frac{\Lambda \nu^2}{2} \eta^2=0\}\,.
\ee
which can be identified with $\mathbb{CP}^1\times \mathbb{CP}^1$. In analogy to $\mathbb{MT}$ in the self-dual Taub NUT case, this $\mathbb{CP}^1\times \mathbb{CP}^1$ again plays the role of a minitwistor space. It is to be viewed as the minitwistor space of $S^3$ \cite{calderbank2001einstein} or the minitwistor space of $\mathbb{H}^3$ after removing the antiholomorphic diagonal. See figure \ref{fig:minitwistoCurved} for an illustration of this in the $\mathbb{H}^3$-case.

\medskip

We saw that the oblate (or prolate) Pedersen metric fibres over $S^3$ (or $\mathbb{H}^3$) through its description as a generalized Gibbons-Hawking metric in equations \eqref{eq:s3_gibbons-hawking} and \eqref{eq:eq:h3_gibbons-hawking}. The full twistor space of the Pedersen metric fibres over the  corresponding minitwistor space $\mathbb{CP}^1\times \mathbb{CP}^1$. We already saw, that our Beltrami differential \eqref{eq:NablaQTN} leads to this correct base space. Let us now analyse the remaining coordinates.

It can be checked immediately that the coordinates 
\be
\label{eq:rhocoord}
\phi^{\pm}=\mu^\pm \Bigg(\frac{\sqrt{\tfrac{2}{\Lambda\nu^2}}+\eta f(\lambda)}{\sqrt{\tfrac{2}{\Lambda\nu^2}}-\eta f(\lambda)}\Bigg)^{\pm \sqrt{ \tfrac{\nu^2}{8\Lambda} }}\,.
\ee
are holomorphic by using that
\be
 \frac{\hat\lambda_\alpha}{\la\lambda\,\hat\lambda\ra}\frac{\p f}{\p\lambda_\alpha}=-f^2
\ee
holds in both patches $\lambda_0\neq 0$ and $\lambda_1 \neq 1$. These patches are somewhat unnatural in the given context.
However, naively working with these patches, gives rise to the transition function
\bea
\label{eq:TransitionFunctionPhi}
g_{12}&= \Bigg(\frac{\sqrt{\tfrac{2}{\Lambda\nu^2}}+\eta \frac{\hat{\lambda}_0}{\lambda_0\langle\lambda\,\hat\lambda\rangle}}{\sqrt{\tfrac{2}{\Lambda\nu^2}}-\eta \frac{\hat{\lambda}_0}{\lambda_0\langle\lambda\,\hat\lambda\rangle}}\Bigg)^{\mp \sqrt{ \tfrac{\nu^2}{8\Lambda} }} 
 \Bigg(\frac{\sqrt{\tfrac{2}{\Lambda\nu^2}}+\eta \frac{\hat{\lambda}_1}{\lambda_1\langle\lambda\,\hat\lambda\rangle}}{\sqrt{\tfrac{2}{\Lambda\nu^2}}-\eta \frac{\hat{\lambda}_1}{\lambda_1\langle\lambda\,\hat\lambda\rangle}}\Bigg)^{\pm \sqrt{ \tfrac{\nu^2}{8\Lambda} }} \\
&=\Bigg(\frac{Z+\sqrt{\tfrac{\Lambda\nu^2}{2}}\eta}{Z-\sqrt{\tfrac{\Lambda\nu^2}{2}}\eta}\Bigg)^{\pm \sqrt{ \tfrac{\nu^2}{8\Lambda} }}\,,
\eea
which precisely matches with the expression found by Pedersen in equation (8.15) of \cite{pedersen1985geometry} when describing the twistor space of the Pedersen metric.  The Pedersen twistor space \cite{pedersen1985geometry,Pedersen:1986vup} is in fact given by a line bundle over $Q_{\Lambda\nu^2}$ with a transition function which is closely related to \eqref{eq:TransitionFunctionPhi}. We conjecture that our deformed twistor space described by the Beltrami differential \eqref{eq:NablaQTN} actually agrees with the full Pedersen twistor space, given by this line bundle. We hope to verify this conjecture more directly in the future by working with deformed holomorphic coordinates that are better adapted to the minitwistor space of $S^3$ or $\mathbb{H}^3$.

Further evidence for this conjecture is given by the fact that our Beltrami differential has the correct limits according to figure \ref{fig:Ped}. For $\Lambda\rightarrow0$, it is immediate that we obtain the twistor space of self-dual Taub-NUT.
For $\Lambda\rightarrow\infty$ and $\nu^2\rightarrow0$ with $\Lambda\nu^2$ fixed, we obtain the holomorphic coordinates of Eguchi-Hanson space with the role of $\lambda_\alpha$ and $\mu^{\da}$ exchanged which was expected from the conformal inversion in equation \eqref{eq:conf}. This can already be seen on the level of the Beltrami differential in \eqref{eq:NablaQTN} where the first term is the Beltrami differential of self-dual Taub-NUT and the second term is that of Eguchi-Hanson space, up to a conformal inversion.

\section{Celestial symmetries from twistor space}
\label{subsec:CelSymmTwistor}

Beyond viewing the twistor space of Eguchi-Hanson space as a backreaction, chapter \ref{chapter4} also obtained the celestial chiral algebra of self-dual gravity from considering the action of $\{\,\,,\,\,\}_0$ on functions in the deformed holomorphic coordinates in section \ref{sec:twistor-algebra}. We will follow the same strategy on the Pedersen twistor space, but since $\Lambda\neq 0$ we will have to use $\{\,\,,\,\,\}_\Lambda$. Acting with $\{\,\,,\,\,\}_\Lambda$ on functions in the undeformed holomorphic coordinates, led to $\LHL$ in chapter \ref{chapter5} \cite{Bittleston:2024rqe, Taylor:2023ajd, Alday:2023jdk}. Acting on holomorphic coordinates of the Pedersen twistor space leads us to a conjectural $2$-parameter deformation of $\mathcal{L}w_\wedge$, which is the unique such deformation that obeys the Jacobi-identity, respects the expected symmetries and has the right limits.

\subsection*{$\{\,\,,\,\,\}_\Lambda$ in deformed coordinates}
Let us now derive the action of $\{\,\,,\,\,\}_\Lambda$ on any pair of holomorphic coordinates  $\phi^\pm,\eta, X,Y,Z$. We derive it in part by acting on holomorphic coordinates with
\bea
\label{eq:LambdaBrack2}
&\{f,g\}_\Lambda=\varepsilon^{\dot\alpha\dot\beta}\frac{\p f}{\p\mu^{\dot\alpha}}\frac{\p g}{\p\mu^{\dot\beta}}+\Lambda\,\varepsilon_{\alpha\beta}\frac{\p f}{\p\lambda_\alpha}\frac{\p g}{\p\lambda_\beta}\,,
\eea
and in part from the two consistency conditions that the Jacobi identity needs to hold and that the subspaces generated by 
\bea
\label{eq:Poissonideals}
\langle \{ \phi^+ \phi^- -\eta\}\rangle\,,\qquad
\langle \{XY-Z^2+\tfrac{\Lambda\nu^2}{2}\eta^2\}\rangle\,,
\eea
need to be Poisson-ideals. 

First, we act with \eqref{eq:LambdaBrack} on a pair of $Y_{\alpha \beta}$ giving
\begin{equation}
    \{Y_{\alpha\beta}, Y_{\gamma\delta}\}_\Lambda=2\Lambda(\varepsilon_{\beta\gamma}Y_{\alpha\delta}+\varepsilon_{\alpha\delta}Y_{\beta\gamma})\,.
\end{equation}
This is equivalent to $X,Y,Z$ obeying the defining relations of $\mathfrak{sl}_2$ similar to equation \eqref{eq:Poisson-bracket} of chapter \ref{chapter4} \be \label{eq:Poisson-bracketXYZ}  
 \{X,Y\}_\Lambda = 4\Lambda Z\,,\qquad \{X,Z\}_\Lambda = 2\Lambda X\,,\qquad \{Y,Z\}_\Lambda = -2 \Lambda Y\,.
\ee
Moreover, we can immediately see
\bea
\{\phi^{\pm}, \eta\}_\Lambda=\pm \phi^{\pm} \,,\qquad 
\{Y_{\alpha \beta},\eta \}_\Lambda=0\,,\qquad 
\{\phi^+,\phi^-\}_\Lambda=\{\mu^+,\mu^-\}_\Lambda=1\,.
\eea
 Demanding that the subspace generated by $\langle \{XY-Z^2+\frac{\Lambda\nu^2}{2}\eta^2\}\rangle$ is a Poisson ideal and the resulting bracket has to obey the Jacobi identity uniquely fixes the remaining brackets (under the additional assumption that $X$ and $Y$ should be treated symmetrically) to be
\be
\label{eq:magicbracket}
\{\phi^\pm,Y\}_\Lambda=\mp \frac{\Lambda\nu^2}{2}  \phi^\pm \frac{\eta}{X}\,, \qquad \{\phi^\pm,X\}_\Lambda=\mp \frac{\Lambda\nu^2}{2}  \phi^\pm \frac{\eta}{Y}\,,\qquad\{\phi^\pm, Z\}_\Lambda=0\,.
\ee
We hope to give a direct derivation of this in the future.
Although a direct derivation is absent, the above discussion leads to strong evidence for the conjectured formula
\bea
\label{eq:PoissonQTN2}
&\{\,\,,\,\,\}_\Lambda= 2\Lambda X \frac{\partial}{\partial X}\wedge\frac{\partial}{\partial Z} -2\Lambda Y \frac{\partial}{\partial Y}\wedge\frac{\partial}{\partial Z} +4\Lambda Z\frac{\partial}{\partial X}\wedge\frac{\partial}{\partial Y}+\frac{\partial}{\partial \phi^+}\wedge\frac{\partial}{\partial \phi^-}\\
& +\phi^-\frac{\partial}{\partial  \phi^+}\wedge\frac{\partial}{\partial \eta}-\phi^+ \frac{\partial}{\partial \phi^-}\wedge\frac{\partial}{\partial \eta}+\frac{\Lambda\nu^2}{2}\,\eta \Bigg(\frac{\phi^-}{X}\frac{\partial}{\partial \phi^-}\wedge\frac{\partial}{\partial Y} +\frac{\phi^-}{Y}\frac{\partial}{\partial \phi^-}\wedge\frac{\partial}{\partial X}\\
&-\frac{\phi^+}{X}\frac{\partial}{\partial \phi^+}\wedge\frac{\partial}{\partial Y} -\frac{\phi^+}{Y}\frac{\partial}{\partial \phi^+}\wedge\frac{\partial}{\partial X}\Bigg)\,.
\eea
Below, we will discuss that the correct limits are all obtained according to figure \ref{fig:Ped}.

\subsection*{Deriving the $2$-parameter deformation of $\mathcal{L}w_\wedge$}

Equation \eqref{eq:PoissonQTN2} can be used to compute the bracket of two generators 
\be
(\phi^+)^a(\phi^-)^b \eta^c X^d Y^e Z^f\,,
\ee
where $a,b,c \in \mathbb{N}$ and $d,e,f \in \mathbb{Z}$. Since $\phi^\pm$ are deformations of the $\mu^\da$ coordinates and $X,Y,Z$ are deformations of the $\lambda_\alpha$ coordinates, these index ranges are the natural generalisation of our treatment in previous chapters based on \cite{Bittleston:2024rqe,Adamo:2021lrv, Bittleston:2023bzp}. We still need to impose the right scaling by demanding that the generators have weight $2$, i.e. $a+b+\tfrac{1}{2}(c+d+e+f)=2$, and impose the relations \eqref{eq:Poissonideals} between the different coordinates. Let us initially only work on the support of $\phi^+\phi^-=\eta$ with no relation between $X,Y,Z,\eta$ imposed. A general polynomial of weight $2$ is then given by
\bea
w[a,b,c,d,e]=(\phi^+)^a (\phi^-)^bX^cY^dZ^e\,,
\eea
where $a,b\in\mathbb{N}_0$ and $c,d,e \in \mathbb{Z}$ and $a+b+\tfrac{1}{2}(c+d+e)=2$. Using \eqref{eq:PoissonQTN2}, leads to the Lie-algebra
\bea
\label{eq:QTNbeforeQuotient}
&\{w[p,q,i,j,k],w[r,s,l,m,n]\}_\Lambda=(ps-qr)\, w[p+r-1,q+s-1,i+l,j+m,k+n]\\
&+2\Lambda\bigg(((i-j)n-(l-m)k)\,w[p+r,q+s,i+l,j+m,k+n-1]
\\
&+2(im-jl)\,w[p+r,q+s,i+l-1,j+m-1,k+n+1]\bigg)\\
&- \frac{\Lambda\nu^2}{2}((p-q)(l+m)-(r-s)(i+j))\,w[p+r+1,q+s+1,i+l-1,j+m-1,k+n]\,,
\eea
which has been explicitly checked to obey the Jacobi identity. Imposing the further constraint $Z^2=XY+\tfrac{\Lambda\nu^2}{2}\eta^2$ means that we can always solve for $X,Y,\eta$ whenever we have $2$ or more powers of $Z$. This leaves us with the generators
\bea
&w[p,q,2i,2j]=(\phi^+)^p(\phi^-)^q X^i Y^j\\
&w[p,q,2i+1,2j+1]=(\phi^+)^p(\phi^-)^q X^i Y^jZ\,,
\eea
where for $w[p,q,i,j]$ the weight $2$ condition reads $p+q+i+j=2$.  Solving for $j=2-p-q-i$ would lead to the standard presentation of generators with $3$ labels but we will not do so to keep the symmetries between $\mu^\da$ and $\lambda_\alpha$ manifest. 

Using the relation  $XY-Z^2=-\tfrac{\Lambda\nu^2}{2}\eta^2$ in  \eqref{eq:QTNbeforeQuotient} then leads us to the final algebra. 
\bea
\label{eq:QTNAlg}
&\{w[p,q,2i,2j],w[r,s,2k,2l]\}_\Lambda =(ps-qr)w[p+r-1,q+s-1,2(i+k),2(j+l)]\\
&+4\Lambda(il-jk) w[p+r,q+s,2(i+k-1)+1,2(j+l-1)+1]\\
&-\frac{\Lambda\nu^2}{2}((p-q)(k+l)-(r-s)(i+j))w[p+r+1,q+s+1,2(i+k-1),2(j+l-1)]
\\
&\{w[p,q,2i,2j],w[r,s,2k+1,2l+1]\}_\Lambda=(ps-qr)w[p+r-1,q+s-1,2(i+k)+1,2(j+l)+1]\\
&+2\Lambda(i(2l+1)-j(2k+1)) w[p+r,q+s,2(i+k),2(j+l)]\\
&+2\Lambda^2\nu^2(il-jk) w[p+r+2,q+s+2,2(i+k-1),2(j+l-1)]\\
&-\frac{\Lambda\nu^2}{2}((p-q)(k+l)-(r-s)(i+j))w[p+r+1,q+s+1,2(i+k-1)+1,2(j+l-1)+1]\\
&\{w[p,q,2i+1,2j+1],w[r,s,2k+1,2l+1]\}_\Lambda\\
&=(ps-qr)w[p+r-1,q+s-1,2(i+k+1),2(j+l+1)]\\
&+\frac{\Lambda\nu^2}{2}(ps-qr)w[p+r+2,q+s+2,2(i+k),2(j+l)]\\
&+\Lambda((2i+1)(2l+1)-(2j+1)(2k+1)) w[p+r,q+s,2(i+k)+1,2(j+l)+1]\\
&+ 2\Lambda^2\nu^2(il-jk) w[p+r+2,q+s+2,2(i+k-1)+1,2(j+l-1)+1]\\
&-\frac{\Lambda\nu^2}{2}((p-q)(k+l)-(r-s)(i+j))w[p+r+1,q+s+1,2(i+k),2(j+l)]\\
&-\Big(\frac{\Lambda\nu^2}{2}\Big)^2((p-q)(k+l)-(r-s)(i+j))w[p+r+3,q+s+3,2(i+k-1),2(j+l-1)]\,.
\eea
Although \eqref{eq:QTNAlg} looks very messy, it is the unique algebra, that respects the expected symmetries, obeys the Jacobi identity and reduces to known algebras in the limits of section \ref{subsec:CelSymmTwistor}. Indeed, when transformed to the generators $w^{p}_{m,a}$, this can be easily seen to be a further deformation of the  $\Lambda$-deformed algebra $\LHL$ considered in chapter \ref{chapter5} and \cite{Bittleston:2024rqe, Taylor:2023ajd}. In fact, we obtain $\LHL^{\mathbb{Z}_2}$, the $\mathbb{Z}_2$-invariant subalgebra of $\LHL$ in the limit $M\rightarrow\infty$. This feature of the $\mathbb{Z}_2$-quotient is expected due to the coordinates \eqref{eq:XYZ} in analogy to equation \eqref{eq:newholomorphic} \cite{Bittleston:2024rqe}. 

After rescaling the generators by $\Lambda$, related to the prefactor of \eqref{eq:EH}, and sending $\Lambda\rightarrow\infty$ and $M\rightarrow\infty$, keeping $c^2=\tfrac{\Lambda}{4M}$ constant, the algebra becomes the expected algebra of Eguchi-Hanson space considered in equation  \eqref{eq:EH-twistor-basis}
of chapter \ref{chapter4} \cite{Bittleston:2023bzp}. The role of $\mu^\da$ and $\lambda_\alpha$ is exchanged as expected from the conformal inversion in equation \eqref{eq:conf}.
In the self-dual Taub-NUT limit, $\Lambda\rightarrow 0$, we find an undeformed $\mathcal{L}w_\wedge$.\footnote{Note again, that due to the coordinates \eqref{eq:XYZ} we only obtain the $\mathbb{Z}_2$-quotient $\mathcal{L}w_\wedge=\mathcal{L}\mathfrak{ham}(\mathbb{C}^2/\mathbb{Z}_2)$ of the full undeformed algebra $\mathcal{L}\mathfrak{ham}(\mathbb{C}^2)$ in the self-dual Taub-NUT limit and the flat space limit.} 

\section{Discussion}

In this chapter, we argued that the twistor space of AdS$_4$, described in chapter \ref{chapter5}, can be further deformed by a backreaction of a defect operator. This conjecturally led to a $2$-parameter twistor space, the Pedersen twistor space, which Penrose transforms to the Pedersen metric on spacetime. The Pedersen metric can be viewed as a Wick-rotation of a self-dual Pleba\'{n}ski-Demia\'{n}ski black hole metric which is, among other reasons, why we refer to it as a \emph{self-dual black hole}. The $2$ parameters of the Pedersen metric are given by a mass parameter and a cosmological constant. In limiting cases of these parameters, previously studied metrics such as Eguchi-Hanson space, AdS$_4$, self-dual Taub-NUT and Burns space (up to a conformal prefactor) arise. The Pedersen twistor space gives rise to a $2$-parameter deformation of $\mathcal{L}w_\wedge$. We derived a conjecture for the explicit form of this $2$-parameter algebra which reduces to previously studied algebras in various limits and respects the expected symmetries. 

We conclude with a brief discussion of some future directions suggested by the results of this chapter.

\medskip

We saw in section \ref{subsec:sdPlebanski} that the Pedersen metric is conformally equivalent to a scalar-flat Kähler manifold which interpolates between Burns space and a double-cover of Eguchi-Hanson space. We hope that these solutions to the equations of Mabuchi-gravity can be engineered from some top-down construction as suggestively displayed in figure \ref{fig:QTNholography}. For instance, it might arise from a generalization of Burns holography with non-trivial boundary conditions or from the more recent top-down construction by Bittleston, Costello and Zeng when simultaneously turning on VEVs for two certain bulk states that were discussed in \cite{Bittleston:2024efo}. From the latter, it might be possible to obtain \emph{self-dual QCD} amplitudes on a Pedersen background. Note that non-trivial tree amplitudes might be present in the theory since the Pedersen metric is not hyperkähler (similar to Burns space).

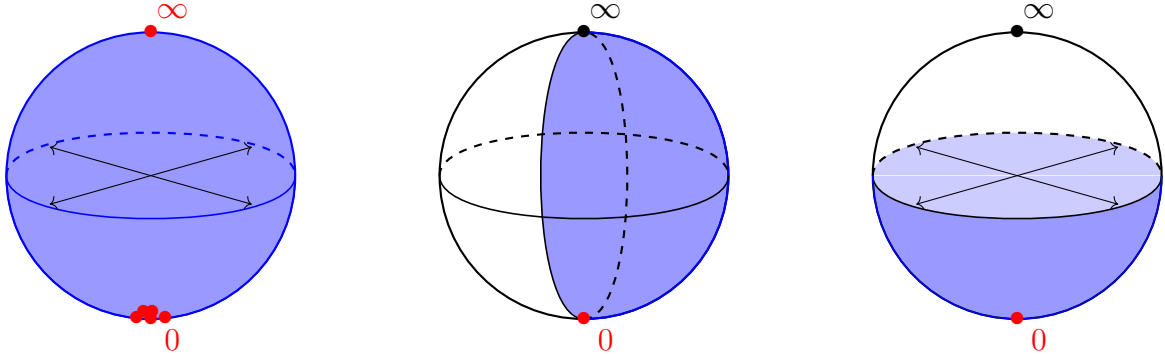
\begin{figure}[t!]
    \centering
\begin{tikzpicture}[scale=1.9]
\filldraw [color= blue, fill=blue!40, thick] (0,0) circle (1);
\draw [name path = A, thick, dashed, blue] (-1,0)  .. controls (-1,0.4) and (1,0.4) ..  (1,0)  ;
\draw [name path = B, semithick, blue] (-1,0)  .. controls (-1,-0.4) and (1,-0.4) ..  (1,0)  ;
\draw [name path = D, semithick, blue] (1,0) arc(0:-180:1);
 \begin{pgfonlayer}{bg}
\fill [blue!20,
          intersection segments={
            of=A and B,
            sequence={L2--R2}
          }];
          \fill [blue!40,
          intersection segments={
            of=B and D,
            sequence={L2--R2}
          }];
\end{pgfonlayer}
 \node[name = a] at (0,-1) {\red{$\bullet$}};
 \node at (-0.1,-0.99) {\red{$\bullet$}};
 \node at (0.1,-0.99) {\red{$\bullet$}};
 \node at (-0.05,-0.95) {\red{$\bullet$}};
 \node at (0.01,-0.95) {\red{$\bullet$}};
  \node at (0.15,-1.15) {\red{$0$}};
\node (b) at (0,1){\red{$\bullet$}}; 
\node at (0.15,1.15) {\red{$\infty$}};

\draw[<->] (-0.7,-0.2)  -- (0.7,0.2);
\draw[<->] (-0.7,0.2)  -- (0.7,-0.2);


\draw [thick, black] (3,0) circle (1);
\draw [name path = A2, thick, dashed] (3,-1)  .. controls (3.4,-1) and (3.4,1) ..  (3,1)  ;
\draw [name path = B2, semithick] (3,-1)  .. controls (2.6,-1) and (2.6,1) ..  (3,1)  ;
\draw [name path = D2, thick, blue] (3,1) arc(90:-90:1);
 \begin{pgfonlayer}{bg}
\fill [blue!20,
          intersection segments={
            of=A2 and B2,
            sequence={L2--R2}
          }];
          \fill [blue!40,
          intersection segments={
            of=B2 and D2,
            sequence={L2--R2}
          }];
\end{pgfonlayer}
\draw [thick, dashed] (2,0)  .. controls (2,0.4) and (4,0.4) ..  (4,0)  ;
\draw [semithick] (2,0)  .. controls (2,-0.4) and (4,-0.4) ..  (4,0)  ;

 \node[name = a] at (3,-1) {\red{$\bullet$}};
  \node at (3.15,-1.15) {\red{$0$}};
\node (b) at (3,1){$\bullet$}; 
\node at (3.15,1.15) {$\infty$};


\draw [thick] (6,0) circle (1);
\draw [name path = A3, thick, dashed] (5,0)  .. controls (5,0.4) and (7,0.4) ..  (7,0)  ;
\draw [name path = B3, semithick] (5,0)  .. controls (5,-0.4) and (7,-0.4) ..  (7,0)  ;
\draw [name path = D3, semithick, blue] (7,0) arc(0:-180:1);
 \begin{pgfonlayer}{bg}
\fill [blue!20,
          intersection segments={
            of=A3 and B3,
            sequence={L2--R2}
          }];
          \fill [blue!40,
          intersection segments={
            of=B3 and D3,
            sequence={L2--R2}
          }];
\end{pgfonlayer}

 \node[name = a] at (6,-1) {\red{$\bullet$}};
  \node at (6.15,-1.15) {\red{$0$}};
\node (b) at (6,1){$\bullet$}; 
\node at (6.15,1.15) {$\infty$};

\draw[<->] (5.3,-0.2)  -- (6.7,0.2);
\draw[<->] (5.3,0.2)  -- (6.7,-0.2);

\end{tikzpicture}
\caption{\emph{Further backreactions are expected to give $A_k$-ALE and $A_k$-ALF spaces (left), \emph{self-dual AdS-Taub-Bolt} (right) and an unknown metric (middle). }}
    \label{fig:moreDefects}
\end{figure}

There are many further configurations of multiple defects some of which are displayed in figure \ref{fig:moreDefects}. We conjecture that these will lead to further known gravitational instantons. In particular, the backreaction on the right of figure \ref{fig:moreDefects} is expected to lead to a $\Lambda\neq 0$ version of Eguchi-Hanson space which is sometimes referred to as \emph{quaternionic Eguchi-Hanson space} or \emph{self-dual AdS-Taub-Bolt}\footnote{The latter terminology is natural to contrast the metric to self-dual AdS-Taub-NUT which is a different name for the Pedersen metric \cite{Hawking:1998ct, Chamblin:1998pz, Martelli:2012sz}.}. It has been known for a long time that there is a Gibbons-Hawking-type phase transition from AdS-Taub-NUT (Pedersen's metric) to AdS-Taub-bolt \cite{Hawking:1998ct, Chamblin:1998pz, Martelli:2012sz}. For oblate squashing with $\nu^2l^2<6+2\sqrt{10}$, self-dual AdS-Taub-NUT \emph{i.e.} the Pedersen metric dominates  but for $\nu^2l^2>6+2\sqrt{10}$ a Taub-bolt AdS$_4$ metric dominates \cite{Chamblin:1998pz}.  It would be very exciting to find a self-dual version of this in the given context, which is expected to be related to \cite{Martelli:2011fu, Martelli:2012sz}. After including the conformal prefactor of section \ref{sec:SDPlebDem}, it might be possible to reproduce some of these
 non-trivial thermodynamic phase structures holographically in the context of Burns holography or related constructions \cite{Bittleston:2024efo, Costello:2023hmi}. This could lead to the first explicit example of a non-trivial thermodynamic phase structures in holography on an asymptotically flat $4$-dimensional spacetime. We hope to be able to provide such a construction in the future.

%% file: Appendix1/appendix1.tex
\chapter{Isomorphic celestial chiral algebras} 
\label{app:isomorphism}

In this appendix, we will discuss two isomorphisms from the  (loop algebras of the) algebras in the natural basis on twistor space --  \eqref{eqs:algebra-in-twistor-basis} for SDGR and \eqref{eq:twistor-S-structure-constants} for SDYM -- to the algebras found
in the scattering basis -- \eqref{eq:sl-infinity} for SDGR and \eqref{eq:SDYM-CCA-twistor} for SDYM.

\medskip

Recall that the polynomials $V[2p,2q]$ and $V[2p+1,2q+1]$ only appear in certain combinations in the scattering states. From their definition in equation \eqref{eq:softbasis}, we see that 
\bea
\label{eq:extractingSoftModes}
W[2p,2q]=&\frac{(2p)!(2q)!}{(2p+2q)!}\sum_{a=0}^{\min(p,q)}\binom{p+q}{p-a,q-a,2a}X^{p-a}Y^{q-a}(2Z)^{2a}\\
=&\frac{(2p)!(2q)!}{(2p+2q)!}\sum_{a=0}^{\min(p,q)}2^{2a} \binom{p+q}{p-a,q-a,2a}  \sum_{\ell=0}^a \binom{a}{\ell} c(\lambda)^{2\ell} X^{p-\ell}Y^{q-\ell}\\
=&\sum_{\ell=0}^{\min(p,q)}  (2c(\lambda))^{2\ell} \left(\frac{(2p)!(2q)!}{(2p+2q)!} \sum_{a=\ell}^{\min(p,q)}2^{2(a-l)} \binom{a}{\ell} \binom{p+q}{p-a,q-a,2a}\right)\\
&V[2(p-\ell),2(q-\ell)]\,,
\eea
where the sum over $a$ can be performed as
\be
\sum_{a=\ell}^{\min(p,q)}2^{2(a-l)} \binom{a}{\ell} \binom{p+q}{p-a,q-a,2a}=\frac{(2p+2q)!}{(2p)!(2q)!} \frac{[p]_\ell\,[q]_\ell\,[p+q]_\ell}{\ell!\,[2(p+q)]_{2\ell}}\,,
\ee
to give
\be
C_0(p,q,\ell)=\frac{[p]_\ell\,[q]_\ell\,[p+q]_\ell}{\ell!\,[2(p+q)]_{2\ell}}\,
\ee
from equation \eqref{eq:c0c1}. For the odd case, we similarly have
\bea
\label{eq:extractingSoftModesOdd}
W[2p+1,2q+1]=&\frac{(2p+1)!(2q+1)!}{(2p+2q+2)!}\sum_{a=0}^{\min(p,q)}\binom{p+q+1}{p-a,q-a,2a+1}X^{p-a}Y^{q-a}(2Z)^{2a+1}\\
=&\frac{(2p+1)!(2q+1)!}{(2p+2q+2)!}\sum_{a=0}^{\min(p,q)}2^{2a+1} \binom{p+q+1}{p-a,q-a,2a}  \sum_{\ell=0}^a \binom{a}{\ell} c(\lambda)^{2\ell} X^{p-\ell}Y^{q-\ell}Z\\
=&\sum_{\ell=0}^{\min(p,q)}  (2c(\lambda))^{2\ell} \left(\frac{(2p+1)!(2q+1)!}{(2p+2q+2)!}\sum_{a=\ell}^{\min(p,q)}2^{2(a-l)+1} \binom{a}{\ell} \binom{p+q+1}{p-a,q-a,2a}\right)\\
&V[2(p-\ell)+1,2(q-\ell)+1]\,,
\eea
where once again the sum over $a$ can be performed as
\be
\sum_{a=\ell}^{\min(p,q)}2^{2(a-l)+1} \binom{a}{\ell} \binom{p+q+1}{p-a,q-a,2a}= \frac{(2p+2q+2)!}{(2p+1)!(2q+1)!} \frac{[p]_{\ell}
\,[q]_{\ell}\,[p+q+1]_{\ell}}{\ell!\,[2(p+q+1)]_{2\ell}}\,,
\ee
to give
\be
C_1(p,q,\ell)= \frac{[p]_{\ell}
\,[q]_{\ell}\,[p+q+1]_{\ell}}{\ell!\,[2(p+q+1)]_{2\ell}}\,.
\ee

These arguments go through line by line to give the form of the soft gluon modes in terms of the polynomials \eqref{eq:GaugePolynomials}
\bea
J_\sfa[2p,2q]&=\sum_{\ell=0}^{\min(p,q)}(2c(\lambda))^{2\ell}\, C_0(p,q,\ell)\, j_\sfa[2(p-\ell),2(q-\ell)]\,,\\
J_\sfa[2p\!+\!1,2q\!+\!1]&= \sum_{\ell=0}^{\min(p,q)}(2c(\lambda))^{2\ell}\,C_1(p,q,\ell)\,j_\sfa[2(p-\ell)\!+\!1,2(q-\ell)\!+\!1]\,.
\eea

To see that the change of basis is actually an isomorphism Lie algebras we expand both sides of 
\bea
\label{eq:appendixW}
&\big[W[p,q],W[r,s]\big] \\
&= \frac{1}{2}\sum_{\ell\geq0} (2c(\lambda))^{2\ell}R_{2\ell+1}(p,q,r,s)\psi_{2\ell+1}\bigg(\frac{p+q}{2},\frac{r+s}{2}\bigg)W[p\!+\!r\!-\!2\ell\!-\!1,q\!+\!s\!-\!2\ell\!-\!1]\,
\eea
in terms of the $V$ basis through equations \eqref{eq:extractingSoftModes} and \eqref{eq:extractingSoftModesOdd}. For, say, two even elements, the left hand side reads 
\bea
\label{eq:IsomLHS}
&[W[2p,2q],W[2r,2s]]=\sum_{\ell=0}^{\min(p,q)+\min(r,s)} (2c(\lambda))^{2\ell}  \, V[2(p\!+\!r\!-\!\ell)\!-\!1,2(q\!+\!s\!-\!\ell)]\\
&\left(\sum_{i=0}^\ell R_1\big(2(p-i),2(q-i),2(r-(\ell-i)),2(s-(\ell-i))\big) 
C_0(p,q,i)C_0(r,s,\ell-i) \right)\,,
\eea
while the right hand side reads
\bea
\label{eq:IsomRHS}
&\sum_{\ell=0}^{\min(p+r,q+s)} (2c(\lambda))^{2\ell}V[2(p\!+\!r\!-\!\ell)\!-\!1,2(q\!+\!s\!-\!\ell)\!-\!1]\\
&\left(\sum_{i=0}^\ell R_{2i+1}(2p,2q,2r,2s)\psi_{2i+1}(2(p+q),2(r+s))C_1(p\!+\!r\!-\!i\!-\!1,q\!+\!s\!-\!i\!-\!1,l\!-\!i) \right)\,.
\eea
While the individual summands in the last lines of \eqref{eq:IsomLHS} and \eqref{eq:IsomRHS} do not match, the whole sum does. This has been verified numerically up to $\ell=8$. Similarly, we have numerically verified up to $\ell=8$ that \eqref{eq:appendixW} also holds for $[W[2p+1,2q+1],W[2r,2s]]$ and for $[W[2p+1,2q+1],W[2r+1,2s+1]]$. Similar checks have also been performed in the case of self-dual Yang-Mills theory.

%% file: Appendix2/appendix2.tex

\chapter{Space-time calculations} \label{app:space-time-calcs}

In these appendices we include derivations of the results used in section \ref{sec:space-time-algebra}.

\section{Self-dual gravity perturbiner} \label{app:SDGR-calcs}

In this appendix we evaluate the contributions of the first and second terms on the right hand side of equation \eqref{eq:SDGR-vertex} to the leading holomorphic collinear singularity in equation \eqref{eq:SDGR-singularity-integral}. These are listed in equation \eqref{eq:SDGR-remaining-terms}. We employ the method outlined in section \ref{subsec:simplify-first-order}, and shall adopt the same notation.

The contribution of the first term in \eqref{eq:SDGR-vertex} to the integral \eqref{eq:SDGR-singularity-integral} is
\be \label{eq:SDGR-term1} -\frac{2}{\pi^2\la\al1\ra^2\la\al2\ra^2}\int_0^1\dif s\,\int_{\bbR^4}\frac{\dif^4y}{(x-y)^2y^6}\,[\vt1][v2]\bigg([12]-\frac{6[v1][\vt2]}{y^2}\bigg)\cos(y\cdot k_1)\cos(s\,y\cdot k_2)\,. \ee
Replacing the factors of $[vi],[\vt i]$ by derivatives with respect to helicity variables, this can written as
\bea \label{eq:SDGR-simplified-term1}
&\frac{1}{\pi^2\la\alpha1\ra^2\la\alpha2\ra^2}\int_0^1\frac{\dif s}{s^2}\,\Big(s[12]\la\al\p_{\lambda_2}\ra\la
\beta\p_{\lambda_1}\ra\big(\cI_2(x,k_-(s)) - \cI_2(x,k_+(s))\big) \\
&+ 6\la\al\p_{\lambda_1}\ra\la\al\p_{\lambda_2}\ra\la
\beta\p_{\lambda_1}\ra\la
\beta\p_{\lambda_2}\ra\big(\cI_3(x,k_-(s)) + \cI_3(x,k_+(s)\big)\Big)\,.
\eea
Differentiating $\cI_k(x,k_\pm(s))$ with respect to spinor helicity variables a total of $k+l$ times generates holomorphic collinear singularities of at worst order $l$. The logarithmic singularities are expected to cancel as they do in self-dual Yang-Mills, so we'll concentrate on the simple pole generated by the second set of terms. Furthermore, recalling that
\be \cI_m(x;k_\pm(s)) = \frac{\pi^2}{k!}\int_0^1\dif t\,(1-t)^m\cos(t\,x\cdot k_\pm(s))\int_0^\infty\dif r\,r^{m-1}e^{-rt(1-t)x^2-k_\pm(s)^2/4r}\,, \ee
the pole of order $l$ is only generated if all $l+m$ derivatives with respect to spinor helicity variables hit the exponential $\exp(-k_\pm(s)^2/4r)$ to bring down a factor of $1/r^{l+m}$. Therefore the singularity in \eqref{eq:SDGR-simplified-term1} is determined by
\bea
&6\la\al\p_{\lambda_1}\ra\la\al\p_{\lambda_2}\ra\la
\beta\p_{\lambda_1}\ra\la
\beta\p_{\lambda_2}\ra\cI_3(x,k_\pm(s)) \\
&\sim \frac{\pi^2s^4\la\al1\ra\la\al2\ra\la1\beta\ra\la2\beta\ra[12]^4}{16}\int_0^1\dif t\,(1-t)^3\cos(t\,x\cdot k_\pm(s))\int_0^\infty\frac{\dif r}{r^2}\,e^{-rt(1-t)x^2-k_\pm(s)^2/4r} \\
&+ \cO(\log\la12\ra) \\
&\sim \pm\frac{\pi^2s^3\la\al1\ra\la\al2\ra\la1\beta\ra\la2\beta\ra[12]^3}{16\la12\ra}\int_0^1\dif t\,(1-t)^3\cos(t\,x\cdot k_\pm(s)) + \cO(\log\la12\ra)\,.
\eea
Hence, in the holomorphic collinear limit equation \eqref{eq:SDGR-term1} has a leading simple pole
\be - \frac{\la1\beta\ra\la2\beta\ra[12]^3}{4\la\al1\ra\la\al2\ra\la12\ra}\int_0^1\dif s\,s\int_0^1\dif t\,(1-t)^3\sin(s\,x\cdot k_1)\sin(st\,x\cdot k_2)\,. \ee
Rescaling $s$ by a factor of $1/t$, so that it now takes values in the range $[0,t]$, gives the first line of equation \eqref{eq:SDGR-remaining-terms}.

Let's move on to the second term in \eqref{eq:SDGR-vertex}, whose contribution to \eqref{eq:SDGR-singularity-integral} is
\bea \label{eq:SDGR-term2}
&-\frac{[12]}{\pi^2\la\al1\ra^2\la\al2\ra}\int_0^1\dif s\,s\int_{\bbR^4}\frac{\dif^4y}{(x-y)^2y^4}\,[\vt2] \\
&\bigg([12]-\frac{2([v1][\vt2] + [\vt1][v2])}{y^2}\bigg)\cos(y\cdot k_1)\sin(s\,y\cdot k_2)\,,
\eea
or equivalently
\bea \label{eq:SDGR-simplified-term2}
&-\frac{[12]}{2\pi^2\la\alpha1\ra^2\la\alpha2\ra}\int_0^1\frac{\dif s}{s}\,\la\beta\p_{\lambda_2}\ra\Big(s[12]\big(\cI_1(x;k_-(s)) + \cI_1(x;k_+(s))\big) \\
&+ 2(\la\al\p_{\lambda_1}\ra\la\beta\p_{\lambda_2}\ra + \la\beta\p_{\lambda_1}\ra\la\al\p_{\lambda_2}\ra)\big(\cI_2(x;k_-(s)) - \cI_2(x;k_+(s))\big)\Big)\,.
\eea
Again, the logarithmic singularities should cancel. The only potential pole is therefore generated by the second set of terms. We have
\bea \label{eq:SDGR-singularity-term2}
&2\la\al\p_{\lambda_2}\ra\la\beta\p_{\lambda_1}\ra\la\beta\p_{\lambda_2}\ra\cI_2(x;k_\pm(s)) \\
&\sim\mp\frac{\pi^2s^3\la\alpha1\ra\la1\beta\ra\la2\beta\ra[12]^3}{8}\int_0^1\dif t\,(1-t)^2\cos(s\,x\cdot k_\pm(s))\int_0^\infty\frac{\dif r}{r^2}\,e^{-rt(1-t)x^2-k_\pm(s)^2/4r} \\
&+ \cO(\log\la12\ra) \\
&\sim - \frac{\pi^2s^2\la\alpha1\ra\la1\beta\ra\la2\beta\ra[12]^2}{4\la12\ra}\int_0^1\dif t\,(1-t)^2\cos(t\,x\cdot k_\pm(s)) + \cO(\log\la12\ra)\,.
\eea
Similarly
\bea
&2\la\al\p_{\lambda_1}\ra\la\beta\p_{\lambda_2}\ra^2\cI_2(x;k_\pm(s)) \\
&\sim - \frac{\pi^2s^2\la\alpha2\ra\la1\beta\ra^2[12]^2}{4\la12\ra}\int_0^1\dif t\,(1-t)^2\cos(t\,x\cdot k_\pm(s)) + \cO(\log\la12\ra)\,.
\eea
Invoking the Schouten identity $\la\alpha2\ra\la1\beta\ra = \la\alpha1\ra\la2\beta\ra + \la12\ra$ and discarding the non-singular piece we find that this simple pole coincides with that in \eqref{eq:SDGR-singularity-term2}. Hence, in the holomorphic collinear limit equation \eqref{eq:SDGR-term2} has a leading simple pole
\be \frac{\la1\beta\ra\la2\beta\ra[12]^3}{2\la\alpha1\ra\la\alpha2\ra\la12\ra}\int_0^1\dif s\,s\int_0^1\dif t\,(1-t)^2\sin(s\,x\cdot k_1)\sin(st\,x\cdot k_2)\,. \ee
Rescaling $s$ by $1/t$ gives the second line  of equation \eqref{eq:SDGR-remaining-terms} in the main text.


\section{Self-dual Yang-Mills perturbiner} \label{app:SDYM-perturbiner}

In this appendix we compute the leading holomorphic collinear singularity in the first order correction to the self-dual Yang-Mills perturbiner at first order in $c^2$, as given in equation \eqref{eq:SDYM-perturbiner-singularity}. We follow the approach used for self-dual gravity, as presented in section \ref{subsec:simplify-first-order} and appendix \ref{app:SDGR-calcs}.

It's sufficient to determine the leading holomorphic collinear singularity in
\be \label{eq:SDYM-singularity-integral} - \frac{1}{4\pi^2}\int_{\bbR^4}\frac{\dif^4y}{(x-y)^2}\,\left[\tilde\p^\da\Phi_{1\sfa}^{(0)}(y),\tilde\p_\da\Phi_{2\sfb}^{(1)}(y)\right]\,. 
\ee
Recalling the first order correction to a null momentum eigenstate on Eguchi-Hanson \eqref{eq:first-order-eigenstate}, we find 
\bea \label{eq:SDYM-vertex}
\left[\tilde\p^\da\Phi_{1\sfa}^{(0)}(x)\,\tilde\p_\da\Phi_{2\sfb}^{(1)}(x)\right] &= -  f^{~~\sfc}_{\sfa\sfb}\,t_\sfc\,\frac{4[\ut2]\sin(x\cdot k_1)}{\la\al1\ra}\bigg(\frac{2[\ut 1][u2]}{x^6}\int_0^1\dif s\,\cos(s\,x\cdot k_2) \\
&\qquad\qquad\qquad + \frac{\la\al2\ra[\ut2][12]}{2x^4}\int_0^1\dif s\,s\sin(s\,x\cdot k_2)\bigg)\,.
\eea
Let's address the two terms on the right hand side separately. Stripping the colour factor, the contribution of the first term to equation \eqref{eq:SDYM-singularity-integral} is
\be \label{eq:SDYM-term1} \frac{2}{\pi^2\la\al1\ra}\int_0^1\dif s\,\int_{\bbR^4}\frac{\dif^4y}{(x-y)^2y^6}\,[\vt1][v2][\vt2]\sin(x\cdot k_1)\cos(s\,x\cdot k_2)\,. \ee
Employing familiar tricks this can be written as
\be \label{eq:SDYM-simplified-term1} \frac{1}{\pi^2\la\al1\ra}\int_0^1\frac{\dif s}{s^2}\,\la\al\p_{\lambda_2}\ra\la\beta\p_{\lambda_1}\ra\la\beta\p_{\lambda_2}\ra\big(\cI_2(x;k_-(s)) + \cI_2(x;k_+(s))\big)\,. \ee
From the discussion in appendix \ref{app:SDGR-calcs}, we know that hitting $\cI_2(x;k_\pm(s))$ with 3 derivatives with respect to spinor helicity variables can generate at worst a simple pole. For now we ignore the subleading logarithmic singularities of the form $\log\la12\ra$, though we'll see that they cancel explicitly in appendix \ref{app:logarithmic-cancellation}. Using equation \eqref{eq:SDGR-singularity-term2}, the leading simple pole in equation \eqref{eq:SDYM-term1} is
\be \label{eq:SDYM-term1-final} - \frac{\la1\beta\ra\la2\beta\ra[12]^2}{4\la12\ra}\int_0^1\dif s\,\int_0^1\dif t\,(1-t)^2\cos(t\,x\cdot k_1)\cos(st\,x\cdot k_2)\,.
\ee

The contribution of the second term in equation \eqref{eq:SDYM-vertex} is
\be \frac{\la\al2\ra[12]}{2\pi^2\la\al1\ra}\int_0^1\dif s\,s\int_{\bbR^4}\frac{\dif^4y}{(x-y)^2y^4}\,[\vt2]^2\sin(x\cdot k_1)\sin(s\,x\cdot k_2)\,. \ee
The usual tricks turn this into
\be \label{eq:SDYM-simplified-term2} \frac{\la\alpha2\ra[12]}{4\pi^2\la\alpha1\ra}\int_0^1\frac{\dif s}{s}\la\beta\p_{\lambda_2}\ra^2\big(\cI_1(x;k_-(s))-\cI_1(x;k_+(s))\big)\,, \ee
which has at worst a simple pole. Ignoring the subleading logarithmic singularity, the leading simple pole was determined in equation \eqref{eq:SDGR-singularity-term3}, giving
\be \label{eq:SDYM-term2-final} \frac{\la\al2\ra\la1\beta\ra^2[12]^2}{4\la\al1\ra\la12\ra}\int_0^1\dif s\,\int_0^1\dif t\,(1-t)\cos(t\,x\cdot k_1)\cos(st\,x\cdot k_2)\,. \ee
The coefficient can be symmetrised using a Schouten.

Combining equations \eqref{eq:SDYM-term1-final} and \eqref{eq:SDYM-term2-final} (after symmetrising the coefficient), and then rescaling $s$ by $1/t$ so that it now takes values in the range $[0,t]$, we arrive at
\be \frac{\la1\beta\ra\la2\beta\ra[12]^2}{4\la12\ra}\int_{0\leq s\leq t\leq1}\dif s\,\dif t\,(1-t)\cos(t\,x\cdot k_1)\cos(s\,x\cdot k_2)\,. \ee
Finally we antisymmetrise in $1\leftrightarrow2$ as indicated in equation \eqref{eq:SDYM-perturbiner-dif}, to get
\be \label{eq:SDYM-perturbiner-final} \frac{\la1\beta\ra\la2\beta\ra[12]^2}{4\la12\ra}\int_0^1\dif s\,\int_0^1\dif t\,(1-\max(s,t))\cos(t\,x\cdot k_1)\cos(s\,x\cdot k_2)\,. \ee
This is the holomorphic collinear singularity in the self-dual Yang-Mills perturbiner on Eguchi-Hanson at first order in $c^2$, minus the term responsible for the shift in the zeroth order perturbiner to its curved counterpart. It appears in equation \eqref{eq:SDYM-perturbiner-singularity} of the bulk manuscript.


\section{Cancellation of logarithmic collinear singularities} \label{app:logarithmic-cancellation}

In this appendix we show that in the holomorphic collinear limit the subleading logarithmic singularities in $\cP_\mathrm{SDYM}^{(1)}(x;k_1,k_2)$ (as defined in equation \eqref{eq:SDYM-perturbiner-dif-1}) vanish. To this end, consider the contributions of the two terms in \eqref{eq:SDYM-vertex} to \eqref{eq:SDYM-singularity-integral} separately.

We already found that the first can be written as \eqref{eq:SDYM-simplified-term1}
\be \label{eq:SDYM-logarithmic-term1} \frac{1}{\pi^2\la\al1\ra}\int_0^1\frac{\dif s}{s^2}\,\la\al\p_{\lambda_2}\ra\la\beta\p_{\lambda_1}\ra\la\beta\p_{\lambda_2}\ra\big(\cI_2(x;k_-(s)) + \cI_2(x;k_+(s))\big)\,. \ee
A careful computation shows that the in the holomorphic collinear limit
\bea
&\la\al\p_{\lambda_2}\ra\la\beta\p_{\lambda_1}\ra\la\beta\p_{\lambda_2}\ra\cI_2(x;k_\pm(s))\sim \mathrm{simple~pole} \\
&+ \frac{s^2\pi^2[12]^2\log\la12\ra}{16}\int_0^1\dif t\,(1-t)^2\Big(2\la\beta1\ra\cos(t\,x\cdot k_\pm(s)) \\
& - 2[\ut1]\la\al1\ra\la\beta1\ra t\sin(t\,x\cdot k_\pm(s)) \mp 2\big([u2]\la\beta1\ra - [\ut2]\la\al1\ra\big)\la\beta2\ra st\sin(t\,x\cdot k_\pm(s)) \\
&\mp \la\al1\ra\la\beta1\ra\la \beta2\ra[12]x^2st(1-t)\cos(t\,x\cdot k_\pm(s))\Big) + \cO(1,\la12\ra\log\la12\ra)\,.
\eea
Therefore, the subleading logarithmic singularity in equation \eqref{eq:SDYM-logarithmic-term1} is
\bea \label{eq:SDYM-logarithmic-singularities1}
&\frac{[12]^2\log\la12\ra}{8\la\al1\ra}\int_0^1\dif s\,\int_0^1\dif t\,(1-t)^2\Big(2\la\beta1\ra\cos(t\,x\cdot k_1)\cos(st\,x\cdot k_2) \\
&- 2[\ut1]\la\al1\ra\la\beta1\ra t\sin(t\,x\cdot k_1)\cos(st\,x\cdot k_2) \\
&- 2\big([u2]\la\beta1\ra - [\ut2]\la\al1\ra\big)\la\beta2\ra st\cos(t\,x\cdot k_1)\sin(st\,x\cdot k_2) \\
&+ \la\al1\ra\la\beta1\ra\la\beta2\ra[12]x^2st(1-t)\sin(t\,x\cdot k_1)\sin(st\,x\cdot k_2)\Big)\,. 
\eea

The contribution of the second term can be written as \eqref{eq:SDYM-simplified-term2}
\be \label{eq:SDYM-logarithmic-term2} \frac{\la\al2\ra[12]}{4\pi^2\la\al1\ra}\int_0^1\frac{\dif s}{s}\la\beta\p_{\lambda_2}\ra^2\big(\cI_1(x;k_-(s))-\cI_1(x;k_+(s))\big)\,. \ee
In the holomorphic collinear limit
\bea
&\la\beta\p_{\lambda_2}\ra^2\cI_1(x,k_\pm(s)) \sim \mathrm{simple~pole} - \frac{s^2\pi^2\la\beta1\ra[12]\log\la12\ra}{4}\int_0^1\dif t\,t(1-t) \\
&\big(4[\ut2]\sin(t\,x\cdot k_\pm(s)) - \la\beta1\ra[12]x^2(1-t)\cos(t\,x\cdot k_\pm(s))\big) + \cO(1,\la12\ra\log\la12\ra)\,.
\eea
Hence, the subleading logarithmic singularity in equation \eqref{eq:SDYM-logarithmic-term2} is
\bea \label{eq:SDYM-logarithmic-singularities2} 
&- \frac{\la\al2\ra\la\beta1\ra[12]^2\log\la12\ra}{8\la\al1\ra}\int_0^1\dif s\,s\int_0^1\dif t\,t(1-t)\big(4[\ut2]\la\beta1\ra\cos(t\,x\cdot k_1)\sin(st\,x\cdot k_2) \\
&+ \la\beta1\ra[12]x^2(1-t)\sin(t\,x\cdot k_1)\sin(st\,x\cdot k_2)\big)\,. 
\eea

We can now proceed by combining like terms in equations \eqref{eq:SDYM-logarithmic-singularities1} and \eqref{eq:SDYM-logarithmic-singularities2}. For example, the terms involving $\sin(t\,x\cdot k_1)\sin(st\,x\cdot k_2)$ can be combined (invoking a Schouten and working modulo non-singular $\la12\ra\log\la12\ra$ terms) to give
\be
- \frac{\la\beta1\ra\la\beta2\ra x^2[12]^3\log\la12\ra}{8}\int_0^1\dif s\,s\int_0^1\dif t\,t^2(1-t)^2\sin(t\,x\cdot k_1)\sin(st\,x\cdot k_2) 
\ee
Exploiting a Schouten, the terms involving $\cos(t\,x\cdot k_1)\sin(st\,x\cdot k_2)$ sum to
\bea
&\frac{[12]^2\log\la12\ra}{4\la\al1\ra}\int_0^1\dif s\,s\int_0^1\dif t\,t(1-t)\cos(t\,x\cdot k_1)\sin(st\,x\cdot k_2) \\
&\qquad\qquad\qquad\big( - (x\cdot k_2)\la\beta1\ra + t([u2]\la\beta1\ra - [\ut2]\la\al1\ra)\la\beta2\ra\big)\,.
\eea
It's then natural to integrate by parts with respect to $s$ in the first of the above terms in order to eliminate $x\cdot k_2$. This gives
\bea
&\frac{\la \beta1\ra[12]^2\log\la12\ra}{4\la\al1\ra}\int_0^1\dif t\,(1-t)\cos(t\,x\cdot k_1)\cos(t\,x\cdot k_2) \\
&- \frac{\la\beta1\ra[12]^2\log\la12\ra}{4\la\al1\ra}\int_0^1\dif s\,\int_0^1\dif t\,(1-t)\cos(t\,x\cdot k_1)\cos(st\,x\cdot k_2)\,. \eea
Having performed these manipulations, and upon rescaling $s$ so that it now takes values in the range $[0,t]$, the total logarithmic singularity is
\bea
&\frac{[12]^2\log\la12\ra}{8\la\al1\ra}\int_{0\leq s\leq t\leq 1}\dif s\,\dif t\,(1-t)\big(- 2\la\beta1\ra\cos(t\,x\cdot k_1)\cos(s\,x\cdot k_2) \\
&\qquad\qquad\qquad\qquad- 2[\ut1]\la\al1\ra\la\beta1\ra(1-t)\sin(t\,x\cdot k_1)\cos(s\,x\cdot k_2) \\
&\qquad\qquad\qquad\qquad+ 2\big([u2]\la\beta1\ra - [\ut2]\la\al1\ra\big)\la\beta2\ra s\cos(t\,x\cdot k_1)\sin(s\,x\cdot k_2) \\
&\qquad\qquad\qquad\qquad- \la\al1\ra\la\beta1\ra\la\beta2\ra x^2[12]s(1-t)\sin(t\,x\cdot k_1)\sin(s\,x\cdot k_2)\big) \\
&+ \frac{\la \beta1\ra[12]^2\log\la12\ra}{4\la\al1\ra}\int_0^1\dif t\,(1-t)\cos(t\,x\cdot k_1)\cos(t\,x\cdot k_2)\,.
\eea
We can now iteratively integrate by parts with respect to $s,t$, so that the integrands become proportional to $\sin(t\,x\cdot k_1)\sin(st\,x\cdot k_2)$. As we do this we must take care to keep track of the boundary terms generated on the diagonal $s=t$. We have
\bea
&\int_{0\leq s\leq t\leq 1}\dif s\,\dif t\,(1-t)\cos(t\,x\cdot k_1)\cos(s\,x\cdot k_2) \\
&= - \frac{1}{2}(x\cdot k_1)(x\cdot k_2)\int_{0\leq s\leq t\leq1}\dif s\,\dif t\,s(1-t)^2\sin(t\,x\cdot k_1)\sin(t\,x\cdot k_2) \\
&+ \frac{1}{2}(x\cdot k_2)\int_0^1\dif t\,t(1-t)^2\cos(t\,x\cdot k_1)\sin(t\,x\cdot k_2) + \int_0^1\dif t\,t(1-t)\cos(t\,x\cdot k_1)\cos(t\,x\cdot k_2)\,, 
\eea
and similarly
\bea
&\int_{0\leq s\leq t\leq 1}\dif s\,\dif t\,(1-t)^2\sin(t\,x\cdot k_1)\cos(s\,x\cdot k_2) \\
&= (x\cdot k_2)\int_{0\leq s\leq t\leq 1}\dif s\,\dif t\,s(1-t)^2\sin(t\,x\cdot k_1)\sin(s\,x\cdot k_2) + \int_0^1\dif t\,t(1-t)^2\sin(t\,x\cdot k_1)\cos(t\,x\cdot k_2)\,,
\eea
as well as
\bea
&\int_{0\leq s\leq t\leq 1}\dif s\,\dif t\,s(1-t)\cos(t\,x\cdot k_1)\sin(s\,x\cdot k_2) \\
&= - \frac{1}{2}(x\cdot k_1)\int_{0\leq s\leq t\leq 1}\dif s\,\dif t\,s(1-t)^2\sin(t\,x\cdot k_1)\sin(s\,x\cdot k_2) + \frac{1}{2}\int_0^1\dif t\,t(1-t)^2\cos(t\,x\cdot k_1)\sin(t\,x\cdot k_2)\,.
\eea
The coefficient of the double integral
\be \int_{0\leq s\leq t\leq1}\dif s\,\dif t\,s(1-t)^2\sin(t\,x\cdot k_1)\sin(s\,x\cdot k_2) \ee
is, after some massaging,
\bea
&(x\cdot k_1)(x\cdot k_2)\la\beta1\ra - 2(x\cdot k_2)[\ut1]\la\al1\ra\la\beta1\ra - (x\cdot k_1)\big([u2]\la\beta1\ra - [\ut2]\la\al1\ra\big)\la\beta2\ra - \la\al1\ra\la\beta1\ra\la\beta2\ra[12]x^2 \\
&= (x\cdot k_1)((x\cdot k_2)\la\beta1\ra - [u2]\la\beta1\ra\la\beta2\ra + [\ut2]\la\al1\ra\la\beta2\ra) - 2(x\cdot k_2)[\ut1]\la\al1\ra\la\beta1\ra - \la\al1\ra\la\beta1\ra\la\beta2\ra[12]x^2 \\
&=  \big([u1]\la\beta1\ra - [\ut1]\la\al1\ra\big)[\ut2]\la12\ra\,,
\eea
so that it's contribution is non-singular. This leaves the boundary terms. The coefficient of
\be \int_0^1\dif t\,t(1-t)^2\cos(t\,x\cdot k_1)\sin(t\,x\cdot k_2) \ee
is proportional to
\bea
&(x\cdot k_2)\la\beta1\ra - [u2]\la\beta1\ra + [\ut2]\la\al1\ra\la\beta2\ra \\
&= [\ut2](\la\al1\ra\la\beta2\ra + \la\al2\ra\la\beta1\ra) = 2[\ut2]\la\al1\ra\la\beta2\ra + [\ut2]\la12\ra\,,
\eea
so that the full logarithmic singularity is
\bea
&\frac{[12]^2\log\la12\ra}{4\la\al1\ra\la\al2\ra}\int_0^1\dif t\,(1-t)^2\Bigg(\la\al2\ra\la\beta1\ra\cos(t\,x\cdot k_1)\cos(t\,x\cdot k_2)
\\
&\qquad\qquad- \la\al1\ra\la\al2\ra t\big([\ut1]\la\beta1\ra\sin(t\,x\cdot k_1)\cos(t\,x\cdot k_2) + (1\leftrightarrow2)\big)\Bigg)\,.
\eea
Upon antisymmetrising to get the perturbiner as indicated in equation \eqref{eq:SDYM-perturbiner-dif-1}, we find that above is non-singular.